%% file: thesis.tex
\documentclass[12pt,oneside,final]{huthesis}
\usepackage{amssymb}
\usepackage{amsmath}
\usepackage{bm}
\usepackage{graphicx}
\usepackage[numbers]{natbib}
\usepackage{listings}
\usepackage{color}

\begin{document}
\def\newblock{\hskip .11em plus .33em minus .07em}%added to get rid of newblock error (PDN 19/04/10)
\input{mathdefs} % my math definitions.

% UNDERLYING SPACING FOR WHOLE DOCUMENT:
% Single spacing: takes place of `draft' mode, without losing figures.
% \ssp

% makes double-spaced: (for GSAS requirement, microfiche):
\dsp

\include{frontmatter}

%start arabic numbering with chaper 1, moved from frontmatter, PDN (10/04/13)
\startarabicpagination
% Chapter 1:
\include{ch1}
\include{ch2}
\include{ch3}
\include{ch4}

% insert other chapters here...

\appendix
\include{ap1}
\include{ap2}

% insert other appendices here...

% bibliography:
\addcontentsline{toc}{chapter}{Bibliography} 
\include{bib}

\end{document}

%% file: mathdefs.tex
%%% mathdefs.tex
%

% Taken from Adam Lupu-Sax ....................................................
%%% derivatives
\newcommand{\deriv}[2]{\frac{d#1}{d#2}}
\newcommand{\derivc}[3]{\left. \frac{d#1}{d#2}\right|_{#3}}
\newcommand{\pd}[2]{\frac{\partial #1}{\partial #2}}
\newcommand{\pdc}[3]{\left. \frac{\partial #1}{\partial #2}\right|_{#3}}

%%% Dirac notation
\newcommand{\bra}[1]{\left\langle #1\right|}
\newcommand{\ket}[1]{\left|#1\right\rangle}
\newcommand{\braket}[2]{\left\langle #1 \left|#2\right.\right\rangle}
\newcommand{\braOket}[3]{\left\langle #1\left|#2\right|#3\right\rangle}

% calligraphic letters in math.
\def\cal#1{\mathcal{#1}}

% Taken from M Haggerty ......................................................
\def\avg#1{\left< #1 \right>}
\def\abs#1{\left| #1 \right|}
\def\recip#1{\frac{1}{#1}}
\def\vhat#1{\hat{{\bf #1}}}
\def\smallfrac#1#2{{\textstyle\frac{#1}{#2}}}
\def\smallrecip#1{\smallfrac{1}{#1}}

% SPSmith's definitions ......................................................
\def\spshalf{{1\over{2}}}
\def\Orabi{\Omega_{\rm rabi}}
\def\btt#1{{\tt$\backslash$#1}}

% My own .....................................................................

%%% Equations
\def\schrod{Schroedinger's Equation}
\def\helm{Helmholtz Equation}

%%% Equation environments
\def\be{\begin{equation}}
\def\ee{\end{equation}}
\def\bea{\begin{eqnarray}}
\def\eea{\end{eqnarray}}
\def\bean{\begin{mathletters}\begin{eqnarray}}
\def\eean{\end{eqnarray}\end{mathletters}}

%%% macros for Feb 2000 PRL
\newcommand{\tbox}[1]{\mbox{\tiny #1}}
\newcommand{\half}{\mbox{\small $\frac{1}{2}$}}
\newcommand{\pit}{\mbox{\small $\frac{\pi}{2}$}}
\newcommand{\sfrac}[1]{\mbox{\small $\frac{1}{#1}$}}
\newcommand{\mbf}[1]{{\mathbf #1}}
% hack to get APS's \text style to work ok:
\def\text{\tbox}

\newcommand{\mV}{{\mathsf{V}}}
\newcommand{\mL}{{\mathsf{L}}}
\newcommand{\mA}{{\mathsf{A}}}
\newcommand{\lB}{\lambda_{\tbox{B}}}  % de Broglie
\newcommand{\ofr}{{(\mbf{r})}}       % (r vec)
\def\ofkr{(k;\mbf{r})}			% (k;r vec)
\def\ofks{(k;\mbf{s})}			% (k;s vec)
\newcommand{\ofs}{{(\mbf{s})}}       % (s vec)
\def\xt{\mbf{x}^{\tbox T}}		% x^T vec

\def\ce{\tilde{C}_{\tbox E}}		% C_E
\def\cew{\tilde{C}_{\tbox E}(\omega)}		% C_E(w)
\def\ceqmw{\tilde{C}^{\tbox{qm}}_{\tbox E}(\omega)}	% C^qm_E(w)
\def\cewqm{\tilde{C}^{\tbox{qm}}_{\tbox E}}	% C^qm_E(w) no omega
\def\ceqm{C^{\tbox{qm}}_{\tbox E}}	% C^qm_E no tilde
\def\cw{\tilde{C}(\omega)}		% C(w)
\def\cfw{\tilde{C}_{\cal F}(\omega)}		% C_F(w)

\def\tcl{\tau_{\tbox{cl}}}		% tau cl
\def\tcol{\tau_{\tbox{col}}}		% tau col
\def\terg{t_{\tbox{erg}}}		% t_erg
\def\tbl{\tau_{\tbox{bl}}}		% tau bl
\def\theis{t_{\tbox{H}}}		% t_Heis

\def\area{\mathsf{A}_D}			% piston effective area A_D 
\def\ve{\nu_{\tbox{E}}}			% nu_E, noise intensity
\def\vewna{\nu_E^{\tbox{WNA}}}		% nu_E for WNA

\def\dxcqm{\delta x^{\tbox{qm}}_{\tbox c}}	% x to mix levels.

%% operators
\newcommand{\rop}{\hat{\mbf{r}}}	% vector valued operators
\newcommand{\pop}{\hat{\mbf{p}}}

%%% Integrals
\newcommand{\sint}{\oint \! d\mbf{s} \,} % surface int
\def\gint{\oint_\Gamma \!\! d\mbf{s} \,} % surface int over Gamma.
\newcommand{\lint}{\oint \! ds \,}	% d=2
\def\infint{\int_{-\infty}^{\infty} \!\!}	% infinite integral
\def\dn{\partial_n}				% d_n
\def\aswapb{a^*\!{\leftrightarrow}b}		% a* <-> b
\def\eps{\varepsilon}				% eps

%%% Dissipation.
\def\dhdxt{\partial {\cal H} / \partial x}
\def\dhdx{\pd{\cal H}{x}}
\def\dhdxnm{\left( \pd{\cal H}{x} \right)_{\!nm}}
\def\dhdxnmsq{\left| \left( \pd{\cal H}{x} \right)_{\!nm} \right| ^2}

%%% vergini
\def\bcs{\stackrel{\tbox{BCs}}{\longrightarrow}}	% apply BCs

%%% DIEL atom project.
\def\wx{\omega_x}
\def\wy{\omega_y}
\newcommand{\ofro}{({\bf r_0})}
\def\Eb{E_{\rm blue,rms}}
\def\Er{E_{\rm red,rms}}
\def\Es2{E_{0,{\rm sat}}^2}
\def\sb{s_{\rm blue}}
\def\sr{s_{\rm red}}

%%% text usefuls
\def\ie{{\it i.e.\ }}
\def\eg{{\it e.g.\ }}
\newcommand{\etal}{{\it et al.\ }}
\newcommand{\ibid}{{\it ibid.\ }}

%%% tables spaces.
\def\gap{\hspace{0.2in}}

%%%%%%%%%%%%%%%%%%%%
%%% mathletters code (from http://www.grad.uiuc.edu/thesis/latexcode.html )
% Modified Alex Barnett 00/9/28 to handle non-arabic \thechapter output
% (ie, now works in Appendices)
% Fails to give correct references in eqnarray (offset by one).
% Still inserts small extra space after equation - unknown reason.
%
%    Original code:
%\newcounter{eqletter}
%\def\mathletters{%
%\setcounter{eqletter}{0}%
%\addtocounter{equation}{1}
%\edef\curreqno{\arabic{equation}}
%\edef\@currentlabel{\theequation}
%\def\theequation{%
%\addtocounter{eqletter}{1}\arabic{chapter}.\curreqno\alph{eqletter}%
%}%
%}
%\def\endmathletters{\setcounter{equation}{\curreqno}}

\newcounter{eqletter}
\def\mathletters{%
\setcounter{eqletter}{0}%
\addtocounter{equation}{1}
\edef\curreqno{\arabic{equation}}
\edef\@currentlabel{\theequation}
\def\theequation{%
\addtocounter{eqletter}{1}\thechapter.\curreqno\alph{eqletter}%
}%
}
\def\endmathletters{\setcounter{equation}{\curreqno}}

%...............................................QPC...................

\newcommand{\bk}{{\bf k}}
\def\kf{k_{\text F}}
\newcommand{\br}{{\bf r}}
\newcommand{\TL}{{\text{(L)}}}
\newcommand{\TR}{{\text{(R)}}}
\newcommand{\TLR}{{\text{L,R}}}
\newcommand{\VSD}{V_{\text{SD}}}
\newcommand{\GT}{\Gamma_{\text{T}}}
\newcommand{\DEL}{\mbox{\boldmath $\nabla$}}
\def\lf{\lambda_{\text F}}
\def\st{\sigma_{\text T}}
\def\stlr{\sigma_{\text T}^{\text{L$\rightarrow$R}}}
\def\strl{\sigma_{\text T}^{\text{R$\rightarrow$L}}}
\def\aeff{a_{\text{eff}}}
\def\aaeff{A_{\text{eff}}}
\def\gat{G_{\text{atom}}}
\newcommand{\LB}{Landauer-B\"{u}ttiker}

% .............. NOTES ..................
% To force linebreak when get overfull hbox from equation in paragraph
% test, use \linebreak (which makes it justify the line it broke).
% Contrast \\ or \newline which leave empty space.
%
%

%% file: frontmatter.tex
%% frontmatter.tex
%%

\title{Quantum Dynamics of Nonlinear Cavity Systems}
\author{Paul David Nation}
\degreemonth{June} % month final submission occurs.
\degreeyear{2010}
\degree{Doctor of Philosophy}
\field{Physics}
\department{Physics}
\advisor{Miles P. Blencowe} % Category I added.

\maketitle
\copyrightpage

\begin{abstract}
\dsp
\ \

\ \
In this work we investigate the quantum dynamics of three different configurations of nonlinear cavity systems.  We begin by carrying out a quantum analysis of a dc superconducting quantum interference device (SQUID) mechanical displacement detector comprising a SQUID with a mechanically compliant loop segment.  The SQUID is approximated by a nonlinear current-dependent inductor, inducing an external flux tunable nonlinear Duffing term in the cavity equation of motion.  Expressions are derived for the detector signal and noise response where it is found that a soft-spring Duffing self-interaction enables a closer approach to the displacement detection standard quantum limit, as well as cooling closer to the ground state.  Next, we consider the use of a superconducting transmission line formed from an array of dc-SQUIDs for investigating analogue Hawking radiation.  We will show that biasing the array with a space-time varying flux modifies the propagation velocity of the transmission line, leading to an effective metric with a horizon.  As a fundamentally quantum mechanical device, this setup allows for investigations of quantum effects such as backreaction and analogue space-time fluctuations on the Hawking process.  Finally, we investigate a quantum parametric amplifier with dynamical pump mode, viewed as a zero-dimensional model of Hawking radiation from an evaporating black hole.  The conditions are derived under which the spectrum of particles generated from vacuum fluctuations deviates from the thermal spectrum predicted for the conventional parametric amplifier.  We find that significant deviation occurs once the pump mode (black hole) has released nearly half of its initial energy in the signal (Hawking radiation) and idler (in-falling particle) modes.  As a model of black hole dynamics, this finding lends support to the view that late-time Hawking radiation contains information about the quantum state of the black hole and is entangled with the black hole's quantum gravitational degrees of freedom.

\end{abstract}

\begin{acknowledgments}
\dsp
\ \

\ \
Although my name is the only one on this thesis, the work done over the last five years, and my maintained sanity during this same period, could not have been accomplished without the help of several individuals whom I would now like to thank.

First and foremost, I am indebted to to my adviser Miles Blencowe for his steadfast support and encouragement over the course of my time at Dartmouth.  Since the beginning, my relationship with Miles has not been that of an advisor and their student, but rather one between colleagues and friends.  Even in the beginning when I knew virtually nothing about our field of research, Miles was always pushing me to think on my own and challenging me to come up with my own research ideas.  This freedom to explore the physics, as opposed to being handed a research topic, has resulted in much of the work presented here as well as our future research endeavors.  Although his methodology has proven fruitful, I recognize that the first few years of my constant questions and comments must have been very trying on Miles.  Therefore, I accept full responsibility for any increase in Miles' red wine consumption over my years at Dartmouth.  As a corollary to our work together, I have also enjoyed many unique traveling experiences with Miles.  From Russian bars and rotisserie chicken in Israel, to red wine tasting in California, outings with Miles have provided for many memorable adventures and valuable life lessons.  I am eager to see what awaits us on our final trip to Japan.  Having the opportunity to reflect on the last five years, I realize that I could not have asked for a better mentor, coworker, and friend.  

In addition to Miles I would also like to thank Alex Rimberg and Eyal Buks for their help on the experimental feasibility and implementation of the superconducting circuit setups considered here.  My minimal knowledge of materials and fabrication methods would have never been enough to complete this work without their help and criticism.  Likewise, I am thankful to Hiroshi Yamaguchi for giving me the opportunity to spend some time in the laboratories of NTT-BRL and for a brief moment experience the life of an experimentalist.  On the theoretical side, I would also like to thank Andrew Armour for helpful discussions over the past several years.  Finally, I have to acknowledge the many great graduate students and postdocs I have had the pleasure of meeting, in particular Baleegh Abdo, Winton Brown, Tatsuo Dougakiuchi, Imran Mahboob, Menno Poot, and Takayuki Watanabe.
 
I have received enormous support over the course of my time at Dartmouth from my wife Hwajung.  Throughout the years in Hanover she has been a constant source of encouragement and happiness.  I can be a very stubborn person and yet somehow she has remained unwavering in her kindness and concern.  As a physicist herself, I have enjoyed our many, often heated, discussions about physics and astronomy and the frequent arguments at the white board.  As I am often the loser of such debates, they have benefited me immensely by opening my mind to new ways of thinking about physics.  However, her greatest assistance has been in getting me to enjoy the world outside of physics.  It is through her relentless prodding that I manage to get out of the office and enjoy the less serious things in life.  I cannot thank her enough for everything she has done.

Life in the office would not have been the same without Danny Milisavljevic who is one of the few people that I have met with a correct understanding of the balance between work and play.  Although many graduate students in physics at Dartmouth tend to idly go through the Ph.D. process, Danny has an excitement and drive about him that shows up not only in his work, but also in his genuine love for science.  Without his motivation I might have fallen in with the rest of the crowd.   Outside of work, our frequent bad-movie nights, parties, and dinners have all been extremely helpful in preventing the downfall of my mental faculties.  Along these same lines, I give thanks to my friend and former roommate Victor de Vries who will be joining me in moving to the Eastern Hemisphere.  Hopefully we can think of something to build that beats our life-size snow bar. 

I am grateful to the faculty and staff of the Dartmouth College Department of Physics and Astronomy whom I have had the great pleasure of getting to know during my time in Hanover.  Furthermore, I am thankful to the graduate students who came with me to Dartmouth in the fall of 2005: Shusa Deng, Ryan Johnson, Hwajung Kang, James Lundberg, Danny Milisavljevic, Dane Owen, David Sicilia, and Sara Walker.  We have managed to maintain a closeness not seen from any other group of students; I hope this continues into the future.  Lastly, I would like give credit to the makers of TeXShop, LaTeXiT, Papers, Keynote, as well as the NumPy, SciPy, and matplotlib packages for Python.  I have used these tools extensively throughout my doctoral work and they are deserving of recognition.

\end{acknowledgments}

\newpage
\thispagestyle{plain}
\ssp
\rule[0pt]{0pt}{0pt} \\
\vspace{2.0in}
\rule[0pt]{0pt}{0pt} \\
\centerline{``Scientific discovery and scientific}
\centerline{}
\centerline{knowledge have been achieved only}
\centerline{}
\centerline{by those who have gone in pursuit}
\centerline{}
\centerline{of it without any practical purpose}
\centerline{}
\centerline{whatsoever in view."}
\centerline{}
\centerline{--Max Planck}
\newpage
\addcontentsline{toc}{section}{Table of Contents}
\tableofcontents
\newpage
\listoffigures
% these are optional in the Jan 2000 Harvard thesis GSAS guide:
%\listoffigures
%\listoftables
%(Cut them for my personal thesis format).

% cccccccccccccccccccccccccccccccccccccccccccccccccccccccccccccccccccccccccc
\begin{citations}

\vspace{0.5in}

\dsp
\noindent
Chapter~\ref{ch:squid} and Appendix~\ref{ap:squid} appear in their entirety as
\begin{quote}\dsp
	``Quantum analysis of a nonlinear microwave-cavity embedded dc-SQUID displacement detector'',
	P. D. Nation, M. P. Blencowe, and E. Buks,
	Phys. Rev. B \textbf{78}, 104516 (2008),
	{\tt arXiv:0806.4171}.
\end{quote}
With minor changes, almost all of Chapter~\ref{ch:black} has been published as
\begin{quote}\dsp
	``Analogue Hawking Radiation in a dc-SQUID Array Transmission Line'',
	P. D. Nation, M. P. Blencowe, A. J. Rimberg, and E. Buks,
	Phys. Rev. Lett. \textbf{103}, 087004 (2009), {\tt arXiv:0904.2589}.
\end{quote}
Finally, Chapter~\ref{ch:trilinear} is to appear without Sec.~\ref{sec:tri-circuit} in the \textit{New Journal of Physics} focus issue: ``Classical and Quantum Analogues for Gravitational Phenomena and Related Effects".
\begin{quote}\dsp
	``The Trilinear Hamiltonian: Zero Dimensional Model for Hawking Radiation from a Quantized Source'',
	P. D. Nation, and M. P. Blencowe, {\tt arXiv:1004.0522} (2010).
\end{quote}
Electronic preprints are available
on the Internet at the following URL:
\begin{quote}
	{\tt http://arXiv.org}
\end{quote}
\end{citations}

%ddddddddddddddddddddddddddddddddddddddddddddddddddddddddddddddddddddddddddd

%%% end

%% file: ch1.tex
%% intro.tex
\dsp
\chapter{Introduction and organization}
%%%%%%%%%%%%%%%%%%%%%%%%%%%%%%%%%%%%%%%%%%%%%%%%
\section{Organization of this thesis}
This thesis is naturally organized into three chapters that are relatively independent from each other:
\begin{itemize}
\item Displacement detection and cooling of a mechanical oscillator using a nonlinear microwave cavity detector (Chapter~\ref{ch:squid})
\item Analogue Hawking radiation in a dc-SQUID array transmission line (Chapter~\ref{ch:black})
\item The Trilinear Hamiltonian: Modeling Hawking radiation from a quantized source (Chapter~\ref{ch:trilinear}) \end{itemize}

Given the self-contained nature of the topics, each chapter includes its own introduction and conclusion sections.  However, the work in the later chapters relies upon the knowledge gained in the previous chapters and is the underlying reason for the ordering of research topics presented here.  To highlight these relations, Chapters \ref{ch:black} and \ref{ch:trilinear} contain ``Motivation" sections which attempt to elucidate the connections between the various research topics.  Given the overall theme of nonlinear cavity systems, in the next section we briefly introduce the field of superconducting circuit cavity systems that provides the foundation for the work presented in this thesis.

\section{Prologue}
Since the early days of quantum mechanics the interaction between two-level systems and harmonic oscillators has played a central role in our understanding of physics below the macroscopic level.  Beginning with the thought experiments of Einstein and Bohr, scientists have devoted significant effort towards understanding the dynamics of these most basic elements of quantum theory.  Unfortunately, as is often the case with physics, it would be another fifty years before these conceptual ideas could be brought to reality.  It wasn't until the 1970's and the advent of Cavity Quantum Electrodynamics (CQED) that physicists could fabricate and control the interaction between an atom (two-level system) and an optical oscillator (harmonic oscillator) while at the same time removing unwanted environmental effects.  Since then, CQED has contributed greatly to our fundamental understanding of the interaction of matter with quantized electromagnetic fields, the physics open quantum systems, and decoherence (see Ref.~\cite{haroche:2006} and references therein).  Additionally, these same systems will undoubtably be utilized in the future realization of a quantum computer.  The experiments made possible by CQED have brought our understanding full circle to where it is possible to test the very foundations upon which quantum mechanics rests.

Building upon the success of CQED, this last decade has seen a large effort devoted to exhibiting complementary effects in solid-state superconducting circuits, a field of research known as circuit-QED\cite{blais:2004,Schoelkopf:2008p1671}.  While similar to CQED, the use of superconducting circuits to fabricate artificial atoms and cavities has several unique advantages.  To begin, the effective two-dimensional coplanar microwave transmission-line cavity\cite{goppl:2008} geometries used in superconducting circuits have mode volumes that are markedly smaller than the corresponding three-dimensional Fabry-P\'{e}rot cavities used in their optical counterparts.  This modal volume is significantly less than the cubic wavelength corresponding to a microwave photon in vacuum ($\lambda=30~\mathrm{cm}$ at $1~\mathrm{GHz}$) and consequently leads to a greatly enhanced electric field strength\cite{Schoelkopf:2008p1671} and correspondingly large amplitude vacuum fluctuations\cite{wallraff:2004,Devoret:2007p1782}.  Additionally, the small size and low-temperature environment of these resonators has routinely produced quality factors of $Q\approx10^{5}$ or higher\cite{goppl:2008} making it possible to reach the ultra-strong coupling regime\cite{abdumalikov:2008,bourassa:2009}.  

The artificial atoms (qubits) to which the cavity resonator interacts are composed of highly nonlinear Josephson junction circuit elements\cite{likharev:1986} that sufficiently modify the harmonic potential of a typical LC-circuit allowing for the isolation of the lower two energy levels\cite{wendin:2007}.  The energy spacing between levels can be fabricated to a desired value and then modified in situ by the application of an external magnetic flux bias.  At present there are a variety of experimental realizations of superconducting atoms based on charge, phase, and flux qubits, each distinguished by the degree of freedom used in generating the approximate two-level potential\cite{you:2005,clarke:2008}.  Since the original demonstration of a solid-state qubit\cite{nakamura:1999} the coherence times of these devices have gone from a few tenths of nanosecond up to $\sim$ one microsecond\cite{clarke:2008}, representing an increase in four orders of magnitude over the last ten years.  This rapid rise in coherence times has helped to positioned solid-state qubits and the circuit-QED architecture at the forefront of the race for a scalable quantum computer. 

To date, circuit-QED has resulted in unparalleled access to the coherent interactions between qubit and cavity resonator.  It is now possible to control and observe photon number states in a microwave resonator using a coupled qubit\cite{schuster:2007,houck:2007,hofheinz:2008}, generate and perform tomography of arbitrary superposition states\cite{hofheinz:2009}, and measure the decay of these states due to the ever present environment\cite{wang:2008}.  Qubits coupled to cavity resonators\cite{majer:2007} can find applications in single-atom lasing\cite{astafiev:2007} and fluorescence\cite{astafiev:2010a}, quantum amplifiers\cite{astafiev:2010b}, coupled qubit dynamics, and the demonstration of quantum computing gates and algorithms (an exhaustive list of references on qubit coupling and computation can be found in Ref.~\cite{clarke:2008}).

In this thesis we will make use of the circuit-QED architecture in exploring the detection and cooling of a mechanical oscillator by a cavity detector, and as analogue models for investigating classical and quantum gravitational physics.  The preparation and manipulation of quantum states of a mechanical resonator has many potential applications in fundamental physics and applied sciences\cite{blencowe:2005,schwab:2005,aspelmeyer:2008}.  This has stimulated many groups to  focus on coupling mechanical resonators to both superconducting\cite{Xue:2007p2711,Blencowe:2007p285,Buks:2007p279,Regal:2008p2560,Teufel:2008p2398,li:2008,elste:2009,hertzberg:2009} and optical cavities\cite{kippenberg:2007,marquardt:2009} with the hope of reaching the quantum ground state in the mechanical element, a goal that has recently been achieved\cite{oconnell:2010}.  However, the push to see quantum effects in progressively larger systems requires ever more advanced methods to reach this goal.  In Chapter~\ref{ch:squid} we will demonstrate the detection and cooling of a mechanical resonator coupled to a nonlinear microwave cavity.  In Chapters~\ref{ch:black} and \ref{ch:trilinear} we will focus on simulating the gravitational physics of black holes using superconducting circuits.  The use of superconducting elements in the investigation of analogue gravitational effects is a previously unexplored area that allows the possibility of observing otherwise elusive quantum gravitational phenomena in the laboratory.

%%%

%% file: ch2.tex
%% intro.tex
\dsp
\chapter{Displacement detection and cooling of a nanomechanical resonator using a nonlinear microwave cavity detector}\label{ch:squid}
%%%%%%%%%%%%%%%%%%%%%%%%%%%%%%%%%%%%%%%%%%%%%%%%

\section{Introduction}\label{sec:squid-intro}
Recently there has been interest in exploiting the nonlinear dynamics of nanoelectromechanical systems (NEMS) for amplification.\cite{Krommer:2000p2946,Aldridge:2005p3614,Almog:2007p2875}  The use of nonlinear mechanical resonators to some extent parallels investigations with systems comprising purely electronic degrees of freedom, such as  nonlinear superconducting devices incorporating  Josephson Junctions (JJ).\cite{Siddiqi:2004p1662,Lupascu:2006p3289,Lee:2007p3227} For example, it was shown that the bistable response of an RF-driven JJ can be employed as a low noise, high-sensitivity amplifier for superconducting qubits.\cite{Siddiqi:2004p1662}  A similar setup consisting of a JJ embedded in a microwave cavity was used to measure the states of a quantronium qubit,\cite{Metcalfe:2007p1060} where the relevant cavity mode was found to obey the Duffing oscillator equation.\cite{Boaknin:2007p3308}

One area of nanomechanics that has yet to fully explore the possibility of exploiting nonlinearities for sensitive detection involves setups in which a nanomechanical resonator couples either capacitively\cite{Xue:2007p2711,Regal:2008p2560,Teufel:2008p2398} or inductively\cite{Blencowe:2007p285,Buks:2007p279} to a superconducting microwave transmission line resonator, combining elements from both the above-described NEMS and superconducting systems.  Such setups are in some sense the solid-state analogues of optomechanical systems, which ponderomotively couple a movable mirror to the optical field inside a cavity using radiation pressure.\cite{Mancini:1994p2815,LAW:1995p3039,Jacobs:1999p2816,Metzger:2004p2871,Gigan:2006p1942,Arcizet:2006p2873,Schliesser:2006p2558,Mavalvala:2007p160801,Thompson:2008p72}  In both areas, the focus has primarily been on operating in the regime where the cavity and resonator behave to a good approximation as harmonic oscillators interacting via their mutual ponderomotive coupling.  However, in the case of microwave cavities, introducing an embedded JJ,\cite{Boaknin:2007p3308} or simply driving the cavity close to the superconducting critical temperature,\cite{Tholen:2007p1916} results in the breakdown of  the harmonic mode approximation; nonlinear dynamical behavior of the cavity must be taken into account. Furthermore, the ponderomotive coupling term between the microwave or optical cavity mode and mechanical mode is by itself  capable of inducing strong, effective nonlinearities in the respective mode equations.  In optical systems, such nonlinearities can manifest themselves in the appearance of  a bistable (or even multistable)  region for the movable mirror.\cite{DORSEL:1983p2818,Marquardt:2006p103901}  By restricting ourselves to linear microwave cavities, we are overlooking a range of nonlinear phenomena that might enable a closer approach to quantum-limited detection, as well as cooling of the mechanical oscillator closer to its ground state. As an illustration, consider the phase sensitive Josephson parametric amplifier,\cite{yurke:1989p2519,Yurke:2006p1911,bergeal:2008,beltran:2008}    which exploits the nonlinear effective inductance of the JJ to perform  (in principle) noiseless amplification and  quantum squeezing of the  respective complimentary quadrature amplitudes of the signal oscillator.

In this chapter, we will go beyond the usually considered ponderomotively-coupled two oscillator system to include a Duffing nonlinearity in the microwave cavity mode equations. The closed system model Hamiltonian describing the nonlinear microwave-coupled mechanical oscillators is given by Eq.~(\ref{eq:hamiltonian-closed}). The nonlinear microwave mode is externally driven with a pump frequency $\omega_p$ that can be detuned from the transmission line mode frequency $\omega_T$. 
Our investigation will focus on the nonlinear amplifier created by embedding a dc-SQUID displacement detector into a superconducting microwave transmission line.\cite{Blencowe:2007p285}  This has the advantage of significantly larger coupling strengths\cite{Devoret:2007p1782} as compared with existing geometrical coupling schemes.\cite{Xue:2007p2711,Regal:2008p2560,Teufel:2008p2398}  The displacement detector comprises a SQUID with one segment consisting of a doubly-clamped mechanical resonator as shown in Fig.~\ref{fig:diagram}.  The net flux, and therefore circulating current, is modulated by the mechanical motion, providing displacement transduction.  The capacitively-coupled  pump/probe feedline both drives and provides readout of the relevant transmission line resonator mode amplitude (or phase). We will assume transmission line losses are predominantly due  to coupling with the feedline, and that the pump drive is coherent.   The main irreducible noise source is therefore microwave photon shot noise from the drive that acts back on the mechanical oscillator via the intermediate nonlinear microwave resonator and SQUID. Environmental influences on the mechanical oscillator other than that due to the SQUID detector are simply modelled as a free oscillator thermal bath.  By operating the amplifier well below the superconducting critical temperature, and with transmission line currents less than the SQUID JJ's critical current threshold, resistive tunneling of electrons and the associated noise is a negligible contribution.  Similar setups involving JJ elements have been considered previously.\cite{Blencowe:2007p285,Buks:2007p279,Xue:2007p2454,Wang:2007p2819}

With JJ plasma frequencies assumed to be larger than both the mechanical and transmission line fundamental mode frequencies, the SQUID can be considered as a passive, effective inductance element that depends on both the externally applied flux and drive current.  The effective inductance can therefore be freely tuned by varying these external parameters.  Previously, we considered only the lowest, zeroth order expansion of the inductance with respect to the current entering (or exiting) the SQUID,\cite{Blencowe:2007p285} yielding the usual ponderomotively-coupled double harmonic oscillator system.  In this companion investigation, we include  the next leading, quadratic order  term, resulting in a nonlinear current dependent inductance. Provided that the current magnitude is small as compared with the JJ's critical current, neglecting higher order terms should not introduce significant errors. The nonlinear inductance induces an effective Duffing (i.e., cubic) self-interaction term in the microwave mode equations of motion.  The results presented here apply  to a broad class of bosonic detector, which includes optomechanical amplifiers with nonlinear cavities\cite{Drummond:1980p3191} that are describable by Hamiltonian~(\ref{eq:hamiltonian-closed}). A related analysis of quantum  noise in a Duffing oscillator amplifier is  given in Ref.~\cite{BabourinaBrooks:2008p3350}.  

The chapter is organized as follows.  In Sec.~\ref{sec:squid-motion}  we first derive the truncated Hamiltonian~(\ref{eq:hamiltonian-closed}) that describes the closed system dynamics of the coupled cavity and mechanical resonator fundamental modes. We then derive the quantum Langevin equations of motion that describe the open system dynamics in the presence of the pump/probe line and mechanical oscillator's external environment.  In Sec.~\ref{sec:squid-response} we find expressions for the detector signal response and noise using a semiclassical treatment of the detector's linear response to the external noise input signal driving the mechanical oscillator.  Section~\ref{sec:squid-bistability} determines the critical drive current for the onset of bistability (not to be confused with the JJ critical current).  Sections \ref{sec:squid-detection} and \ref{sec:squid-cooling} discuss the effects of the microwave mode Duffing nonlinearity on mechanical mode displacement detection and cooling, respectively, giving illustrative examples  assuming achievable device parameters.  Section~\ref{sec:squid-conclusion} briefly concludes, while the more technical aspects of the signal and noise term derivations are relegated to Appendix~\ref{ap:squid}.  Source code for the numerical analysis in Secs.~\ref{sec:squid-detection} and \ref{sec:squid-cooling} is given in Appendix~\ref{sec:ap2-squid}.

\section{Equations of Motion}\label{sec:squid-motion}
\subsection{Closed System Hamiltonian}\label{subsec:closed}
\begin{figure}[htbp]
\begin{center}
\includegraphics[width=4in]{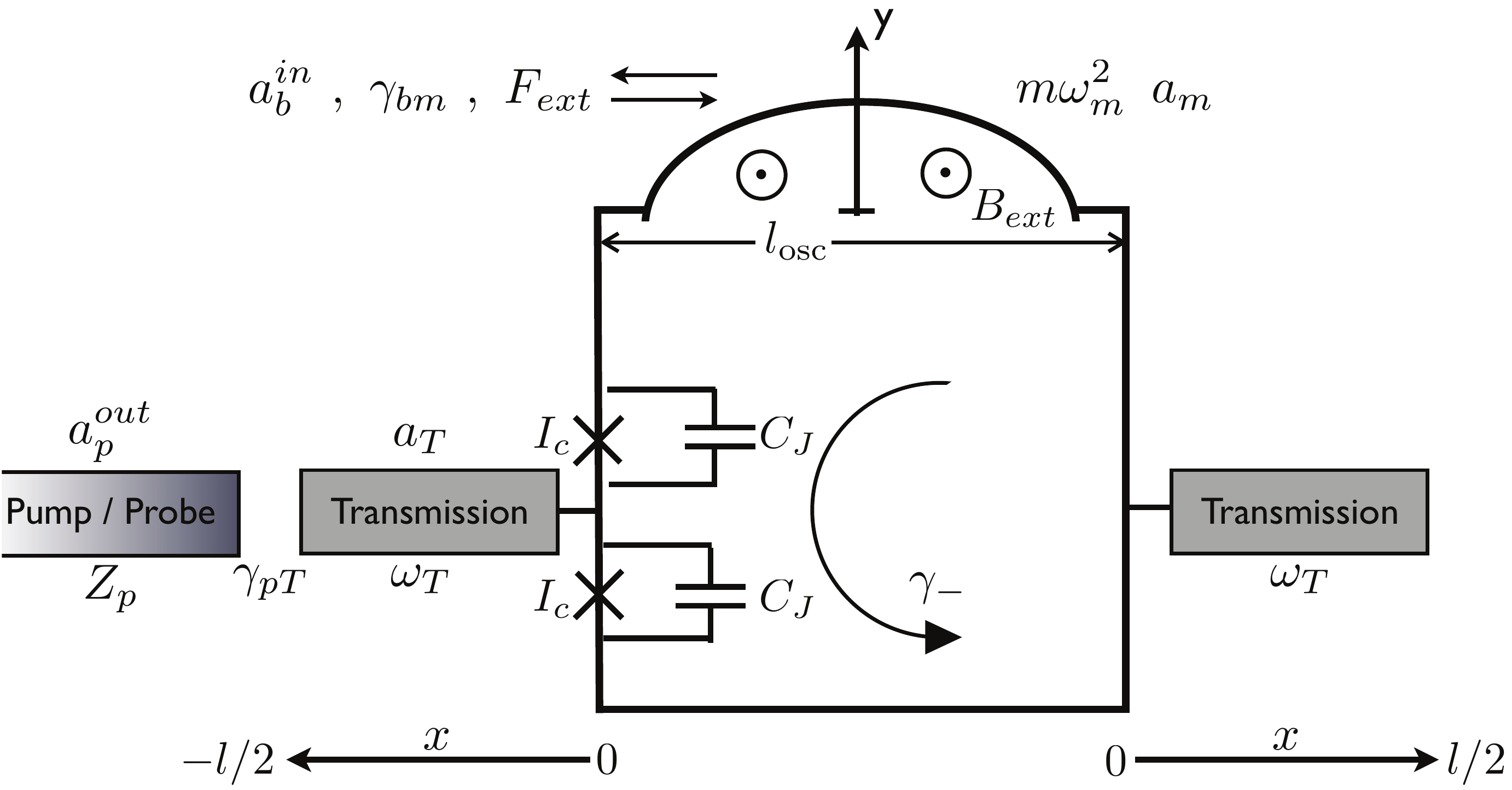}
\caption{Layout of the dc SQUID displacement detector.  The dimensions of the SQUID have been enlarged relative to the transmission line to show the key characteristics employed in the analysis.}
\label{fig:diagram}
\end{center}
\end{figure}

The displacement detector scheme is shown in Fig. \ref{fig:diagram}.  The device consists of a stripline resonator (transmission line) of length $l$ bisected by a SQUID.  The transmission line is characterized by an inductance and capacitance per unit length $L_{T}$ and $C_{T}$ respectively.  The SQUID comprises two identical Josephson junctions with critical current $I_{C}$ and capacitance $C_{J}$.  A segment of the SQUID loop is mechanically compliant, forming a doubly clamped resonator of length $l_{\mathrm{osc}}$.  A similar setup without the microwave cavity has recently been constructed\cite{etaki:2008}. We only take into account mechanical fundamental mode displacements in the plane of the loop and assume that the resonator can be modeled effectively as a harmonic oscillator with the $y$ coordinate giving the center of mass displacement.  The magnetic flux threading the loop is given by $\Phi_{\mathrm{ext}}(y)=\Phi_{\mathrm{ext}}(0)+\lambda B_{\mathrm{ext}}l_{\mathrm{osc}}y$,   where $\Phi_{\mathrm{ext}}(0)\equiv\Phi_{\mathrm{ext}}$ is the flux with the mechanical oscillator fixed at $y=0$, $B_{\mathrm{ext}}$ is the externally applied field in the vicinity of the resonator, and $\lambda$ is a geometrical factor that corrects for the non-uniform displacement of the oscillator along its length.  The coupling between mechanical oscillator and external heat bath is characterized by the oscillator amplitude damping rate $\gamma_{bm}$, while the pump-probe line-transmission line coupling is characterized by the transmission line amplitude damping rate  $\gamma_{pT}$.  In what follows, we will assume weak couplings (i.e., large quality factors for the transmission line and mechanical oscillator) and also that the dominant dissipation mechanism for the transmission line is due to its coupling to the pump-probe line, $\gamma_{pT}$.

In analyzing the SQUID dynamics, an appropriate choice of variables is $\gamma_{\pm}=(\phi_{1}\pm\phi_{2})/2$, where $\phi_{1}$ and $\phi_{2}$ are the gauge invariant phases across the Josephson junctions,\cite{orlando}  while for the transmission line we choose  the phase field $\phi(x,t)$. The transmission line current and voltage are described in terms of $\phi(x,t)$ using the standard telegraphic relations:
\begin{eqnarray}
I(x,t)&=&-\frac{\Phi_{0}}{2\pi L_{T}}\frac{\partial \phi(x,t)}{\partial x}, \\
V(x,t)&=&\frac{\Phi_{0}}{2\pi}\frac{\partial \phi(x,t)}{\partial t},
\label{eq:currentvoltage}
\end{eqnarray}
where $\Phi_{0}=h/(2e)$ is the flux quantum.  Assuming that the SQUID can be lumped at the midpoint $x=0$ of the transmission line, the equations of motion for the closed system comprising the SQUID, transmission line and mechanical oscillator are given by\cite{Buks:2006p346}
\begin{equation}\label{eq:squid-wave}
\frac{\partial^{2}\phi}{\partial t^{2}}=\frac{1}{L_{T}C_{T}}\frac{\partial^{2}\phi}{\partial x^{2}},
\end{equation}
\begin{equation}\label{eq:current-circ}
{\omega_{J}^{-2}}\ddot{\gamma}_{-}+\cos(\gamma_{+})\sin(\gamma_{-})+{2}{\beta^{-1}_{L}}\left[\gamma_{-}-\pi\left(n+\frac{\Phi_{\mathrm{ext}}+\lambda B_{\mathrm{ext}}l_{\mathrm{osc}} y}{\Phi_{0}}\right)\right]=0,
\end{equation}
\begin{equation}\label{eq:current-avg}
{\omega_{J}^{-2}}\ddot{\gamma}_{+}+\sin(\gamma_{+})\cos(\gamma_{-})+\frac{\Phi_{0}}{4\pi L_{T}I_{c}}\frac{\partial \phi(0,t)}{\partial x}=0,
\end{equation}
and
\begin{equation}\label{eq:force}
m\ddot{y}+m\omega_{m}^{2}y-\frac{\Phi_{0}}{\pi L}\lambda B_{\mathrm{ext}}l_{\mathrm{osc}}\gamma_{-}=0,
\end{equation}
where $\omega_{J}=\sqrt{2\pi I_{c}/(C_{J}\Phi_{0})}$ is the plasma frequency of the Josephson junctions, $\beta_{L}\equiv 2\pi L I_{c}/\Phi_{0}$ is a dimensionless parameter with $L$ the SQUID loop self-inductance and $I_c$ the Josephson junction critical current, and where $n$ takes on integer values arising from the requirement that the phase around the loop be single-valued.  Equation (\ref{eq:squid-wave}) is the wave equation for the transmission line,  equation~(\ref{eq:current-circ}) describes the current circling the loop, which depends on the external flux and oscillator position,  equation~(\ref{eq:current-avg}) describes the average current in the loop, and  Eq.~(\ref{eq:force}) is Newton's second law for the mechanical oscillator with  Lorentz force acting on the oscillator.   The  current and voltage across the SQUID must also obey the boundary conditions
\begin{eqnarray}
\frac{\partial \phi(\pm l/2,t)}{\partial x}=0;\ \ \frac{\partial \phi(0^{-},t)}{\partial x}=\frac{\partial \phi(0^{+},t)}{\partial x},\label{eq:currentbc}\\
\dot{\gamma}_{+}-\frac{L}{4 L_{T}}\frac{\partial^2 \phi(0,t)}{\partial t\partial x}=\frac{\partial \phi(0^{-},t)}{\partial t}-\frac{\partial \phi(0^{+},t)}{\partial t}.\label{eq:voltagebc}
\end{eqnarray}

Using Eqs.~(\ref{eq:squid-wave})-(\ref{eq:voltagebc}), we shall now derive approximate equations of motion describing a single mode of the transmission line interacting with the mechanical oscillator, where the form of  the interaction between the two oscillators is governed by the SQUID parameters and boundary conditions.  Assume that the following conditions are satisfied: (a)~$\omega_{J}\gg\omega_{T}\gg\omega_{m}$. (b)~$\beta_{L}\ll 1$. (c)~$|I(0,t)/I_{c}|\ll1$.  (d)~$|\lambda B_{\mathrm{ext}}l_{\mathrm{osc}}y/{\Phi_{0}}|\ll 1$.  Condition~(a) states that the SQUID plasma frequency is much larger than the transmission line mode frequency of interest, $\omega_T$, and consequently we shall ignore the SQUID inertia terms in (\ref{eq:current-circ}) and (\ref{eq:current-avg}). Condition~(b) allows us to neglect the SQUID loop self inductance and, together with (a), eliminate $\gamma_{\pm}$ from the equations by expressing them in terms of the transmission line and oscillator coordinates as series expansions in $\beta_{L}$. Conditions (c) and (d) allow us  to expand the above equations in the transmission line current $I(0,t)\equiv I(t)$ at $x=0$ and in the oscillator displacement $y$. Keeping terms to first order in $y$ and to leading, second order in $I$, Eq.~(\ref{eq:force}) for the mechanical oscillator becomes approximately 
\begin{equation}
m\ddot{y}+m\omega_{m}^{2}y-L_{01}I^2/2 =0.
\label{eq:force-approx}
\end{equation}
The voltage boundary condition (\ref{eq:voltagebc}) can be expressed as
\begin{equation}
\frac{\partial}{\partial t}[L(I,y)I]=\frac{\Phi_{0}}{2\pi}\left[\frac{\partial \phi(0^{-},t)}{\partial t}-\frac{\partial \phi(0^{+},t)}{\partial t}\right]\label{eq:voltage},
\end{equation}
where $L(I,y)$ is the effective inductance, which expanded to second order in $I$ takes the form
\begin{equation}\label{eq:squid-inductance}
 L(I,y)=L_{00}+L_{20}\left({I}/{I_{c}}\right)^{2}+L_{01}y,
\end{equation}
where the $L_{ij}$ coefficients are defined as
\begin{eqnarray}
L_{00}&=&\frac{\Phi_{0}}{4\pi I_{c}}\sec\left({\pi\Phi_{\mathrm{ext}}}/{\Phi_{0}}\right)\label{eq:L00}\\
L_{20}&=&\frac{\Phi_{0}}{96\pi I_{c}}\sec^{3}\left({\pi\Phi_{\mathrm{ext}}}/{\Phi_{0}}\right)\label{eq:L20}\\
L_{01}&=&\frac{\lambda B_{\mathrm{ext}}l_{\mathrm{osc}}}{4I_{c}}\sec\left({\pi\Phi_{\mathrm{ext}}}/{\Phi_{0}}\right)\tan\left({\pi\Phi_{\mathrm{ext}}}/{\Phi_{0}}\right).\label{eq:L01}
\end{eqnarray}
Note that we have neglected the $I^2 y$ term in (\ref{eq:squid-inductance}), restricting ourselves to the leading order coupling only between the transmission line and mechanical oscillator, as already stated. The above equations differ from those of the prequel~\cite{Blencowe:2007p285} through the inclusion of the nonlinear, leading order current-dependent contribution [$L_{20}(I/I_c)^2$] to the effective inductance $L(I,y)$.

The nonlinear voltage boundary condition~(\ref{eq:voltage}) with inductance given by Eq.~(\ref{eq:squid-inductance}) generates frequency tripling harmonics of the transmission line resonator mode. Omitting for the time being the mechanical oscillator degree of freedom, a trial perturbative mode solution to the wave equation~(\ref{eq:squid-wave}) that includes the leading harmonic and solves the current boundary conditions~(\ref{eq:currentbc}) is the following:
\begin{equation}\label{eq:phase-field}
\phi(x,t)=\left\{\begin{array}{ccc}
+A\cos(\omega t +\varphi)\cos\left[k(x-l/2)\right]+a A^3 \cos(3\omega t +3\varphi)\cos\left[3k(x-l/2)\right]; x>0\\
-A\cos(\omega t +\varphi)\cos\left[k(x+l/2)\right]-a A^3 \cos(3\omega t +3\varphi)\cos\left[3k(x+l/2)\right]; x<0,
\end{array}\right.
\end{equation}
where $k=k^{(0)} + k^{(1)}$ and $\omega=|k|/\sqrt{L_T C_T}$. The coefficients $a$, $k^{(0)}$ and $k^{(1)}$ are determined by substituting Eq.~(\ref{eq:phase-field}) into the voltage boundary condition~(\ref{eq:voltage}) and solving perturbatively to order $A^3$, with $k^{(1)}$ scaling as $A^2$. We obtain: $a=-1/48$,
\begin{equation}
({k^{(0)}l}/{2})\tan\left({k^{(0)}l}/{2}\right)=\zeta^{-1}
\label{eq:k0}
\end{equation} 
and
\begin{equation}
k^{(1)} l=-\frac{1}{8}\zeta^3 A^2 (k^{(0)}l)^3 \sin^2 \left({k^{(0)}l}/{2}\right),
\label{eq:k1}
\end{equation}
where
\begin{equation}
\zeta=\frac{L_{00}}{L_Tl}=\frac{\Phi_0}{4\pi L_T l I_c} \sec\left({\pi\Phi_{\mathrm{ext}}}/{\Phi_0}\right).
\label{eq:zeta}
\end{equation}
Considering the transmission line phase field at the location $x=-l/2$,  where the field is pumped and probed (see Fig.~\ref{fig:diagram}), the perturbative solution (\ref{eq:phase-field}) can be obtained from the following single mode equation for $\phi(-l/2,t)\equiv\phi(t)$:
\begin{eqnarray}
\frac{d^2\phi}{dt^2}+\omega_T^2 \phi &+&\frac{1}{12}\omega_T^2 \left(1-18\zeta^3\left[\left({k^{(0)}l}/{2}\right)\sin\left({k^{(0)}l}/{2}\right)\right]^2\right)\phi^3\cr
&&+\frac{1}{12}\left(1-2\zeta^3\left[\left({k^{(0)}l}/{2}\right)\sin\left({k^{(0)}l}/{2}\right)\right]^2\right)\frac{d^2(\phi^3)}{dt^2}=0,
\label{eq:nonlinearmode}
\end{eqnarray}
 where $\omega_T=|k^{(0)}|/\sqrt{L_TC_T}$.
The awkward nonlinear term $\ddot{\phi^3}$ can be eliminated by redefining the phase mode coordinate as $\phi=\psi (1+\Gamma\psi^2)$, provided $|\Gamma|\phi^2\ll 1$, where 
\begin{equation}
\Gamma=-\frac{1}{12}\left(1-2\zeta^3\left[\left({k^{(0)}l}/{2}\right)\sin\left({k^{(0)}l}/{2}\right)\right]^2\right).
\label{eq:Gamma}
\end{equation}
The mode equation~(\ref{eq:nonlinearmode}) in terms of the redefined phase coordinate $\psi$ then becomes
\begin{equation}
\ddot{\psi}+\omega^2_T \psi -\frac{4}{3}\omega_T^2\zeta^3 \left[\left({k^{(0)}l}/{2}\right)\sin\left({k^{(0)}l}/{2}\right)\right]^2 \psi^3 =0.
\label{eq:psimode}
\end{equation}
Thus, embedding a SQUID in a microwave transmission line induces a cubic nonlinearity in the effective single mode equations (under the conditions of small currents as compared with the Josephson junction critical current), resulting in the  familiar (undamped) Duffing oscillator. 

We now restore the mechanical degree of freedom $y(t)$ by assuming that for small and slow displacements [conditions (a) and (d) above], the interaction with $\psi$ can be obtained by expanding $\omega_T$ [through its dependence on $\Phi_{\mathrm{ext}}(y)$] to first order in $y$ in Eq.~(\ref{eq:psimode}) to obtain
\begin{eqnarray}
\ddot{\psi}+\omega^2_T \psi &-&\frac{4}{3}\omega_T^2\zeta^3 \left[\left({k^{(0)}l}/{2}\right)\sin\left({k^{(0)}l}/{2}\right)\right]^2 \psi^3\cr
&&=\frac{\lambda B_{\mathrm{ext}} l_{\mathrm{osc}} y}{\left(\Phi_0/\pi\right)}\frac{\Phi_{0}}{4\pi L_Tl I_c}\tan\left({\pi\Phi_{\mathrm{ext}}}/{\Phi_0}\right)\sec\left({\pi\Phi_{\mathrm{ext}}}/{\Phi_0}\right)\omega^2_T\psi.
\label{eq:mass-psimode}
\end{eqnarray}  
Equation~(\ref{eq:force-approx}) for the mechanical oscillator, together with Eq.~(\ref{eq:mass-psimode}) for the phase coordinate, follow from the Lagrangian:
\begin{eqnarray}
 &&\mathcal{L}(\psi,y,\dot{\psi},\dot{y})=\frac{1}{2}m\dot{y}^{2}-\frac{1}{2}m\omega_{m}^{2}y^{2}+\frac{1}{2}C_{T}l\left(\frac{\Phi_{0}}{2\pi}\right)^{2}\sin^{2}(k_{0}l/2)\cr
 &&\times\left\{\frac{1}{2}\dot{\psi}^{2}(t)-\frac{1}{2}\left[1-\frac{\lambda B_{\mathrm{ext}}l_{\mathrm{osc}}y}{(\Phi_{0}/\pi)}\frac{\Phi_{0}}{4\pi L_{T}lI_{c}}\tan\left({\pi\Phi_{\mathrm{ext}}}/{\Phi_{0}}\right)\sec\left({\pi\Phi_{\mathrm{ext}}}/{\Phi_{0}}\right)\right]\omega_{T}^{2}\psi^{2}(t)\right.\cr
&&\left.+\frac{1}{3}\omega^2_T\zeta^{3}\left[\left({k^{(0)}l}/{2}\right)\sin\left({k^{(0)}l}/{2}\right)\right]^2\psi^{4}\right\}.
\end{eqnarray}
We now introduce the phase momentum coordinate $p_{\psi}=\partial{\mathcal{L}}/\partial{\dot{\psi}}=m_{\psi}\dot{\psi}$ and raising (lowering) operators
\begin{eqnarray}
 \hat{a}_{T}^{\pm}&=&\frac{1}{\sqrt{2m_{\psi}\hbar\omega_{T}}}\left(m_{\psi}\omega_{T}\hat{\psi}\mp i\hat{p}_{\psi}\right)\label{eq:aT}\\
  \hat{a}_{m}^{\pm}&=&\frac{1}{\sqrt{2m\hbar\omega_{m}}}\left(m\omega_{m}\hat{y}\mp i\hat{p}_{y}\right)\label{eq:am}
\end{eqnarray}
satisfying the usual commutation relations, where the effective phase mass is $m_{\psi}=\frac{1}{2}C_{T}l\left(\Phi_{0}/2\pi\right)^{2}\sin^{2}(k_{0}l/2)$. In terms of the raising (lowering) operators, the Hamiltonian operator is
 \begin{equation}\label{eq:hamiltonian-closed}
 H=\hbar\omega_{T}{a}_{T}^{+}{a}_{T}+\frac{1}{12}\hbar\omega_{T}K_{d}({a}_{T}^{+}+{a}_{T})^{4}+\hbar\omega_{m}{a}_{m}^{+}{a}_{m}
+\frac{1}{2}\hbar\omega_{T}K_{Tm}({a}_{T}^{+}+{a}_{T})^2({a}_{m}^{+}+{a}_{m}),
 \end{equation}
 where, for notational convenience, hats on the operators and the minus superscript on the lowering operators will be suppressed from now on. The parameter characterizing the strength of the interaction between the transmission line mode and mechanical oscillator mode is
 \begin{equation}
 K_{Tm}=\frac{\lambda B_{\mathrm{ext}}l_{\mathrm{osc}}\Delta_{zp}}{\left(\Phi_{0}/\pi\right)}\frac{\Phi_{0}}{4\pi L_{T}lI_{c}}\tan\left(\pi\Phi_{\mathrm{ext}}/\Phi_{0}\right)\sec\left(\pi\Phi_{\mathrm{ext}}/\Phi_{0}\right),
\label{eq:ktm}
\end{equation}
where $\Delta_{zp}=\sqrt{\hbar/(2m\omega_m)}$ is the zero-point displacement uncertainty.
 The parameter $K_d$ characterizing the strength of the Duffing nonlinear term takes the form
 \begin{equation}
  K_d = -\left(k^{(0)} l\right)^2 \left(\frac{L_{00}}{L_Tl}\right)^3 \left[\frac{(2e)^2/(2C_Tl)}{\hbar\omega_T}\right],
  \label{eq:Kd}
  \end{equation}
which has been written in such a way as to make clear its various dependencies. In particular, $K_d$ depends essentially on the cube of the ratio of the linear SQUID effective inductance $L_{00}$ to transmission line inductance $L_Tl$, as well as on the ratio of the single Cooper pair charging energy to the microwave mode photon energy of the transmission line. Since the strength and sign of the  linear SQUID inductance depends on the external flux  $\Phi_{\mathrm{ext}}$  [see Eq.~(\ref{eq:L00})], it is possible to vary the strength as well as the sign of the Duffing constant by tuning the external flux either side of  $\Phi_0/2$.  Thus, we can have either spring hardening or spring softening of the transmission line oscillator mode.  Previously this flux tunability was observed in the readout of a persistent current qubit.\cite{Lee:2007p3227} Note, however, that the perturbative approximations that go into deriving the above Hamiltonian~(\ref{eq:hamiltonian-closed}) do not allow too close an approach to the singular half-integer flux quantum point. In particular, the validity of the expansions in $I_{c}$ and $\beta_L$ properly require the following conditions to hold:
\begin{eqnarray}
\left|\frac{I}{I_{c}}\sec\left({\pi\Phi_{\mathrm{ext}}}/{\Phi_{0}}\right)\right|&\ll&1\label{eq:Iccondition}\\
\left|\beta_{L}\sec\left({\pi\Phi_{\mathrm{ext}}}/{\Phi_{0}}\right)\right|&\ll&1\label{eq:betalcondition}.
\end{eqnarray}

As already noted, Eq.~(\ref{eq:hamiltonian-closed}) without the Duffing nonlinearity coincides with the Hamiltonian commonly used to describe the single mode of an optical cavity interacting with a mechanical mirror via the radiation pressure.  However, we have just seen that embedding a SQUID within a microwave transmission line cavity induces a tunable Duffing self-interaction term as well; it is not so easy to achieve a similar, tunable nonlinearity in the optical cavity counterpart. 

\subsection{Open System Dynamics}\label{subsec:squid-open}
Up until now we have considered the transmission line, SQUID and mechanical oscillator  as an isolated system. It is straightforward to couple the transmission line to an external pump-probe feedline and mechanical oscillator to a thermal bath using the `input-output' formalism of Gardiner and Collett.\cite{Gardiner:1985p1483} Assuming weak system-bath couplings justify making the rotating wave approximation (RWA), and furthermore making a Markov approximation for the bath dynamics, the following Langevin equations can be derived for the system mode operators in the Heisenberg picture:
\begin{eqnarray}\label{eq:motionm}	
\frac{d{a}_{m}}{dt}&=&-i\omega_{m}{a}_{m}+\frac{i}{\hbar}\sqrt{\frac{\hbar}{2m\omega_{m}}}F_{\mathrm{ext}}(t)-i\omega_{T}K_{Tm}{a}^{+}_{T}{a}_{T}\cr
&&-\gamma_{bm}{a}_{m}-i\sqrt{2\gamma_{bm}}e^{i\phi_{bm}}{a}_{b}^{\mathrm{in}}(t)
\end{eqnarray}
and
\begin{eqnarray}\label{eq:motiont}
\frac{d{a}_{T}}{dt}&=&-i\omega_{T}{a}_{T}-i\omega_{T}K_{d}{a}_{T}^{+}{a}_{T}{a}_{T}-i\omega_{T}K_{Tm}{a}_{T}({a}^{+}_{m}+{a}_{m})\cr
&&-\gamma_{pT}{a}_{T}-i\sqrt{2\gamma_{pT}}e^{i\phi_{pT}}{a}_{p}^{\mathrm{in}}(t),
\end{eqnarray}
where $\gamma_{bm}$ is the mechanical oscillator amplitude damping rate due to coupling to the bath, $\gamma_{pT}$ is the transmission line mode damping rate due to coupling to the pump-probe line, and we have also assumed that the small Duffing coupling $K_d$ and transmission line-mechanical oscillator coupling $K_{Tm}$ justify applying the RWA to the transmission line mode operator terms.
The `in'  bath  and probe line operators are defined as
\begin{equation}\label{eq:inoperators}
	{a}_{i}^{\mathrm{in}}(t)=\frac{1}{\sqrt{2\pi}}\int d\omega e^{-i\omega(t-t_{0})}{a}_{i}(\omega,t_{0}),
\end{equation}
where $t>t_{0}$, with the states of the pump-probe line and oscillator bath assigned at $t_0$, interpreted as the initial time in the past before the measurement commences. For completeness, we have also included a classical, external time-dependent force $F_{\mathrm{ext}}$(t) acting on the mechanical oscillator, although we shall not address the force detection sensitivity in the present work.

It will be convenient to work with the Fourier transformed Langevin equations. With $O(\omega)=\frac{1}{\sqrt{2\pi}}\int_{-\infty}^{\infty}d\omega e^{i\omega t} O(t)$, Eqs.~(\ref{eq:motionm}) and (\ref{eq:motiont}) become 
\begin{eqnarray}\label{eq:amw}
	{a}_{m}(\omega)&=&\frac{1}{\omega-\omega_{m}+i\gamma_{bm}}\left\{\sqrt{2\gamma_{bm}}e^{i\phi_{bm}}{a}_{b}^{\mathrm{in}}(\omega)-\frac{1}{\sqrt{2m\hbar\omega_{m}}}F_{\mathrm{ext}}(\omega)\right.\cr
	&&\left.+\frac{\omega_{T}K_{Tm}}{2\sqrt{2\pi}}\int_{-\infty}^{\infty}d\omega'
\left[{a}_{T}(\omega'){a}_{T}^{+}(\omega'-\omega)+{a}_{T}^{+}(\omega'){a}_{T}(\omega'+\omega)\right]\right\}
 \end{eqnarray}
 and
 \begin{eqnarray}\label{eq:atw}
{a}_{T}(\omega)&=&\frac{1}{\omega-\omega_{T}+i\gamma_{pT}}\left\{\frac{}{}\sqrt{2\gamma_{pT}}e^{i\phi_{pT}}{a}_{p}^{\mathrm{in}}(\omega)+\frac{\omega_{T}K_{d}}{2\pi}\right.\cr
&&\left.\times \int_{-\infty}^{\infty}d\omega'\int_{-\infty}^{\infty}d\omega''{a}_{T}^{+}(\omega''){a}_{T}(\omega'){a}_{T}(\omega+\omega''-\omega')\right.\cr
&&\left.+\frac{\omega_{T}K_{Tm}}{\sqrt{2\pi}}\int_{-\infty}^{\infty}d\omega'{a}_{T}(\omega') \left[{a}_{m}(\omega-\omega')+{a}_{m}^{+}(\omega'-\omega)\right]\right\}.
\end{eqnarray} 

\section{Detector Response}\label{sec:squid-response}
The probe line observables are expressed in terms of the `out' mode operator:
\begin{equation}\label{eq:outoperator}
	{a}_{p}^{\mathrm{out}}(t)=\frac{1}{\sqrt{2\pi}}\int d\omega e^{-i\omega(t-t_1)}{a}_{p}(\omega,t_1),
\end{equation}
where $t_1>t$. The  `out' and `in' probe operators are related via the following useful identity:\cite{Gardiner:1985p1483}
\begin{equation}\label{eq:inoutidentity}
 {a}_{p}^{\mathrm{out}}(t)=-i\sqrt{2\gamma_{pT}}e^{-i\phi_{pT}}{a}_{T}(t)+{a}_{p}^{\mathrm{in}}(t), 
\end{equation}  
which allows us to obtain the expectation value of  a given observable once $a_T(t)$ is determined. 
As an illustrative expectation value, we shall consider the variance in the probe line reflected current in a given bandwidth $\delta\omega$ centered about the signal frequency of interest $\omega_s$:\cite{Blencowe:2007p285}
\begin{eqnarray}
\overline{\langle\left[{\delta I}^{\mathrm{out}}(\omega_{s},\delta\omega)\right]^{2}\rangle}&=&\frac{1}{Z_{p}}\int_{\omega_{s}-\delta\omega/2}^{\omega_{s}+\delta\omega/2}\frac{d\omega_{1}d\omega_{2}}{2\pi}\hbar\omega_{1}\left(\frac{2\sin\left[(\omega_{1}-\omega_{2})T_{M}/2\right]}{(\omega_{1}-\omega_{2})T_{M}}\right)\cr
&&\times\frac{1}{2}\langle{a}_{p}^{\mathrm{out}}(\omega_{1}){a}_{p}^{{\mathrm{out}}+}(\omega_{2})+{a}_{p}^{{\mathrm{out}}+}(\omega_{2}){a}_{p}^{\mathrm{out}}(\omega_{1})\rangle,
\label{eq:currentvariance}
\end{eqnarray}
where, in addition to the ensemble average, there is also a time average denoted by the overbar, with the averaging time  taken to be the duration of the measurement  $T_M$, assumed much longer than all other timescales associated with the detector dynamics. In particular, time averaging is required when $F_{\mathrm{ext}}(t)$ has a deterministic time dependence.\cite{Blencowe:2007p285} Expectation values of other observables, such as the reflected voltage variance and reflected power are simply obtained from Eq.~(\ref{eq:currentvariance}) with appropriate inclusions of the probe line impedance $Z_p=\sqrt{L_p/C_p}$: $P^{\mathrm{out}}=\overline{\langle [\delta V^{\mathrm{out}}]^2\rangle}/Z_p=\overline{\langle [\delta I^{\mathrm{out}}]^2\rangle}  Z_p$.  

From the form of the $K_{Tm}$ coupling term in Eq.~(\ref{eq:motiont}), we can see that the motion of the mechanical resonator modulates the transmission line frequency, and thus a complimentary way to transduce displacements besides measuring the current amplitude, is to measure the frequency-dependent, relative phase shift between the `in' pump drive current  and `out' probe current using the homodyne detection procedure.\cite{gardiner} While we shall focus on amplitude detection, the homodyne method can be straighforwardly addressed and is expected to give similar results for the quantum limited detection sensitivity.

Substituting Eq.~(\ref{eq:amw}) into (\ref{eq:atw}), we obtained the following single equation for the transmission line mode operator $a_T$:
\begin{eqnarray}\label{eq:atwonly}
	{a}_{T}(\omega)&=&\int_{-\infty}^{\infty}d\omega'{a}_{T}(\omega-\omega')A(\omega,\omega')+\int_{-\infty}^{\infty}d\omega'B(\omega,\omega'){a}_{T}(\omega-\omega')\cr
	&&\times\int_{-\infty}^{\infty}d\omega''\left[{a}_{T}(\omega''){a}_{T}^{+}(\omega''-\omega')+{a}_{T}^{+}(\omega''){a}_{T}(\omega''+\omega')\right]\cr
&&+D(\omega)\int_{-\infty}^{\infty}d\omega''\int_{-\infty}^{\infty}d\omega'{a}^{+}_{T}(\omega''){a}_{T}(\omega'){a}_{T}(\omega+\omega''-\omega')+C(\omega),
\end{eqnarray}
where
\begin{equation}\label{eq:A}	
	A(\omega,\omega')=\frac{\omega_T K_{Tm}}{\sqrt{2\pi}}\frac{1}{\omega-\omega_{T}+i\gamma_{pT}}
	\left[\frac{S_{m}(\omega')}{\omega'-\omega_{m}+i\gamma_{bm}}+\frac{S_{m}^{+}(-\omega')}{-\omega'-\omega_{m}-i\gamma_{bm}}\right],
\end{equation}
\begin{equation}\label{eq:B}	
	B(\omega,\omega')=\frac{(\omega_T K_{Tm})^{2}}{4\pi}\frac{1}{\omega-\omega_{T}+i\gamma_{pT}} \left[\frac{1}{\omega'-\omega_{m}+i\gamma_{bm}}+\frac{1}{-\omega'-\omega_{m}-i\gamma_{bm}}\right],
\end{equation}
\begin{equation}\label{eq:C}
	C(\omega)=\frac{S_{T}(\omega)}{\omega-\omega_{T}+i\gamma_{pT}}
\end{equation}
and
\begin{equation}\label{eq:D}
	D(\omega)=\frac{\omega_T K_d}{2\pi}\frac{1}{\omega-\omega_{T}+i\gamma_{pT}},
\end{equation}
with mechanical signal operator
\begin{equation}
	S_{m}(\omega)=\sqrt{2\gamma_{bm}}e^{i\phi_{bm}}{a}_{b}^{\mathrm{in}}(\omega)-\frac{1}{\sqrt{2m\hbar\omega_{m}}}F_{\mathrm{ext}}(\omega)\label{eq:sm}
\end{equation}
and noise operator
\begin{equation}
	S_{T}(\omega)=\sqrt{2\gamma_{pT}}e^{i\phi_{pT}}{a}_{p}^{\mathrm{in}}(\omega).\label{eq:st}
\end{equation}
For small signal strength, it is assumed that Eq.~(\ref{eq:atwonly}) can be solved as a series expansion up to first order in $A(\omega,\omega')$, giving the usual linear-response approximation. I.e., $a_T(\omega)\approx a_T^{(0)}(\omega)+a_T^{(1)}(\omega)$, where the noise component  $a_T^{(0)}(\omega)$ is the solution to Eq.~(\ref{eq:atwonly}) with the mechanical signal source term $A(\omega,\omega')$ set to zero, while the signal component $a_T^{(1)}(\omega)$ is the part of the solution to Eq.~(\ref{eq:atwonly}) that depends linearly on $A(\omega,\omega')$. Thus, from Eq.~(\ref{eq:inoutidentity}) we can express the `out' probe mode operator as follows:
\begin{equation}
a_p^{\mathrm{out}}(\omega)=\left[-i\sqrt{2\gamma_{pT}}e^{-i\phi_{pT}} a^{(1)}_T (\omega)\right]+\left[-i\sqrt{2\gamma_{pT}}e^{-i\phi_{pT}} a^{(0)}_T (\omega)+a_p^{\mathrm{in}}(\omega)\right],
\label{eq:asignalnoise}
\end{equation}
where the first square bracketed term gives the signal contribution to the detector response and the second square bracketed term gives the noise contribution. 

As `in' states, we consider the mechanical oscillator bath to be in a thermal state at temperature $T$ and the pump line to be in a coherent state centered about the pump frequency $\omega_p$:\cite{Johansson:2006p2429}
\begin{equation}\label{eq:coherent}
\left|\{\alpha(\omega)\}\right.\rangle_{p}=\exp\left(\int d\omega\alpha(\omega)\left[{a}_{p}^{\mathrm{in}+}(\omega)-{a}_{p}^{\mathrm{in}}(\omega)\right]\right)\left|0\right.\rangle_{p},
\end{equation}
where $\left|0\right.\rangle_{p}$ is the vacuum state and 
\begin{equation}\label{eq:alpha}
\alpha(\omega)=-I_{0}\sqrt{\frac{Z_{p}T^{2}_{M}}{2\hbar}}\frac{e^{-(\omega-\omega_{p})^{2}T^{2}_{M}/2}}{\sqrt{\omega}}.
\end{equation}
The coherent state coordinate $\alpha(\omega)$ is parametrized such that the expectation value of the right-propagating `in' current $I^{\mathrm{in}}(x,t)$ with respect to this coherent state has amplitude $I_0$, where
\begin{equation}
I^{\mathrm{in}}(x,t)=i\sqrt{\frac{\hbar}{4\pi Z_p}}\int_0^{\infty} d\omega \sqrt{\omega} \left[e^{i\omega (x/v_p -t)} a_p^{\mathrm{in}}(\omega)-e^{-i\omega (x/v_p -t)} a_p^{\mathrm{in}+}(\omega)\right],
\label{eq:incurrentoperator}
\end{equation}
with $v_p=1/\sqrt{L_pC_p}$ the wave propagation velocity in the pump probe line.

With the pump probe line in a coherent state, we assume that for large drive currents  Eq.~(\ref{eq:atwonly}) can be approximately solved  using a semiclassical, `mean field' approximation, where the quantum fluctuation $\delta a_T^{(0)}(\omega)$ in $a_T^{(0)}(\omega)=\langle a_T^{(0)}(\omega)\rangle +\delta a_T^{(0)}(\omega)$ is kept to first order only. However,  the nonlinear Duffing  and transmission line-mechanical oscillator interaction terms can give rise to a bistability in the transmission line oscillator dynamics and one must be careful when interpreting the results from the mean field approximation when operating close to a bifurcation point; large fluctuations can occur in the oscillator amplitude as it jumps between the two metastable amplitudes, which are not accounted for in the mean field approximation. (See Refs. \cite{Dykman:1980p480} and~\cite{Dykman:2007p1864} for respective analyses of the classical and quantum oscillator fluctuation dynamics near a bifurcation point).  This issue will be further discussed in the following sections. 

The solutions to the signal $a_T^{(1)}(\omega)$ and noise $a_T^{(0)}$ terms parallel closely our previous calculations, which omitted the Duffing nonlinearity;\cite{Blencowe:2007p285} the Duffing ($D$) term in Eq.~(\ref{eq:atwonly}) has a very similar form to the transmission line oscillator coupling ($B$) term, both involving $a_T^2 a_T^+$ operator combinations. We therefore relegate the solution details to the appendix, presenting only the essential results in this section.

The solution to $\langle a _T^{(0)}(\omega)\rangle$ is sharply peaked about the pump frequency $\omega_p$ for large $T_M$   and so can be approximately expressed as a delta function:  $\langle a _T^{(0)}(\omega)\rangle=\chi \delta (\omega-\omega_p)$. Substituting this expression into Eq.~(\ref{eq:zeroth-atw-coherent}), we obtain for the  amplitude $\chi$: 
\begin{equation}\label{eq:mean-field0}
	\chi=c+\left[2B(\omega_{p},0)+D(\omega_{p})\right]\chi\left|\chi\right|^{2},
\end{equation}
with
\begin{equation}
	c=\frac{i\sqrt{2\pi}e^{i\phi_{pT}}}{\gamma_{pT}-i\Delta\omega}\sqrt{\frac{I_{0}^{2}Z_{p}\gamma_{pT}}{\hbar\omega_{p}}}, \label{eq:c}
\end{equation}
where $\Delta\omega=\omega_{p}-\omega_{T}$ is the detuning of the pump frequency $\omega_p$ from the transmission line resonance frequency $\omega_T$. Using the expressions for $B(\omega_p,0)$ and $D(\omega_p)$, Eq.~(\ref{eq:mean-field0}) can be written as
\begin{equation}
(\omega_T-\omega_p -i \gamma_{pT})\chi+\frac{\omega_T}{2\pi} {\mathcal{K}}\chi |\chi|^2 = e^{i\phi_{pT}}\sqrt{{{2\pi}I_0^2 Z_p \gamma_{pT}}/{(\hbar\omega_p})},\label{eq:meanfield} 
\end{equation}
where the effective Duffing coupling is defined as
\begin{equation}
{\mathcal{K}}=K_d -\frac{2 \omega_T \omega_m}{\omega_m^2 +\gamma_{bm}^2} K^2_{Tm}.
\label{eq:k}
\end{equation}
Notice that the interaction between the transmission line and mechanical oscillator induces an additional Duffing nonlinearity (the second term involving $K_{Tm}$ in ${\mathcal{K}}$) in the transmission line mode amplitude effective equations of motion (\ref{eq:meanfield}).  However, in contrast with the $K_d$ nonlinearity, which can be tuned to have either sign, the former mechanically-induced nonlinearity is always negative and thus has a ``spring-softening" effect on the transmission line mode. Interestingly, by choosing an appropriate compensating ``spring hardening" $K_d>0$, the effective Duffing constant ${\mathcal{K}}$ can in principle be completely suppressed so that the next non-vanishing higher order nonlinearity would govern the mode amplitude dynamics.

Once we have the solution for $\langle a _T^{(0)}(\omega)\rangle$, the solutions for the quantum signal $a_T^{(1)}(\omega)$ and quantum noise $\delta a_T^{(0)}(\omega)$ are obtained from Eqs.~(\ref{eq:first-atw}) and (\ref{eq:zeroth-atw-quantum}), respectively. These solutions can be expressed as follows:
\begin{equation}
{a}_{T}^{(1)}(\omega)=\alpha_{1}(\omega)A(\omega,\omega-\omega_{p})+\alpha_{2}(\omega)A(\omega-2\Delta\omega,\omega-\omega_{p})\label{eq:a1solution}
\end{equation}
and
\begin{equation}\label{eq:a0solution}
\delta{a}_{T}^{(0)}(\omega)=\beta_{1}(\omega)\delta C(\omega)+\beta_{2}(\omega)\delta C^{+}(2\omega_{p}-\omega),
\end{equation}
where the $\alpha_{i}(\omega)$ and $\beta_{i}(\omega)$ functions are defined in Eqs.~(\ref{eq:alpha1}), (\ref{eq:alpha2}), (\ref{eq:beta1}), and (\ref{eq:beta2}).

Substituting Eqs.~(\ref{eq:a1solution}) and (\ref{eq:a0solution}) into the expression (\ref{eq:asignalnoise}) for $a^{\mathrm{out}}(\omega)$ and then evaluating the signal component of the detector response~(\ref{eq:currentvariance}), we obtain\cite{Blencowe:2007p285}
\begin{eqnarray}\label{eq:current-signal}
&&\left.\overline{\langle\left[{\delta I}^{\mathrm{out}}(\omega_{s},\delta\omega)\right]^{2}\rangle}\right|_{\mathrm{signal}}=\left(\frac{I_{0}K_{Tm}\omega_{T}}{\gamma_{pT}}\right)^{2}\frac{\gamma_{pT}^{2}}{\gamma_{pT}^{2}+\Delta\omega^{2}}\cr
&&\times\int_{\omega_{s}-\delta\omega/2}^{\omega_{s}+\delta\omega/2}\frac{d\omega}{2\pi}\left(\frac{\omega}{\omega_{p}}\frac{\gamma_{pT}^{2}}{(\omega-\omega_{p}+\Delta\omega)^{2}+\gamma_{pT}^{2}}\right)\left|\frac{\alpha_{1}(\omega)}{c}+\frac{\alpha_{2}(\omega)}{c}\left(\frac{\omega-\omega_{p}+\Delta\omega+i\gamma_{pT}}{\omega-\omega_{p}-\Delta\omega+i\gamma_{pT}}\right)\right|^{2}\cr
&&\times\left(\frac{2\gamma_{bm}}{(\omega-\omega_{p}-\omega_{m})^{2}+\gamma_{bm}^{2}}[2n(\omega-\omega_{p})+1]+\frac{2\gamma_{bm}}{(\omega_{p}-\omega-\omega_{m})^{2}+\gamma_{bm}^{2}}[2n(\omega_{p}-\omega)+1]\right)\cr
&&+\left(\frac{I_{0}K_{Tm}\omega_{T}}{\gamma_{pT}}\right)^{2}\frac{\gamma_{pT}^{2}}{\gamma_{pT}^{2}+\Delta\omega^{2}}\frac{1}{2m\hbar\omega_{m}\gamma_{bm}}\int_{\omega_{s}-\delta\omega/2}^{\omega_{s}+\delta\omega/2}\frac{d\omega d\omega'}{2\pi}\left(\frac{\omega}{\omega_{p}}\frac{\gamma_{pT}^{2}}{(\omega-\omega_{p}+\Delta\omega)^{2}+\gamma_{pT}^{2}}\right)\cr
&&\times\left|\frac{\alpha_{1}(\omega)}{c}+\frac{\alpha_{2}(\omega)}{c}\left(\frac{\omega-\omega_{p}+\Delta\omega+i\gamma_{pT}}{\omega-\omega_{p}-\Delta\omega+i\gamma_{pT}}\right)\right|^{2}\frac{\sin\left[(\omega-\omega')T_{m}/2\right]}{(\omega-\omega')T_{m}/2}\cr
&&\times\left(\frac{2\gamma_{bm}}{(\omega-\omega_{p}+\Delta\omega)^{2}+\gamma_{bm}^{2}}F_{\mathrm{ext}}(\omega-\omega_{p})F^{*}_{\mathrm{ext}}(\omega'-\omega_{p})\right.\cr
&&\left.+\frac{2\gamma_{bm}}{(\omega_{p}-\omega-\omega_{m})^{2}+\gamma_{bm}^{2}}F_{\mathrm{ext}}(\omega_{p}-\omega)F^{*}_{\mathrm{ext}}(\omega_{p}-\omega')\right),	
\end{eqnarray}	
where $n(\omega)=(e^{\hbar\omega/k_BT} -1)^{-1}$ is the thermal average occupation number for bath mode $\omega$. In the limit of small drive current amplitude $I_0\rightarrow 0$, we have $\alpha_1 (\omega)/c \rightarrow 1$, $\alpha_2 (\omega)/c \rightarrow 0$, and we see that the signal spectrum comprises two Lorentzian peaks centered at $\omega_p\pm\omega_m$. The $\omega_p+\omega_m$ peak corresponds to  phase preserving detection, in the sense that $a_p^{\mathrm{out}}$ gives the amplified $a_b^{\mathrm{in}}$ signal, while the  $\omega_p-\omega_m$ peak corresponds to phase conjugating detection, with $a_p^{\mathrm{out}}$ amplifying the $a^{\mathrm{in}+}_b$ signal.\cite{Caves:1982p1311}    Increasing the drive current amplitude causes the peaks to shift, and the peak widths relative to their height to change, signifying renormalization of the mechanical oscillator frequency and damping rate. The noise component of the detector response is
\begin{eqnarray}\label{eq:current-noise}
&&\left.\overline{\langle\left[{\delta I}^{\mathrm{out}}(\omega_{s},\delta\omega)\right]^{2}\rangle}\right|_{\mathrm{noise}}=\frac{1}{Z_{p}}\int_{\omega_{s}-\delta\omega/2}^{\omega_{s}+\delta\omega/2}\frac{d\omega}{2\pi}\hbar\omega\frac{2\gamma_{pT}^{2}}{(\omega-\omega_{p}+\Delta\omega)^{2}+\gamma_{pT}^{2}}\cr
&&\times\left(\left|\beta_{1}(\omega)\right|^{2}+\frac{(\omega-\omega_{p}+\Delta\omega)^{2}+\gamma_{pT}^{2}}{(\omega-\omega_{p}-\Delta\omega)^{2}+\gamma_{pT}^{2}}\left|\beta_{2}(\omega)\right|^{2}
-{\mathrm{Re}}\left[\beta_{1}(\omega)\right]+\frac{(\omega-\omega_{p}+\Delta\omega)}{\gamma_{pT}}{\mathrm{Im}}\left[\beta_{1}(\omega)\right]\right)\cr
&&+Z_{p}^{-1}\frac{\hbar\omega_{s}}{2}\frac{\delta\omega}{2\pi},
\end{eqnarray}
where the integral term involving the $\beta_i (\omega)$ functions includes the back reaction noise  on the mechanical oscillator and the term involving $Z_p$ describes the probe line zero-point fluctuations added at the output.

In Sec.~\ref{sec:squid-detection} we will numerically evaluate Eqs.~(\ref{eq:current-signal}) and (\ref{eq:current-noise}) and in particular compare the detector noise with the minimum noise bound discussed by Caves\cite{{Blencowe:2007p285},{Caves:1982p1311}}  that follows from the Heisenberg uncertainty principle for the detector:
\begin{eqnarray}\label{eq:current-min}
&&\left.\overline{\langle\left[\delta I^{\mathrm{out}}(\omega_{s},\delta\omega)\right]^{2}\rangle}\right|_{\mathrm{min-noise}}=\left|Z_p^{-1}\frac{\hbar\omega_{s}}{2}\frac{\delta\omega}{2\pi}-\left(\frac{I_{0}K_{Tm}\omega_{T}}{\gamma_{pT}}\right)^{2}\frac{\gamma^{2}_{pT}}{\gamma^{2}_{pT}+\Delta\omega^{2}}\right.\cr
&&\left.\times\int_{\omega_{s}-\delta\omega/2}^{\omega_{s}+\delta\omega/2}\frac{d\omega}{2\pi}\left(\frac{\omega}{\omega_{p}}\frac{\gamma_{pT}^{2}}{(\omega-\omega_{p}+\Delta\omega)^{2}+\gamma_{pT}^{2}}\right)\left|\frac{\alpha_{1}(\omega)}{c}+\frac{\alpha_{2}(\omega)}{c}\left(\frac{\omega-\omega_{p}+\Delta\omega+i\gamma_{pT}}{\omega-\omega_{p}-\Delta\omega+i\gamma_{pT}}\right)\right|^{2}\right.\cr
&&\left.\times
\left(\frac{2\gamma_{bm}}{(\omega-\omega_{p}-\omega_{m})^{2}+\gamma^{2}_{bm}}-\frac{2\gamma_{bm}}{(\omega_{p}-\omega-\omega_{m})^{2}+\gamma^{2}_{bm}}\right)\right|.
\end{eqnarray}

\section{Bistability Conditions}\label{sec:squid-bistability}

We have seen [Hamiltonian  (\ref{eq:hamiltonian-closed})]  that the current-dependent SQUID effective inductance gives rise to a transmission line Duffing type nonlinearity with strength $K_d$. Furthermore, there is a nonlinear coupling with strength $K_{Tm}$ between the transmission line and mechanical oscillator.  These two nonlinearities correspond respectively to the cubic terms  proportional  to $K_{d}$ and $K^2_{Tm}$ in the mean transmission line coordinate amplitude  $\chi$ equation (\ref{eq:mean-field0}). For sufficiently large drive current amplitude $I_0$ and/or coupling strengths $K_{Tm}$, $K_d$,  the cubic term $\chi\left|\chi\right|^2$ term in Eq.~(\ref{eq:meanfield})  becomes appreciable, resulting in three real solutions over a certain pump frequency range $\omega_p$.  This parameter regime defines the bistable region of the detector phase space (the intermediate amplitude solution is unstable and cannot be realized in practice).  In the following, we determine the conditions on the parameters for the bistable region  employing the analysis of Ref.~\cite{Yurke:2006p1911}.

\begin{figure}[htbp]
\begin{center}
\includegraphics[width=3.0in]{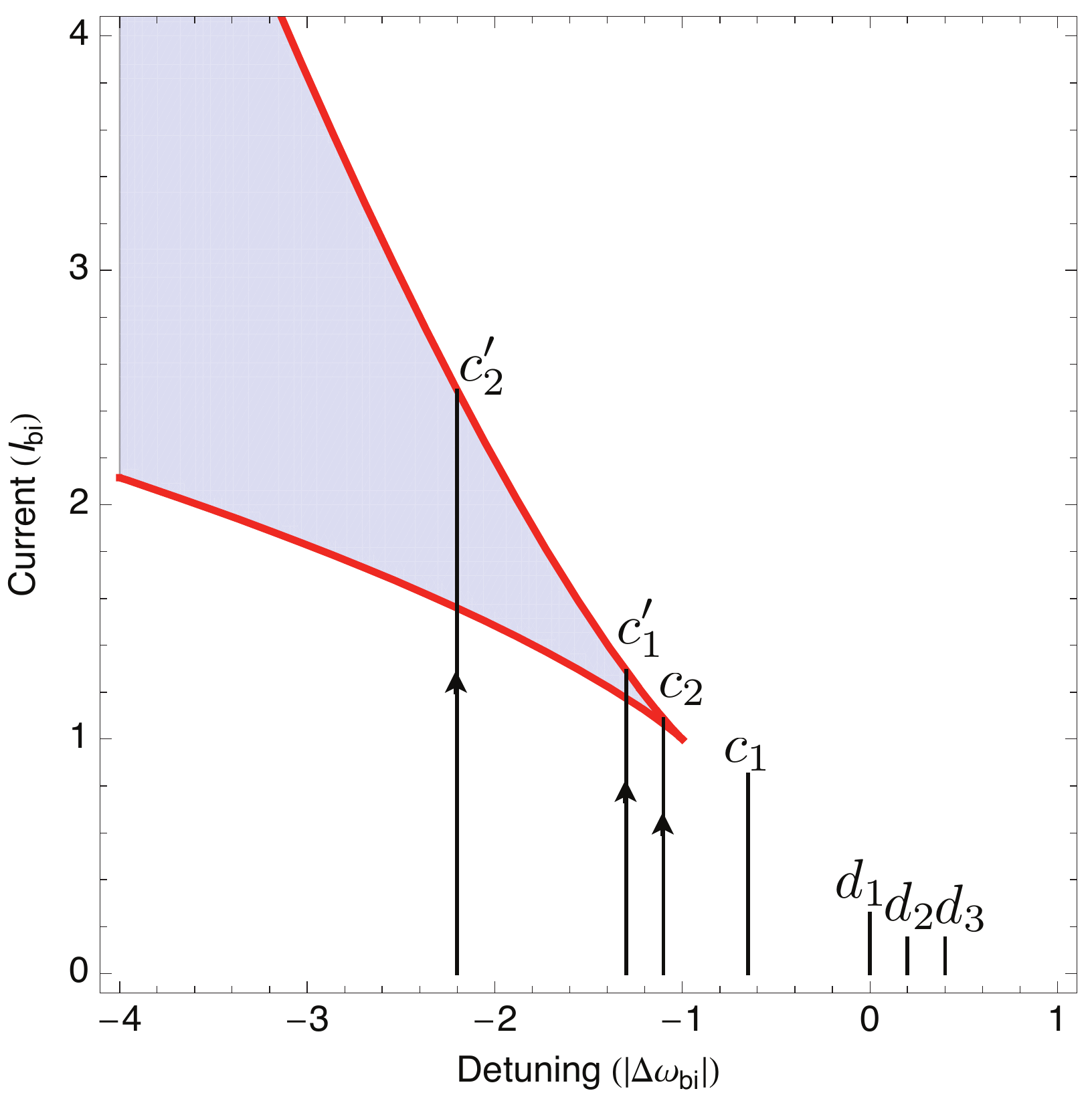}
\caption{Bistable region (shaded) of the cavity-oscillator system for negative, spring softening Duffing nonlinearity.  The drive current and detuning are expressed in units of the bistability onset critical values $I_{bi}$ and $|\Delta\omega_{bi}|$. The labelled straight line traces correspond to  detection ($d$) and cooling ($c$) current drive-detuning parameter examples considered in Secs~\ref{sec:squid-detection} and \ref{sec:squid-cooling}. The arrows give the direction in which the drive current is varied in order to enter the bistable region on the small amplitude branch.}
\label{fig:region}
\end{center}
\end{figure}

We first express the transmission line mode coordinate in terms of its phase and amplitude:
\begin{eqnarray}\label{eq:chi_B}
 \langle a_T^{(0)}(t)\rangle&=&Me^{-i(\omega_{p}t+\phi_{M})},\cr
 \chi&=&\sqrt{2\pi}M e^{-i\phi_M},
 \end{eqnarray}
where the amplitude $M$ is a positive real constant and recall $\chi$ is defined through the relation $ \langle a_T^{(0)}(\omega)\rangle=\chi \delta(\omega-\omega_p)$. Equation~(\ref{eq:meanfield}) then becomes
\begin{equation}\label{eq:kerr-eq}
\left(\omega_{T}-\omega_{p}-i\gamma_{pT}\right)M+\mathcal{K}\omega_T M^{3}=\sqrt{2\gamma_{pT}}\langle b_{pT}^{\mathrm{in}}\rangle e^{i(\phi_{pT}+\phi_{M})},
\end{equation}
where
\begin{equation}\label{eq:bin}
\langle b_{pT}^{\mathrm{in}}\rangle=\sqrt{\frac{I_{0}^{2}Z_{p}}{2\hbar\omega_{p}}}.
\end{equation}
Multiplying both sides of Eq.~(\ref{eq:kerr-eq}) by their complex conjugates and substituting $E=M^{2}$, we obtain  the following third-order polynomial in $E$:
\begin{equation}\label{eq:third_poly}
	E^{3}+\frac{2(\omega_{T}-\omega_{p})}{\omega_T\mathcal{K}}E^{2}+\frac{(\omega_{T}-\omega_{p})^{2}+\gamma_{pT}^{2}}{\omega_T^2\mathcal{K}^{2}}E=\frac{2\gamma_{pT}\langle b_{pT}^{\mathrm{in}}\rangle^{2}}{\omega_T^2\mathcal{K}^{2}}.
\end{equation}
The bifurcation line in current drive and detuning parameter space that delineates between the single solution and bistable solution regions occurs where the susceptibility $\partial E/\partial\omega_{p}$ diverges. If we further impose the condition that the transition between the two regions is continuous, i.e. $\partial^{2}\omega_{p}/\partial^{2}E=0$, we obtain the bistability onset critical point.  From Eq.~(\ref{eq:third_poly}), these two requirements can be written
\begin{eqnarray}
&3\mathcal{K}^{2}E^{2}+4(\omega_{T}-\omega_{p})\mathcal{K}E+(\omega_{T}-\omega_{p})^{2}+\gamma_{pT}^{2}=0,\label{eq:susceptibility} \\
&6\mathcal{K}^{2}E+4(\omega_{T}-\omega_{p})\mathcal{K}=0.
\end{eqnarray}
Solving these equations simultaneously for $E$ and $\Delta\omega=\omega_p-\omega_T$ yields the following bistability onset critical values:
\begin{eqnarray}
E_{{bi}}&=&\frac{2\gamma_{pT}}{\sqrt{3}\omega_T|\mathcal{K}|}\label{eq:bi}, \cr
\Delta\omega_{{bi}}&=&\sqrt{3}\gamma_{pT}\frac{\mathcal{K}}{|\mathcal{K}|}.\label{eq:detune}
\end{eqnarray}
Substituting these critical values into Eq.~(\ref{eq:third_poly}) gives 
\begin{equation}
\langle b_{pT}^{\mathrm{in}}\rangle_{{bi}}^{2}=\frac{4\gamma_{pT}^{2}}{3\sqrt{3}\omega_T|\mathcal{K}|}.
\end{equation}
Finally, using Eq.~(\ref{eq:bin}) we obtain the driving current critical amplitude:
\begin{equation}\label{eq:Ibi}
I_{{bi}}=2\gamma_{pT}\sqrt{\frac{2\hbar\omega_{p}}{3\sqrt{3}\omega_T|\mathcal{K}|Z_{p}}}.
\end{equation}
Note, the requirement that we operate below the Josephson critical current, $I_{0}<I_{c}$, gives a lower limit on the value of $|\mathcal{K}|$ for which our system can approach the bistability onset. 
The boundary of the bistable region that is given by the diverging susceptibility equation (\ref{eq:susceptibility}) can be expressed in units of the bistability onset critical current $I_{bi}$ and detuning value $\Delta\omega_{bi}$  using Eqs.~(\ref{eq:Ibi}) and~(\ref{eq:detune}) to obtain\cite{Dykman:1980p480}
\begin{equation}\label{eq:boundary}
\frac{I}{I_{bi}}=\frac{1}{2}\left(\frac{\Delta\omega}{\Delta\omega_{bi}}\right)^{3/2}\left\{1+3\left(\frac{\Delta\omega_{bi}}{\Delta\omega}\right)^{2}\pm\left[1-\left(\frac{\Delta\omega_{bi}}{\Delta\omega}\right)^{2}\right]^{3/2}\right\}^{1/2},
\end{equation}
where the $\pm$ roots give the upper and lower boundaries of the bistable region, respectively (see Fig.~\ref{fig:region}).  As mentioned in the preceding section, care must be taken when applying our semiclassical, mean field approximations to the detector signal and noise response when approaching closely the  bifurcation boundary lines. Fluctuation-induced jumps between the small and large amplitude solutions of the transmission line  mode can occur that are not accounted for in the mean field approximation. Nevertheless, in the next two sections we shall in some instances evaluate the detector response close the boundaries of the bistability region. For example, we shall see that significant improvements in cooling can be achieved provided a way is found to keep the transmission line mode on the low amplitude solution branch when operating in the bistable region.  

\section{Displacement Detection}\label{sec:squid-detection}

Assuming that $\gamma_{bm}\ll\gamma_{pT}$, i.e., the unrenormalized mechanical oscillator amplitude damping rate is much smaller  than the transmission line oscillator amplitude damping rate, then the detector spectral noise and response in the mechanical signal bandwidth is approximately white over a large range of drive current and detuning parameter space. The mechanical signal and  noise response spectra are therefore approximately Lorentzian and Eqs.~(\ref{eq:current-signal}) and (\ref{eq:current-noise}) can be parametrized as
\begin{eqnarray}
&&\left.\overline{\langle\left[{\delta I}^{\mathrm{out}}(\omega_{s}=\omega_p\pm R_{\omega}\omega_m,\delta\omega)\right]^{2}\rangle}\right|_{\mathrm{signal}} Z_p\cr
&&=G_{\pm} \frac{\hbar}{2m R_{\omega}\omega_m}\int_{\omega_{s}-\delta\omega/2}^{\omega_{s}+\delta\omega/2}\frac{d\omega}{2\pi} \frac{2\gamma_{bm}[2n(R_{\omega}\omega_m)+1]}{(\omega-\omega_p\mp R_{\omega}\omega_m)^2 +(R_{\gamma} \gamma_{bm})^2}
\label{eq:paramcurrentsignal}
\end{eqnarray} 
and
\begin{eqnarray}
&&\left.\overline{\langle\left[{\delta I}^{\mathrm{out}}(\omega_{s}=\omega_p\pm R_{\omega}\omega_m,\delta\omega)\right]^{2}\rangle}\right|_{\mathrm{noise}} Z_p\cr
&&=G_{\pm} \frac{\hbar}{2m R_{\omega}\omega_m}\int_{\omega_{s}-\delta\omega/2}^{\omega_{s}+\delta\omega/2}\frac{d\omega}{2\pi} \frac{2\gamma_{\mathrm{back}}[2n^{\pm}_{\mathrm{back}}+1]}{(\omega-\omega_p\mp R_{\omega}\omega_m)^2 +(R_{\gamma} \gamma_{bm})^2}\cr
&&+\left.\overline{\langle\left[{\delta I}^{\mathrm{out}}(\omega_{s}=\omega_p\pm R_{\omega}\omega_m,\delta\omega)\right]^{2}\rangle}\right|_{\mathrm{added}~\mathrm{noise}} Z_p,
\label{eq:paramcurrentnoise}
\end{eqnarray} 
where $G_{\pm}$ is the phase preserving (conjugating) gain (in W$\cdot$m$^{-2}$), $n(R_{\omega}\omega_m)$ is the mechanical oscillator's external bath occupation number at its renormalized frequency $R_{\omega} \omega_m$,  $R_{\gamma} \gamma_{bm}$ is the renormalized (i.e., net) mechanical oscillator  damping rate, and the detector back reaction noise on the oscillator is effectively that of a thermal bath  with damping rate $\gamma_{\mathrm{back}}=\gamma_{bm} (R_{\gamma}-1)$ and thermal average occupation number $n^{\pm}_{\mathrm{back}}$. Note, from here on we do not consider an external classical force driving the mechanical oscillator; the focus is on displacement detection rather than force detection. The added noise term in Eq.~(\ref{eq:paramcurrentnoise}) comprises output noise that is not due to the action  of the detector on the mechanical oscillator; the added noise is present even when there is no coupling to the mechanical oscillator, i.e., when $K_{Tm}=0$.  In the absence of the  transmission line Duffing nonlinearity, the added noise simply consists of the probe line zero-point fluctuations $\hbar\omega_s\delta\omega/(4\pi Z_p)$. However, with the Duffing nonlinearity present, the added noise will be in excess of the probe line zero-point fluctuations.

The convenient  Lorentzian parametrization approximations of the mechanical signal~(\ref{eq:paramcurrentsignal}) and noise response spectra~(\ref{eq:paramcurrentnoise}) that provide the above-described  effective thermal description of the back reaction noise will break down as one approaches arbitrarily closely the jump points at the ends of the small or large amplitude transmission line oscillator solution branches occuring at the boundaries of the bistable region indicated in Fig.~\ref{fig:region}. This is a consequence of the diverging damping (i.e., ring-down) time of transmission line mode.\cite{Yurke:2006p1911} Thus, when numerically solving ~(\ref{eq:current-signal}) and (\ref{eq:current-noise}) to extract the effective thermal properties of the detector back reaction, it is important to always check the accuracy of the Lorentzian spectrum approximation.

For sufficiently large gain (i.e., large current drive amplitude), we can neglect the added noise contribution and we have for the noise-to-signal response ratio when the mechanical oscillator external bath is at absolute zero [i.e., $n(R_{\omega} \omega_m)=0$]:
\begin{equation} 
\frac{\langle\left[{\delta I}^{\mathrm{out}}\right]^{2}\rangle_{\mathrm{noise}}}{\langle\left[{\delta I}^{\mathrm{out}}\right]^{2}\rangle_{\mathrm{signal}}}=(2n^{\pm}_{\mathrm{back}}+1)\frac{ \gamma_{\mathrm{back}}}{\gamma_{bm}}.
\label{eq:noisetosignal}
\end{equation}
On the other hand,  in the large gain limit the Caves noise lower bound (\ref{eq:current-min}) gives a noise-to-signal ratio of one. For large gain, we typically have $|2n^{\pm}_{\mathrm{back}}+1|\gg 1$ and thus to approach the Caves bound necessarily requires $|\gamma_{\mathrm{back}}|\ll \gamma_{bm}$.\cite{Clerk:2004p245306}

As an example, we numerically solve  for the signal and noise contributions of the detector response, Eqs.~(\ref{eq:current-signal}) and (\ref{eq:current-noise}) respectively, as well as the Caves lower bound on the quantum noise (\ref{eq:current-min}). We consider Duffing nonlinearities $K_d=-3.4\times 10^{-6}$  and  $K_d=0$ (i.e., no nonlinearity). The integrated signal and noise bandwidth is taken to be $\delta\omega=2R_{\gamma}\gamma_{bm}$.  The corresponding parameter values are: probe line impedance $Z_p=50~{\mathrm{Ohms}}$, transmission line mode angular frequency $\omega_T/(2\pi)=5\times 10^9~{\mathrm{s}}^{-1}$, transmission line mode quality factor $Q_T=\omega_T/(2\gamma_{pT})=300$, mechanical frequency $\omega_m/(2\pi)=4\times 10^6~{\mathrm{s}}^{-1}$, mechanical quality factor $Q_m=\omega_m /(2\gamma_{bm})=10^3$, oscillator mass $m=10^{-16}~{\mathrm{kg}}$, Josephson junction critical current $I_{c}=4.5\times10^{-6}~\mathrm{A}$,  junction capacitance $C_{J}=10^{-14}~\mathrm{F}$, external flux bias $\Phi_{\mathrm{ext}}=0.442\ \Phi_0$, and external field in the vicinity of the mechanical resonator $B_{\mathrm{ext}}=0.05~\mathrm{T}$. These values give a zero-point uncertainty $\Delta_{zp}=1.45 \times 10^{-13}~{\mathrm{m}}$ and transmission line-oscillator coupling $K_{Tm}=1.1\times10^{-5}$.  

\begin{figure}[htbp]
\begin{center}
\includegraphics[width=4in]{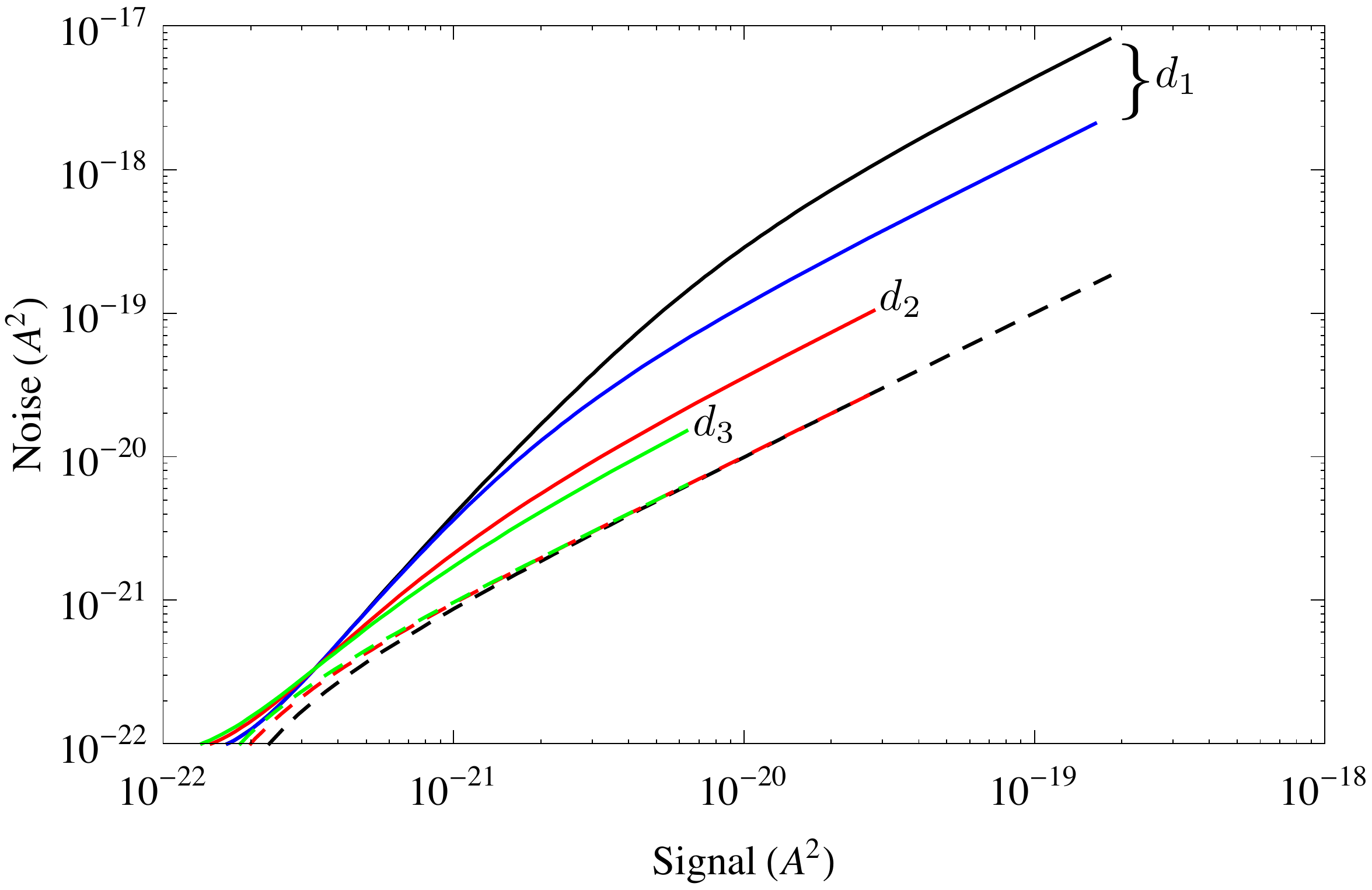}
\caption{Detector noise versus signal response at $\Delta\omega=0$ for harmonic ($K_d=0$) transmission line, Duffing ($K_d<0$) transmission line ($d_1$), and Caves' bound (black-dashed).  Noise for the nonlinear transmission line is also evaluated for blue detunings: $\Delta\omega=+0.2|\Delta\omega_{bi}|$ ($d_2$), and $+0.4|\Delta\omega_{bi}|$ ($d_3$). The labeled curves correspond to the  traces in Fig.~\ref{fig:region}.  The dashed, colored lines  give Caves' bound for the corresponding detuning values.}
\label{fig:detection}
\end{center}
\end{figure}

The advantage of using a spring softening nonlinearity, $K_d<0$, is clearly evident in Fig. \ref{fig:detection}, where we plot the noise versus response signal under increasing current drive for a transmission line both with and without Duffing term driven on resonance, $\Delta\omega=\omega_p-\omega_T=0$. We also plot the response of the nonlinear transmission line for several positively detuned values, $\Delta\omega=\omega_p-\omega_T>0$.  Termination of the curves indicates the signal value at which the damping renormalization $R_{\gamma}= 0$, beyond which the derived solutions become unphysical due to the net mechanical damping rate becoming negative and hence the motion unstable about the original fixed point.
Note that the same criterion, namely $R_{\gamma}>0$, is employed throughout the paper in order to ensure stability of the system. Again, the semiclassical, mean field approximation is expected to break down in the vicinity of termination points, where large fluctuations in the mechanical oscillator amplitude occur.  In Fig.~\ref{fig:detection}, we see that with positive detuning,  we can further approach the Caves bound. However, this is at the expense of reduced gain; depending on one's point of view, large renormalizations of the mechanical oscillator damping rate (and frequency) due to detector back action may or may not be allowed in detector displacement sensitivity figures of merit, affecting the maximum achievable gain as one approaches more closely the Caves bound.

\begin{figure}[htbp]
\begin{center}
\includegraphics[width=5in]{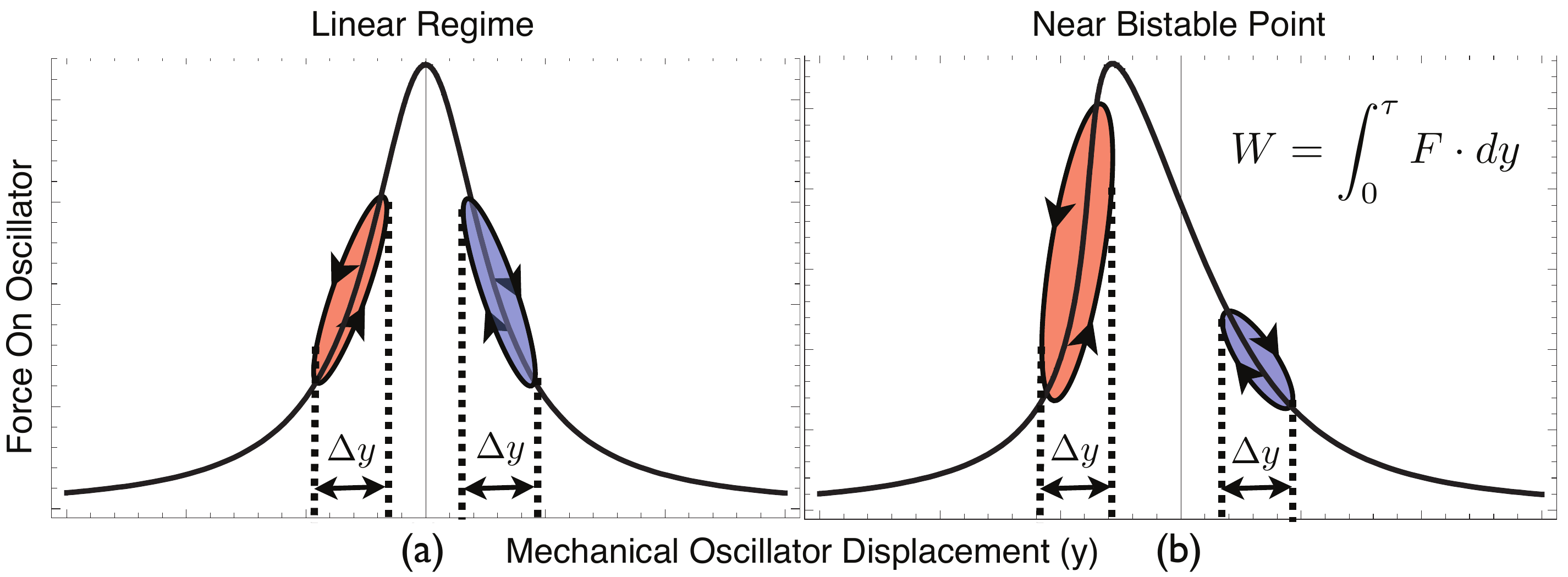}
\caption{Cartoon indicating the `radiation pressure' force exerted on the mechanical oscillator by the transmission line mode during one cycle of mechanical motion: (a) the harmonic transmission line mode approximation;  (b) approaching the onset of bistability.  The work done on the oscillator is proportional to the area swept out during each cycle, considerably exaggerated here for clarity.  Positive mechanical damping (red detuning) results on the positive slope side of the curves. Negative mechanical damping (blue detuning) results on the negative slope side. A spring softening nonlinearity can result in improved cooling for red detuning and improved signal-to-noise amplification  for blue detuning. }
\label{fig:curves}
\end{center}
\end{figure}

The trends displayed in Fig.~\ref{fig:detection} can be partly explained by invoking Fig.~\ref{fig:curves}, which indicates qualitatively the force on the mechanical oscillator due to the microwave transmission line `ponderomotive radiation pressure' force, both with vanishing  and with nonzero Duffing nonlinearity  and also for `red' and `blue'  pump frequency detunings.  
The work done on the mechanical oscillator by the radiation pressure force during one period of motion, due to the delayed transmission line resonator response, is given by the area enclosed within the hysteresis loop\cite{Kippenberg:2007p1995,Marquardt:2008p1309} and can be related to the steady-state back action damping rate through
\begin{equation}
\gamma_{\mathrm{back}}=-\frac{W}{\bar{E}}\frac{1}{\tau},
\label{eq:hysteresis}
\end{equation}
where $W$ is the work done on the mechanical oscillator, $\bar{E}$ is the average oscillator energy and $\tau$ is the period of motion.
When frequency pulling is taken into account,  the usual notions of red-detuned ($\Delta\omega<0$) or blue-detuned ($\Delta\omega>0$)  hold only in the weak drive limit.  We will assume red (blue)-detuned  to correspond to drive and detuning values $\Delta\omega$ where the net work done on the oscillator is negative (positive) as seen in Fig.~\ref{fig:curves}.
For a harmonic transmission line and for low drive powers, the frequency pulling effects can be ignored, since the effective Duffing coupling Eq.~(\ref{eq:k}) is proportional to the square of the transmission line-mechanical oscillator coupling $K_{Tm}$, which contributes only weakly for the considered parameter values.  Conversely, the Duffing term causes frequency pulling even at low input power and can significantly alter the slope of the response curve.  From Eq.~(\ref{eq:hysteresis}), the decreased slope on the blue detuned side leads to a decrease in the  damping rate magnitude which, through Eq.~(\ref{eq:noisetosignal}), leads as demonstrated above to a closer approach to the Caves' limit. As mentioned above, benefits in lower noise-to-signal resulting from further detuning deep into the blue region are offset by diminished achievable signal gain levels.   

Tuning the sign of the Duffing coupling $\mathcal{K}$ (\ref{eq:k}) to be positive, so that we have a hardening spring, results in an increased back action damping rate for blue detuning, and hence  a corresponding decrease in signal to noise relative to the harmonic transmission line resonator detector case.   

\section{Cooling}\label{sec:squid-cooling}

Referring to the parametrizations (\ref{eq:paramcurrentsignal}) and (\ref{eq:paramcurrentnoise}) of the  signal and noise components of the detector response,  we define the mechanical oscillator's net occupation number through the following equation:
\begin{equation}
\gamma_{\mathrm{net}}(2n^{\pm}_{\mathrm{net}}+1)=\gamma_{bm}\left[2n(R_{\omega}\omega_{m})+1\right]+\gamma_{\mathrm{back}}(2n^{\pm}_{\mathrm{back}}+1),
\label{eq:gammanetnnet}
\end{equation}
 where the net damping rate is $\gamma_{\mathrm{net}}=\gamma_{bm}+\gamma_{\mathrm{back}}=R_{\gamma}\gamma_{bm}$. 
 The oscillator's net occupation number is then
 \begin{equation}
 2n_{\mathrm{net}}+1=R_{\gamma}^{-1}\left[2n(R_{\omega}\omega_{m})+1\right] +(1-R_{\gamma}^{-1}) (2n^{\pm}_{\mathrm{back}}+1).
 \label{eq:nnet}
 \end{equation}  
In order to  cool a mechanical oscillator to its ground state using detector back action, we therefore require a large detector back action damping rate, equivalently large damping rate renormalization $R_{\gamma}\gg 1$, together with a small detector back action effective occupation number $n^{\pm}_{\mathrm{back}}\ll 1$.   
 
\begin{figure}[htbp]
\begin{center}
\includegraphics[width=4.0in]{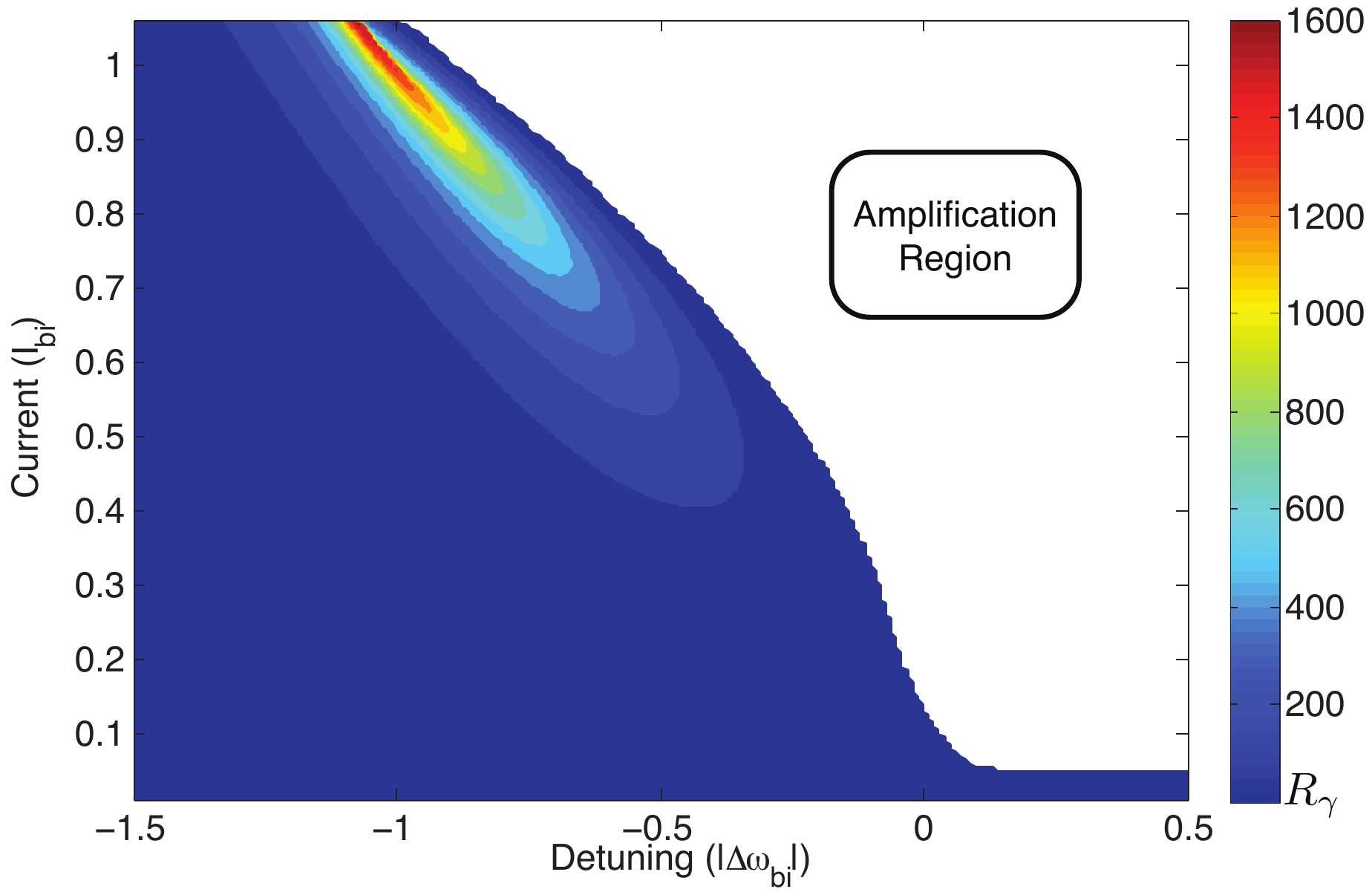}
\caption{Mechanical oscillator damping renormalization factor $R_{\gamma}$ for detunings both above and below the bistable detuning $\Delta\omega_{bi}$. The amplification region corresponds to negative back action damping, i.e., $R_{\gamma}<1$.}
\label{fig:gamma}
\end{center}
\end{figure}

Referring to Fig.~\ref{fig:curves}, operating closer to the bistability increases the negative work done per cycle on the oscillator by the cavity  and hence  increases the back action damping rate for given current drive.  In Fig.~\ref{fig:gamma}, we plot the mechanical oscillator damping rate renormalization factor $R_{\gamma}$, using the same parameter values as in  Sec.~\ref{sec:squid-detection} (e.g., Duffing coupling $K_{d}=-3.4\times10^{-6}$), but with a larger yet still feasible mechanical quality factor $Q_m=10^4$ (which we shall adopt throughout this section).  We clearly see the enhanced damping as one approaches the onset of bistability given by $I_{{bi}}$ (\ref{eq:Ibi}) and $\Delta\omega_{{bi}}<0$ (\ref{eq:detune}).  

For the  example parameter choices of  Sec.~\ref{sec:squid-detection}, we have $\omega_{m}/\gamma_{pT}\approx0.5$ and thus we are operating in the so-called bad cavity limit, where cooling close to the ground state (i.e., $n_{\mathrm{net}}\ll 1$) is not possible.\cite{Blencowe:2007p285,Marquardt:2007p978,WilsonRae:2007p502}
 While it is not difficult to achieve the good cavity limit $\omega_m>\gamma_{pT}$ simply by realizing sufficiently large quality factor superconducting microwave resonators, together with high frequency mechanical resonators,\cite{Teufel:2008p2398} it is nevertheless worthwhile to address how nonlinearities can improve on the cooling limits in the bad-cavity case. With the fundamental motivation to demonstrate macroscopic quantum behavior, the anticipated trend is to work with increasingly massive and hence lower frequency oscillators, making it progressively more difficult to achieve the good cavity limit.   

\begin{figure}[htbp]
\begin{center}
\includegraphics[width=6.0in]{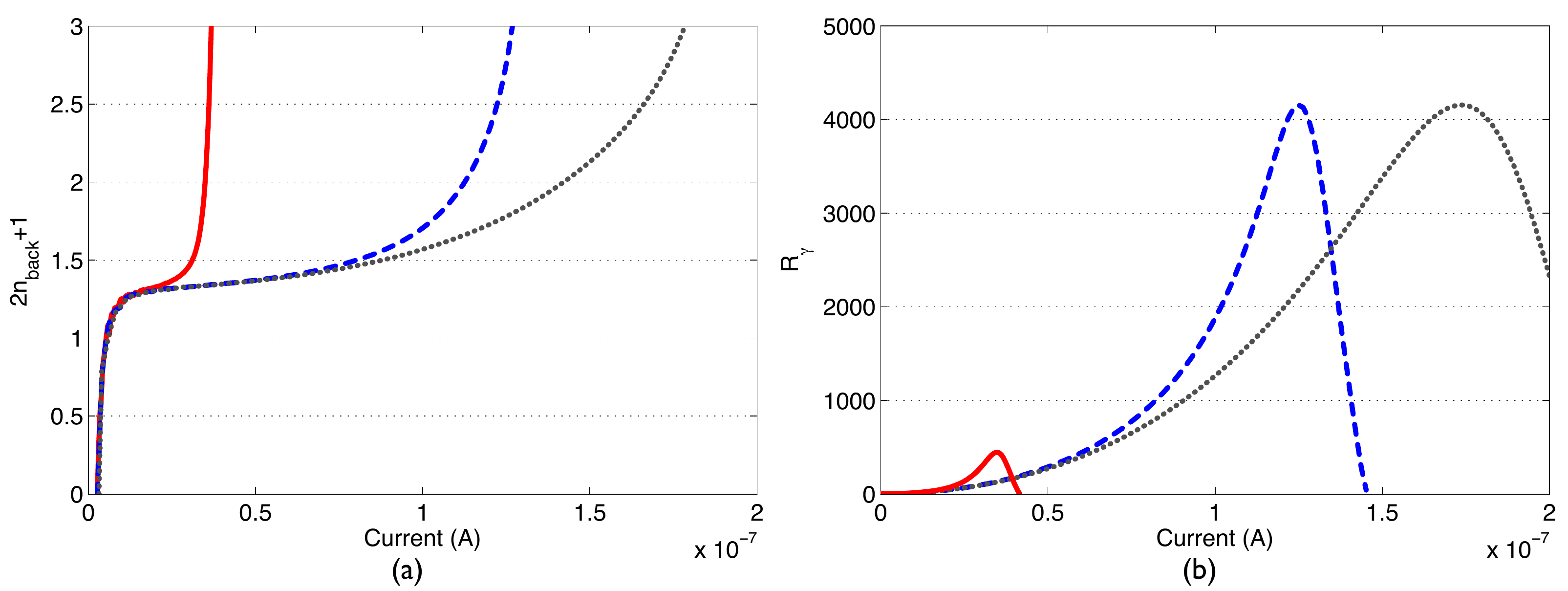}
\caption{(a) Detector noise effective back action occupation number versus current drive when red-detuned at $\Delta\omega=-\sqrt{\omega_{m}^{2}+\gamma_{pT}^{2}}$, $|\Delta\omega|<|\Delta\omega_{bi}|$, with a Duffing nonlinearity (solid line),  without  a Duffing nonlinearity (dashed line), and without both Duffing and ponderomotive nonlinearities (dotted line). (b) Oscillator coupling renormalization factor $R_{\gamma}$ for the corresponding back-action occupation number curves. These plots are obtained for the straight line trace labeled $c_1$ in Fig.~\ref{fig:region}.}
\label{fig:optimal}
\end{center}
\end{figure}

In Fig.~\ref{fig:optimal}, we plot the dependence of detector's noise effective back action occupation number $n_{\mathrm{back}}$ on microwave drive current amplitude  at the detuning bias $\Delta\omega=-\sqrt{\omega_{m}^{2}+\gamma_{pT}^{2}}$, where  $|\Delta\omega|<|\Delta\omega_{bi}|$. This is the optimum detuning in the harmonic, transmission line oscillator approximation, i.e.,  when nonlinear effects are ignored.    The noise effective occupation number is indicated  for both a nonzero ($K_d=-3.4\times10^{-6}$) as well as zero ($K_d=0$)  Duffing nonlinearity transmission line. We also show for comparison the effective back action occupation number when the frequency pulling effects of both the ponderomotive coupling $K_{Tm}$ and Duffing coupling  $K_{d}$ are neglected.  The latter case is obtained by dropping the nonlinear microwave mode amplitude term in the mean field equation~(\ref{eq:mean-field0}).  The sharp rise in occupation number and associated sharp drop in damping renormalization at larger current drives is a consequence of  crossing over into the amplification region due to negative frequency pulling of the cavity response relative to the fixed detuning.  The decrease in  occupation number as $I \rightarrow 0$ is accompanied by weak back action damping,  which prevents cooling the mechanical oscillator to  such occupation numbers. Note that at smaller current drives the damping renormalization in the presence of a Duffing nonlinearity peaks above the corresponding damping renormalization without the Duffing nonlinearity. This damping enhancement can be qualitatively explained with the aid of Fig.~\ref{fig:curves}. In the presence of the nonlinearity then, improved cooling can be achieved for smaller current drives.        

\begin{figure}[htbp]
\begin{center}
\includegraphics[width=3.2in]{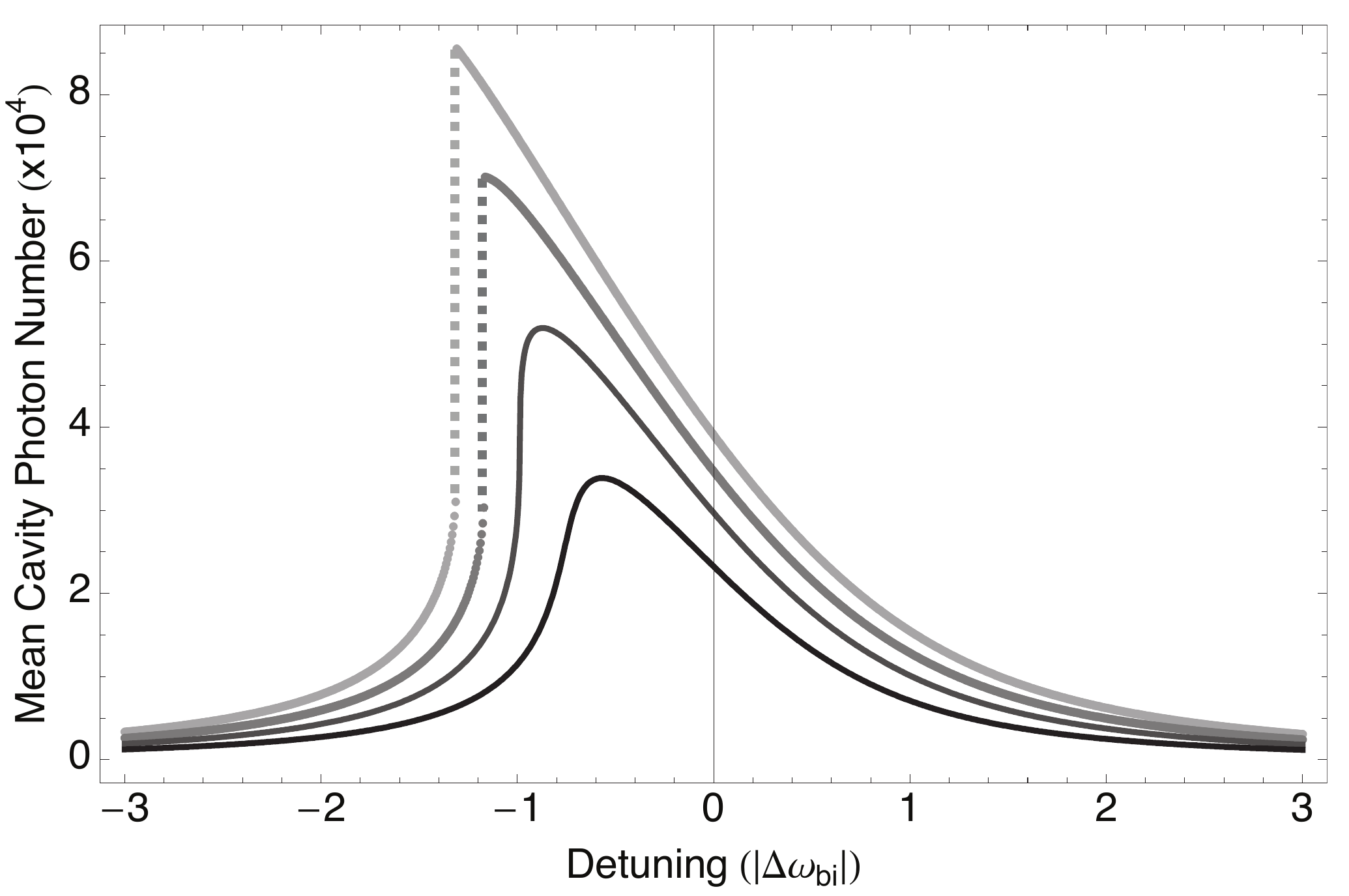}
\caption{Transmission line resonator response curve for $Q_{T}=300$ restricted to the small amplitude solution branch.  The example drive currents from bottom-to-top: $I/I_{bi}=0.8$, 0.95, 1.15, and 1.3.  The  jump between small (red-detuned) and large (blue-detuned) amplitude solutions is indicated by the dotted lines.}
\label{fig:response}
\end{center}
\end{figure}

According to the above discussion, any  improvements in mechanical oscillator cooling are due solely to enhancements in the detector's back action damping rate for given drive; as can be seen from Fig.~\ref{fig:optimal}, the absolute minimum attainable detector effective occupation number is the same both in the presence and absence of the transmission line resonator Duffing nonlinearity.  While the effects of enhanced back action damping may be beneficial in situations where one is facing constraints on the maximum achievable drive power,\cite{Teufel:2008p2398} it would nevertheless be more significant if reductions in detector effective occupation number  could similarly  be achieved through nonlinear effects.   To see how this might be possible, we consider detunings corresponding to the pump frequency being to the left and away from the cavity resonance, i.e., $|\Delta\omega| > |\Delta\omega_{bi}|$, $\Delta\omega<0$. For such detunings, the mechanical oscillator `sees' a transmission line resonator effective quality factor that is determined by the steeper slope on the left side of the  response curve  (see Fig.~\ref{fig:curves}).  As we drive the transmission line resonator towards the lower bistable boundary (see Fig.~\ref{fig:region}), the slope of the response curve increases sharply and mimics a resonator with larger quality factor, effectively getting closer to the good cavity limit and hence resulting in a lower detector occupation number.\cite{Blencowe:2007p285,Marquardt:2007p978,WilsonRae:2007p502} Continuing to drive the transmission line resonator into the bistable region, and assuming that the resonator can be maintained on the low amplitude, red-detuned solution branch,\cite{Naaman:2008} the detector effective occupation number further decreases while the back action damping rate on the mechanical resonator increases (as explained by Fig.~\ref{fig:curves}). Eventually, the transmission line resonator becomes unstable at  the upper bistable boundary indicated in Fig.~\ref{fig:region}, and the oscillator jumps to the larger amplitude, blue-detuned solution (see Fig.~\ref{fig:response}).

\begin{figure}[htbp]
\begin{center}
\includegraphics[width=6.0in]{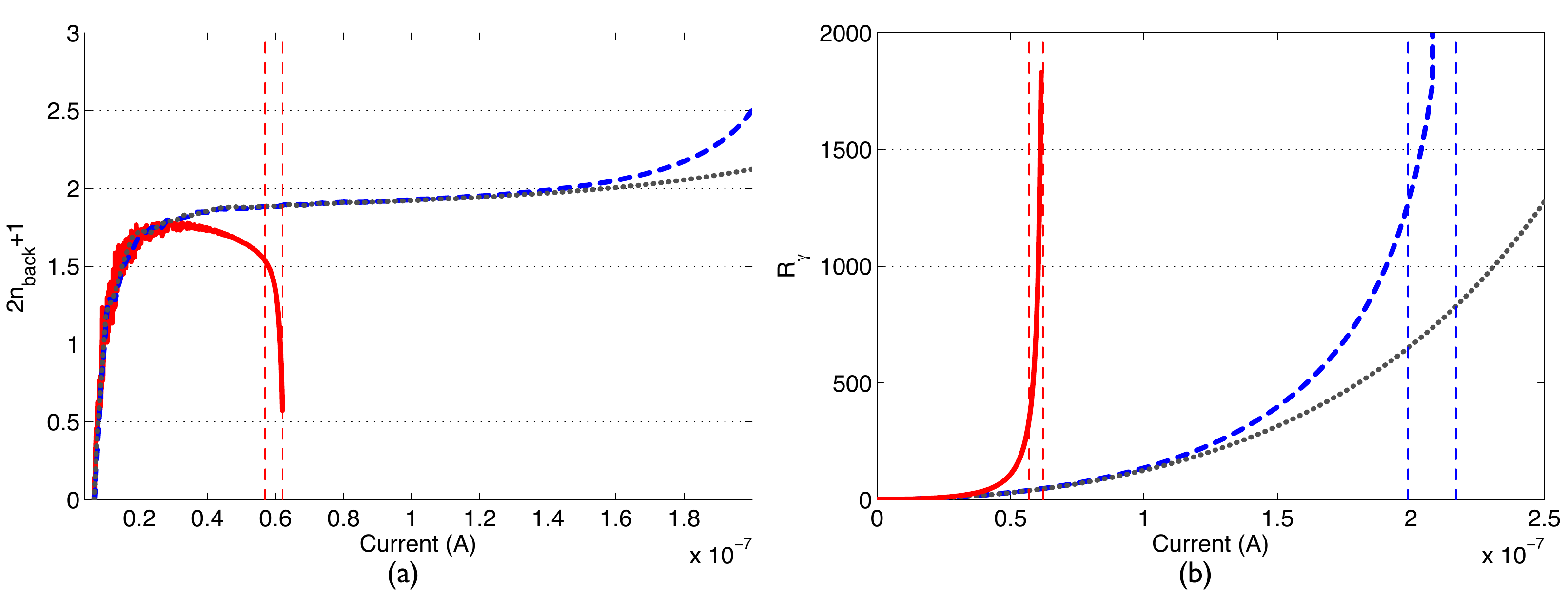}
\caption{(a) Detector noise effective occupation number versus current drive when red detuned at $\Delta\omega=1.3\ \Delta\omega_{bi}$, corresponding to straight line trace $c'_1$ in Fig.~\ref{fig:region}.  The Duffing nonlinear transmission line resonator occupation number (solid line) rapidly decreases as the resonator is driven towards the upper bistable boundary, assuming the resonator can be maintained on the small amplitude metastable stable solution branch. In contrast, a harmonic transmission line resonator (dashed line) or a cavity with neither Duffing nor ponderomotive nonlinearities in its mean field microwave mode equations (dotted line)  shows no such decrease in the occupation number. (b) Mechanical oscillator damping renormalization factor for the  same fixed detuning and drive current range.  The dashed vertical lines indicate the boundaries of the bistable region for the given transmission line resonator parameters.}
\label{fig:goodcavity}
\end{center}
\end{figure}

\begin{figure}[htbp]
\begin{center}
\includegraphics[width=6.0in]{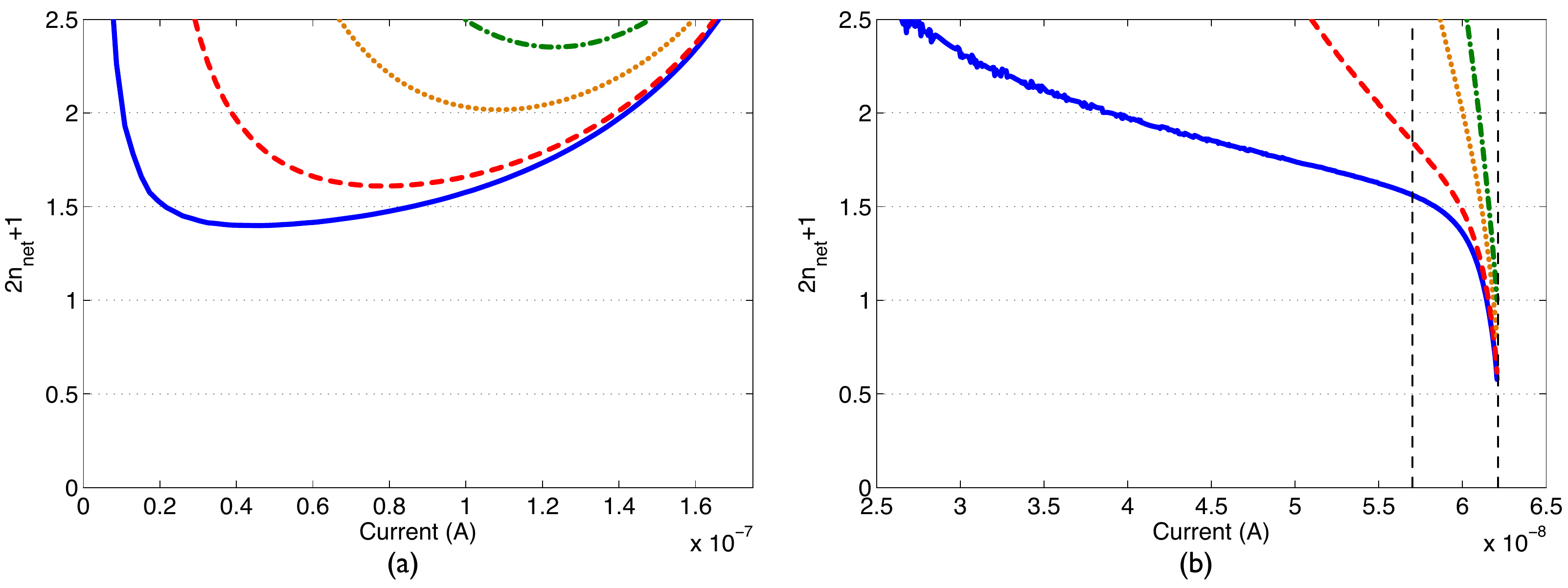}
\caption{(a) Net mechanical occupation number at $\Delta\omega=-\sqrt{\omega_{m}^{2}+\gamma_{pT}^{2}}$ for a harmonic transmission line resonator.  External bath temperatures: $T=1$ (solid line), 10 (dashed line), 50 (dotted line) and 100 (dot-dashed line) $\mathrm{mK}$. (b) Dependence of the net mechanical oscillator occupation number on current drive for a Duffing transmission line with detuning $\Delta\omega=1.3\ \Delta\omega_{bi}$.  The bistable region boundaries are indicated by the dashed vertical lines.}
\label{fig:cooling}
\end{center}
\end{figure}

In Fig.~\ref{fig:goodcavity}, we plot the dependence of the detector effective occupation number $n_{\mathrm{back}}$ on current drive for an example detuning value of $\Delta\omega=-2\sqrt{\omega_{m}^{2}+\gamma_{pT}^{2}}=1.3\Delta\omega_{bi}$.  
Driving the nonlinear transmission line resonator towards the upper boundary of the bistable region (see Fig.~\ref{fig:curves}) produces a sharp decrease in detector occupation number, and an occupation number value of $(2n_{\mathrm{back}}+1)\approx 0.55$ can be obtained,  well below that achievable when ignoring frequency pulling effects.  The harmonic cavity shows no such decrease in occupation number, indicating the qualitatively different quantum dynamical dependencies on $K_{d}$ and $K_{Tm}$ and the necessity of the former.  We can quantify the effect of frequency pulling by comparing with a harmonic transmission line resonator with a quality factor  $Q_{T}^{\mathrm{eff}}$ value chosen so as  to give the same detector effective occupation number.  For  the occupation number value $(2n_{\mathrm{back}}+1)\approx 0.55$, we have $Q_{T}^{\mathrm{eff}}\approx 600$,  corresponding to $\omega_{m}/\gamma^{\mathrm{eff}}_{pT}=0.95$, and therefore the mechanical oscillator behaves as if it is coupled to a cavity with double the quality factor.  This translates into lower net mechanical temperatures as shown in Fig.~\ref{fig:cooling}, where we give the net oscillator occupation number $n_{\mathrm{net}}$ (\ref{eq:nnet}) for various external bath temperatures.
The combination of nonlinearly-enhanced coupling $R_{\gamma}\gamma_{bm}$ and enhanced transmission line effective quality factor can be seen to significantly affect cooling of the mechanical motion, even for relatively large external temperatures.

In the numerical solutions to Eqs.~(\ref{eq:current-signal}) and (\ref{eq:current-noise}), the Lorentzian parametrizations (\ref{eq:paramcurrentsignal}) and (\ref{eq:paramcurrentnoise}) were found to give good approximations even when the upper bistable boundary is  approached quite closely. This is a consequence of the wide separation in the relaxation rates that determine the line widths of the harmonic transmission line resonator and unrenormalized mechanical oscillator modes, i.e., $\gamma_{bm}\ll\gamma_{pT}$. The upper bistable boundary has to be approached pretty closely in order for the nonlinear transmission line resonator ring-down time to exceed the renormalized mechanical oscillator damping time,  resulting in the breakdown of the  effective thermal description of the detector back reaction.  The deviation of Eqs.~(\ref{eq:current-signal}) and (\ref{eq:current-noise}) from the assumed Lorentzian response near the upper bistable boundary gives a general criteria for measuring how close this limit can be approached.   In all of the plots shown in this section, the Lorentzian approximation is a good one over the resolvable scale of the plots. The actual minimum temperature  that can be achieved depends on the upper drive threshold where the Lorentzian approximation breaks down, as well as on the ability to keep the transmission line resonator on the small amplitude solution branch; the latter condition becomes progressively more difficult to satisfy as the upper boundary is approached, owing to the increasing probability of noise-induced jumps to the large amplitude branch.

\begin{figure}[htbp]
\begin{center}
\includegraphics[width=6.0in]{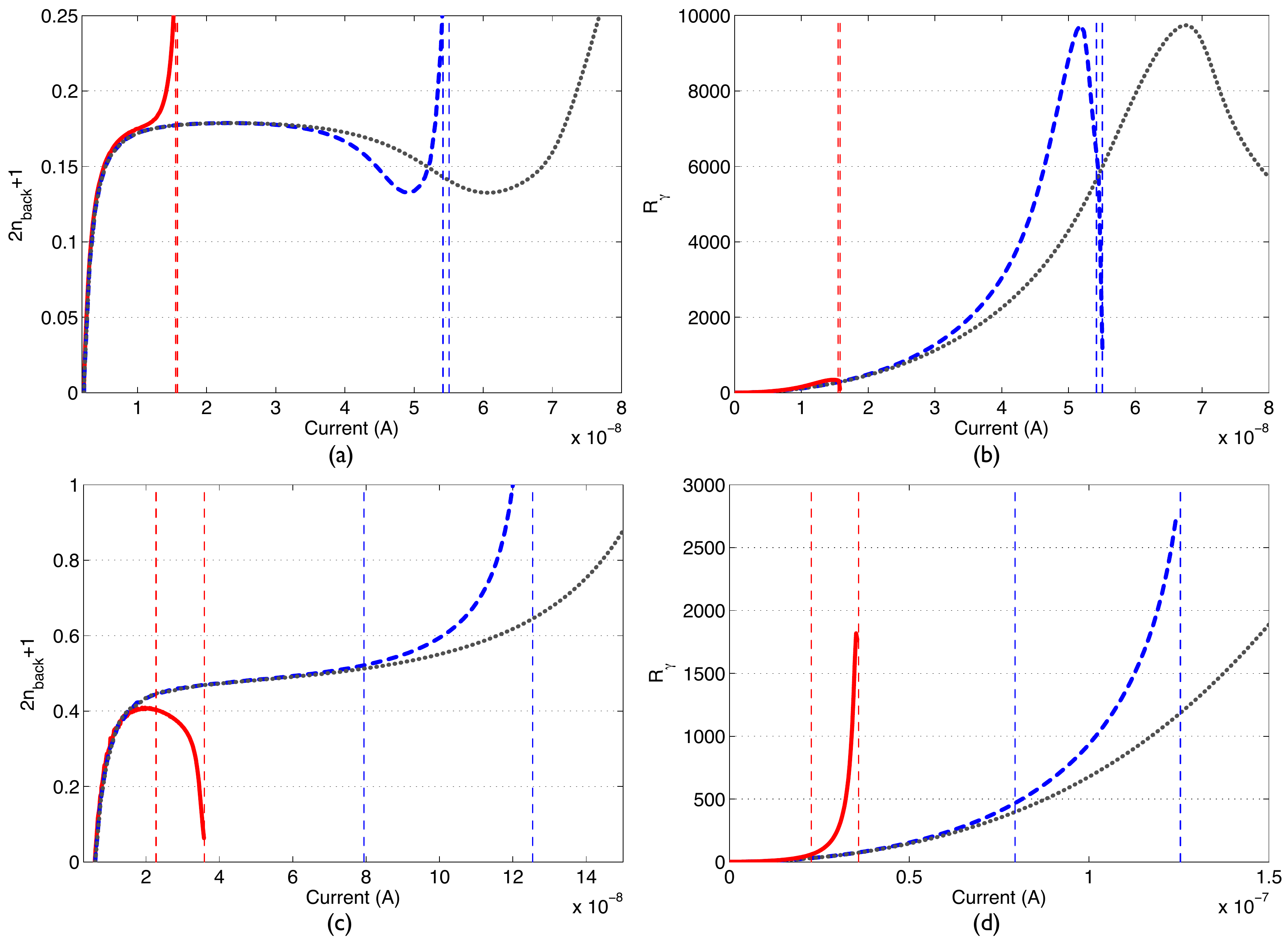}
\caption{(a) Detector noise effective occupation number versus current drive when red-detuned at $\Delta\omega=-\sqrt{\omega_{m}^{2}+\gamma_{pT}^{2}}$, $|\Delta\omega|<|\Delta\omega_{bi}|$, for a Duffing nonlinear (solid line),  harmonic (dashed line) transmission line, and with the effects of frequency pulling due to both ponderomotive and Duffing nonlinearities neglected (dotted line).  The vertical dashed lines give the bistable region boundaries for the Duffing and harmonic transmission line resonators. This plot is obtained for the straight line trace labeled $c_2$ in Fig.~\ref{fig:region} (b) Oscillator coupling renormalization factor $R_{\gamma}$ for optimal harmonic detuning.  (c) Detector occupation number detuned at twice the harmonic optimum, $\Delta\omega=2.2\Delta\omega_{bi}$ and locked to the lower stable amplitude solution. This plot is obtained for the straight line trace labeled $c'_2$ in Fig.~\ref{fig:region} (d) Corresponding back-action damping rate when driven to the upper bistable boundary.}
\label{fig:1000}
\end{center}
\end{figure}

\begin{figure}[htbp]
\begin{center}
\includegraphics[width=6.0in]{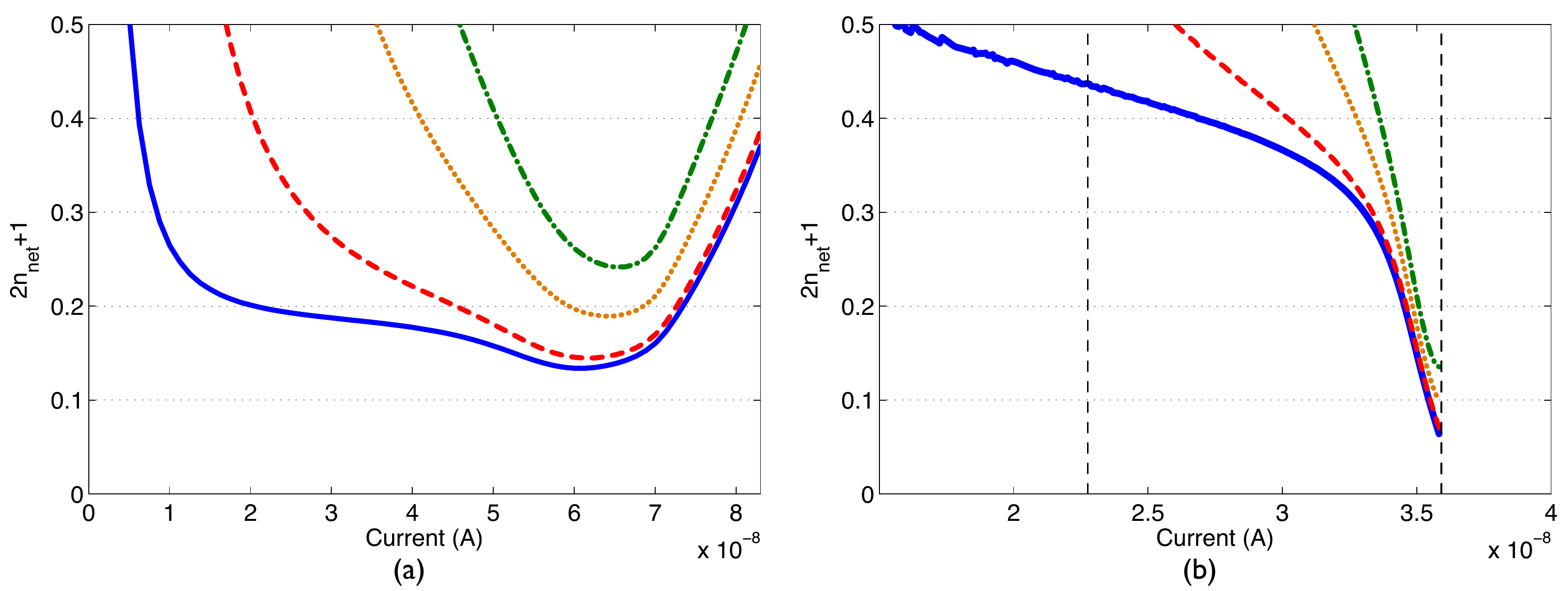}
\caption{(a) Net mechanical occupation number at $\Delta\omega=-\sqrt{\omega_{m}^{2}+\gamma_{pT}^{2}}$ for a harmonic transmission line resonator.  External bath temperatures: $T=1$ (solid line), 10 (dashed line), 50 (dotted line) and 100(dot-dashed line) $\mathrm{mK}$. (b) Dependence of the net mechanical oscillator occupation number on current drive for a Duffing transmission line with detuning $\Delta\omega=2.2\Delta\omega_{bi}$.}
\label{fig:cooling1000}
\end{center}
\end{figure}

A Duffing transmission line resonator nonlinearity can also produce cooling gains in the good cavity limit.  In Fig.~\ref{fig:1000}, we consider a transmission line resonator with $Q_{T}=1000$, giving $\omega_{m}/\gamma_{pT}=1.6$, and compare the  nonlinear transmission line resonator with the harmonic resonator approximation at optimal harmonic detuning.  Again, by detuning to twice the optimal harmonic resonator  value, $\Delta\omega=-2\sqrt{\omega_{m}^{2}+\gamma_{pT}^{2}}\approx2.2 \Delta\omega_{bi}$, we see that the effective back action occupation number decreases, while the  back action damping increases as the system is driven towards the upper boundary of the bistable region. 
Driving a Duffing transmission line resonator at twice the optimal harmonic detuning can yield a detector occupation number $(2n_{\mathrm{back}}+1)\approx 0.06$ just below the upper boundary of the bistable region, which is equivalent to an effective harmonic resonator quality factor of $Q_{T}^{\mathrm{eff}}\approx 1400$ or $\omega_{m}/\gamma^{\mathrm{eff}}_{pT}=2.2$.  In comparison, the minimum effective detector occupation number ignoring nonlinear effects is $2n_{\mathrm{back}}+1=0.13$.  In Fig.~\ref{fig:cooling1000},  we plot the net mechanical occupation number for the good cavity transmission line resonator both in the presence and absence of the Duffing nonlinearity.
Again, we see the strong cooling effects provided by frequency pulling of the cavity response. As discussed above, the minimum achievable net occupation number will depend on the threshold drive for which the Lorentzian approximation breaks down, as well as on the ability to lock the transmission line resonator onto the small amplitude solution branch in the bistable region.

\section{Conclusions}\label{sec:squid-conclusion}

We have provided a quantum analysis of a nonlinear microwave amplifier for displacement detection and cooling of a mechanical oscillator. The amplifier comprises a microwave stripline resonator with embedded dc SQUID.  The SQUID gives rise to an effective, Duffing-type nonlinearity in the fundamental microwave mode equations, as well as a ponderomotive-type coupling between the microwave and fundamental mechanical modes.  It was found that  a spring-softening Duffing nonlinearity enables  a closer approach to the standard quantum limit for position detection as expressed by the Caves bound, as well as cooling closer to the  mechanical mode ground state. These findings can be qualitatively explained by considering the effects of frequency pulling in the response curve of the transmission line resonator `ponderomotive force' acting on the mechanical oscillator   (see Fig.~\ref{fig:curves}).  With blue detuning, the decrease in  damping allows for a closer approach to the quantum limit with large amplifier gain.  Conversely,  red detuning towards the bistable point of the force response curve increases the back action damping, improving the thermal contact to the detector `cold load'. Furthermore, effectively increasing the cavity quality factor due to the nonlinearity mimics the so-called good cavity limit in the harmonic case,  allowing cooling closer to the ground state. 

The present investigation has by no means exhaustively searched the large parameter space of  the transmission line resonator-embedded SQUID-mechanical resonator system for establishing the optimal displacement detection sensitivity and cooling parameters. Rather, our intention has been to point out general trends, using specific parameter values as illustrative examples. It may be that other choices of parameters (e.g., using a mechanical resonator with a smaller quality factor) lead to a closer approach to the standard quantum limit, or to cooling closer to the ground state.  

The  semiclassical, mean field methods  employed in the present work do not take into account  classical or quantum noise-induced jumps between the small and large amplitude metastable solutions that  become more likely as the bistability region boundaries are approached. Unless ways can be found to keep the transmission line resonator locked onto the smaller amplitude solution branch, the predicted effects of nonlinearity-induced cooling will be less substantial, as it will be necessary to operate deeper in the bistability region to avoid jumps. The driven microwave mode amplitude dynamics in the vicinity of the bistable region boundaries is still a relatively unexplored area that requires more sophisticated theoretical techniques in order to elucidate the fluctuations  between the small and large amplitude metastable solution branches.\cite{Dykman:1980p480,Dykman:2004p061102,Dykman:2005p021102,Dykman:2007p1864,Serban:2007p3199,Lifshitz:2007p040404,Kogan:2008p0972} This will be the subject of a future investigation.

%%%

%% file: ch3.tex
%% intro.tex
\dsp
\chapter{Analogue Hawking radiation in a dc-SQUID array transmission line}\label{ch:black}
%%%%%%%%%%%%%%%%%%%%%%%%%%%%%%%%%%%%%%%%%%%%%%%%

\section{Motivation}\label{sec:black-motivation}
In the previous chapter we saw how the embedding of a dc-SQUID into a microwave cavity results in a nonlinear Duffing oscillator with the strength of the nonlinearity governed by the amplitude of an externally applied magnetic flux, Eq.~(\ref{eq:Kd}).  One can therefore modulate the dynamical properties of the microwave resonator by applying a time-dependent flux through the SQUID loop.  In this chapter we wish to extend the SQUID-embedded cavity model to allow for control of the dynamics in \textit{both} time and space.   To this end, we now consider an effective 1+1 dimensional system involving a superconducting coplanar waveguide with centerline formed from an array of dc-SQUIDs\footnote{A 1+1 dimensional system is one that can be described using a field theory comprising a single time and space dimension.}.  In this configuration a dc-SQUID approximated as a lumped inductor forms an LC oscillator together with the geometric capacitance of the transmission line ground planes.  Therefore this setup is essentially an array of coupled oscillators each with a nonlinear flux-dependent frequency.  Quite remarkably, the propagation of free-fields\footnote{A free field interacts only with the classical gravitational background.} in curved space-times is very similar to a set of coupled oscillators with time-dependent frequencies\cite{jacobson:2003,mukhanov:2007} suggesting we should be able to create a 1+1 dimensional analogue of photons propagating in an effective curved geometry in this device.  In this chapter we will show that this is indeed the case by considering perhaps the most important example of quantum field dynamics in curved spacetime: the Hawking effect\cite{hawking:1974}.

\section{Introduction}\label{sec:black-introduction}
The possibility of observing Hawking radiation \cite{hawking:1974} in a condensed matter system was first suggested by Unruh who uncovered the analogy between sound waves in a fluid and a scalar field in curved space-time \cite{unruh:1981}.  In particular, the fluid equations of motion can formally be expressed in terms of an effective metric matching that of a gravitating spherical, non-rotating massive body in Painlev\'e-Gullstrand coordinates \cite{painleve:1921}
\begin{equation}\label{eq:painleve}
ds^{2}=-\left[c_{s}^{2}-v(r)^{2}\right]dt^{2}+2v(r)drdt+dr^{2}+r^{2}d\Omega^{2},
\end{equation}
where $c_{s}$ is the speed of sound and $v(r)$ is the spatially varying velocity of the fluid.  For a sound wave excitation in the fluid, with velocity $c_{s}$, the horizon occurs where $v^{2}(r)=c_{s}^{2}$ and the excitation is incapable of surmounting the fluid flow.   Since Unruh's original proposal, Hawking radiation analogues have been proposed using Bose-Einstein condensates \cite{garay:2000}, liquid Helium \cite{volovik:1999}, electromagnetic transmission lines \cite{schutzhold:2005}, and fiber-optic setups \cite{philbin:2008}.  Estimated Hawking temperatures in these systems vary from a few nano-Kelvin to $10^{3}\mathrm{K}$ respectively, far above temperatures predicted for astronomical black holes and thus usher in the possibility of experimental observation.  Additionally, the understanding of the physics associated with laboratory system analogues may provide clues as to resolving unanswered questions associated with Hawking's original calculation such as the trans-Planckian problem \cite{jacobson:1991}.

\begin{figure}[htbp]
\begin{center}
\includegraphics[width=3.1in]{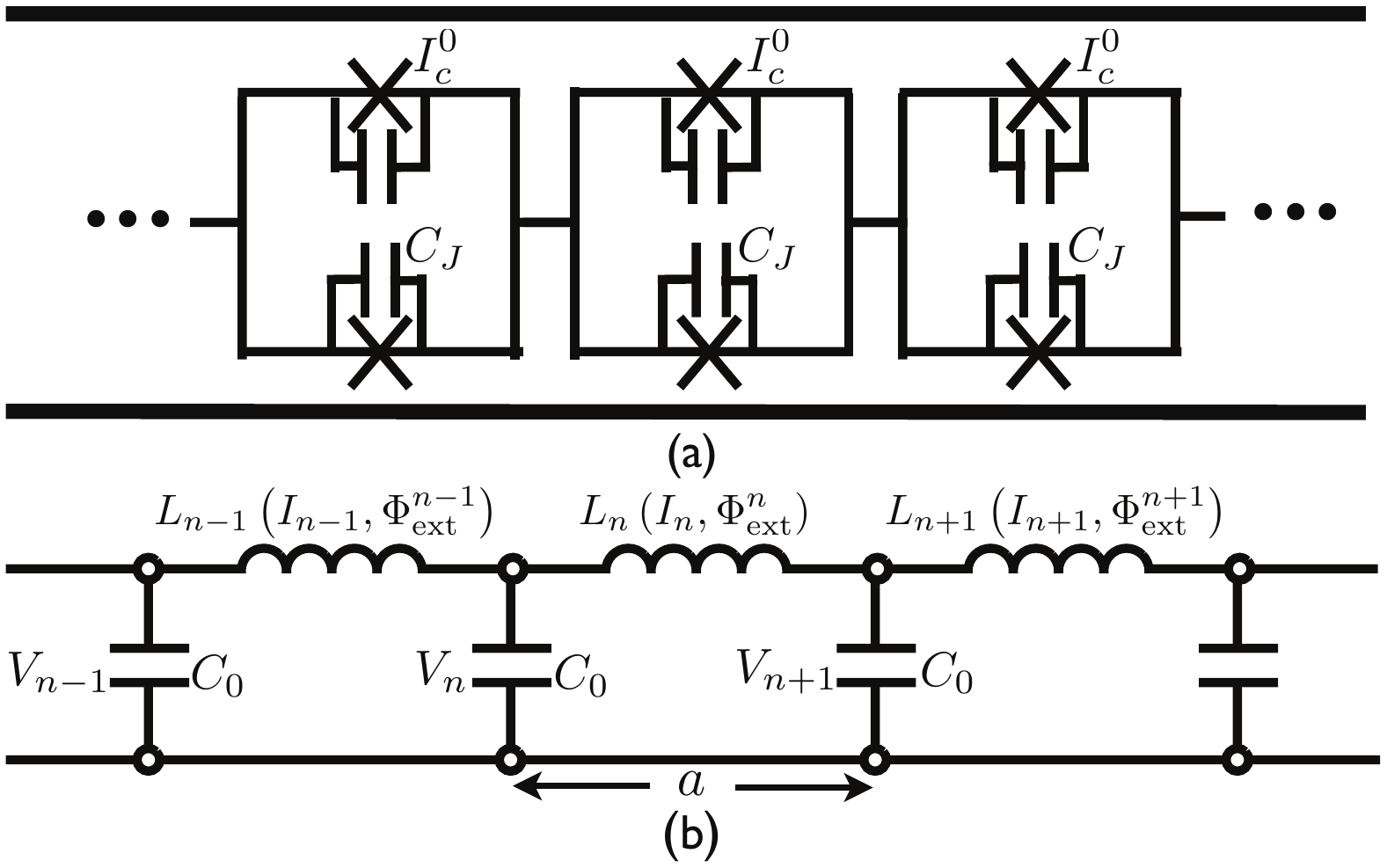}
\caption{a) Layout of the dc-SQUID transmission line. We assume each SQUID element is formed from identical tunnel junctions with critical current $I_{c}$ and capacitance $C_{J}$. b) Effective lumped circuit model valid for frequencies below the plasma frequency and negligible SQUID self-inductance.}
\label{fig:setup}
\end{center}
\end{figure}

In this chapter, we propose using a metamaterial formed from an array of direct-current superconducting quantum interference devices (dc-SQUID's).  Modulation of the propagation velocity, necessary for the formation of an horizon, is accomplished through application of an external flux bias through the SQUID loops as indicated in Fig.~\ref{fig:setup}a.
Under appropriate conditions, this configuration provides the superconducting realization of Ref. \cite{schutzhold:2005}, with the benefit of available fabrication methods.  Indeed, arrays of SQUID's with parameters near those required to observe the Hawking effect have already been constructed \cite{beltran:2007,beltran:2008}.  Furthermore, as a quantum device, the SQUID array goes beyond the capabilities of previously proposed systems, allowing the possibility to probe the effect on Hawking radiation of quantum fluctuations in the space-time metric. Thus, in principle, this setup enables the exploration of analogue quantum gravitational effects.

\section{Model}\label{sec:black-model}
We consider a coplanar transmission line composed of a centerline conductor formed by a long, $N\gg1$, series array of dc-SQUID's indicated in Fig.~\ref{fig:setup}a.  For simplicity, we assume that all Josephson junctions (JJ) have identical critical current $I_{c}$ and capacitance $C_{J}$ values.  For an individual dc-SQUID, with $\phi_{1}$ and $\phi_{2}$ representing the gauge invariant phases across the JJ's, the equations of motion for $\gamma_{\pm}=\left(\phi_{1}\pm\phi_{2}\right)/2$ take the form
\begin{align}\label{eq:black-motion}
\frac{1}{\omega_{p}^{2}}\frac{d^{2}\gamma_{+}}{dt^{2}}+\frac{1}{\omega_{c}}\frac{d\gamma_{+}}{dt}+\cos(\gamma_{-})\sin(\gamma_{+})&=\frac{I}{2I_{c}} \notag \\
\frac{1}{\omega_{p}^{2}}\frac{d^{2}\gamma_{-}}{dt^{2}}+\frac{1}{\omega_{c}}\frac{d\gamma_{-}}{dt}+\cos(\gamma_{+})\sin(\gamma_{-})+\frac{2\gamma_{-}}{\beta_{L}}&=\frac{1}{\beta_{L}}\frac{2\pi\Phi_{\mathrm{ext}}}{\Phi_{0}},
\end{align}
with plasma frequency $\omega_{p}=(2\pi I_{c}/C_{J}\Phi_{0})^{1/2}$, characteristic frequency $\omega_{c}=2\pi I_{c}R_{N}/\Phi_{0}$, and normalized self-inductance $\beta_{L}=2\pi L I_{c}/\Phi_{0}$.  The parallel, normal current resistance of the junction is denoted $R_{N}$, while $\Phi_{0}=h/2e$ is the flux quantum and $\Phi_{\mathrm{ext}}$ is the external flux through the SQUID loop.    If $\beta_{L}\ll1$ then the SQUID dynamics can be approximated by a JJ with a flux-tunable critical current, $I^{s}_{c}=2I_{c}\cos(\pi\Phi_{\mathrm{ext}}/\Phi_{0})$, the dynamics of which can be written
\begin{equation}
\frac{1}{(\omega^{s}_{p})^{2}}\frac{d^{2}\gamma_{+}}{dt^{2}}+\sin\left(\gamma_{+}\right)=\frac{I}{I_{c}^{s}},
\end{equation}
where we have dropped the damping term, assuming the  temperature is well below the superconducting critical temperature, and
where the effective plasma frequency is given by $\omega^{s}_{p}=\sqrt{2\pi I^{s}_{c}/(2C_{J}\Phi_{0})}$. We will assume the validity of this approximation and consider a flux-tunable array of Josephson Junctions (JJA).  If we additionally restrict ourselves to frequencies well below the plasma frequency and currents below the critical current, then a JJ behaves as a passive, flux and current dependent inductance given by
\begin{equation}\label{eq:inductance}
L_{n}(I_{n},\Phi^{n}_{\mathrm{ext}})=\frac{\Phi_{0}}{2\pi I_{n}}\arcsin\left(I_{n}/I^{s}_{c}\right)
\end{equation}
for the \textit{n}th JJ in the array. The equivalent circuit is given in Fig.~\ref{fig:setup}b where we have labeled the length and capacitance to ground of each JJ by $a$ and $C_{0}$, respectively.  Using Kirchoff's laws, we can write the discrete equations of motion as
\begin{equation}\label{eq:discrete}
V_{n+1}-V_{n}=-\frac{d L_{n}I_{n}}{dt}\ \ ;\ \ I_{n+1}-I_{n}=-C_{0}\frac{d V_{n+1}}{dt}.
\end{equation}
 From (\ref{eq:inductance}), we see that by controlling the external flux bias, or by creating a varying current in the transmission line, we are able to modify the inductance and thus propagation velocity inside the transmission line.  Here, we focus on using the flux degree of freedom as our tunable parameter.  Creating a space-time varying current pulse, as in Ref.~\cite{philbin:2008}, can also be accomplished in our device.  However, our simplified model does not admit the correct dispersion relation to support the required stable nonlinear solitonic localized pulses in the parameter region of interest.  Charge solitons can however be produced in the high impedance regime of our device \cite{haviland:1996}.

\section{Effective Geometry and Hawking Temperature}\label{sec:black-effective}
\begin{figure}[htbp]
\begin{center}
\includegraphics[width=3.5in]{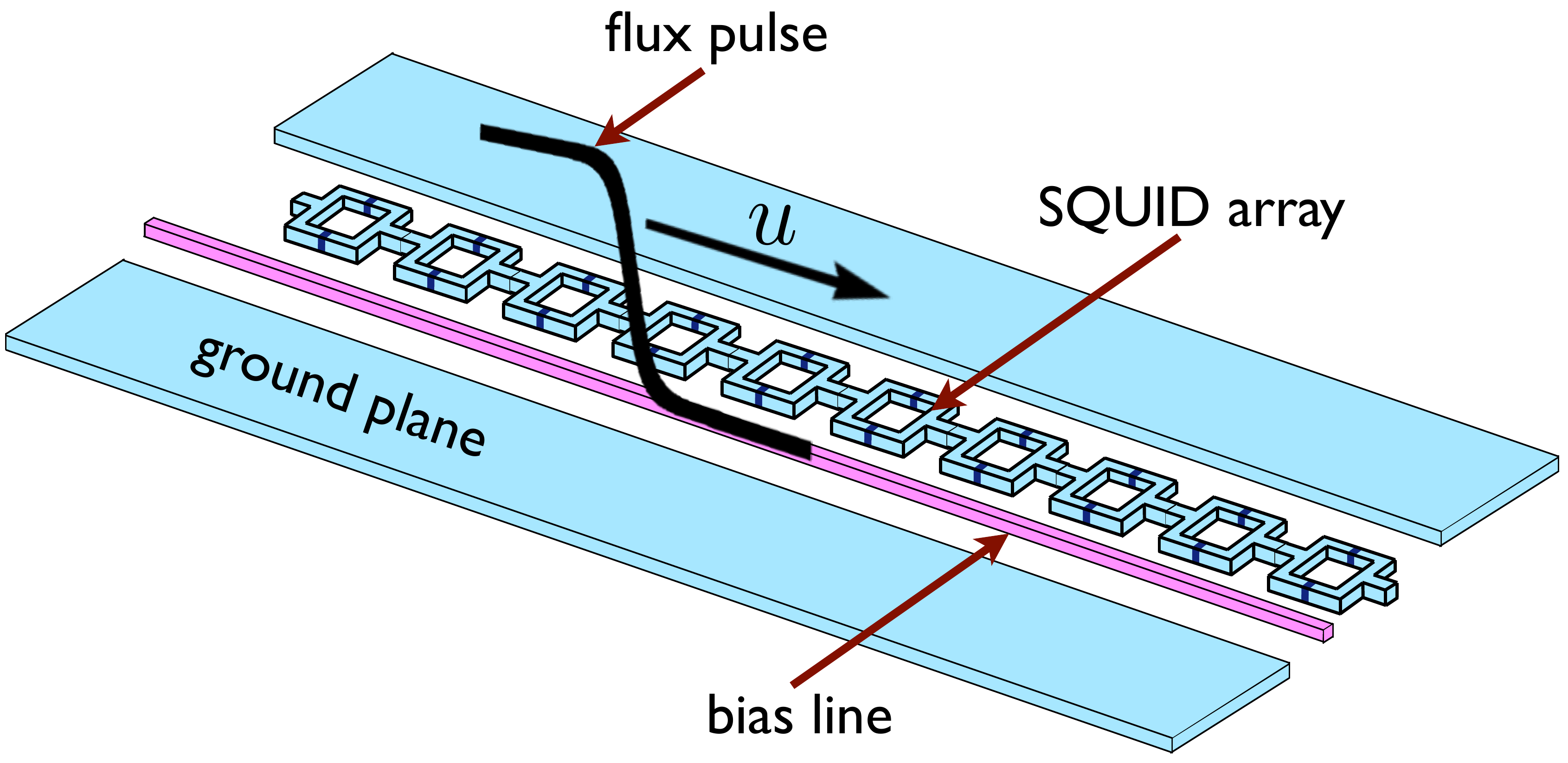}
\caption{Layout of the dc-SQUID array transmission line used in generating analogue Hawking radiation.  The flux pulse creates a space-time varying magnetic field that modulates the speed of light in the SQUID array leading to an effective horizon in the comoving frame where $u^{2}=c^{2}(x)$.}
\label{fig:array}
\end{center}
\end{figure}

\begin{figure}[htbp]
\begin{center}
\includegraphics[width=3.1in]{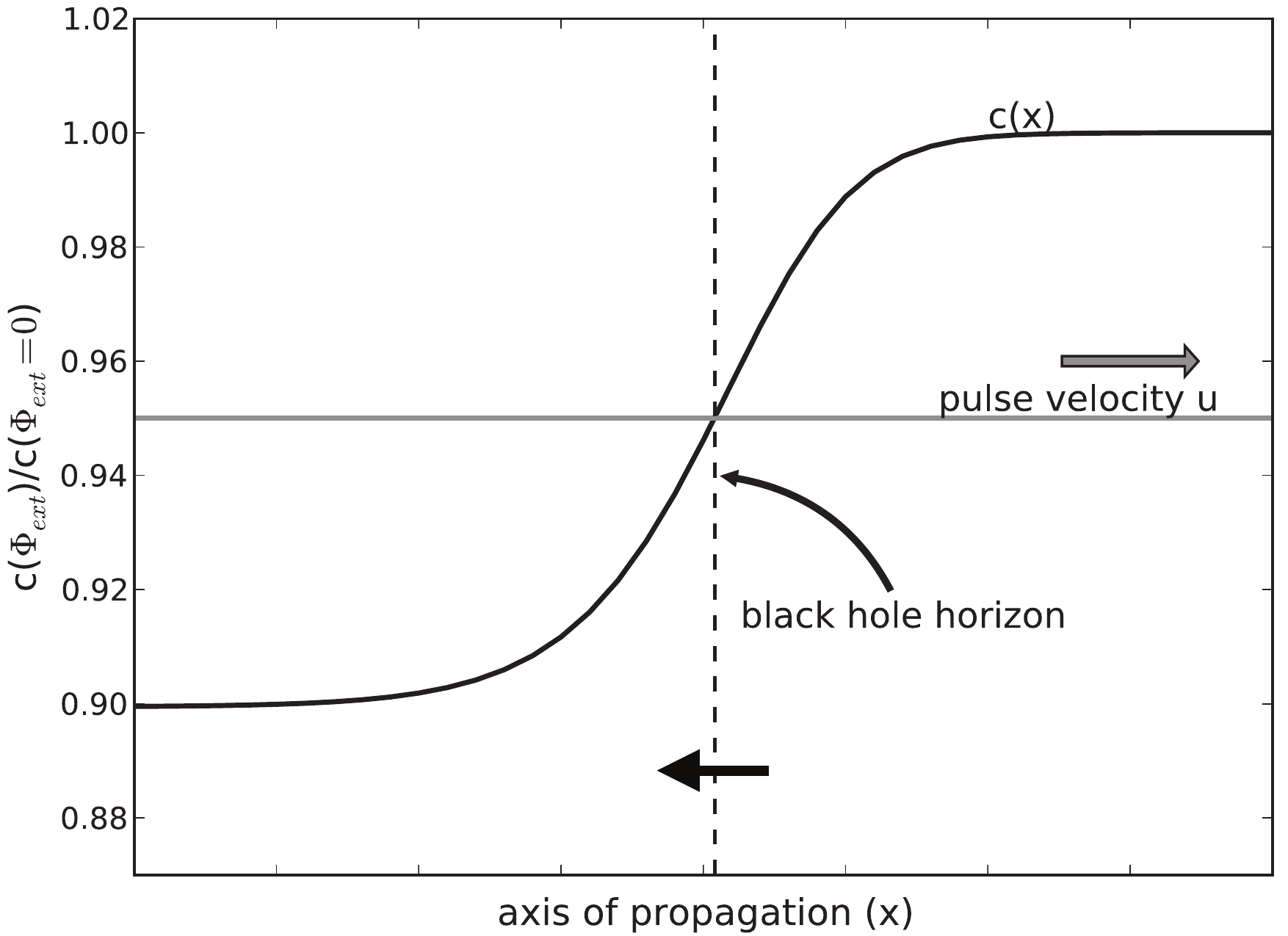}
\caption{Effect of a step-like flux pulse on the propagation velocity of a JJA as seen in the comoving frame.  The pulse velocity was chosen to be $u=0.95c\left(\Phi_{\mathrm{ext}}=0\right)$.  The black hole horizon occurs where $c(x)=u$.  Arrow indicates the only permissible direction of travel across the horizon.}
\label{fig:regions}
\end{center}
\end{figure}

By defining potentials $A_{n}$ such that $I_{n}=-C_{0}dA_{n}/dt$ and $V_{n}=A_{n}-A_{n-1}$ \cite{schutzhold:2005}, the equations of motion (\ref{eq:discrete}) can be combined to yield the discretized wave equation, 
\begin{equation}
\frac{d}{dt}L_{n}C_{0}\frac{d}{dt}A_{n}=A_{n+1}-2A_{n}+A_{n-1} .
\end{equation}
For wavelengths much longer than the dimensions of a single SQUID the dispersion relation becomes to lowest order in $k$:
\begin{equation}\label{eq:dispersion}
\omega^{2}(k)=\frac{4}{LC_{0}}\sin^{2}\left(\frac{ka}{2}\right)\approx c^{2}k^{2},
\end{equation}
where we have defined the velocity of propagation as $c=a/\sqrt{LC_{0}}$, which in practice is well below the vacuum speed of light $c_{0}$.  In this limit, the wave equation approaches the continuum
\begin{equation}
\left(\frac{\partial}{\partial t}\frac{1}{c^{2}}\frac{\partial}{\partial t}-\frac{\partial^{2}}{\partial x^{2}}\right)A=0.
\end{equation}
By ignoring higher-order terms in Eq.~(\ref{eq:dispersion}), we effectively remove the discreteness of the array which, along with dispersion from JJ inertia terms, can play the role of Planck scale physics in our system \cite{unruh:2005,philbin:2008,jacobson:1991}.  For parameter values considered below, the relevant short distance scale is $c/\omega_{p}~(>a)$.  Requiring the propagation speed to vary in \textit{both} space and time,
\begin{equation}\label{eq:c}
c^{2}\rightarrow c^{2}(x-ut),
\end{equation}
with fixed velocity $u$ set by an external flux bias pulse, Fig.~\ref{fig:array},
 the wave equation in the comoving frame becomes
\begin{equation}\label{eq:wave}
\left[\left(\frac{\partial}{\partial t}-u\frac{\partial}{\partial x}\right)\frac{1}{c^{2}}\left(\frac{\partial}{\partial t}-u\frac{\partial}{\partial x}\right)-\frac{\partial^{2}}{\partial x^{2}}\right]A=0,
\end{equation}
where $x$ and $t$ now label the comoving coordinates.  This wave equation can be re-expressed in terms of an effective space-time metric,
\begin{equation}\label{eq:metric}
g^{\mathrm{eff}}_{\mu\nu}=\begin{pmatrix}
 c^{2}-u^{2} & -u  \\ 
-u & -1
\end{pmatrix}.
\end{equation}
Comparing this metric with Eq.~(\ref{eq:painleve}), we see that our system contains a horizon located wherever $u^{2}=c^{2}(x)$.  In Fig.~\ref{fig:regions} we plot the effect of a step-like hyperbolic tangent flux bias pulse, similar to that shown in Fig.~\ref{fig:array}, with amplitude $\Phi_{\mathrm{ext}}=0.2\Phi_{0}$ on a JJA with inductances given by Eq.~(\ref{eq:inductance}), where we have kept only the lowest term in the $I_{c}/I^{s}_{c}$ expansion.
Additionally, since $\Phi_{\mathrm{ext}}$ can only increase the inductance, the flux-bias pulse velocity $u$ must be below the unbiased transmission line propagation velocity $c$ in order to establish a horizon.  We do not consider Gaussian or similar pulse shapes as they generate both black hole and white hole horizons \cite{hawking:1976} which complicates interpretation of the emission process.

So far, we have focused on demonstrating a classical effective background geometry with an event horizon. The next step is to quantize small perturbations in the potential field $A$ about this background. The correct commutation relations between quantum field operators are required for conversion of vacuum fluctuations into photons \cite{unruh:2003}.  These relations have been verified in the systems to which ours is analogous \cite{schutzhold:2005}.  The resulting Hawking temperature  is determined by the gradient of the JJA velocity at the horizon
\begin{equation}\label{eq:temp}
T_{H}=\frac{\hbar}{2\pi k_{b}}\left|\frac{\partial c(x)}{\partial x}\right|_{c^{2}=u^{2}}.
\end{equation}
The radiated power in the comoving frame coincides with the optimal rate for single-channel bosonic heat flow in one-dimension \cite{schutzhold:2005,blencowe:2000,meschke:2006}
\begin{equation}\label{eq:power}
\frac{dE}{dt}=\frac{\pi}{12\hbar}\left(k_{b}T_{H}\right)^{2}.
\end{equation}
Eq.~(\ref{eq:power}) is universal for bosons since the channel-dependent group velocity and density of states cancel each other in one dimension \cite{blencowe:2000}. For a detector at the end of the transmission line, the radiation emitted by an incoming bias pulse will be doppler shifted yielding higher power compared to Eq.~(\ref{eq:power}).  However, the rate of emitted photons remains approximately unchanged.

\section{Model Validity}\label{sec:black-model}
\begin{figure}[htbp]
\begin{center}
\includegraphics[width=3.1in]{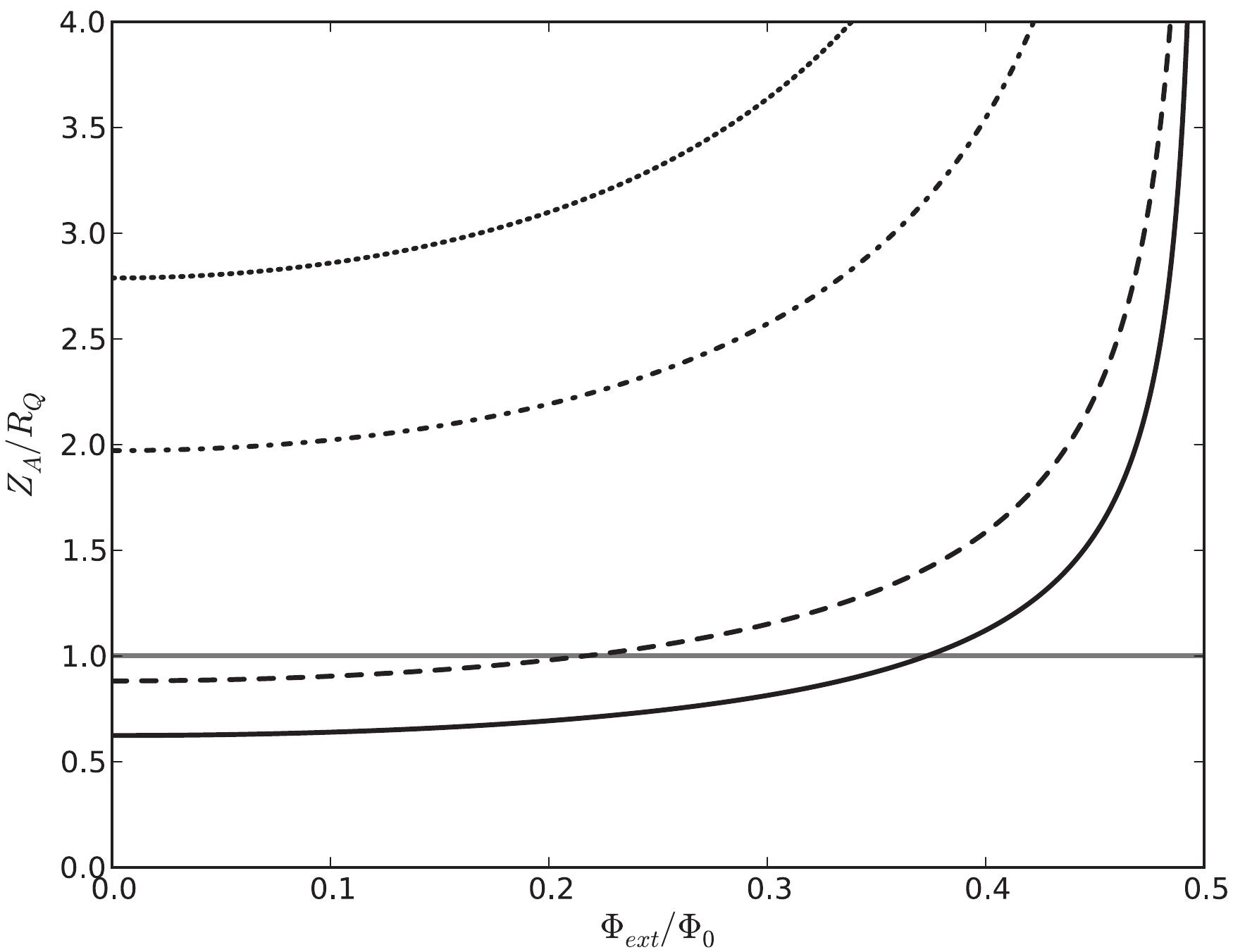}
\caption{Ratio of array impedance $Z_{A}$ to the resistance quantum $R_{Q}$ as a function of the external flux bias for a critical current $I_{c}=2~\mathrm{\mu A}$ and example ground capacitor values: $C_{0}$= $10^{-16}$(solid), $5\times10^{-17}$(dashed), $10^{-17}$(dash-dot), and $5\times10^{-18}\mathrm{F}$(dotted).  High and low impedance regions are defined above and below $Z_{A}/R_{Q}=1$ respectively.}
\label{fig:impedance}
\end{center}
\end{figure}

For a single effective JJ, the magnitude of quantum fluctuations in the phase variable $\gamma_{+}$ depends on both the ratio of Josephson energy, $E_{J}=\Phi_{0}I^{s}_{c}/2\pi$, to charging energy, $E_{C}=e^{2}/4C_{J}$, as well as on the impedance of the junction's electromagnetic environment.  These energy scales give a representation of the phase-charge uncertainty relation $\Delta\gamma\Delta Q\ge e$, and relate the amplitude of quantum fluctuations between these variables \cite{likharev:1986}.   When $E_{J}/E_{C}>1$ and the impedance seen by the junction is less than the resistance quantum, $R_{Q}=h/4e^{2}\approx6.45\mathrm{k}\Omega$, the phase operator behaves as a semiclassical quantity, i.e. the $\gamma_{+}$ quantum fluctuations are small with respect to its average, and the JJ is in the superconducting state, allowing for a lumped inductor approximation.  In the majority of experimental configurations, a single JJ is connected to probe leads with impedance $\sim50\Omega$ and as such is in the low-impedance regime $Z/R_{Q}\ll 1$.  In contrast, a JJA has an environment that comprises not only the input and output ports, but also all the other JJ's in the array.  In this case, we can define an effective impedance as seen by a single junction to be $Z_{J}=Z_{E}+Z_{A}$ where $Z_{E}$ is the environmental impedance of the leads and $Z_{A}$ is the array impedance that, for frequencies below the plasma frequency, can be written as \cite{haviland:2000}
\begin{equation}\label{eq:impedance}
Z_{A}=R_{Q}\sqrt{\frac{4E_{C}}{E_{J}}}\sqrt{\frac{C_{J}}{C_{0}}}=R_{Q}\sqrt{\frac{2\pi e^{2}}{\Phi_{0}C_{0}I_{c}}\sec\left(\pi\Phi_{\mathrm{ext}}/\Phi_{0}\right)},
\end{equation}
where the last equality explicitly shows the dependence on the external flux bias and single junction parameters.  Thus, even for a small energy ratio $E_{C}/E_{J}$, the lumped inductor model applies only when $Z_{A}/R_{Q}\lesssim 1$.   In Fig.~\ref{fig:impedance} we show the dependence of array impedance $Z_{A}$ on the external bias for fixed critical current $I_{c}=2~\mu\mathrm{A}$ and a range of experimentally valid capacitances to ground.
As $\Phi_{\mathrm{ext}}\rightarrow \Phi_{0}/2$, high impedance causes large phase fluctuations, indicating a breakdown of our semiclassical description; the array undergoes a quantum phase transition from superconducting to insulating Coulomb blockade behavior \cite{chow:1998}.  Note, the small JJ parameter variability in actual arrays \cite{beltran:2007,beltran:2008,chow:1998} will prevent the divergence in Fig.~\ref{fig:impedance}, as well as cause some transmission line scattering in the low impedance superconducting state.

The dependence of the Josephson energy $E_{J}$ on external flux $\Phi_{\mathrm{ext}}$ as described above allows for the systematic introduction of quantum fluctuations in our model.  With the phase variable governing the circuit inductance, these fluctuations manifest themselves in the effective metric (\ref{eq:metric}) through the propagation velocity $c$. As the amplitude of fluctuations increases, the metric becomes a quantum dynamical variable which must be included in the description of the Hawking process.  Thus, consequences of back-reaction from the Hawking process as well as quantum dynamical space-time can be probed by this configuration.  Both processes, not included in the original Hawking derivation, represent analogue quantum gravitational effects present in our system \cite{thiemann:2007}.

\section{Experimental Realization}\label{sec:black-experiment}
\begin{figure}[htbp]
\begin{center}
\includegraphics[width=4.5in]{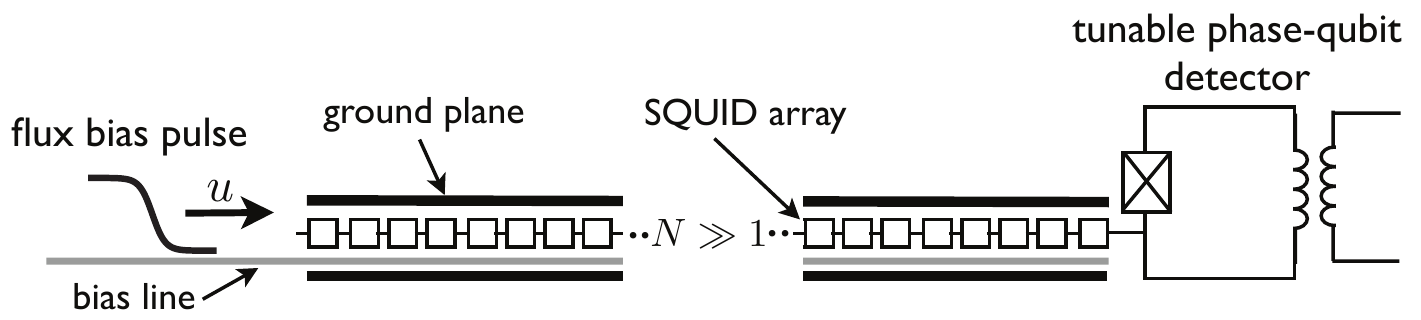}
\caption{Possible transmission line and detector realization.  Current pulses in the bias-line provide external flux necessary to modify the SQUID array propagation velocity.  A phase-qubit at the end of the JJA functions as the photon detector.}
\label{fig:experiment}
\end{center}
\end{figure}

A possible realization of the JJA is shown in Fig.~\ref{fig:experiment}, which consists of the JJA transmission line as well as an additional conducting line producing the space-time varying external flux bias $\Phi_{\mathrm{ext}}$.  To provide a space-time changing velocity, the JJA is modulated by generating current pulses in the bias-line, the propagation velocity of which are assumed to be slightly below that of the unbiased JJA.  The required bias pulse velocities $u$ can be achieved by similarly employing individual JJ's in series as the bias line.  Additionally, a dc-external flux can be used to fine-tune the transmission line velocity closer to that of the bias-line, eliminating the need for large amplitude current bias pulses.  Unavoidable current pulse dispersion in the bias-line, resulting in a decrease in Hawking temperature, can be minimized with appropriate choice of pulse shape and transmission line length.

Unambiguous verification of the Hawking process will require frequency-tunable, single-shot photon detection at the  end of the JJA opposite to that of the bias pulse origin.  Although not presently available, microwave single-photon detectors based on superconducting qubits are under active investigation \cite{wang:2008,romero:2009}.  We will assume a phase-qubit as our model detector \cite{wang:2008}.  By repeatedly injecting current pulses down the bias-line, the predicted blackbody spectrum associated with the Hawking process can be probed by tuning the qubit resonant frequency.  Correlations across the horizon between the emitted photon pairs can be established through coincidence detection.  We emphasize the essential need for correlation information in order to establish that a photon is produced by the Hawking effect rather than some other ambient emission process, or spuriously generated via capacitive coupling to the bias-line.  Unwanted directional coupling can be minimized with proper engineering of the transmission line.

To estimate the Hawking temperature we will assume parameters similar to those of Ref.~\cite{beltran:2007}, with SQUID's composed of tunnel junctions with $I_{c}=2~\mathrm{\mu A}$ and an upper bound achievable plasma frequency $\omega_{p}=2\pi\times10^{12}~\mathrm{Hz}$.  The capacitance to ground is assumed to be $C_{0}=5\times10^{-17}~\mathrm{F}$ (dashed line in Fig.~\ref{fig:impedance}).  Using a SQUID length $a=0.25~\mu\mathrm{m}$ gives an unbiased transmission line velocity $c\sim c_{0}/100$.  Equation~(\ref{eq:temp}) gives the temperature as determined by the rate at which the JJA transmission line velocity varies that, in our case, is limited by the plasma frequency $\omega^{s}_{p}$.  Assuming the maximum rate is an order of magnitude below $\omega^{s}_{p}\left(\Phi_{\mathrm{ext}}=0\right)/2\pi$, then the Hawking temperature is $\sim 120~\mathrm{mK}$; identical to a black hole with a mass of $10^{24}~\mathrm{kg}$, or equivalently a Schwarzschild radius of $1.5~\mathrm{mm}$.  This temperature can be a factor of ten larger than the ambient temperature set by a dilution refrigerator and therefore should be visible above the background thermal spectrum.  Using Eq.~(\ref{eq:power}) and the sample pulse in Fig.~(\ref{fig:regions}) gives an initial Hawking temperature $120~\mathrm{mK}$, which decreases $\sim10\%$ every $1000$ JJA elements due to bias-line dispersion.  Applying the power expression (\ref{eq:power}) yields an average emission rate of one photon per pulse for $\sim 4800$ SQUID's. Of course, the transmission line can be made considerably shorter at the expense of an increase in number of pulse repetitions in order to accumulate sufficient photon counts to verify the Hawking radiation.  The parameters and pulse shapes chosen here illustrate feasibility of this setup, but do not represent the only available configuration.  These values can likely be improved upon and optimized in terms of both performance and fabrication of this proposal. 

\section{Conclusion}\label{sec:black-conclusion}
We have demonstrated that an array of dc-SQUID's in a coplanar transmission line, when biased by a space-time dependent flux, creates an effective space-time metric with a horizon.  As a quantum device, the superconducting transmission line allows for the possibility of observing not only the Hawking effect, but also the effects of quantum fluctuations in an analogue gravitational system.

%%%

%% file: ch4.tex
%% intro.tex
\dsp
\chapter{The Trilinear Hamiltonian: Modeling Hawking radiation from a quantized source}\label{ch:trilinear}
%%%%%%%%%%%%%%%%%%%%%%%%%%%%%%%%%%%%%%%%%%%%%%%%
\section{Motivation}
We have shown that a magnetic field-pulsed microwave transmission line comprising an array of superconducting quantum interference devices, or SQUID's, not only reproduces physics analogous to that of a radiating black hole, but does so in a system where the high-energy and quantum mechanical properties are well understood and can be directly manipulated in the laboratory.  This may help in addressing the validity of several assumptions made by Hawking as to the ultra-high energy physics where quantum gravity effects are important.  Missing from the analysis presented in Chapter \ref{ch:black} is the energy loss due to the emission of analogue Hawking radiation, and therefore this setup does not address the process of black hole evaporation.  Indeed, the number of SQUIDs needed to produce a single photon per flux pulse is already pushing the current limits on experimental feasibility.  However, hidden in conventional derivations of black hole evaporation is a breakdown of the unitary evolution in quantum mechanics known as the Information Loss Paradox\cite{mathur:2009}.  Such a severe violation of the laws of quantum mechanics has stimulated a vast amount of literature devoted to solving this dilemma.  Given that the evaporation process requires understanding the laws of physics at the Planck scale, any solution must wait for a complete formulation of quantum gravity and is therefore likely out of reach for the foreseeable future.  Building on our previous work of condensed matter analogues for black hole dynamics, in this chapter we will examine a simple nonlinear quantum optics Hamiltonian as a toy model for the emission of Hawking radiation by an evaporating black hole.  Although the system presented here is highly simplified, we will show that it is capable of reproducing many of the dynamical features thought to be present in the original problem.  In addition, this work suggests that Hawking radiation may be entangled with gravitational degrees of freedom leading to nonclassical states of the black hole.

\section{Introduction}\label{sec:tri-intro}
In the 35 years since Hawking's seminal paper on the quantum emission of radiation from a black hole\cite{hawking:1974}, a large body of work has been devoted to solving the so called information loss problem.  For a black hole of fixed mass $M$, this emission process yields a black body spectrum with characteristic temperature $T_{\mathrm{H}}=\hbar c^{3}/ 8\pi k_{\mathrm{B}}GM$, irrespective of the initial state of the matter from which the black hole is formed.  The inability to reconstruct the initial, possibly pure state of the black hole from the total emitted radiation signals the apparent breakdown of unitary evolution and the $S$-matrix description of the Hawking process.  Although Hawking's calculations have since been verified in a number of ways\cite{hawking:1975,hartle:1976,boulware:1976,gibbons:1977,parentani:2000}, this breakdown at the foundation of quantum mechanics suggests that our understanding of black hole dynamics is not yet complete. 

The information loss problem rests on two key assumptions made in the standard derivation of Hawking radiation: (1) the perfectly thermal (i.e. mixed) character of the outgoing radiation; (2) the validity of this emission process over the lifetime of the black hole.  The notion of thermal spectrum considered here is not that of a blackbody frequency spectrum, but rather the quantum thermal probability distribution defined by the temperature $T_{H}$ of a single mode (single frequency) of Hawking radiation.  The traditional picture of the Hawking process leaves no room for deviations from this thermal distribution and thus breaks the requirement of pure-state$\rightarrow$pure-state evolution enforced by unitarity.  By itself, this process need not lead to information loss since the information content of the black hole may be stored in entanglement between particle pairs created on opposite sides of the horizon\cite{bombelli:1986,srednicki:1993}.  Results from many-body theory suggesting that entanglement across a boundary scales with the area of the boundary lends credence to this view\cite{eisert:2010}.  However, with the second assumption, the black hole causes a loss of entanglement and thus information.

With the expectation that information must be conserved, many suggestions for resolving the information loss problem have been put forward.  Current proposed solutions include long-lived and stable Planck-scale remnants\cite{aharanov:1987,giddings:1992}, baby universes\cite{hawking:1988,frolov:1990}, and the possibility of information escaping as non-thermal Hawking radiation \cite{page:1993,parikh:2000}.  In all of these proposals, corrections to the Hawking process manifest themselves when the black hole has evaporated to a size near that of the Planck length, $l_{p}=\sqrt{\hbar G/c^{3}}\approx 10^{-35}~\mathrm{m}$, at which quantum gravitational effects, neglected in Hawking's original analysis, are expected to play a role.  In considering quantum states of the gravitational field, there is the possibility of back-reaction and entanglement of the radiating matter degrees of freedom with those of gravity\cite{hawking:2005,terno:2005}.  Although it is natural to consider a quantized gravitational field for the Hawking process, the current lack of a full quantum mechanical description of gravity severely limits progress in directly addressing this scenario.  In fact, exactly which degrees of freedom, if any, should be quantized is still subject to debate\cite{jacobson:1995,carlip:2008}.  In this paper we investigate a simple, zero-dimensional quantum optics model of Hawking radiation that mimics some of the essential physics present in the original information loss problem.

\begin{figure}[htbp]
\begin{center}
\includegraphics[width=4.0in]{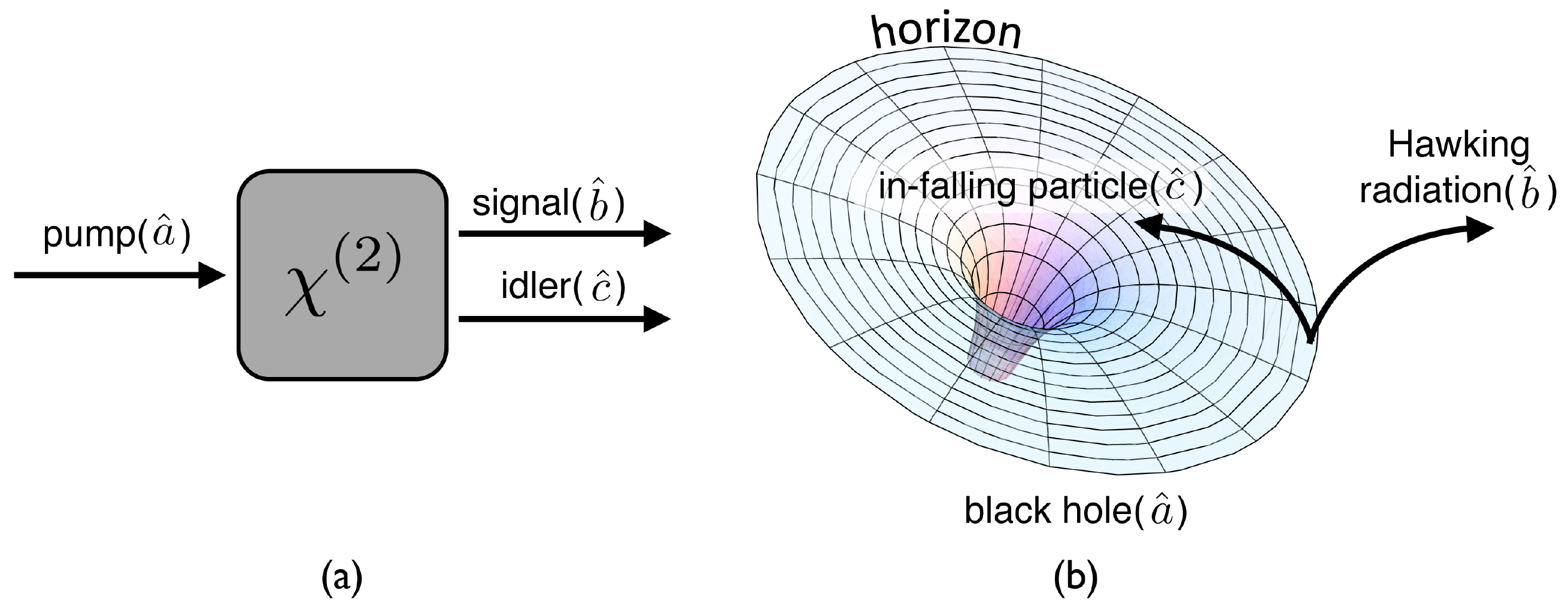}
\caption{a.)Diagram depicting the dynamics of the trilinear Hamiltonian.  Initial vacuum modes are not shown for clarity.  The nonlinear interaction is generated by a system possessing a second-order susceptibility $\chi^{(2)}$.  b.)Equivalent dynamical elements involved in the Hawking process.}
\label{fig:relation}
\end{center}
\end{figure}

As a zero-dimensional model we consider the following trilinear Hamiltonian:
\begin{equation}\label{eq:trilinear}
\hat{H}=\hbar\omega_{a}\hat{a}^{+}\hat{a}+\hbar\omega_{b}\hat{b}^{+}\hat{b}+\hbar\omega_{c}\hat{c}^{+}\hat{c}+i\hbar\chi\left(\hat{a}\hat{b}^{+}\hat{c}^{+}-\hat{a}^{+}\hat{b}\hat{c}\right),
\end{equation}
consisting of three harmonic oscillator modes with the frequency relation, $\omega_{a}=\omega_{b}+\omega_{c}$.  We will designate the modes as pump ($\hat{a}$), signal ($\hat{b}$), and idler ($\hat{c}$) respectively. This Hamiltonian describes several quantum optics processes including frequency conversion, Raman and Brillouin scattering, the interaction of two-level atoms with a single mode resonant EM field, and is the full quantum generalization of the parametric amplifier\cite{dicke:1953,mollow:1967,travis:1968,tucker:1969,walls:1970,lu:1973,agrawal:1974,mcneil:1983}.  The connection between black hole radiance and  parametric amplification was appreciated shortly after Hawking's discovery\cite{gerlach:1976}. Both processes amplify vacuum fluctuations resulting in the production of correlated photon pairs.  Tracing over one of the two subsystems (i.e., signal and idler) yields statistics that are identical to that of a thermal distribution\cite{barnett:1985,yurke:1987}.  The energy source in the parametric amplifier is assumed to be a classical pump such as a laser or microwave generator with fixed amplitude driving a system with $\chi^{(2)}$ nonlinearity, the first nonlinear susceptibility in a medium without inversion symmetry\cite{boyd:2003}.  Viewed as a black hole model, the pump plays the role of black hole mass $M$ while the signal and idler modes of the parametric amplifier represent the escaping and trapped Hawking photons, respectively, as depicted schematically in Fig.~\ref{fig:relation}.
The trilinear Hamiltonian (\ref{eq:trilinear}) generalizes the parametric amplifier by quantizing the pump mode and allowing for energy loss to the signal and idler modes.  The expectation value of the pump mode energy is analogous to the mass $M$ of a quantum mechanical black hole.  We will explore this model by establishing the conditions under which the signal mode spectrum of the trilinear Hamiltonian deviates from the predicted  thermal spectrum of the conventional parametric amplifier.  We shall see that the quantization of the pump mode degree of freedom results over time in entanglement with the signal and idler modes and dynamics that become markedly different from the parametric approximation. In particular, the signal mode develops a strongly non-thermal spectrum that is dependent on the initial pump mode state. The corresponding entropy is reduced relative to that of a thermal (maximally mixed) state indicating the presence of information.  These model system results lend support to the view that late-time Hawking radiation contains an increasing amount of information about the initial quantum state of the black hole and is composed of particles entangled with quantized gravitational states. 

The outline of this chapter is as follows:  In Sec.~\ref{sec:tri-parametric} we review the derivation of the amplification of vacuum fluctuations under the parametric approximation.  Sec.~\ref{sec:tri-semiclassical} considers the semi-classical approximation whereby the back-reaction from the quantized radiation onto the classical pump is accounted for and derives the self-consistent equations of motion for the signal and idler modes.  In Sec.~\ref{sec:tri-quantum} we consider the full quantum dynamics Eq.~(\ref{eq:trilinear}) under the short time analytical approximation and compare this result along with the full quantum numerical solution to the parametric and semi-classical approximations from the previous sections.  Sec.~\ref{sec:tri-entanglement} investigates the role of entropy and entanglement in the production of non-thermal states.  In Sec.~\ref{sec:tri-circuit} we give a possible microwave cavity realization of the trilinear Hamiltonian.  Finally, Sec.~\ref{sec:tri-conclusion} concludes with a brief discussion of the results and consequences for black hole evaporation.  Details of the numerical calculations are presented in Appendix~\ref{sec:ap2-trilinear}.

\section{Parametric Amplifier and Hawking Emission}\label{sec:tri-parametric}
In the following,  we  derive the well-known thermal spectrum of the signal mode under the parametric assumption of a fixed amplitude pump mode.  Replacing the pump mode in Eq.~(\ref{eq:trilinear}) with a fixed amplitude drive $A$ results in the interaction frame Hamiltonian
\begin{equation}\label{eq:parametric}
H_{I}=i\hbar\chi A\left(b^{+}c^{+}-bc\right).
\end{equation}
The Heisenberg equations of motion for the signal and idler mode operators are,
\begin{equation}
\frac{db(t)}{dt}=A\chi c(t)^{+};\qquad \frac{dc(t)}{dt}=A\chi b(t)^{+}.
\end{equation}
These can be readily solved yielding the Bogoliubov transformations,
\begin{eqnarray}\label{eq:semi-motion}
b(\tau)&=&b(0)\cosh\left(A\tau\right)+c(0)^{+}\sinh\left(A\tau\right) \\
c(\tau)&=&c(0)\cosh\left(A\tau\right)+b(0)^{+}\sinh\left(A\tau\right)\nonumber,
\end{eqnarray}
where we have expressed the dynamics in dimensionless time $\tau=\chi t$.   If the system starts with both signal and idler modes in the ground state, $\left|\Psi(0)\right>=\left| 0,0\right>_{bc}$, then Eq.~(\ref{eq:semi-motion}) gives for the number operators $N_{b}$ and $N_{c}$:
\begin{equation}\label{eq:occupation}
N_{b}(\tau)=N_{c}(\tau)=\sinh^{2}\left(A\tau\right).
\end{equation}
Additionally, we are interested in the probability distribution of the individual signal and idler subsystems in the number state basis.  With the system initially in the ground state, the unitary evolution corresponding to Eq.~(\ref{eq:parametric}) can be expressed as,
\begin{equation}
\left| \Psi(\tau)\right>=\exp\left[A\tau\left(b^{+}c^{+}-bc\right)\right]\left| 0,0\right>_{bc}.
\end{equation}
Making use of the disentangling theorem\cite{truax:1985},
\begin{equation}\label{eq:disentangle}
\exp\left[A\tau\left(b^{+}c^{+}-bc\right)\right]=e^{\Gamma b^{+}c^{+}}e^{-g\left(b^{+}b+c^{+}c+1\right)}e^{-\Gamma bc},
\end{equation} 
with $\Gamma=\tanh\left(A\tau\right)$ and $g=\ln\cosh\left(A\tau\right)$,  where the last term in Eq.~(\ref{eq:disentangle}) vanishes for the ground state and the middle term reduces to $e^{-g}$, we have
\begin{equation}
e^{\Gamma b^{+}c^{+}}\left| 0,0\right>_{bc}=\sum_{n=0}^{\infty}\frac{\Gamma^{n}\left(b^{+}c^{+}\right)^{n}}{n!}\left| 0,0\right>_{bc}=\sum_{n=0}^{\infty}\tanh^{n}\left(A\tau\right)\left| n,n\right>_{bc},
\end{equation}
with the evolving state vector $\left| \Psi(\tau)\right>$ given by 
\begin{equation}\label{eq:squeezed}
\left| \Psi(\tau)\right>=\mathrm{sech}\left(A\tau\right)\sum_{n}\tanh^{n}\left(A\tau\right)\left| n,n\right>_{bc}.
\end{equation}
Let us now focus on operators acting on a subsystem spanned by the states of either the signal or idler mode individually.  Assuming we are interested in the signal mode only, tracing over the idler subsystem in Eq.~(\ref{eq:squeezed}) gives the operator expectation value
\begin{equation}\label{eq:semi-spectrum}
\left< O(\tau)\right>_{b}=\mathrm{sech}^{2}\left(A\tau\right)\sum_{n}\tanh^{2n}\left(A\tau\right)\left< n|O| n\right>_{b}
\end{equation}
for an arbitrary signal mode operator $O$.  Comparing Eq.~(\ref{eq:semi-spectrum}) with the spectrum of a thermal state defined by temperature $T$,
\begin{equation}\label{eq:semi-thermal}
\left< O\right>=\sum_{n}P_{n}\left< n_{b}|O|n_{b}\right>=\left(1-e^{-\hbar\omega_{b}/k_{\mathrm{B}}T}\right)\sum_{n}e^{-n\hbar\omega_{b}/k_{\mathrm{B}}T}\left< n_{b}|O|n_{b}\right>,
\end{equation}
indicates that the signal mode is in a thermal state provided we define the temperature as
\begin{equation}\label{eq:temperature}
T\left(\tau\right)=\frac{\hbar\omega_{b}}{2k_{\mathrm{B}}\ln\left[\coth\left(A\tau\right)\right]},
\end{equation}
where the time dependence is a consequence of the rapid increase in the occupation number of the modes, Eq.(\ref{eq:occupation}).  As in the Hawking process\cite{fuentes:2005}, the particle pairs generated by the parametric amplifier form an entangled two-mode squeezed state given by Eq.~(\ref{eq:squeezed})\cite{walls:2008}.  The bipartite structure of this system allows calculating the entanglement between particle pairs via the Von-Neumann entropy $S_{i}$, also referred to as entanglement entropy. For mode $i=b,c$,  using the reduced density matrix $\rho_{i}$ we have
\begin{equation}\label{eq:VNentropy}
S_{i}=-\mathrm{Tr}\left(\rho_{i}\ln\rho_{i}\right),
\end{equation}
where we drop the usual $k_{\mathrm{B}}$ factor.  For both the signal and idler modes, this entropy is given by
\begin{equation}\label{eq:thermal-entropy}
S^{\mathrm{th}}=-\ln\left[1-e^{-\hbar\omega/k_{\mathrm{B}}T(\tau)}\right]-\frac{\hbar\omega}{k_{\mathrm{B}}T(\tau)}\left[1-e^{\hbar\omega/k_{\mathrm{B}}T(\tau)}\right]^{-1},
\end{equation}
which is the thermal entropy of a quantum harmonic oscillator with temperature defined by Eq.~(\ref{eq:temperature}).  Thus, as for the Hawking thermal radiation,  we see that the temperature of  the parametric oscillator is determined by the entropy generated from tracing over one of the two modes in a particle pair squeezed state. 

\section{Semi-Classical Analysis: Backreaction on a Classical Pump Mode}\label{sec:tri-semiclassical}
We now go beyond the just-considered parametric amplifier model and incorporate backreaction effects due to the emission process using a semi-classical approximation where the pump mode is treated as a classical variable, while the signal and idler modes are quantized.  In order to self consistently solve for the system evolution, we first work out the dynamics of the pump mode assuming it behaves as a classical variable affected by the expectation values of the signal and idler modes, after which we substitute the resulting c-number expressions for the pump operators into the Hamiltonian and calculate the signal and idler evolution.  A similar procedure arises for the semi-classical Einstein equations:
\begin{equation}
G_{\mu\nu}=\frac{8\pi G}{c^{4}}\left<\Psi |\hat{T}_{\mu\nu}|\Psi\right>,
\end{equation}
where the mass $M$ black hole spacetime geometry is modeled classically through the usual Einstein tensor $G_{\mu\nu}$,  interacting with the quantum matter/radiation fields via the expectation value of the stress-energy-momentum tensor operator $\hat{T}_{\mu\nu}$ with respect to a suitable incoming quantum field state $\left| \Psi\right>$.  Solving these semiclassical equations using an analogous procedure to that outlined above results in a nonlinear Schr\"{o}dinger equation\cite{kibble:1980} and also a possible loss of spacetime stability\cite{horowitz:1980}. In addition, issues such as non-causal dynamics\cite{anselmi:2007} and  possible inconsistencies with wavefunction collapse\cite{unruh:1984} occur.  In contrast, the semi-classical model presented here can be readily solved by exploiting the symmetries present in the Hamiltonian.  

To begin, we consider Eq.~(\ref{eq:trilinear}) in the interaction frame,
\begin{equation}\label{eq:interaction}
H_{I}=i\hbar\chi\left(ab^{+}c^{+}-a^{+}bc\right)
\end{equation}
and obtain the following mode equations,
\begin{equation}\label{eq:semiQ_dt}
\frac{da}{dt}=-\chi bc;\qquad  \frac{db}{dt}=\chi ac^{+}; \qquad  \frac{dc}{dt}=\chi ab^{+}
\end{equation}
that lead to the evolution of the pump mode number operator,
\begin{equation}
\frac{dN_{a}}{dt}=-\chi\left(ab^{+}c^{+}+a^{+}bc\right).
\end{equation}
It is easy to show the corresponding number operators for signal and idler modes are given by $dN_{b}/dt=dN_{c}/dt=-dN_{a}/dt$. To proceed, we will use the following Manley-Rowe constants of motion\cite{manley:1956},
\begin{equation}\label{eq:MR}
M_{ab}=N_{a}+N_{b};\  \ M_{ac}=N_{a}+N_{c};\  \ M_{bc}=N_{b}-N_{c},
\end{equation}
expressing the underlying SU(2) and SU(1,1) symmetries in our model\cite{yurke:1986,brif:1996}. Differentiating Eqs.~(\ref{eq:semiQ_dt}) again results in decoupled equations of motion containing only commuting operators.
\begin{eqnarray}\label{eq:number}
\frac{d^{2}N_{a}}{dt^{2}}&=&2\chi^{2}\left[3N_{a}^{2}-N_{a}\left(2M_{ab}+2M_{ac}+1\right)+M_{ab}M_{ac}\right]\nonumber \\
\frac{d^{2}N_{b}}{dt^{2}}&=&-2\chi^{2}\left[3N_{b}^{2}-N_{b}\left(4M_{ab}-2M_{ac}-1\right)+M_{ab}\left(M_{ab}-M_{ac}-1\right)\right]\\
\frac{d^{2}N_{c}}{dt^{2}}&=&-2\chi^{2}\left[3N_{c}^{2}-N_{c}\left(4M_{ac}-2M_{ab}-1\right)+M_{ac}\left(M_{ac}-M_{ab}-1\right)\right]\nonumber
\end{eqnarray}
We now take the expectation value of the pump number operator equation~(\ref{eq:number}) for $N_a$ and make the approximation $\left<N_a^2\right>\approx\left<N_a\right>\left< N_a\right>$, the validity of which can be checked using full quantum numerical simulations of Eq.~(\ref{eq:semiQ_dt}) [see Sec.~\ref{sec:tri-quantum}].  These approximations, along with the condition that both signal and idler mode start in the vacuum state, lead to the semiclassical evolution for the pump mode:
\begin{equation}\label{eq:semiNa}
N_{a}(\tau)=\beta_{+}+\left[N_{a}(0)-\beta_{+}\right]\mathrm{dn}\left[\sqrt{\beta_{+}-\beta_{-}}\tau,\frac{N_{a}(0)-\beta_{-}}{\beta_{+}-\beta_{-}}\right]^{-2},
\end{equation}
where $dn(u,k)$ is the Jacobi elliptic function and
\begin{equation}\label{eq:semibeta}
\beta_{\pm}=\frac{1}{4}\left[ 1+2N_{a}(0)\pm\sqrt{1+12N_{a}(0)+4N_{a}(0)^{2}}\right].
\end{equation}
It is important to note that both Eqs.~(\ref{eq:semiNa}) and (\ref{eq:semibeta}) are expressed in terms of the initial conditions of the pump mode only, a consequence of  the relations (\ref{eq:MR}) and the fact that the signal and idler are initially in the vacuum (ground) state. Equations of motion for both signal and idler can then be obtained by substitution of the c-number expressions for the pump mode with amplitude given by Eq.~(\ref{eq:semiNa}),
\begin{equation}\label{eq:cnum}
a=\sqrt{N_{a}(\tau)}e^{-i\phi(t)};\ \
a^{+}=\sqrt{N_{a}(\tau)}e^{i\phi(t)},
\end{equation}
where $\phi(t)$ is a slowly varying function of time, into Eq.~(\ref{eq:interaction}):
\begin{equation}
\tilde{H}_{I}=i\hbar\chi N_{a}(t)^{1/2}\left(b^{+}c^{+}e^{-i\phi(t)}-bc e^{i\phi(t)}\right).
\end{equation}
The time evolution of this now bilinear Hamiltonian can be straightforwardly calculated if we assume \cite{lu:1973}
\begin{equation}\label{eq:phase1}
\sqrt{N_{a}(t)}\frac{d\phi(t)}{dt}=\mathrm{const},
\end{equation}
which is the most general condition under which the Hamiltonian at different times commutes, $\left[\tilde{H}_{I}(t),\tilde{H}_{I}(t')\right]=0$.  Combining Eqs.~(\ref{eq:semiQ_dt}) and (\ref{eq:cnum}) results in the phase relation
\begin{equation}
a^{+}\frac{d a}{dt}-a\frac{d a^{+}}{dt}=-2i N_{a}\frac{d\phi(t)}{dt},
\end{equation}
which, expressing the left-hand side as Eq.~(\ref{eq:interaction}) through the Heisenberg equations for the operators Eq.~(\ref{eq:semiQ_dt}), yields a second condition for the pump phase:
\begin{equation}\label{eq:phase2}
2N_{a}(t)\frac{d\phi(t)}{dt}=\left<\tilde{H}_{I}\right>.
\end{equation}
As long as the initial state contains at least one mode in the vacuum state, we have the constant of motion $\left<H_{I}\right>=0$. Thus, Eqs.~(\ref{eq:phase1}) and (\ref{eq:phase2}) are only consistant if $\phi(t)=0$.  This leads to a bilinear Hamiltonian of the form,
\begin{equation}
\tilde{H}_{I}=i\hbar\chi N_{a}(t)^{1/2}\left(b^{+}c^{+}-bc\right).
\end{equation}
Using a similar analysis to that employed in Sec.~\ref{sec:tri-parametric} gives,
\begin{eqnarray}
\frac{d b(\tau)}{d\tau}&=&b(0)\cosh\left(\theta(\tau)\right)+c^{+}(0)\sinh\left(\theta(\tau)\right)\\
\frac{d c(\tau)}{d\tau}&=&c(0)\cosh\left(\theta(\tau)\right)+b^{+}(0)\sinh\left(\theta(\tau)\right)\nonumber,
\end{eqnarray}
where,
\begin{equation}
\theta(\tau)=\int_{0}^{\tau}\sqrt{N_{a}\left(\tau'\right)}d\tau'.
\end{equation}
Therefore, we see that each mode behaves similarly as in the parametric approximation but with the number of particles dependent on the pump dynamics:
\begin{equation}
N_{b}(\tau)=N_{c}(\tau)=\sinh^{2}\left(\theta(\tau)\right).
\end{equation}
However, most importantly the modes are again in a thermal state with distribution
\begin{equation}
\left< O(
\tau)\right>=\mathrm{sech}^{2}\left(\sqrt{N_{a}(\tau)}\tau\right)\sum_{n}\tanh^{2n}\left(\sqrt{N_{a}(\tau)}\tau\right)\left<n| O| n\right>.
\end{equation}
Thus, the spectrum of the signal mode (Hawking radiation) remains thermal when pump mode energy loss due to signal and idler particle creation is taken into account within the semi-classical approximation.   

\section{Full Quantum Description}\label{sec:tri-quantum}

\subsection{Short-time Approximation}\label{sec:short-time}
\begin{figure}[htbp]
\begin{center}
\includegraphics[width=5.5in]{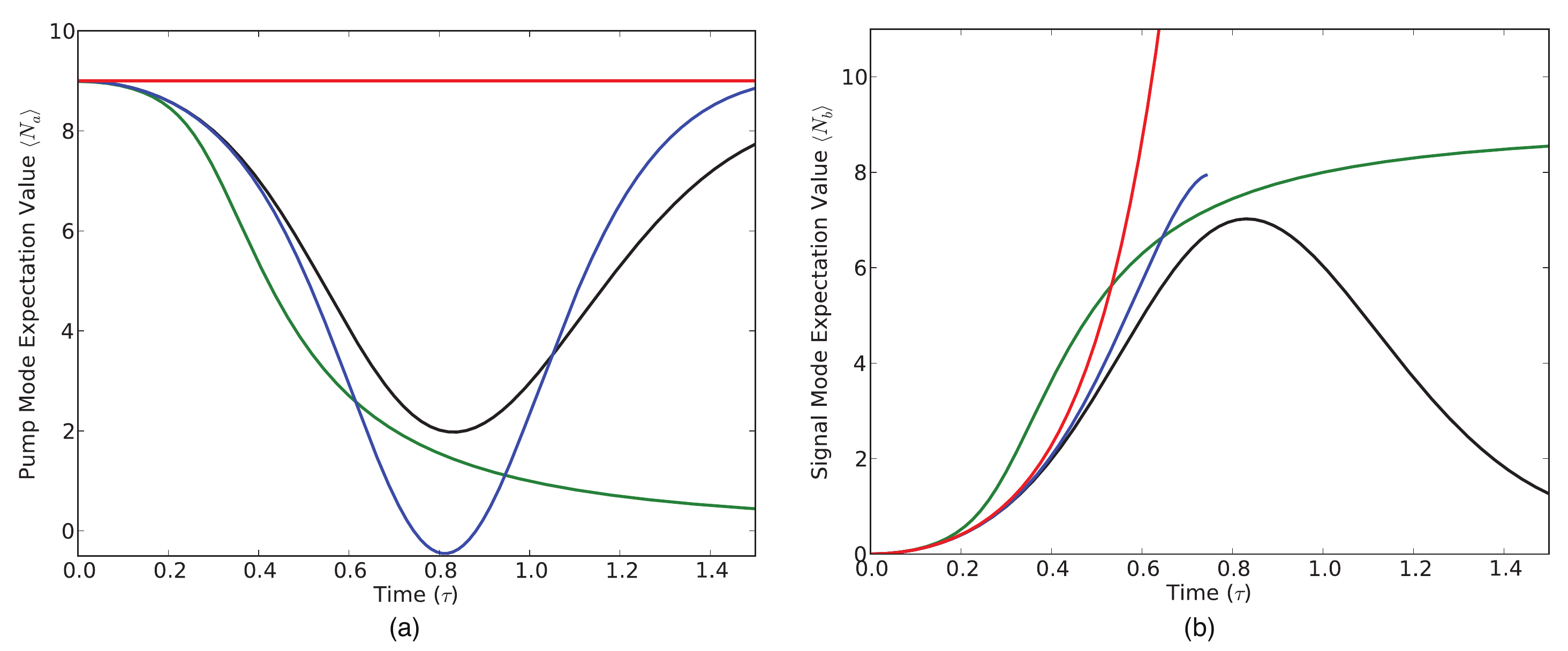}
\caption{a.)Evolution of pump mode initially in coherent state $\left< N_{a}(0)\right>=9$ for the parametric (red), semi-classical (blue),  short-time quantum approximation (green), and full quantum numerical solution (black), as a function of dimensionless time $\tau=\chi t$. b.)Corresponding population of the signal mode.  Note that the semi-classical analysis ceases to be valid when the pump mode becomes depleted.}
\label{fig:averages}
\end{center}
\end{figure}

In this section we consider the full quantum dynamics of the trilinear Hamiltonian Eq.~(\ref{eq:trilinear}), (i.e. quantize the pump mode as well).  Although the trilinear system does not allow for a general analytical solution, we can obtain approximate equations in the short time limit $\tau=\chi t\ll 1$.  Using the condition $\omega_{b}=\omega_{c}=\omega_{a}/2$, appropriate for modeling particle generation by a black hole, we can rewrite Eq.~(\ref{eq:trilinear}) ignoring constant terms as
\begin{equation}\label{eq:fullquantumH}
H=H_{0}+H_{\mathrm{int}}=\hbar\omega_{a}\left(a^{+}a+K_{z}\right)+i\hbar\chi\left(aK_{+}-a^{+}K_{-}\right),
\end{equation}
where
\begin{equation}
K_{+}=b^{+}c^{+},\ \ K_{-}=bc,\ \ K_{z}=\frac{1}{2}\left(b^{+}b+c^{+}c+1\right).
\end{equation}
We now construct the Casimir operator,
\begin{equation}
K^{2}=K_{z}^{2}-\frac{1}{2}\left(K_{+}K_{-}+K_{-}K_{+}\right)=\frac{1}{4}\left(M^{2}_{bc}-1\right),
\end{equation}
where the second equality comes from  definition (\ref{eq:MR}).  This operator obeys the eigenvalue equation
\begin{equation}
K^{2}\left|\Psi\right>=k(k-1) \left|\Psi\right>,
\end{equation}
where $k=1/2\left(\left| M_{bc}\right|+1\right)$.  We now look for simultaneous eigenstates of both $K^{2}$ and $K_{z}$: $\left| k,n_{c}\right>$, where $n_{c}$ is the number of idler particles.  These states can also be decomposed using the number state basis
\begin{equation}
\left| k,n_{c}\right>=\left| n_{c}+2k-1\right>_{b}\left| n_{c}\right>_{c}=\left| n_{b}\right>\left| n_{c}\right>,
\end{equation}  
with $n_{b}$ representing the number of particles in the signal.  Operating on this state with $K_{z}$ gives
\begin{equation}
K_{z}\left| k,n_{c}\right>=\left(k+n_{c}\right) \left| k,n_{c}\right>=\left(\frac{n_{b}+n_{c}+1}{2}\right)\left| n_{b}\right>\left| n_{c}\right>,
\end{equation}
valid only for initial conditions where both signal and idler modes are in pure states satisfying $n_{b}\ge n_{c}$.  Including the pump mode state in the full state vector and expressing the idler mode population as a function of the pump amplitude
\begin{equation}\label{eq:quantum-state}
\left| n_{a}\right>\left| k,M_{ac}-n_{a}\right>,
\end{equation}
one may switch to the interaction frame where the evolution depends only on $H_{\mathrm{int}}$ (\ref{eq:fullquantumH})
\begin{equation}\label{eq:quantum-evolution}
\left| \tau;k,M_{ac}\right>=e^{\tau\left(aK_{+}-a^{+}K_{-}\right)}\left|0;k,M_{ac}\right>,
\end{equation}
with the initial state with $M_{ac}=N_{a}(0)$ is given by
\begin{equation}\label{eq:initial}
\left| 0;k,M_{ac}\right>=\left| M_{ac}\right>\left|k,0\right>.
\end{equation}
The short time approximation $\tau=\chi t\ll1$ to Eq.~(\ref{eq:quantum-evolution}) can be calculated using the Baker-Campbell-Hausdorff formula truncated to $O\left(\tau^{2}\right)$
\begin{equation}\label{eq:truncate}
e^{\tau\left(aK_{+}-a^{+}K_{-}\right)}\approx e^{\tau aK_{+}}e^{-\tau a^{+}K_{-}}e^{-\frac{\tau^2}{2}\left[aK_{+},a^{+}K_{-}\right]}.
\end{equation}
Acting with the third exponential operator term on our initial state gives
\begin{equation}
e^{-\frac{\tau^2}{2}\left[aK_{+},a^{+}K_{-}\right]}\left| M_{ac}\right>\left|k,0\right>\approx \left(1-kM_{ac}\tau^{2}\right)\left| M_{ac}\right>\left|k,0\right>
\end{equation}
indicating clearly that the time over which the approximation is valid $t=\tau/\chi<1/(\chi\sqrt{kM_{ac}})$ decreases with pump amplitude and coupling strength.  Evaluating the BCH approximated expression~(\ref{eq:truncate}) on the initial state~(\ref{eq:initial}),  we obtain for the full evolution of the state:
\begin{equation}\label{eq:zstate}
\left|\tau;k,M_{ac}\right>=\frac{1}{\sqrt{N_{M_{ac}}(\tau)}}\sum_{n=0}^{M_{ac}}f_{n}\left(k,M_{ac}\right){\tau}^{n}\left|M_{ac}-n\right>\left|k,n\right>,
\end{equation}
where $N_{M_{ac}}(\tau)$ is the time-dependent normalization factor
\begin{equation}
N_{M_{ac}}(\tau)=e^{\tau^{-2}}\tau^{2M_{ac}}\Gamma\left(M_{ac}+1,\tau^{-2}\right),
\end{equation}
with $\Gamma\left(a,b\right)$ the reduced gamma function and
\begin{equation}
f_{n}\left(k,M_{ac}\right)=\left[\frac{M_{ac}!\Gamma\left(2k+n\right)}{n!\left(M_{ac}-n\right)!\Gamma\left(2k\right)}\right]^{1/2}.
\end{equation}
Our interest in the crossover from classical to quantum dynamics for the pump mode suggests that we use coherent states built from linear combinations of Eq.~(\ref{eq:zstate}) for the initial pump state.  To this end, we consider a general initial state
\begin{equation}\label{eq:initial-state}
\left|\Psi(0)\right>=\sum_{s=0}^{\infty}a_{s}\frac{f_{0}(s)}{\sqrt{N_{s}(0)}}\left|s\right>\left|0\right>\left|0\right>,
\end{equation}
with the pump mode in an as yet unspecified initial pure state with probability $P_{s}=\left|a_{s}\right|^2$ of being in state $s$ and both signal and idler modes in the vacuum state ($k=1/2$).  Here we implicitly assume the probabilities are normalized and add to unity.  The density matrix at some later time $\tau$ resulting from (\ref{eq:initial-state}) is given by
\begin{equation}
\rho_{abc}^{\Psi}(\tau)=\sum_{s=0}^{\infty}\sum_{r=0}^{\infty}\sum_{i=0}^{s}\sum_{j=0}^{r}a_{s}a_{r}^{*}\frac{f_{i}(s)\tau^{i}}{\sqrt{N_{s}(\tau)}}\frac{f_{j}(r)\tau^{j}}{\sqrt{N_{r}(\tau)}}\left|s-i\right>\left|i,i\right>\left<r-j\right|\left<j,j\right|,
\end{equation} 
which, after performing a partial trace, leads to the reduced density operators
\begin{equation}\label{eq:partiala}
\rho_{a}^{\Psi}(\tau)=\sum_{s=0}^{\infty}\sum_{r=0}^{\infty}\sum_{i=0}^{s}\sum_{j=0}^{r}a_{s}a_{r}^{*}\frac{f_{i}(s)\tau^{i}}{\sqrt{N_{s}(\tau)}}\frac{f_{j}(r)\tau^{j}}{\sqrt{N_{r}(\tau)}}\delta_{i,j}\left|s-i\right>\left<r-j\right|,
\end{equation}
where $\delta_{i,j}$ is the Kronecker delta, and
\begin{equation}\label{eq:partialb}
\rho_{b}^{\Psi}(\tau)=\sum_{s=0}^{\infty}\sum_{i=0}^{s}P_{s}\frac{f_{i}^{2}(s)}{N_{s}(\tau)}\tau^{2i}\left|i\right>\left<i\right|
\end{equation}
for the pump and signal modes respectively.  Fig.~\ref{fig:averages} compares the expectation values of the pump and signal modes for the parametric, semi-classical, and short-time quantum approximations, as well as the full numerical solution to Eqs.~(\ref{eq:semiQ_dt}) where the pump mode is initially in a coherent state with amplitude $\left<N_{a}(0)\right>=9$ corresponding to classical pump modes with amplitude $A=3$, and the signal and idler modes are initially in their vacuum (ground) states.

\subsection{Mode Spectrum Dynamics Under the Short-time Approximation}\label{sec:tri-short-dynamics}
The main limitation of the short-time approximation is the inability to account for backreaction effects resulting from the build up of quanta in the signal and idler modes.  For particles produced in the Hawking process, however,  the entangled pairs generated early in the evolution do not, to first-order, effect those created later from the black hole\cite{mathur:2009}. Furthermore, the emitted radiation does not build up in the vicinity of the black hole, but  escapes to spatial infinity. Therefore, in this section we will suppose that the expressions derived above for the evolving pump and signal states, Eqs.~(\ref{eq:partiala}) and (\ref{eq:partialb}) respectively, in fact provide a more relevant zero-dimensional model of a radiating black-hole when extrapolated beyond their original short-time domain of validity.   The pump and signal modes contain in general a large number of coefficients for each number state and cannot be easily evaluated.  However, in the long-time limit where the pump is nearly depleted, these equations can be considerably simplified by noting that a general element of Eq.~(\ref{eq:partialb}) for fixed $s$ is given by
\begin{equation}
P_{s}\frac{f_{i}^{2}(s)}{N_{s}(\tau)}\tau^{2i}=P_{s}\frac{e^{-\tau^{-2}}}{u!\tau^{2u}}\frac{\Gamma(s+1)}{\Gamma\left(s+1,\tau^{-2}\right)},\ \ u=s-i.
\end{equation}
As such, only those elements where $i=s$ remain nonzero, leading to the signal mode density matrix
\begin{equation}\label{eq:long-time-state}
\rho^{\Psi}_{b}=\sum_{s=0}^{\infty}P_{s}\left|s\right>\left<s\right|,
\end{equation}
which is in general a mixed state with number-state probability distribution $P_{s}$ identical to that of the initial pump state.  Therefore, by measuring the late-time signal or idler mode distribution we recover all but the phase information associated with the initial pure state of the pump mode.  Additionally, Eq.~(\ref{eq:long-time-state}) shows that the number of quanta in the signal mode is equal to the initial number of quanta in the pump mode, indicating that the pump (\ref{eq:partiala}) ends up in the vacuum (ground) state.  

\begin{figure}[htbp]
\begin{center}
\includegraphics[width=5.5in]{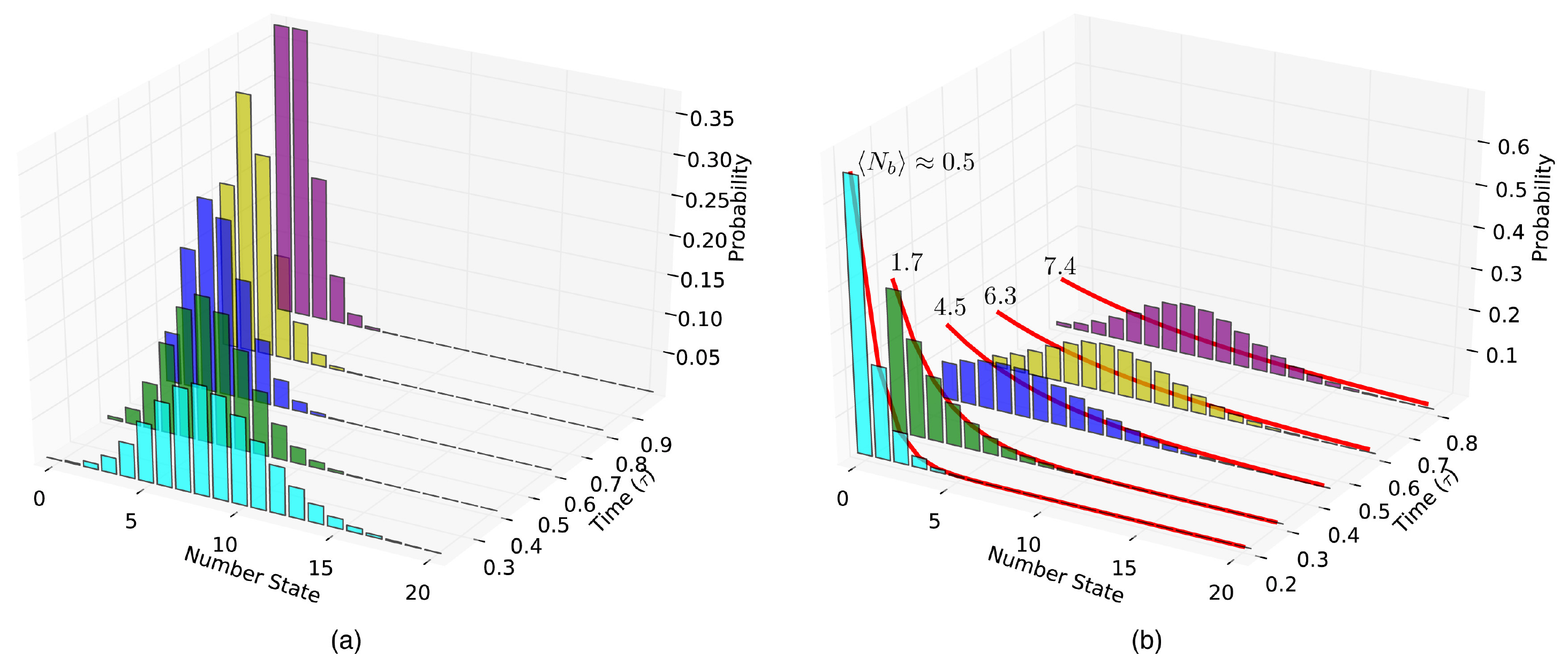}
\caption{a) Spectrum of the pump mode at several time-steps when the pump is initially in a coherent state with $\left<N_{a}(0)\right>=9$. b) Corresponding spectrum of the signal mode.  Red lines indicate what the distribution would be at each time-step if in a thermal state corresponding to occupation number $\left<N_{b}(\tau)\right>$, Eq.~(\ref{eq:bose}).}
\label{fig:spectrum}
\end{center}
\end{figure}

The most important knowledge gained from Eq.~(\ref{eq:long-time-state}) is that the signal mode spectrum will no longer be that of a thermal state, in contrast to the parametric and semi-classical approximations in Sec.~\ref{sec:tri-parametric} and Sec.~\ref{sec:tri-semiclassical}, respectively.  Focusing on coherent states, in Fig.~\ref{fig:spectrum} we give an example of the evolution of Eqs.~(\ref{eq:partiala}) and (\ref{eq:partialb}) by plotting the probability distributions for both modes as a function of time $\tau$ for a pump initially in a coherent state with amplitude $\left<N_{a}(0)\right>=9$ and initial vacuum state for both signal and idler.
Additionally, in Fig.~\ref{fig:spectrum}b we highlight what the thermal distribution would be at each time step by equating $\left<N_{b}(\tau)\right>$ with a Bose-Einstein distribution to extract an effective temperature:
\begin{equation}\label{eq:bose}
\left<N_{b}(\tau)\right>=\left[e^{\hbar\omega_{b}/k_{\mathrm{b}}T_{\mathrm{eff}}\left(\tau\right)}-1\right]^{-1}.
\end{equation}
In Fig.~\ref{fig:spectrum}a see we the evolution of the initial coherent state as it loses quanta to the signal and idler modes and evolves towards the ground state represented by a peak in the probability distribution at the $n=0$ number state value.  The corresponding evolution of the signal mode in Fig.~\ref{fig:spectrum}b starts with the vacuum state and progresses towards the state with probability distribution identical to that of the initial pump coherent state, Eq.~(\ref{eq:long-time-state}).  By comparison with the effective thermal state~(\ref{eq:bose}), we can see that the initial probability distribution for the signal mode is nearly that of a thermal state up until the pump mode has transferred nearly half of its initial energy corresponding to $\left<N_{b}(\tau)\right>=4.5$.  As the evolution continues, Fig.~\ref{fig:spectrum}b shows the increasing deviation from the effective thermal description for the signal mode distribution as expected from Eq.~(\ref{eq:long-time-state}).

\subsection{Non-thermal Spectra and Information}\label{sec:tri-info}
\begin{figure}[htbp]
\begin{center}
\includegraphics[width=5.5in]{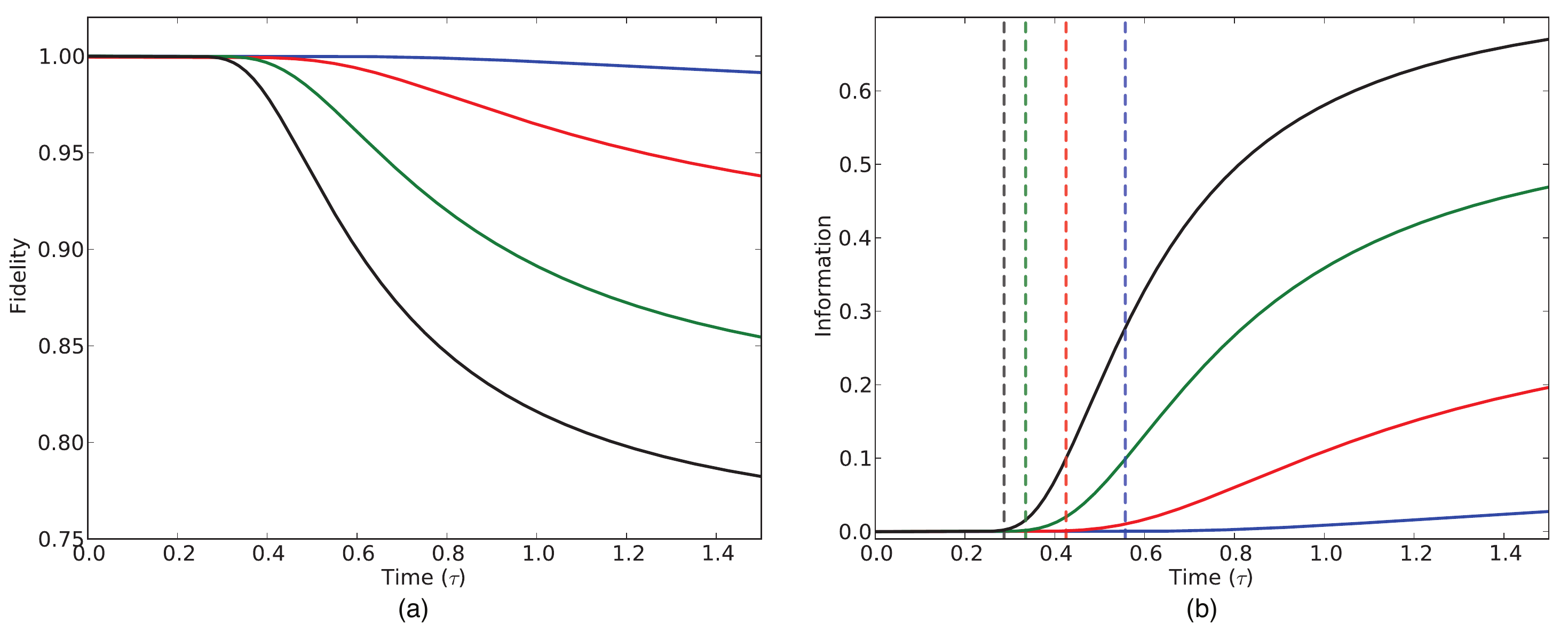}
\caption{a) Fidelities of the signal mode given a pump mode initially in a coherent state with occupation number $\left<N_{a}\right>=1$ (blue), $3$ (red), $6$ (green), and $9$ (black) under the short-time approximation.  b)  Information present in the signal spectrum calculated from Eq.~(\ref{eq:information}).  Equivalently colored vertical lines denote the time at which the effective subspace dimensions satisfy, $d^{\mathrm{eff}}_{a}=d^{\mathrm{eff}}_{bc}$.}
\label{fig:fig4}
\end{center}
\end{figure}

We now quantify the deviations of the signal mode spectrum Eq.~(\ref{eq:partialb}) from that of a thermal state using the fidelity\cite{nielson:2000}
\begin{equation}\label{eq:fidelity}
F_{b}(\tau)=\mathrm{Tr}\sqrt{\rho_{b}(\tau)^{1/2}\sigma(\tau)\rho_{b}(\tau)^{1/2}},
\end{equation}
where $\rho_{b}(\tau)$ is the density matrix of the signal mode and $\sigma(\tau)$ is a thermal density matrix with effective temperature determined by Eq.~(\ref{eq:bose}) using the occupation number $\left<N_{b}\left(\tau\right)\right>$.  The fidelity provides a measure of the distance between these two states in distribution space with range $0\le F\le 1$, where unity indicates that the two density matrices are identical.  In Fig.~\ref{fig:fig4}a we plot the fidelity of the signal mode assuming a pump mode initially in a coherent state starting with several occupation numbers ranging from one to nine.  By comparison with Fig.~\ref{fig:averages}, we can see that for the initial coherent state $\left<N_{a}(0)\right>=9$ the fidelity remains essentially unity, indicating that the signal mode density matrix Eq.~(\ref{eq:partialb}) is nearly thermal, until the pump has transferred close to half of its initial energy into the signal and idler modes beyond which point there is a strong deviation from the thermal state as the signal mode asymptotically approaches the state given by Eq.~(\ref{eq:long-time-state}).  This is in agreement with the qualitative description presented in Fig.~\ref{fig:spectrum}b and remains true for all the initial states considered in Fig.~\ref{fig:fig4}a. 

This deviation from a thermal distribution also indicates that the signal mode spectrum contains information defined as\cite{page:1993}
\begin{equation}\label{eq:information}
I_{b}(\tau)=S_{b}^{\mathrm{th}}(\tau)-S_{b}(\tau),
\end{equation}
where $S_{b}^{\mathrm{th}}(\tau)$ is the Von Neumann entropy of the signal mode in a thermal state with equal average occupation number, Eq.~(\ref{eq:thermal-entropy}), and $S_{b}(\tau)$ is the actual entropy of the mode calculated using Eq.~(\ref{eq:VNentropy}) and the reduced density matrix of the signal mode $\rho_{b}(\tau)$.  In Fig.~4b we plot the information contained in the signal mode for the initial pump coherent states considered in Fig.~4a.  Given that the fidelity is nearly unity up until the pump mode transfers half  of its original quanta to the signal and idler modes, it follows that the information content of the signal or idler mode is close to zero over this same time interval before increasing as the signal spectrum becomes identical to that of the initial coherent states.  

In order to account for the dynamics of the signal mode information, we first consider a general bipartite pure state of a system with fixed total energy that is composed of subsystems $A$ and $B$, each with finite Hilbert space dimensions $d_{A}$ and $d_{B}$, respectively.  It is known that subsystem $B$ will be nearly thermal and thus contain approximately no information as long as $d_{A}\gg d_{B}$, with the information content of subsystem $B$ becoming apparent only after $d_{A}\approx d_{B}$\cite{page:1993,page:1993-2,popescu:2006}.  Similar dynamics for the information content of the signal mode is shown in Fig.~\ref{fig:fig4}. However, the three harmonic oscillator modes from which a pure state of the trilinear Hamiltonian Eq.~(\ref{eq:trilinear}) is composed all contain an infinite set of states, preventing the direct application of dimensional arguments to our model.  One can, however, define an effective subspace dimension for each mode\cite{popescu:2006}
\begin{equation}\label{eq:effective}
d^{\mathrm{eff}}_{i}(\tau)=\frac{1}{\mathrm{Tr}\left[\sigma_{i}^{2}(\tau)\right]},\ \ i=a,b,c
\end{equation}
determined by the purity of the effective thermal state $\sigma_{i}(\tau)$ with temperature given by Eq.~(\ref{eq:bose}).  This definition is motived by the fact that the purity of a mixed state density matrix is proportional to the number of basis states over which the fractional population of the mixed state is nonzero.  For a state with an energy (quanta) constraint, using the thermal state $\sigma(\tau)$ gives the minimal value for the purity, and as such Eq.~(\ref{eq:effective}) represents an effective maximum number of states constrained by the number of quanta in the mode at time $\tau$.  Therefore, if we partition our initial pure state into bipartite subsystems $d_{a}$ and $d_{bc}$ composed of pump and combined signal and idler modes respectively, then the initial effective subspace dimensions satisfy $d^{\mathrm{eff}}_{a}\gg d^{\mathrm{eff}}_{bc}=\left(d^{\mathrm{eff}}_{b}\right)^{2}$, where the equality is due to the symmetry between signal and idler modes.  In Fig.~\ref{fig:fig4}b we plot the times at which $d^{\mathrm{eff}}_{a}=d^{\mathrm{eff}}_{bc}$ for each of the initial pump mode coherent states considered in Fig.~\ref{fig:fig4}a.  Just as for finite dimensional pure states, the information contained in the signal or idler subsystems remains nearly zero until after $d^{\mathrm{eff}}_{a}(\tau)\approx d^{\mathrm{eff}}_{bc}(\tau)$, provided we define the effective subspace dimension using Eq.~(\ref{eq:effective}).  Our results are in agreement with those of Page\cite{page:1993} where a similar argument was put forth for the evolution of information in the Hawking radiation from a finite dimensional black hole.

\subsection{Numerical Results for the Full Trilinear Evolution}\label{sec:tri-num}
\begin{figure}[htbp]
\begin{center}
\includegraphics[width=5.5in]{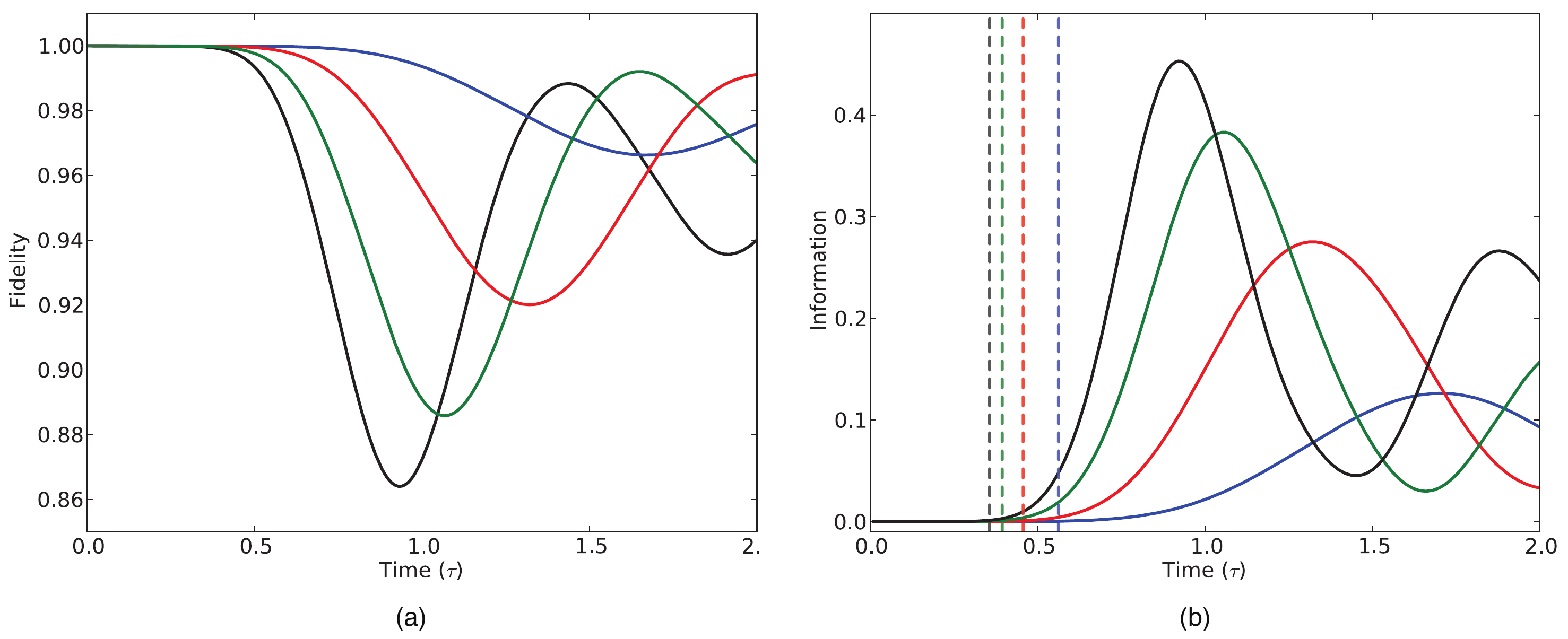}
\caption{a) Fidelities of the signal mode using numerical simulations of Eqs.~(\ref{eq:semiQ_dt}) with a pump mode initially in a coherent state with occupation number $\left<N_{a}\right>=1$ (blue), $3$ (red), $6$ (green), and $9$ (black).  b)  Corresponding information present in the signal mode spectrum and times at which $d^{\mathrm{eff}}_{a}(\tau)=d^{\mathrm{eff}}_{bc}(\tau)$ (dashed lines).}
\label{fig:fig5}
\end{center}
\end{figure}

Unlike Hawking radiation, which is well-modeled using the quantum short-time approximation, backaction due to the build up of quanta in the signal and idler modes has a strong effect on the evolution of the trilinear Hamiltonian (\ref{eq:interaction}).  In Fig.~\ref{fig:averages} we see that backreaction effects quickly lead to differing evolutions between the short-time and full quantum dynamics; backreaction prevents the full transfer of quanta from the pump mode\cite{drobny:1994}, and results in the oscillation of quanta between pump and signal/idler modes. To account for these backaction terms we repeat the analysis in Sec.~\ref{sec:tri-info} by numerically calculating the full dynamics of Eq.~(\ref{eq:semiQ_dt}).  Fig.~\ref{fig:fig5} shows the modifications to both the fidelity and information content of the signal mode when backaction is included in the dynamics, as well as the subspace dimension condition $d^{\mathrm{eff}}_{a}(\tau)=d^{\mathrm{eff}}_{bc}(\tau)$.  
As expected, the short-time dynamics in Fig.~\ref{fig:fig5} are nearly identical to those in Fig.~\ref{fig:fig4} up until $d^{\mathrm{eff}}_{a}(\tau)\approx d^{\mathrm{eff}}_{bc}(\tau)$ which occurs later than in the short-time approximation as the backaction from particles in the signal mode begins to impede the transfer of quanta from the pump.  The peaks in the information content of the signal mode correspond to the times at which backaction from the pump mode has completely stopped the transfer of energy between modes.  With the majority of quanta now in the signal and idler modes, the flow of energy reverses directions as the signal and idler now drive the pump mode leading to the oscillations seen in Fig.~\ref{fig:fig5}. As a model for black hole evaporation these oscillations represent the unphysical process of Hawking radiation flowing back into the black hole; the connection between the trilinear Hamiltonian and Hawking emission is valid only for the initial transfer of quanta from pump to signal/idler modes.

\section{Tripartite Entanglement}\label{sec:tri-entanglement}
\begin{figure}[htbp]
\begin{center}
\includegraphics[width=5.5in]{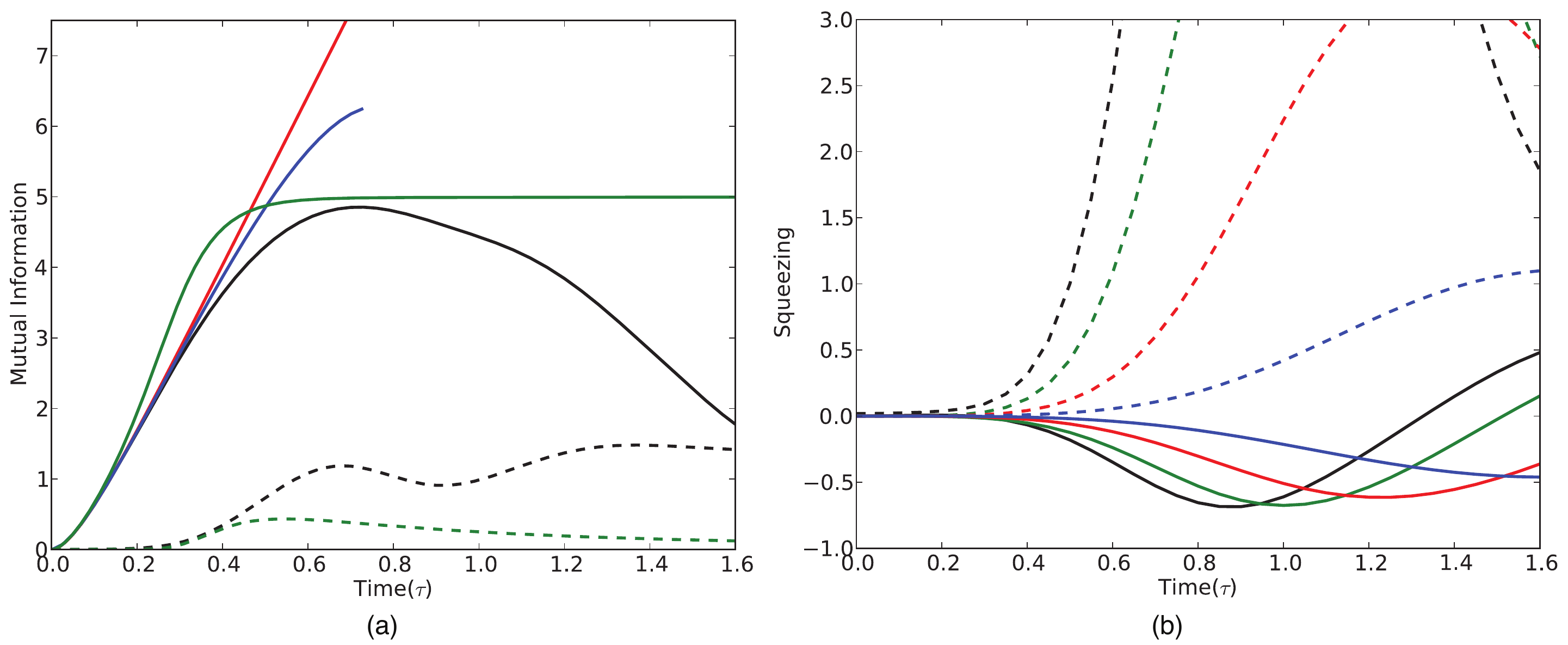}
\caption{a) Signal-idler mutual information $I_{b-c}$ for pump with $\left<N_{a}(0)\right>=9$ in the parametric (red), semi-classical (blue), quantum short-time (green), and full numerical solution (black).  Dashed lines represent mutual information $I_{a-bc}$ indicating entanglement with the pump mode.  b) Quadrature squeezing parameters $q_{\pm}$ of the pump mode given the pump is initially in a coherent state with occupation number $\left<N_{a}\right>=1$ (blue), $3$ (red), $6$ (green), and $9$ (black).  Negative values indicating squeezing of the $q_{-}$ quadrature.  Dashed lines represent the corresponding $q_{+}$ quadrature.}
\label{fig:fig6}
\end{center}
\end{figure}

With the pump mode quantized, one can consider the entanglement between pump and signal/idler modes.  In Sec.~\ref{sec:tri-parametric}, it was shown that ignoring this entanglement (pump mode treated classically) and tracing over one mode of a two-mode squeezed state leads directly to the thermal characteristics of the remaining system.  In the full quantum description the statistics of the signal mode is obtained by tracing over both idler and pump modes and as such the distribution of entanglement between modes of this now tripartite system is important for characterizing the spectrum of the signal mode.  Additionally, entanglement with the pump mode may appreciably alter the state dynamics of the pump compared to the classical approximations in Secs.~\ref{sec:tri-parametric} and \ref{sec:tri-semiclassical}.  Given that multipartite entanglement is not as well understood as for the bipartite case, we begin by again considering the system to be partitioned into two subsystems consisting of the pump mode and the combined signal-idler.  This bipartite separation into pump and signal-idler subsystems allows us to use the mutual information\cite{nielson:2000}
\begin{equation}\label{eq:mutual}
I_{a-bc}=S_{a}+S_{bc}-S_{abc}
\end{equation}
as a measure of total correlations, both classical and quantum, between subsystems\cite{groisman:2005}.  Since our tripartite system state remains pure throughout its evolution, the total entropy $S_{abc}=0$ and the subsystem entropies satisfy $S_{a}=S_{bc}$.  The mutual information is therefore twice the entropy of the pump mode subsystem, $I_{a-bc}=2S_{a}$.  Thus, the subsystem entropy of the pump is a direct measure of entanglement with the signal and idler modes, which is not taken into account in either the parametric or semi-classical solutions.  Of course the signal and idler subsystems are also entangled with each other as in Secs.~\ref{sec:tri-parametric} and \ref{sec:tri-semiclassical}.  With both signal and pump modes in identical states, the signal-idler mutual information is given by
\begin{equation}\label{eq:bc-info}
I_{b-c}=S_{b}+S_{c}-S_{bc}=2S_{b}-S_{a}
\end{equation}
indicating the tradeoff between the entanglement of the signal/idler and that of the pump mode Eq.~(\ref{eq:mutual}). In Fig.~\ref{fig:fig6}a we plot the mutual information Eqs.~(\ref{eq:mutual}) and (\ref{eq:bc-info}) for a pump initially in a coherent state of amplitude $\left<N_{a}(0)\right>=9$ for the parametric, semi-classical and short-time approximations along with the full numerical solution obtained using Eqs.~(\ref{eq:semiQ_dt}).  We see the mutual information (pump entanglement) $I_{a-bc}$ begins to become appreciable around the same time as the information content of the signal mode becomes apparent in Figs.~\ref{fig:fig4}b and \ref{fig:fig5}b indicating the increasing role of the pump mode in the dynamics.  The numerical solution shows that this increase in pump entanglement reduces the signal-idler mutual information $I_{b-c}$ with respect to the semi-classical and parametric approximations as expected from Eq.~(\ref{eq:bc-info}).  In the full quantum dynamics the pump is never allowed to be fully depleted and is therefore always entangled with the signal and idler.  However in the short-time approximation the late-time pump mode is depleted and approaches the ground state where $S_{a}=0$.

Finally, we show that the entanglement buildup with the pump mode given by $I_{a-bc}$ not only affects the signal and idler modes, but also results in amplitude dependent squeezing of the pump.  In Fig.~\ref{fig:fig6}b we plot the amplitude of the squeezing parameters\cite{walls:2008},
\begin{equation}
q_{\pm}=4\left<\Delta X_{\pm}^{2}\right>-1, \ \ X_{\pm}=\frac{a(\tau)\pm a^{+}(\tau)}{2}
\end{equation}
 for the pump mode, where $\left<\Delta X_{\pm}^{2}\right>$ is the variance of the quadratures.  As can be seen, the entanglement with the signal and idler modes leaves the pump mode in a non-classical squeezed state over the time range of interest.

\section{Circuit QED realization of the trilinear Hamiltonian}\label{sec:tri-circuit}
\begin{figure}[htbp]
\begin{center}
\includegraphics[width=4.5in]{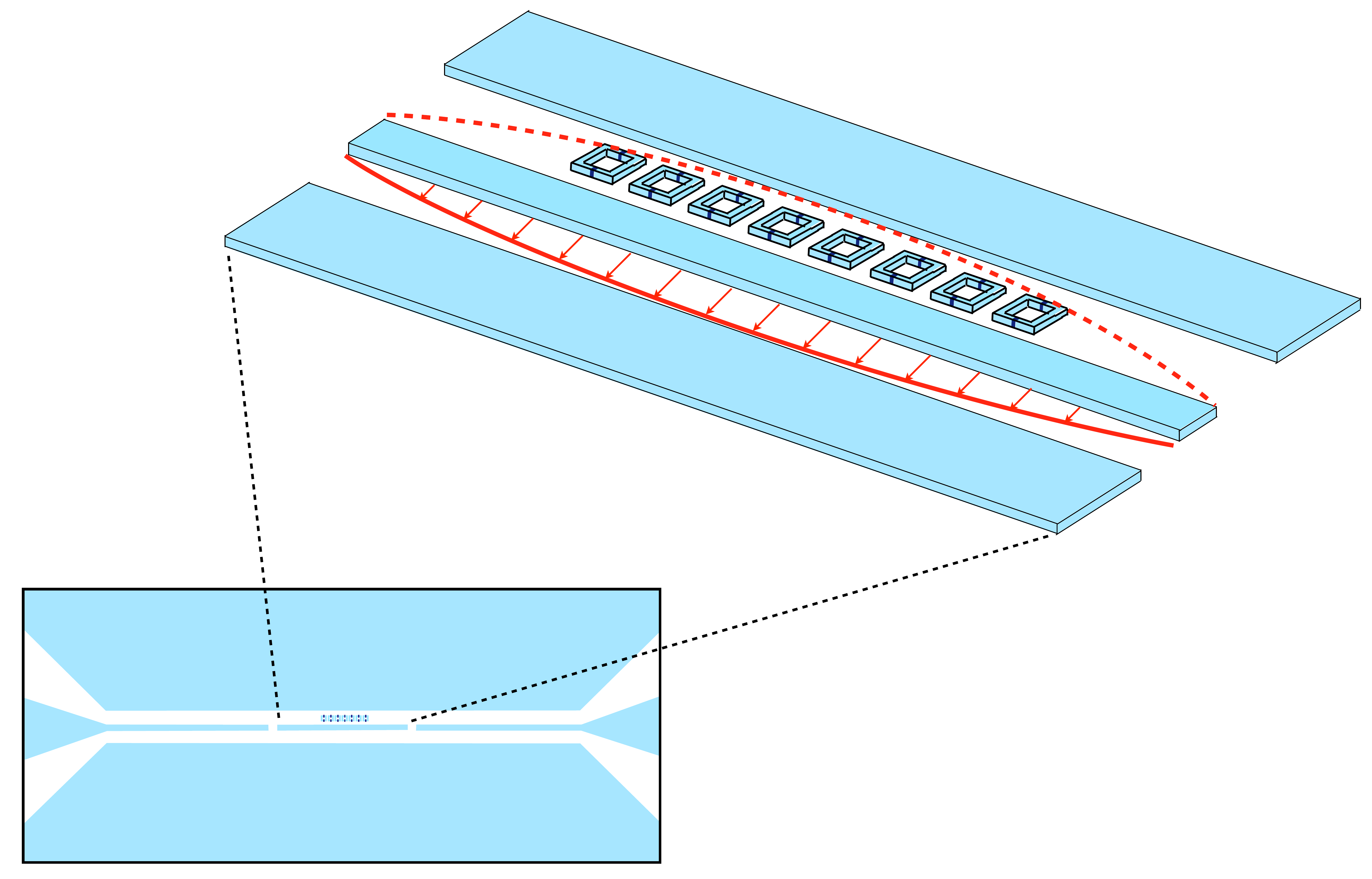}
\caption{Cartoon of a typical coplanar transmission line resonator with flux qubits (SQUIDs) situated in between the centerline conductor and ground plane.  In modeling the Dicke Hamiltonian, Eq.~(\ref{eq:tri-dicke}) the amplitude of the resonant cavity mode (red arrows) is assumed to be constant over the length of the qubit ensemble.}
\label{fig:tri-fig7}
\end{center}
\end{figure}

In attempting to fabricate a direct realization of the trilinear Hamiltonian Eq.~(\ref{eq:trilinear}) we run into difficulty when generating the specific nonlinear coupling between oscillators.  This coupling must not only obey the symmetries of Eqs.~(\ref{eq:MR}), but must also be tunable so that the signal and idler modes may be decoupled while the pump mode is driven to its initial energy state.  Although in principle nothing prohibits us from creating such a coupling, at present we do not have a suitable model for this interaction.  Instead we will make use of the connection between the trilinear Hamiltonian and the Dicke model\cite{dicke:1953} in realizing our system as an array of $N$ superconducting qubits interacting with a single resonant mode of a microwave cavity\cite{chen:2007,lambert:2009,fink:2009}.  In keeping with the theme of this thesis we will assume the use of flux qubits formed by biasing a SQUID with an external magnetic flux with amplitude near a half-flux quantum, $\Phi_{\mathrm{ext}}=\Phi_{0}/2$.  An illustration of this setup is given in Fig.~\ref{fig:tri-fig7}.  Of course, one may readily substitute one of the many other qubit varieties\cite{clarke:2008}.

The Dicke Hamiltonian for a collection of qubits interacting with a single mode of a microwave cavity resonator is given as\cite{dicke:1953}
\begin{equation}\label{eq:tri-dicke}
H=\hbar\omega_{c} c^{+}c+\hbar\omega_{c} J_{z}-i\hbar\chi\left(c^{+}+c\right)\left(J_{+}-J_{-}\right),
\end{equation}
where $\omega_{c}$ is the frequency of the field mode, assumed to be resonant with the level splitting of the qubits, $\chi$ is the coupling strength of the qubits to the cavity field, and $\left\{J_{\pm},J_{z}\right\}$ are collective qubit operators satisfying the standard angular momentum commutation relations.
\begin{equation}
\left[J_{+},J_{-}\right]=2\hbar J_{z}, \ \ \left[J_{\pm},J_{z}\right]=\mp J_{\pm}
\end{equation}
It is implicit in this Hamiltonian that the only qubit interactions are with the cavity field and that the wavelength of this field is much larger than the size of an individual qubit; each atom experiences the same amplitude of the resonant cavity field.  In the absence of any qubit-cavity interaction, the operators $c(t),J_{-}(t)$ and $c^{+}(t),J_{+}(t)$ rotate in the Heisenberg frame as $\exp\left(-i\omega t\right)$ and $\exp\left(i\omega t\right)$ respectively.  Therefore, in making the rotating-wave-approximation (RWA) we can ignore terms containing $cJ_{-}$ and $c^{+}J_{+}$ which gives the simplified Hamiltonian
\begin{equation}\label{eq:tri-rwa}
H=\hbar\omega_{c} c^{+}c+\hbar\omega_{c} J_{z}+i\hbar\chi\left(cJ_{+}-c^{+}J_{+}\right).
\end{equation}
This Hamiltonian can be rewritten using the Schwinger representation\cite{schwinger:1965} where the angular momentum operators are replaced by products of bosonic operators
\begin{equation}\label{eq:tri-bosonic}
J_{+}=a^{+}b, \ \ J_{-}=ab^{+}, \ \ J_{z}=\frac{1}{2}\left(a^{+}a-b^{+}b\right),
\end{equation}
with the boson operators $a,a^{+}$ and $b,b^{+}$ satisfying the usual boson commutation relations.  These operators correspond to the annihilation and creation of excited and ground-level qubits respectively.  The angular momentum operator  $J_{+}$, Eq.~(\ref{eq:tri-bosonic}), now represents the switching of a qubit from the ground to excited state while the reverse process is governed by $J_{-}$.  Upon substitution of Eqs.~(\ref{eq:tri-bosonic}) into Eq.~(\ref{eq:tri-rwa}) we obtain the Hamiltonian
\begin{equation}
H=\hbar\omega_{a}a^{+}a+\hbar\omega_{b}b^{+}b+\hbar\omega_{c}c^{+}c+i\hbar\chi\left(a^{+}bc+ab^{+}c^{+}\right)
\end{equation}
satisfying the frequency relation $\omega_{a}=-\omega_{b}=\omega_{c}/2$, which is equivalent to our trilinear Hamiltonian Eq.~(\ref{eq:trilinear}).  The number operators $N_{1}=a^{+}a$ and $N_{2}=b^{+}b$ corresponding to the excited and ground state modes do not represent the actual mode population values but rather give the effective occupations obeying $N_{1}+N_{2}=2J\le N$, where $J$ is the total angular momentum of the $N$ qubits.  These effective values coincide with the actual level populations when all of the qubits are in either the ground or excited state, $2J=N$.

In constructing the trilinear Hamiltonian from a simplified Dicke model we have replaced the conceptually simple coupled cavity system with a much more abstract realization based on the effective oscillatory modes of the collective dynamics of a qubit ensemble.  While at first this may appear disadvantageous, the connection to the Dicke model suggests that the dynamics of the information content in the signal and idler modes presented in Secs.~\ref{sec:tri-info} and \ref{sec:tri-num} may be related to the well-known super-radiant phase transition of the Dicke Hamiltonian\cite{hepp:1973,lambert:2004,baumann:2010}.  This connection will be addressed in a later work.

\section{Conclusion}\label{sec:tri-conclusion}
In this chapter we have investigated the effect of a dynamical quantized pump mode on the generation of quanta in a parametric amplifier.  We have shown that a quantized pump mode leads naturally to a non-thermal spectrum for the signal and idler modes once the pump has released nearly half of its initial energy, such that the effective subspace dimensions of the pump and signal/idler mode systems approximately coincide.  The departure from a thermal state indicates that the signal spectrum contains information that may be used to partially reconstruct the initial state of the pump mode. Once quantized, the pump mode becomes entangled with the signal and idler leading to non-classical squeezed states of the pump. As a simple, zero-dimensional model of the Hawking effect, the present findings lend support to  the possibility for non-thermal emission from a quantum (as opposed to semiclassical) black hole; the  emitted radiation is entangled with the quantized gravitational degrees of freedom and yields information concerning initial formation of the black hole.

%%%

%% file: ap1.tex
%%%

\chapter{Derivation of signal and noise expressions for SQUID detector}\label{ap:squid}
In this appendix, we give the derivation of the signal $a_T^{(1)}(\omega)$ and noise $a_T^{(0)}(\omega)$ terms of Eq.~(\ref{eq:asignalnoise}) in Chapter~\ref{ch:squid}.  Suppressing the signal dependent term $A(\omega,\omega')$ in  Eq.~(\ref{eq:atwonly}), we obtain the noise equation
\begin{eqnarray}\label{eq:zeroth-atw}
	{a}_{T}^{(0)}(\omega)&=&\int_{-\infty}^{\infty}d\omega'B(\omega,\omega'){a}_{T}^{(0)}(\omega-\omega')\cr
	&\times&\int_{-\infty}^{\infty}d\omega''\left[{a}_{T}^{(0)}(\omega''){a}_{T}^{{(0)}+}(\omega''-\omega')+{a}_{T}^{{(0)}+}(\omega''){a}_{T}^{(0)}(\omega''+\omega')\right]\cr
&+&D(\omega)\int_{-\infty}^{\infty}d\omega''\int_{-\infty}^{\infty}d\omega'
{a}_{T}^{{(0)}+}(\omega''){a}_{T}^{(0)}(\omega'){a}_{T}^{(0)}(\omega+\omega''-\omega')\cr
&+&C(\omega).
\end{eqnarray}
Keeping only terms to first order in $A(\omega,\omega')$ or $a_T^{(1)}$ in Eq.~(\ref{eq:atwonly}), we obtain the signal equation
 \begin{eqnarray}\label{eq:first-atw}
{a}_{T}^{(1)}(\omega)&=&\int_{-\infty}^{\infty}d\omega'{a}_{T}^{(0)}(\omega-\omega')A(\omega,\omega')+\int_{-\infty}^{\infty}d\omega'B(\omega,\omega'){a}_{T}^{(1)}(\omega-\omega')\cr
&\times&\int_{-\infty}^{\infty}d\omega''\left[{a}_{T}^{(0)}(\omega''){a}_{T}^{{(0)}+}(\omega''-\omega')+{a}_{T}^{{(0)}+}(\omega''){a}_{T}^{(0)}(\omega''+\omega')\right]\cr
&+&\int_{-\infty}^{\infty}d\omega'B(\omega,\omega'){a}_{T}^{(0)}(\omega-\omega')\int_{-\infty}^{\infty}d\omega''\left[{a}_{T}^{(1)}(\omega''){a}_{T}^{{(0)}+}(\omega''-\omega')\right.\cr
&+&\left.{a}_{T}^{{(1)}+}(\omega''){a}_{T}^{(0)}(\omega''+\omega')
+{a}_{T}^{(0)}(\omega''){a}_{T}^{{(1)}+}(\omega''-\omega')+{a}_{T}^{{(0)}+}(\omega''){a}_{T}^{(1)}(\omega''+\omega')\right]\cr
&+&D(\omega)\int_{-\infty}^{\infty}d\omega''\int_{-\infty}^{\infty}d\omega'\left[{a}_{T}^{{(0)}+}(\omega''){a}_{T}^{(0)}(\omega'){a}_{T}^{(1)}(\omega+\omega''-\omega')\right.\cr
&+&\left.{a}_{T}^{{(0)}+}(\omega''){a}_{T}^{(1)}(\omega'){a}_{T}^{(0)}(\omega+\omega''-\omega')\right.\cr
&+&\left.{a}_{T}^{{(1)}+}(\omega''){a}_{T}^{(0)}(\omega'){a}_{T}^{(0)}(\omega+\omega''-\omega')\right].
\end{eqnarray}
Decomposing $a_T^{(0)}(\omega)=\langle a_T^{(0)}(\omega)\rangle +\delta a_T^{(0)}(\omega)$ and expanding Eq.~(\ref{eq:zeroth-atw}) to first order in the quantum noise fluctuation $\delta a_T^{(0)}(\omega)$, we obtain the following two equations:
\begin{eqnarray}\label{eq:zeroth-atw-coherent}
\langle{a}_{T}^{(0)}(\omega)\rangle&=&\int_{-\infty}^{\infty}d\omega'B(\omega,\omega')\langle{a}_{T}^{(0)}(\omega-\omega')\rangle\cr
&\times&\int_{-\infty}^{\infty}d\omega''\left[\langle{a}_{T}^{(0)}(\omega'')\rangle\langle{a}_{T}^{{(0)}+}(\omega''-\omega')\rangle+\langle{a}_{T}^{{(0)}+}(\omega'')\rangle\langle{a}_{T}^{(0)}(\omega''+\omega')\rangle\right]\cr
&+&D(\omega)\int_{-\infty}^{\infty}d\omega''\int_{-\infty}^{\infty}d\omega'\langle{a}_{T}^{{(0)}+}(\omega'')\rangle\langle{a}_{T}^{(0)}(\omega')\rangle\langle{a}_{T}^{(0)}(\omega+\omega''-\omega')\rangle\cr
&+&\langle C (\omega)\rangle
\end{eqnarray}
and
 \begin{eqnarray}\label{eq:zeroth-atw-quantum}
	&&\delta{a}_{T}^{(0)}(\omega)=\int_{-\infty}^{\infty}d\omega'B(\omega,\omega')\delta{a}_{T}^{(0)}(\omega-\omega')\cr
&\times&\int_{-\infty}^{\infty}d\omega''\left[\langle{a}_{T}^{(0)}(\omega'')\rangle\langle{a}_{T}^{{(0)}+}(\omega''-\omega')\rangle+\langle{a}_{T}^{{(0)}+}(\omega'')\rangle\langle{a}_{T}^{(0)}(\omega''+\omega')\rangle\right]\cr
&+&\int_{-\infty}^{\infty}d\omega'B(\omega,\omega')\langle{a}_{T}^{(0)}(\omega-\omega')\rangle
\int_{-\infty}^{\infty}d\omega''\left[\delta{a}_{T}^{(0)}(\omega'')\langle{a}_{T}^{{(0)}+}(\omega''-\omega')\rangle\right.\cr
&+&\left.\delta{a}_{T}^{{(0)}+}(\omega''-\omega')\langle{a}_{T}^{(0)}(\omega'')\rangle+\delta{a}_{T}^{{(0)}+}(\omega'')\langle{a}_{T}^{(0)}(\omega''+\omega')\rangle\right.\cr
&+&\left.\delta{a}_{T}^{(0)}(\omega''+\omega')\langle{a}_{T}^{{(0)}+}(\omega'')\rangle\right]\cr 
&+&D(\omega)\int_{-\infty}^{\infty}d\omega''\int_{-\infty}^{\infty}d\omega'\left[\delta{a}_{T}^{{(0)}+}(\omega'')\langle{a}_{T}^{(0)}(\omega')\rangle\langle{a}_{T}^{(0)}(\omega+\omega''-\omega')\rangle\right.\cr
&+&\left.\delta{a}_{T}^{(0)}(\omega')\langle{a}_{T}^{{(0)}+}(\omega'')\rangle\langle{a}_{T}^{(0)}(\omega+\omega''-\omega')\rangle+\delta{a}_{T}^{(0)}(\omega+\omega''-\omega')\langle{a}_{T}^{{(0)}+}(\omega'')\rangle\langle{a}_{T}^{(0)}(\omega')\rangle\right]\cr
&+&\delta C(\omega).
\end{eqnarray}

Assuming $\langle a_T^{(0)}(\omega)\rangle$ can be expressed approximately as a delta function, i.e., $\langle a_T^{(0)}(\omega)\rangle=\chi\delta(\omega-\omega_p)$, Eq.~(\ref{eq:zeroth-atw-coherent}) reduces to Eq.~(\ref{eq:mean-field0}) for $\chi$. The semiclassical approximation to Eq.~(\ref{eq:first-atw}) for $a_T^{(1)}(\omega)$, with  $a_T^{(0)}(\omega)$ replaced by $\langle a_T^{(0)}(\omega)\rangle=\chi\delta(\omega-\omega_p)$, then becomes 
\begin{eqnarray}\label{eq:atw_chi}
&&\left\{1-2\left|\chi\right|^{2}\left[B(\omega,0)+B(\omega,\omega-\omega_{p})+D(\omega)\right]\right\}{a}_{T}^{(1)}(\omega)\cr
&&-\chi^{2}\left[2B(\omega,\omega-\omega_{p})+D(\omega)\right]{a}_{T}^{{(1)}+}(2\omega_{p}-\omega)=\chi A(\omega,\omega-\omega_{p}).
\end{eqnarray}
In order to invert and obtain ${a}_{T}^{(1)}(\omega)$, we require a second, linearly independent equation that also depends on ${a}_{T}^{{(1)}+}(2\omega_{p}-\omega)$. Such an equation is obtained by making the replacement $\omega\rightarrow 2\omega_{p}-\omega$ in Eq.~(\ref{eq:atw_chi}) and then taking the adjoint:
\begin{eqnarray}\label{eq:atw_one_chi}
&\left\{1+2\left|\chi\right|^{2}\left[B(\omega-2\Delta\omega,0)+B(\omega-2\Delta\omega,\omega-\omega_{p})+D(\omega-2\Delta\omega)\right]\right\}{a}_{T}^{{(1)}+}(2\omega_p-\omega)\cr
&+\chi^{*2}\left[2B(\omega-2\Delta\omega,\omega-\omega_{p})+D(\omega-2\Delta\omega)\right]{a}_{T}^{{(1)}}(\omega)=-\chi^* A(\omega-2\Delta\omega,\omega-\omega_{p}).
\end{eqnarray}
Now inverting, we obtain
\begin{equation}
{a}_{T}^{(1)}(\omega)=\alpha_{1}(\omega)A(\omega,\omega-\omega_{p})+\alpha_{2}(\omega)A(\omega-2\Delta\omega,\omega-\omega_{p}),
\end{equation}
where
\begin{equation}
\alpha_{1}(\omega)=\mathcal{D}(\omega)^{-1}\left\{1+2\left|\chi\right|^{2}\left[B(\omega-2\Delta\omega,0)+B(\omega-2\Delta\omega,\omega-\omega_{p})+D(\omega-2\Delta\omega)\right]\right\}\chi,\label{eq:alpha1}
\end{equation}
\begin{equation}
\alpha_{2}(\omega)=-\mathcal{D}(\omega)^{-1}\left[2B(\omega,\omega-\omega_{p})+D(\omega)\right]\left|\chi\right|^{2}\chi\label{eq:alpha2}
\end{equation}
and the determinant is given by,
\begin{eqnarray}\label{eq:determinant}
\mathcal{D}(\omega)&=&\left\{1-2\left|\chi\right|^{2}\left[B(\omega,0)+B(\omega,\omega-\omega_{p})+D(\omega)\right]\right\}\cr
&\times&\left\{1+2\left|\chi\right|^{2}\left[B(\omega-2\Delta\omega,0)+B(\omega-2\Delta\omega,\omega-\omega_{p})+D(\omega-2\Delta\omega)\right]\right\}\cr
&+&\left|\chi\right|^{4}\left[2B(\omega,\omega-\omega_{p})+D(\omega)\right]
\left[2B(\omega-2\Delta\omega,\omega-\omega_{p})\right.\cr
&+&\left.D(\omega-2\Delta\omega)\right].
\end{eqnarray}
A similar approach is used to obtain $\delta{a}_{T}^{(0)}(\omega)$ from Eq.~(\ref{eq:zeroth-atw-quantum}),  giving
\begin{equation}
	\delta{a}_{T}^{(0)}(\omega)=\beta_{1}(\omega)\delta C(\omega)+\beta_{2}(\omega)\delta C^{+}(2\omega_{p}-\omega),
\end{equation}
with
\begin{equation}
\beta_{1}(\omega)=\mathcal{D}(\omega)^{-1}\left\{1+2\left[B(\omega-2\Delta\omega,0)+B(\omega-2\Delta\omega,\omega-\omega_{p})+D(\omega-2\Delta\omega)\right]\left|\chi\right|^{2}\right\}\label{eq:beta1}
\end{equation}
and
\begin{equation}
\beta_{2}(\omega)=\mathcal{D}(\omega)^{-1}\left[2B(\omega,\omega-\omega_{p})+D(\omega)\right]\chi^{2}.\label{eq:beta2}
\end{equation}

 %%%

%% file: ap2.tex
%%%

\ssp
\chapter{Simulation source codes}\label{ap:code}
\definecolor{listinggray}{gray}{0.95}
Included in this appendix are the source codes for the numerical simulations used in Chapters~\ref{ch:squid} and \ref{ch:trilinear}.  All functions are coded for use in the MATLAB computing environment.  In addition, the functions used in simulating the trilinear Hamiltonian, Sec.~\ref{sec:ap2-trilinear}, require the Quantum Optics Toolbox by Sze M. Tan\cite{tan:toolbox}.
\section{Signal, Noise, and Cooling Codes}\label{sec:ap2-squid}
%---------------------------------------------------------------
\lstset{language=matlab}
\lstset{basicstyle=\footnotesize}
%\lstset{backgroundcolor=listinggray,framerulecolor=blue}
%\lstset{backgroundcolor=listinggray,rulecolor=blue}
\lstset{backgroundcolor=\color{listinggray},rulecolor=\color{blue}}
\lstset{linewidth=6in}
%\lstset{labelstep=10}
%\lstset{commentstyle=\textit, stringstyle=\upshape,stringspaces=false}
\lstset{commentstyle=\textit, stringstyle=\upshape,showspaces=false}
\lstset{frame=single,frameround=tttt}
\begin{lstlisting}[]
%Simulation codes for results obtained in
%Phys. Rev. B 78, 104516 (2008)
%Paul D. Nation, Dartmouth College, 2008
%alpha 1-----------------------------------------
function [out]=alpha1(w,I0,dw,Ktm,Kd,vars)
%alpha1 in Eq.56
out=deter(w,I0,dw,Ktm,Kd,vars).^-1.*(1+2.*(B(w-2*dw,0,Ktm,vars)+...
    B(w-2*dw,w-(dw+vars(2)),Ktm,vars)+...
    D(w-2*dw,Kd,vars)).*abs(ChiNumeric(dw,I0,Ktm,Kd,vars)).^2).*...
    ChiNumeric(dw,I0,Ktm,Kd,vars);
%alpha 2------------------------------------------
function [out]=alpha2(w,I0,dw,Ktm,Kd,vars)
%alpha2 in Eq.56
out=-deter(w,I0,dw,Ktm,Kd,vars).^-1.*(2.*B(w,w-(dw+vars(2)),Ktm,vars)+...
    D(w,Kd,vars)).*...
    abs(ChiNumeric(dw,I0,Ktm,Kd,vars))^2*ChiNumeric(dw,I0,Ktm,Kd,vars);
%B-----------------------------------------------
function [out] = B(w,wprime,Ktm,vars)
%Eq.41
out=((vars(2)*Ktm/sqrt(2*pi))^2/2)./(w-vars(2)+i*vars(4)).*...
    (1./(wprime-vars(3)+i*vars(5))-1./(wprime+vars(3)+i*vars(5)));
%beta1--------------------------------------------
function [out]=beta1(w,I0,dw,Ktm,Kd,vars)
%beta1 in Eq.57
out=deter(w,I0,dw,Ktm,Kd,vars).^-1.*(1+2.*(B(w-2*dw,0,Ktm,vars)+...
    B(w-2*dw,w-(dw+vars(2)),Ktm,vars)+D(w-2*dw,Kd,vars)).*...
    abs(ChiNumeric(dw,I0,Ktm,Kd,vars))^2);
%beta2--------------------------------------------
function [out]=beta2(w,I0,dw,Ktm,Kd,vars)
%beta2 in Eq.57
out=deter(w,I0,dw,Ktm,Kd,vars).^-1.*(2.*B(w,w-(dw+vars(2)),Ktm,vars)+...
    D(w,Kd,vars)).*ChiNumeric(dw,I0,Ktm,Kd,vars)^2;
%C------------------------------------------------
function [out]=C(I0,dw,vars)
%Eq.42
out=abs(i*sqrt(2*pi)./(vars(4)-i.*dw).*...
    sqrt((I0.^2*vars(6)*vars(4))./(vars(1).*(dw+vars(2)))));
%ChiNumeric---------------------------------------
function [out]=ChiNumeric(dw,I0,Ktm,Kd,vars)
%Upper or lower solution to Eq.50 in bistable region
%determined by input vars=NLswitch
if vars(9)==0
    out=abs(C(I0,dw,vars));
else
    p3=1;
    p2=-2.*dw./(vars(2)*K(Ktm,Kd,vars));
    p1=(dw.^2+vars(4)^2)./(vars(2)^2*K(Ktm,Kd,vars)^2);
    p0=-vars(4)/(vars(2)^2*K(Ktm,Kd,vars)^2).*...
        (I0.^2*vars(6))/(vars(1)*(dw+vars(2)));
    poly=cell(1,length(I0));
    for j=1:length(I0)
        poly{1,j}=[p3,p2,p1,p0(1,j)];
    end
    r=cell(1,length(I0));
    for j=1:length(I0)
        r{1,j}=roots(poly{1,j});
    end
    reals=zeros(1,length(I0));
    for j=1:length(I0)
        reals(1,j)=min(r{1,j}(imag(r{1,j})==0));
    end
    out=sqrt(2*pi)*sqrt(reals);
end
%D-----------------------------------------------
%Eq.43
function [out]=D(w,Kd,vars)
out=(vars(2)*Kd)/(2*pi).*1./(w-vars(2)+i*vars(4));
%deter-------------------------------------------
function [out]=deter(w,I0,dw,Ktm,Kd,vars)
%Determinant, Eq.A10
out=(1-2*abs(ChiNumeric(dw,I0,Ktm,Kd,vars))^2.*(B(w,0,Ktm,vars)+...
    B(w,w-(dw+vars(2)),Ktm,vars)+D(w,Kd,vars))).*...
    (1+2*abs(ChiNumeric(dw,I0,Ktm,Kd,vars))^2.*(B(w-2*dw,0,Ktm,vars)+...
    B(w-2*dw,w-(dw+vars(2)),Ktm,vars)+D(w-2*dw,Kd,vars)))+...
    abs(ChiNumeric(dw,I0,Ktm,Kd,vars))^4.*...
    (2.*B(w,w-(dw+vars(2)),Ktm,vars)+D(w,Kd,vars)).*...
    (2*B(w-2*dw,w-(dw+vars(2)),Ktm,vars)+D(w-2*dw,Kd,vars));
%dwc---------------------------------------------
function [out]=dwc(Ktm,Kd,vars)
%Eq.65
out=sqrt(3)*vars(4)*sign(K(Ktm,Kd,vars));
%G-----------------------------------------------
function [out]=G(w,I0,dw,Ktm,Kd,vars)
%Gain function defined by Eq.56
out=I0.^2./(2*pi).*(Ktm*vars(2)/vars(4))^2*vars(4)^2/(vars(4)^2+dw^2).*...
    (w./(dw+vars(2)).*vars(4)^2./((w-(dw+vars(2))+dw).^2+vars(4)^2)).*...
    abs(alpha1(w,I0,dw,Ktm,Kd,vars)./C(I0,dw,vars)+...
    alpha2(w,I0,dw,Ktm,Kd,vars)./C(I0,dw,vars).*((w-(dw+vars(2))+dw+...
    i*vars(4))./(w-(dw+vars(2))-dw+i*vars(4)))).^2;
%Ibi---------------------------------------------
function [out]=Ibi(dw,Ktm,Kd,vars)
%Bistable current Eq.67
out=2*vars(4)*sqrt((2*vars(1)*(dw+vars(2)))/(3*sqrt(3)*...
    vars(2)*abs(K(Ktm,Kd,vars))*vars(6)));
%intminnoise--------------------------------------
function [out]=intminnoise(w,dw,I0,Ktm,Kd,vars)
%integrand of min. noise (caves bound) in Eq.58
out=1/(2*pi).*(w/(dw+vars(2)).*vars(4)^2./((w-(dw+vars(2))+dw).^2+...
    vars(4)^2)).*abs(alpha1(w,I0,dw,Ktm,Kd,vars)./C(I0,dw,vars)+...
    alpha2(w,I0,dw,Ktm,Kd,vars)./C(I0,dw,vars).*((w-(dw+vars(2))+dw+...
    i*vars(4))./(w-(dw+vars(2))-dw+i*vars(4)))).^2.*...
    (2*vars(5)./((w-(dw+vars(2))-vars(3)).^2+vars(5)^2)-...
    2*vars(5)./(((dw+vars(2))-w-vars(3)).^2+vars(5)^2));
%intnoise-----------------------------------------
function [out]=intnoise(w,dw,I0,Ktm,Kd,vars)
%integrand in Eq.57
out=1/vars(6)*(vars(1).*w)./(2*pi)*2*vars(4)^2./((w-(dw+vars(2))+dw).^2+...
    vars(4)^2).*(abs(beta1(w,I0,dw,Ktm,Kd,vars)).^2+...
    ((w-(dw+vars(2))+dw).^2+vars(4)^2)./((w-(dw+vars(2))-dw).^2+...
    vars(4)^2).*abs(beta2(w,I0,dw,Ktm,Kd,vars)).^2-...
    real(beta1(w,I0,dw,Ktm,Kd,vars))+(w-(dw+vars(2))+dw)./vars(4).*...
    imag(beta1(w,I0,dw,Ktm,Kd,vars)));
%K-----------------------------------------------
function [out] = K(Ktm,Kd,vars)
%Effective nonlinearity, Eq.53
out=Kd-(2*vars(2)*vars(3)*Ktm^2)/(vars(3)^2+vars(5)^2);
%lowbnd------------------------------------------
function [out]=lowbnd(dw)
%lower boundary of bistable region, Eq. 68
out=(1/2).*dw.^(3/2).*(1+3.*dw.^(-2)-(1-(1./dw).^(2)).^(3/2)).^(1/2);
%minnoise----------------------------------------
function [out]=minnoise(ws,gamma,I0,dw,Ktm,Kd,vars)
%min. noise (caves bound), Eq.58
int=@(x)intminnoise(x,dw,I0,Ktm,Kd,vars);
out=abs(1/vars(6)*(vars(1)*ws*2*gamma)/(4*pi)-...
    (I0*Ktm*vars(2)/vars(4))^2*vars(4)^2/(dw^2+vars(4)^2)*...
    quadgk(int,ws-gamma,ws+gamma));
%n------------------------------------------------
function [out]=n(w,vars)
%Bose-Einstein occupation number
out=(exp((vars(1).*w)./(vars(7)*vars(10)))-1).^-1;
%noise--------------------------------------------
function [out]=noise(ws,gamma,I0,dw,Ktm,Kd,vars)
%noise, Eq.57
int=@(x)intnoise(x,dw,I0,Ktm,Kd,vars);
val=quadgk(int,ws-gamma,ws+gamma);
out=val+(vars(1)*ws*2*gamma)/(4*pi*vars(6));
%num----------------------------------------------
function [out]=num(w,dw,vars)
%effective occupation number with coupling, in Eq.56
out=(2*vars(5)./((w-(dw+vars(2))-vars(3)).^2+vars(5)^2)).*...
    (2.*n(w-(dw+vars(2)),vars)+1)+(2*vars(5)./(((dw+vars(2))-w-...
    vars(3)).^2+vars(5)^2)).*(2.*n((dw+vars(2))-w,vars)+1);
%params-------------------------------------------
function [out]=params(I0,dw,Ktm,Kd,vars)
%Numerical solution for width of signal and noise distribution 
%properties, i.e central frequency, width,...
out=zeros(length(I0),4);
opts=optimset('MaxFunEvals',1E5,'TolX',1E-25);
for k=1:length(I0)
    [xmax,b]=fminbnd((@(x)-G(x,I0(k),dw,Ktm,Kd,vars).*...
        num(x,dw,vars)),dw+vars(2)+vars(3)-9000*vars(8),dw+vars(2)+...
        vars(3)+12000*vars(8),opts);
    ymax=-b;
    [nmax,nb]=fminbnd((@(x)-intnoise(x,dw,I0(k),Ktm,Kd,vars)),dw+...
        vars(2)+vars(3)-9000*vars(8),dw+vars(2)+...
        vars(3)+12000*vars(8),opts);
    nsemax=-nb;
    hmax=fzero((@(x)G(x,I0(k),dw,Ktm,Kd,vars).*...
        num(x,dw,vars)-ymax/2),xmax,opts);
    nsmax=fzero((@(x)intnoise(x,dw,I0(k),Ktm,Kd,vars)-...
        nsemax/2),nmax,opts);
    g=abs(xmax-hmax);
    h=abs(nmax-nsmax);
    out(k,1)=xmax;
    out(k,2)=g;
    out(k,3)=nmax;
    out(k,4)=h;
end
%signal---------------------------------------------
function [out]=signal(ws,gamma,I0,dw,Ktm,Kd,vars)
%signal, Eq.56
int=(@(x)G(x,I0,dw,Ktm,Kd,vars).*num(x,dw,vars));
out=quadgk(int,ws-gamma,ws+gamma);
%upbnd----------------------------------------------
function [out]=upbnd(dw)
%upper boundary for bistable region, Eq.68
out=(1/2).*dw.^(3/2).*(1+3.*dw.^(-2)+(1-(1./dw).^(2)).^(3/2)).^(1/2);
\end{lstlisting}
%---------------------------------------------------------------

\section{Simulation of Full Trilinear Evolution}\label{sec:ap2-trilinear}
%---------------------------------------------------------------
\lstset{language=matlab}
\lstset{basicstyle=\footnotesize}
%\lstset{backgroundcolor=listinggray,framerulecolor=blue}
%\lstset{backgroundcolor=listinggray,rulecolor=blue}
\lstset{backgroundcolor=\color{listinggray},rulecolor=\color{blue}}
\lstset{linewidth=6in}
%\lstset{labelstep=10}
%\lstset{commentstyle=\textit, stringstyle=\upshape,stringspaces=false}
\lstset{commentstyle=\textit, stringstyle=\upshape,showspaces=false}
\lstset{frame=single,frameround=tttt}
\begin{lstlisting}[]
function [] = run_new_mc()
%Monte-Carlo simulation of a non-degenerate parametric amplifier
%driven by a harmonic oscillator initially in a coherent state.
%Coupling is assumed to be to a zero temperature bath.
%Requires Quantum Optics Toolbox
%Paul D. Nation, Dartmouth College
%Last modified April 2010 (code cleanup for thesis)
N0=11; %number of basis states for zeroth mode
N1=11; %number of basis states for first mode
N2=N1; %number of basis states for second mode
K=1; %value of parametric coupling strength (rate)
gamma0=0.0;
gamma1=0.0;
gamma2=0.0;
alpha=sqrt(3);%coherent state amplitude
epsilon=0.5i; %squeezing parameter
tfinal=4.0;
dt=0.02;
tlist=0:dt:tfinal; %evaluation times for evaluating differential equation 
taulist=K.*tlist;
ntraj=1; %number of trajectories to run
%defining lowering operators
a0=tensor(destroy(N0),identity(N1),identity(N2));
a1=tensor(identity(N0),destroy(N1),identity(N2)); 
a2=tensor(identity(N0),identity(N1),destroy(N2));
%define quadratures
X0=(a0+a0')/2;
Y0=(a0-a0')/2i;
X1=(a1+a1')/2;
Y1=(a1-a1')/2i;
X2=(a2+a2')/2;
Y2=(a2-a2')/2i;
p=(a1-1i*a2);
%define number oeprators for modes 0->2
num0=a0'*a0;
num1=a1'*a1;
num2=a2'*a2;
nump=p'*p;
%dissipative operators for zero-temp. bath
C0=sqrt(2*gamma0)*a0;
C1=sqrt(2*gamma1)*a1;
C2=sqrt(2*gamma2)*a2;
%inital state for system: coherent state for mode 0 and vacuum for 1&2
vacuum=tensor((basis(N0,1)),basis(N1,1),basis(N2,1));
D=expm(alpha*a0'-conj(alpha)*a0); %mode 0 displacement operator
%S=expm(0.5*conj(epsilon)*a0^2-0.5*epsilon*(a0')^2);
%D=1;
S=1;
init_state=D*S*vacuum;
%interaction picture Hamiltonian
H=1i*K*(a0*a1'*a2'-a0'*a1*a2);
%effective non-unitary Hamiltonian (includes losses)
Heff=H-0.5*1i*((C0'*C0)+(C1'*C1)+(C2'*C2));
% %options for solver (needed to prevent error accumulation)
options.lmm='BDF';
options.iter='NEWTON';
options.mxstep=100000;
options.reltol=1e-6;
options.abstol=1e-6;
%call to monte-carlo solver
mc2file('test.dat',-1i*Heff,{C0,C1,C2},{},init_state,tlist,ntraj,options)
mcsolve('test.dat','out.dat','clix.dat');
fid=fopen('out.dat','rb'); %open data file
%init. arrays for expectation values of num. oper. at tlist times
photon_num0=zeros(1,length(tlist));
photon_num1=zeros(1,length(tlist));
photon_num2=zeros(1,length(tlist));
photon_nump=zeros(1,length(tlist));
%init matricies for prob. of nth num. state at times tlist
Pn0=zeros(N0,length(tlist));
Pn1=zeros(N1,length(tlist));
Pn2=zeros(N2,length(tlist));
%init von-Nuemann entropy
S0=zeros(length(tlist));
S1=zeros(length(tlist));
S12=zeros(length(tlist));
S2=zeros(length(tlist));
Sth=zeros(length(tlist));
Sth0=zeros(length(tlist));
Sth2=zeros(length(tlist));
%initilize matricies
Fid1=zeros(length(tlist));
Fid0=zeros(length(tlist));
Tdist=zeros(length(tlist));
Tdist0=zeros(length(tlist));
thermal=zeros(N1);
thermal0=zeros(N0);
variance=zeros(1,length(tlist));
statepump=zeros(1,length(tlist));
statesignal=zeros(1,length(tlist));
expec0=zeros(1,N0);
V0X=zeros(length(tlist));
V0Y=zeros(length(tlist));
for k=1:ntraj
    if gettraj(fid)~=k, error('Unexpected data in file'); end
    psi=qoread(fid,dims(init_state),size(tlist)); %readout state vectors 
    %expectation values of number operator in modes 0->2
    photon_num0=photon_num0+expect(num0,psi)./norm(psi).^2;
    photon_num1=photon_num1+expect(num1,psi)./norm(psi).^2;
    photon_num2=photon_num2+expect(num2,psi)./norm(psi).^2;
    photon_nump=photon_nump+expect(nump,psi)./norm(psi).^2;
    %calculate avg. probability of being in n-th number state at tlist
    for j=1:length(tlist)
        %partial trace over modes
        p0=ptrace(psi{j},1)/trace(ptrace(psi{j},1));
        p1=ptrace(psi{j},2)/trace(ptrace(psi{j},2));
        p2=ptrace(psi{j},3)/trace(ptrace(psi{j},3));
        %probability distributions over number state of each mode
        elems0=diag(full(p0(:,:)));
        elems1=diag(full(p1(:,:)));
        elems2=diag(full(p2(:,:)));
        Pn0(:,j)=Pn0(:,j)+elems0(:); %distribution of pump states
        Pn1(:,j)=Pn1(:,j)+elems1(:); %distribution of mode 1 states
        Pn2(:,j)=Pn2(:,j)+elems2(:); %distribution of mode 2 states
        %expectation values
        n1=expect(num1,psi{j});
        n0=expect(num0,psi{j});
        n2=expect(num2,psi{j});
        variance(j)=(expect(num1^2,psi{j})-n1^2)/n1;
        [A0,B0]=eig(full(p0(:,:)));
        b0=diag(B0);
        ex=zeros(1,N0);
        for q=1:N0
            for p=1:N0
                ex(q)=ex(q)+A0(p,q)*conj(A0(p,q))*(p-1);
            end
            expec0(q)=ex(q);
        end
        %calculates log of matrix (in proper way)
        [A1,B1]=eig(full(p1(:,:)));
        ln0=A0*diag(log(b0))*A0^-1;
        ln1=A1*diag(log(diag(B1)))*A1^-1;
        S0(j)=S0(j)-trace(full(p0(:,:))*ln0);
        S1(j)=S1(j)-trace(full(p1(:,:))*ln1);
        S2(j)=S2(j)+(1-trace(p0*p0));
        for i=1:N1
            %populates diagonal elements of thermal matrix with <n1> 
            thermal(i,i)=(1+n1)^-1*(n1/(1+n1))^(i-1); 
        end
        for i=1:N0
            %populates diagonal elements of thermal matrix with <n1>
            thermal0(i,i)=(1+n0)^-1*(n0/(1+n0))^(i-1);
        end
        rho=p1;
        rho0=p0;
        %sqrt of density matrix
        sqrt_rho0=(full(rho0(:,:)))^(1/2);
        sqrt_rho=(full(rho(:,:)))^(1/2);
        %fidelity calculation
        Fid1(j)=Fid1(j)+trace((sqrt_rho*thermal*sqrt_rho)^(1/2));
        Fid0(j)=Fid0(j)+trace((sqrt_rho0*thermal0*sqrt_rho0)^(1/2));
        xx=log(1/n1+1);
        xy=log(1/n0+1);
        %thermal entropies
        Sth(j)=Sth(j)-(log(1-exp(-xx))+xx*(1-exp(xx))^(-1));
        Sth0(j)=Sth2(j)-(log(1-exp(-xy))+xy*(1-exp(xy))^(-1));
        %quadrature variance
        V0X(j)=V0X(j)+(expect(X0^2,psi{j})-expect(X0,psi{j})^2);
        V0Y(j)=V0Y(j)+(expect(Y0^2,psi{j})-expect(Y0,psi{j})^2); 
    end
end
fclose(fid); %close data file (prevents errors)
%normalize by number of traj. taken
%avg. num. of photons in each mode at times tlist
avg_photon0=photon_num0/ntraj;
avg_photon1=photon_num1/ntraj;
avg_photon2=photon_num2/ntraj;
%probability of mode 0,1,2 being in the ith-# state for the jth elem. in
%tlist
Pn0=Pn0./ntraj;
Pn1=Pn1./ntraj;
Pn2=Pn2./ntraj;
Fid1=Fid1./ntraj;
Fid0=Fid0./ntraj;
S0=S0./ntraj;
S1=S1./ntraj;
S12=S12./ntraj;
S2=S2./ntraj;
Sth=Sth./ntraj;
Sth2=Sth2./ntraj;
V0X=V0X./ntraj;
V0Y=V0Y./ntraj;
%quadrature claculationws
q1=4*V0X-1;
q2=4*V0Y-1;
end %ends program---------------------------------

\end{lstlisting}
%---------------------------------------------------------------
 %%%

%% file: bib.tex
%%% bibliography for thesis.
%\nocite{}

{
\ssp % single-spacing

\bibliography{bibs}
}

%% file: thesis.bbl
\begin{thebibliography}{159}%
\makeatletter
\providecommand \@ifxundefined [1]{%
 \@ifx{#1\undefined}
}%
\providecommand \@ifnum [1]{%
 \ifnum #1\expandafter \@firstoftwo
 \else \expandafter \@secondoftwo
 \fi
}%
\providecommand \@ifx [1]{%
 \ifx #1\expandafter \@firstoftwo
 \else \expandafter \@secondoftwo
 \fi
}%
\providecommand \natexlab [1]{#1}%
\providecommand \enquote  [1]{``#1''}%
\providecommand \bibnamefont  [1]{#1}%
\providecommand \bibfnamefont [1]{#1}%
\providecommand \citenamefont [1]{#1}%
\providecommand \href@noop [0]{\@secondoftwo}%
\providecommand \href [0]{\begingroup \@sanitize@url \@href}%
\providecommand \@href[1]{\@@startlink{#1}\@@href}%
\providecommand \@@href[1]{\endgroup#1\@@endlink}%
\providecommand \@sanitize@url [0]{\catcode `\\12\catcode `\$12\catcode
  `\&12\catcode `\#12\catcode `\^12\catcode `\_12\catcode `\%12\relax}%
\providecommand \@@startlink[1]{}%
\providecommand \@@endlink[0]{}%
\providecommand \url  [0]{\begingroup\@sanitize@url \@url }%
\providecommand \@url [1]{\endgroup\@href {#1}{\urlprefix }}%
\providecommand \urlprefix  [0]{URL }%
\providecommand \Eprint [0]{\href }%
\providecommand \doibase [0]{http://dx.doi.org/}%
\providecommand \selectlanguage [0]{\@gobble}%
\providecommand \bibinfo  [0]{\@secondoftwo}%
\providecommand \bibfield  [0]{\@secondoftwo}%
\providecommand \translation [1]{[#1]}%
\providecommand \BibitemOpen [0]{}%
\providecommand \bibitemStop [0]{}%
\providecommand \bibitemNoStop [0]{.\EOS\space}%
\providecommand \EOS [0]{\spacefactor3000\relax}%
\providecommand \BibitemShut  [1]{\csname bibitem#1\endcsname}%
\let\auto@bib@innerbib\@empty
%</preamble>
\bibitem [{\citenamefont {Haroche}\ and\ \citenamefont
  {Raimond}(2006)}]{haroche:2006}%
  \BibitemOpen
  \bibfield  {author} {\bibinfo {author} {\bibfnamefont {S.}~\bibnamefont
  {Haroche}}\ and\ \bibinfo {author} {\bibfnamefont {J.~M.}\ \bibnamefont
  {Raimond}},\ }\href@noop {} {\emph {\bibinfo {title} {Exploring the
  Quantum}}}\ (\bibinfo  {publisher} {Oxford University Press},\ \bibinfo
  {year} {2006})\BibitemShut {NoStop}%
\bibitem [{\citenamefont {Blais}\ \emph {et~al.}(2004)\citenamefont {Blais},
  \citenamefont {Huang}, \citenamefont {Wallraff}, \citenamefont {Girvin},\
  and\ \citenamefont {Schoelkopf}}]{blais:2004}%
  \BibitemOpen
  \bibfield  {author} {\bibinfo {author} {\bibfnamefont {A.}~\bibnamefont
  {Blais}}, \bibinfo {author} {\bibfnamefont {R.-S.}\ \bibnamefont {Huang}},
  \bibinfo {author} {\bibfnamefont {A.}~\bibnamefont {Wallraff}}, \bibinfo
  {author} {\bibfnamefont {S.~M.}\ \bibnamefont {Girvin}}, \ and\ \bibinfo
  {author} {\bibfnamefont {R.~J.}\ \bibnamefont {Schoelkopf}},\ }\href@noop {}
  {\bibfield  {journal} {\bibinfo  {journal} {Phys. Rev. A}\ }\textbf {\bibinfo
  {volume} {69}},\ \bibinfo {pages} {062320} (\bibinfo {year}
  {2004})}\BibitemShut {NoStop}%
\bibitem [{\citenamefont {Schoelkopf}\ and\ \citenamefont
  {Girvin}(2008)}]{Schoelkopf:2008p1671}%
  \BibitemOpen
  \bibfield  {author} {\bibinfo {author} {\bibfnamefont {R.~J.}\ \bibnamefont
  {Schoelkopf}}\ and\ \bibinfo {author} {\bibfnamefont {S.~M.}\ \bibnamefont
  {Girvin}},\ }\href {\doibase 10.1038/451664a} {\bibfield  {journal} {\bibinfo
   {journal} {Nature}\ }\textbf {\bibinfo {volume} {451}},\ \bibinfo {pages}
  {664} (\bibinfo {year} {2008})}\BibitemShut {NoStop}%
\bibitem [{\citenamefont {G\"{o}ppl}\ \emph {et~al.}(2008)\citenamefont
  {G\"{o}ppl}, \citenamefont {Fragner}, \citenamefont {Baur}, \citenamefont
  {Bianchetti}, \citenamefont {Filipp}, \citenamefont {Fink}, \citenamefont
  {Leek}, \citenamefont {Puebla}, \citenamefont {Steffen},\ and\ \citenamefont
  {Wallraff}}]{goppl:2008}%
  \BibitemOpen
  \bibfield  {author} {\bibinfo {author} {\bibfnamefont {M.}~\bibnamefont
  {G\"{o}ppl}}, \bibinfo {author} {\bibfnamefont {A.}~\bibnamefont {Fragner}},
  \bibinfo {author} {\bibfnamefont {M.}~\bibnamefont {Baur}}, \bibinfo {author}
  {\bibfnamefont {R.}~\bibnamefont {Bianchetti}}, \bibinfo {author}
  {\bibfnamefont {S.}~\bibnamefont {Filipp}}, \bibinfo {author} {\bibfnamefont
  {J.~M.}\ \bibnamefont {Fink}}, \bibinfo {author} {\bibfnamefont {P.~J.}\
  \bibnamefont {Leek}}, \bibinfo {author} {\bibfnamefont {G.}~\bibnamefont
  {Puebla}}, \bibinfo {author} {\bibfnamefont {L.}~\bibnamefont {Steffen}}, \
  and\ \bibinfo {author} {\bibfnamefont {A.}~\bibnamefont {Wallraff}},\
  }\href@noop {} {\bibfield  {journal} {\bibinfo  {journal} {J. App. Phys.}\
  }\textbf {\bibinfo {volume} {104}},\ \bibinfo {pages} {113904} (\bibinfo
  {year} {2008})}\BibitemShut {NoStop}%
\bibitem [{\citenamefont {Wallraff}\ \emph {et~al.}(2004)\citenamefont
  {Wallraff}, \citenamefont {Schuster}, \citenamefont {Blais}, \citenamefont
  {Frunzio}, \citenamefont {Huang}, \citenamefont {Majer}, \citenamefont
  {Kumar}, \citenamefont {Girvin},\ and\ \citenamefont
  {Schoelkopf}}]{wallraff:2004}%
  \BibitemOpen
  \bibfield  {author} {\bibinfo {author} {\bibfnamefont {A.}~\bibnamefont
  {Wallraff}}, \bibinfo {author} {\bibfnamefont {D.~I.}\ \bibnamefont
  {Schuster}}, \bibinfo {author} {\bibfnamefont {A.}~\bibnamefont {Blais}},
  \bibinfo {author} {\bibfnamefont {L.}~\bibnamefont {Frunzio}}, \bibinfo
  {author} {\bibfnamefont {R.~S.}\ \bibnamefont {Huang}}, \bibinfo {author}
  {\bibfnamefont {J.}~\bibnamefont {Majer}}, \bibinfo {author} {\bibfnamefont
  {S.}~\bibnamefont {Kumar}}, \bibinfo {author} {\bibfnamefont {S.~M.}\
  \bibnamefont {Girvin}}, \ and\ \bibinfo {author} {\bibfnamefont {R.~J.}\
  \bibnamefont {Schoelkopf}},\ }\href@noop {} {\bibfield  {journal} {\bibinfo
  {journal} {Nature}\ }\textbf {\bibinfo {volume} {431}},\ \bibinfo {pages}
  {162} (\bibinfo {year} {2004})}\BibitemShut {NoStop}%
\bibitem [{\citenamefont {Devoret}\ \emph {et~al.}(2007)\citenamefont
  {Devoret}, \citenamefont {Girvin},\ and\ \citenamefont
  {Schoelkopf}}]{Devoret:2007p1782}%
  \BibitemOpen
  \bibfield  {author} {\bibinfo {author} {\bibfnamefont {M.~H.}\ \bibnamefont
  {Devoret}}, \bibinfo {author} {\bibfnamefont {S.~M.}\ \bibnamefont {Girvin}},
  \ and\ \bibinfo {author} {\bibfnamefont {R.~J.}\ \bibnamefont {Schoelkopf}},\
  }\href@noop {} {\bibfield  {journal} {\bibinfo  {journal} {Ann. Phys-Berlin}\
  }\textbf {\bibinfo {volume} {16}},\ \bibinfo {pages} {767} (\bibinfo {year}
  {2007})}\BibitemShut {NoStop}%
\bibitem [{\citenamefont {Abdumalikov}\ \emph {et~al.}(2008)\citenamefont
  {Abdumalikov}, \citenamefont {Astafiev}, \citenamefont {Nakamura},
  \citenamefont {Pashkin},\ and\ \citenamefont {Tsai}}]{abdumalikov:2008}%
  \BibitemOpen
  \bibfield  {author} {\bibinfo {author} {\bibfnamefont {A.~A.}\ \bibnamefont
  {Abdumalikov}}, \bibinfo {author} {\bibfnamefont {O.}~\bibnamefont
  {Astafiev}}, \bibinfo {author} {\bibfnamefont {Y.}~\bibnamefont {Nakamura}},
  \bibinfo {author} {\bibfnamefont {Y.~A.}\ \bibnamefont {Pashkin}}, \ and\
  \bibinfo {author} {\bibfnamefont {J.~S.}\ \bibnamefont {Tsai}},\ }\href@noop
  {} {\bibfield  {journal} {\bibinfo  {journal} {Phys. Rev. B}\ }\textbf
  {\bibinfo {volume} {78}},\ \bibinfo {pages} {180502} (\bibinfo {year}
  {2008})}\BibitemShut {NoStop}%
\bibitem [{\citenamefont {Bourassa}\ \emph {et~al.}(2009)\citenamefont
  {Bourassa}, \citenamefont {Gambetta}, \citenamefont {Abdumalikov},
  \citenamefont {Astafiev}, \citenamefont {Nakamura},\ and\ \citenamefont
  {Blais}}]{bourassa:2009}%
  \BibitemOpen
  \bibfield  {author} {\bibinfo {author} {\bibfnamefont {J.}~\bibnamefont
  {Bourassa}}, \bibinfo {author} {\bibfnamefont {J.~M.}\ \bibnamefont
  {Gambetta}}, \bibinfo {author} {\bibfnamefont {A.~A.}\ \bibnamefont
  {Abdumalikov}}, \bibinfo {author} {\bibfnamefont {O.}~\bibnamefont
  {Astafiev}}, \bibinfo {author} {\bibfnamefont {Y.}~\bibnamefont {Nakamura}},
  \ and\ \bibinfo {author} {\bibfnamefont {A.}~\bibnamefont {Blais}},\
  }\href@noop {} {\bibfield  {journal} {\bibinfo  {journal} {Phys. Rev. A}\
  }\textbf {\bibinfo {volume} {80}},\ \bibinfo {pages} {032109} (\bibinfo
  {year} {2009})}\BibitemShut {NoStop}%
\bibitem [{\citenamefont {Likharev}(1986)}]{likharev:1986}%
  \BibitemOpen
  \bibfield  {author} {\bibinfo {author} {\bibfnamefont {K.~K.}\ \bibnamefont
  {Likharev}},\ }\href@noop {} {\emph {\bibinfo {title} {Dynamics of Josephson
  Junctions and Circuits}}}\ (\bibinfo  {publisher} {Gordon and Breach
  Science},\ \bibinfo {year} {1986})\BibitemShut {NoStop}%
\bibitem [{\citenamefont {Wendin}\ and\ \citenamefont
  {Shumeiko}(2007)}]{wendin:2007}%
  \BibitemOpen
  \bibfield  {author} {\bibinfo {author} {\bibfnamefont {G.}~\bibnamefont
  {Wendin}}\ and\ \bibinfo {author} {\bibfnamefont {V.~S.}\ \bibnamefont
  {Shumeiko}},\ }\href@noop {} {\bibfield  {journal} {\bibinfo  {journal} {Low
  Temp. Phys.}\ }\textbf {\bibinfo {volume} {33}},\ \bibinfo {pages} {724}
  (\bibinfo {year} {2007})}\BibitemShut {NoStop}%
\bibitem [{\citenamefont {You}\ and\ \citenamefont {Nori}(2005)}]{you:2005}%
  \BibitemOpen
  \bibfield  {author} {\bibinfo {author} {\bibfnamefont {J.~Q.}\ \bibnamefont
  {You}}\ and\ \bibinfo {author} {\bibfnamefont {F.}~\bibnamefont {Nori}},\
  }\href@noop {} {\bibfield  {journal} {\bibinfo  {journal} {Physics Today}\
  }\textbf {\bibinfo {volume} {58}},\ \bibinfo {pages} {42} (\bibinfo {year}
  {2005})}\BibitemShut {NoStop}%
\bibitem [{\citenamefont {Clarke}\ and\ \citenamefont
  {Wilhelm}(2008)}]{clarke:2008}%
  \BibitemOpen
  \bibfield  {author} {\bibinfo {author} {\bibfnamefont {J.}~\bibnamefont
  {Clarke}}\ and\ \bibinfo {author} {\bibfnamefont {F.~K.}\ \bibnamefont
  {Wilhelm}},\ }\href@noop {} {\bibfield  {journal} {\bibinfo  {journal}
  {Nature}\ }\textbf {\bibinfo {volume} {453}},\ \bibinfo {pages} {1031}
  (\bibinfo {year} {2008})}\BibitemShut {NoStop}%
\bibitem [{\citenamefont {Nakamura}\ \emph {et~al.}(1999)\citenamefont
  {Nakamura}, \citenamefont {Pashkin},\ and\ \citenamefont
  {Tsai}}]{nakamura:1999}%
  \BibitemOpen
  \bibfield  {author} {\bibinfo {author} {\bibfnamefont {Y.}~\bibnamefont
  {Nakamura}}, \bibinfo {author} {\bibfnamefont {Y.~A.}\ \bibnamefont
  {Pashkin}}, \ and\ \bibinfo {author} {\bibfnamefont {J.~S.}\ \bibnamefont
  {Tsai}},\ }\href@noop {} {\bibfield  {journal} {\bibinfo  {journal} {Nature}\
  }\textbf {\bibinfo {volume} {398}},\ \bibinfo {pages} {786} (\bibinfo {year}
  {1999})}\BibitemShut {NoStop}%
\bibitem [{\citenamefont {Schuster}\ \emph {et~al.}(2007)\citenamefont
  {Schuster}, \citenamefont {Houck}, \citenamefont {Schreier}, \citenamefont
  {Wallraff}, \citenamefont {Gambetta}, \citenamefont {Blais}, \citenamefont
  {Frunzio}, \citenamefont {Majer}, \citenamefont {Johnson}, \citenamefont
  {Devoret}, \citenamefont {Girvin},\ and\ \citenamefont
  {Schoelkopf}}]{schuster:2007}%
  \BibitemOpen
  \bibfield  {author} {\bibinfo {author} {\bibfnamefont {D.~I.}\ \bibnamefont
  {Schuster}}, \bibinfo {author} {\bibfnamefont {A.~A.}\ \bibnamefont {Houck}},
  \bibinfo {author} {\bibfnamefont {J.~A.}\ \bibnamefont {Schreier}}, \bibinfo
  {author} {\bibfnamefont {A.}~\bibnamefont {Wallraff}}, \bibinfo {author}
  {\bibfnamefont {J.~M.}\ \bibnamefont {Gambetta}}, \bibinfo {author}
  {\bibfnamefont {A.}~\bibnamefont {Blais}}, \bibinfo {author} {\bibfnamefont
  {L.}~\bibnamefont {Frunzio}}, \bibinfo {author} {\bibfnamefont
  {J.}~\bibnamefont {Majer}}, \bibinfo {author} {\bibfnamefont
  {B.}~\bibnamefont {Johnson}}, \bibinfo {author} {\bibfnamefont {M.~H.}\
  \bibnamefont {Devoret}}, \bibinfo {author} {\bibfnamefont {S.~M.}\
  \bibnamefont {Girvin}}, \ and\ \bibinfo {author} {\bibfnamefont {R.~J.}\
  \bibnamefont {Schoelkopf}},\ }\href@noop {} {\bibfield  {journal} {\bibinfo
  {journal} {Nature}\ }\textbf {\bibinfo {volume} {445}},\ \bibinfo {pages}
  {515} (\bibinfo {year} {2007})}\BibitemShut {NoStop}%
\bibitem [{\citenamefont {Houck}\ \emph {et~al.}(2007)\citenamefont {Houck},
  \citenamefont {Schuster}, \citenamefont {Gambetta}, \citenamefont {Schreier},
  \citenamefont {Johnson}, \citenamefont {Chow}, \citenamefont {Frunzio},
  \citenamefont {Majer}, \citenamefont {Devoret}, \citenamefont {Girvin},\ and\
  \citenamefont {Schoelkopf}}]{houck:2007}%
  \BibitemOpen
  \bibfield  {author} {\bibinfo {author} {\bibfnamefont {A.~A.}\ \bibnamefont
  {Houck}}, \bibinfo {author} {\bibfnamefont {D.~I.}\ \bibnamefont {Schuster}},
  \bibinfo {author} {\bibfnamefont {J.~M.}\ \bibnamefont {Gambetta}}, \bibinfo
  {author} {\bibfnamefont {J.~A.}\ \bibnamefont {Schreier}}, \bibinfo {author}
  {\bibfnamefont {B.~R.}\ \bibnamefont {Johnson}}, \bibinfo {author}
  {\bibfnamefont {J.~M.}\ \bibnamefont {Chow}}, \bibinfo {author}
  {\bibfnamefont {L.}~\bibnamefont {Frunzio}}, \bibinfo {author} {\bibfnamefont
  {J.}~\bibnamefont {Majer}}, \bibinfo {author} {\bibfnamefont {M.~H.}\
  \bibnamefont {Devoret}}, \bibinfo {author} {\bibfnamefont {S.~M.}\
  \bibnamefont {Girvin}}, \ and\ \bibinfo {author} {\bibfnamefont {R.~J.}\
  \bibnamefont {Schoelkopf}},\ }\href@noop {} {\bibfield  {journal} {\bibinfo
  {journal} {Nature}\ }\textbf {\bibinfo {volume} {449}},\ \bibinfo {pages}
  {328} (\bibinfo {year} {2007})}\BibitemShut {NoStop}%
\bibitem [{\citenamefont {amd E.~M.~Weig}\ \emph {et~al.}(2008)\citenamefont
  {amd E.~M.~Weig}, \citenamefont {Ansmann}, \citenamefont {Bialczak},
  \citenamefont {Lucero}, \citenamefont {Neeley}, \citenamefont {O'Connell},
  \citenamefont {Wang}, \citenamefont {Martinis},\ and\ \citenamefont
  {Cleland}}]{hofheinz:2008}%
  \BibitemOpen
  \bibfield  {author} {\bibinfo {author} {\bibfnamefont {M.~H.}\ \bibnamefont
  {amd E.~M.~Weig}}, \bibinfo {author} {\bibfnamefont {M.}~\bibnamefont
  {Ansmann}}, \bibinfo {author} {\bibfnamefont {R.~C.}\ \bibnamefont
  {Bialczak}}, \bibinfo {author} {\bibfnamefont {E.}~\bibnamefont {Lucero}},
  \bibinfo {author} {\bibfnamefont {M.}~\bibnamefont {Neeley}}, \bibinfo
  {author} {\bibfnamefont {A.~D.}\ \bibnamefont {O'Connell}}, \bibinfo {author}
  {\bibfnamefont {H.}~\bibnamefont {Wang}}, \bibinfo {author} {\bibfnamefont
  {J.~M.}\ \bibnamefont {Martinis}}, \ and\ \bibinfo {author} {\bibfnamefont
  {A.~N.}\ \bibnamefont {Cleland}},\ }\href@noop {} {\bibfield  {journal}
  {\bibinfo  {journal} {Nature}\ }\textbf {\bibinfo {volume} {454}},\ \bibinfo
  {pages} {310} (\bibinfo {year} {2008})}\BibitemShut {NoStop}%
\bibitem [{\citenamefont {Hofheinz}\ \emph {et~al.}(2009)\citenamefont
  {Hofheinz}, \citenamefont {Wang}, \citenamefont {Ansmann}, \citenamefont
  {Bialczak}, \citenamefont {Lucero}, \citenamefont {Neeley}, \citenamefont
  {O'Connell}, \citenamefont {Sank}, \citenamefont {Wenner}, \citenamefont
  {Martinis},\ and\ \citenamefont {Cleland}}]{hofheinz:2009}%
  \BibitemOpen
  \bibfield  {author} {\bibinfo {author} {\bibfnamefont {M.}~\bibnamefont
  {Hofheinz}}, \bibinfo {author} {\bibfnamefont {H.}~\bibnamefont {Wang}},
  \bibinfo {author} {\bibfnamefont {M.}~\bibnamefont {Ansmann}}, \bibinfo
  {author} {\bibfnamefont {R.~C.}\ \bibnamefont {Bialczak}}, \bibinfo {author}
  {\bibfnamefont {E.}~\bibnamefont {Lucero}}, \bibinfo {author} {\bibfnamefont
  {M.}~\bibnamefont {Neeley}}, \bibinfo {author} {\bibfnamefont {A.~D.~O.}\
  \bibnamefont {O'Connell}}, \bibinfo {author} {\bibfnamefont {D.}~\bibnamefont
  {Sank}}, \bibinfo {author} {\bibfnamefont {J.}~\bibnamefont {Wenner}},
  \bibinfo {author} {\bibfnamefont {J.~M.}\ \bibnamefont {Martinis}}, \ and\
  \bibinfo {author} {\bibfnamefont {A.~N.}\ \bibnamefont {Cleland}},\
  }\href@noop {} {\bibfield  {journal} {\bibinfo  {journal} {Nature}\ }\textbf
  {\bibinfo {volume} {469}},\ \bibinfo {pages} {546} (\bibinfo {year}
  {2009})}\BibitemShut {NoStop}%
\bibitem [{\citenamefont {Wang}\ \emph
  {et~al.}(2008{\natexlab{a}})\citenamefont {Wang}, \citenamefont {Hofheinz},
  \citenamefont {Ansmann}, \citenamefont {Bialczak}, \citenamefont {Lucero},
  \citenamefont {Neeley}, \citenamefont {O'Connell}, \citenamefont {Sank},
  \citenamefont {Wenner}, \citenamefont {Cleland},\ and\ \citenamefont
  {Martinis}}]{wang:2008}%
  \BibitemOpen
  \bibfield  {author} {\bibinfo {author} {\bibfnamefont {H.}~\bibnamefont
  {Wang}}, \bibinfo {author} {\bibfnamefont {M.}~\bibnamefont {Hofheinz}},
  \bibinfo {author} {\bibfnamefont {M.}~\bibnamefont {Ansmann}}, \bibinfo
  {author} {\bibfnamefont {R.~C.}\ \bibnamefont {Bialczak}}, \bibinfo {author}
  {\bibfnamefont {E.}~\bibnamefont {Lucero}}, \bibinfo {author} {\bibfnamefont
  {M.}~\bibnamefont {Neeley}}, \bibinfo {author} {\bibfnamefont {A.~D.}\
  \bibnamefont {O'Connell}}, \bibinfo {author} {\bibfnamefont {D.}~\bibnamefont
  {Sank}}, \bibinfo {author} {\bibfnamefont {J.}~\bibnamefont {Wenner}},
  \bibinfo {author} {\bibfnamefont {A.~N.}\ \bibnamefont {Cleland}}, \ and\
  \bibinfo {author} {\bibfnamefont {J.~M.}\ \bibnamefont {Martinis}},\
  }\href@noop {} {\bibfield  {journal} {\bibinfo  {journal} {Phys. Rev. Lett.}\
  }\textbf {\bibinfo {volume} {101}},\ \bibinfo {pages} {240401} (\bibinfo
  {year} {2008}{\natexlab{a}})}\BibitemShut {NoStop}%
\bibitem [{\citenamefont {Majer}\ \emph {et~al.}(2007)\citenamefont {Majer},
  \citenamefont {Chow}, \citenamefont {Gambetta}, \citenamefont {Koch},
  \citenamefont {Johnson}, \citenamefont {Schreier}, \citenamefont {Frunzio},
  \citenamefont {Schuster}, \citenamefont {Houck}, \citenamefont {Wallraff},
  \citenamefont {Blais}, \citenamefont {Devoret}, \citenamefont {Girvin},\ and\
  \citenamefont {Schoelkopf}}]{majer:2007}%
  \BibitemOpen
  \bibfield  {author} {\bibinfo {author} {\bibfnamefont {J.}~\bibnamefont
  {Majer}}, \bibinfo {author} {\bibfnamefont {J.~M.}\ \bibnamefont {Chow}},
  \bibinfo {author} {\bibfnamefont {J.~M.}\ \bibnamefont {Gambetta}}, \bibinfo
  {author} {\bibfnamefont {J.}~\bibnamefont {Koch}}, \bibinfo {author}
  {\bibfnamefont {B.~R.}\ \bibnamefont {Johnson}}, \bibinfo {author}
  {\bibfnamefont {J.~A.}\ \bibnamefont {Schreier}}, \bibinfo {author}
  {\bibfnamefont {L.}~\bibnamefont {Frunzio}}, \bibinfo {author} {\bibfnamefont
  {D.~I.}\ \bibnamefont {Schuster}}, \bibinfo {author} {\bibfnamefont {A.~A.}\
  \bibnamefont {Houck}}, \bibinfo {author} {\bibfnamefont {A.}~\bibnamefont
  {Wallraff}}, \bibinfo {author} {\bibfnamefont {A.}~\bibnamefont {Blais}},
  \bibinfo {author} {\bibfnamefont {M.~H.}\ \bibnamefont {Devoret}}, \bibinfo
  {author} {\bibfnamefont {S.~M.}\ \bibnamefont {Girvin}}, \ and\ \bibinfo
  {author} {\bibfnamefont {R.~J.}\ \bibnamefont {Schoelkopf}},\ }\href@noop {}
  {\bibfield  {journal} {\bibinfo  {journal} {Nature}\ }\textbf {\bibinfo
  {volume} {449}},\ \bibinfo {pages} {443} (\bibinfo {year}
  {2007})}\BibitemShut {NoStop}%
\bibitem [{\citenamefont {Astafiev}\ \emph {et~al.}(2007)\citenamefont
  {Astafiev}, \citenamefont {Inomata}, \citenamefont {Niskanen}, \citenamefont
  {Yamamoto}, \citenamefont {Pashkin}, \citenamefont {Nakamura},\ and\
  \citenamefont {Tsai}}]{astafiev:2007}%
  \BibitemOpen
  \bibfield  {author} {\bibinfo {author} {\bibfnamefont {O.}~\bibnamefont
  {Astafiev}}, \bibinfo {author} {\bibfnamefont {K.}~\bibnamefont {Inomata}},
  \bibinfo {author} {\bibfnamefont {A.~O.}\ \bibnamefont {Niskanen}}, \bibinfo
  {author} {\bibfnamefont {T.}~\bibnamefont {Yamamoto}}, \bibinfo {author}
  {\bibfnamefont {Y.~A.}\ \bibnamefont {Pashkin}}, \bibinfo {author}
  {\bibfnamefont {Y.}~\bibnamefont {Nakamura}}, \ and\ \bibinfo {author}
  {\bibfnamefont {J.~S.}\ \bibnamefont {Tsai}},\ }\href@noop {} {\bibfield
  {journal} {\bibinfo  {journal} {Science}\ }\textbf {\bibinfo {volume}
  {449}},\ \bibinfo {pages} {588} (\bibinfo {year} {2007})}\BibitemShut
  {NoStop}%
\bibitem [{\citenamefont {Astafiev}\ \emph
  {et~al.}(2010{\natexlab{a}})\citenamefont {Astafiev}, \citenamefont
  {Zagoskin}, \citenamefont {Abdumalikov}, \citenamefont {Pashkin},
  \citenamefont {Yamamoto}, \citenamefont {Inomata}, \citenamefont {Nakamura},\
  and\ \citenamefont {Tsai}}]{astafiev:2010a}%
  \BibitemOpen
  \bibfield  {author} {\bibinfo {author} {\bibfnamefont {O.}~\bibnamefont
  {Astafiev}}, \bibinfo {author} {\bibfnamefont {A.~M.}\ \bibnamefont
  {Zagoskin}}, \bibinfo {author} {\bibfnamefont {A.~A.}\ \bibnamefont
  {Abdumalikov}}, \bibinfo {author} {\bibfnamefont {Y.~A.}\ \bibnamefont
  {Pashkin}}, \bibinfo {author} {\bibfnamefont {T.}~\bibnamefont {Yamamoto}},
  \bibinfo {author} {\bibfnamefont {K.}~\bibnamefont {Inomata}}, \bibinfo
  {author} {\bibfnamefont {Y.}~\bibnamefont {Nakamura}}, \ and\ \bibinfo
  {author} {\bibfnamefont {J.~S.}\ \bibnamefont {Tsai}},\ }\href@noop {}
  {\bibfield  {journal} {\bibinfo  {journal} {Science}\ }\textbf {\bibinfo
  {volume} {327}},\ \bibinfo {pages} {840} (\bibinfo {year}
  {2010}{\natexlab{a}})}\BibitemShut {NoStop}%
\bibitem [{\citenamefont {Astafiev}\ \emph
  {et~al.}(2010{\natexlab{b}})\citenamefont {Astafiev}, \citenamefont
  {Abdumalikov}, \citenamefont {Zagoskin}, \citenamefont {Pashkin},
  \citenamefont {Nakamura},\ and\ \citenamefont {Tsai}}]{astafiev:2010b}%
  \BibitemOpen
  \bibfield  {author} {\bibinfo {author} {\bibfnamefont {O.}~\bibnamefont
  {Astafiev}}, \bibinfo {author} {\bibfnamefont {A.~A.}\ \bibnamefont
  {Abdumalikov}}, \bibinfo {author} {\bibfnamefont {A.~M.}\ \bibnamefont
  {Zagoskin}}, \bibinfo {author} {\bibfnamefont {Y.~A.}\ \bibnamefont
  {Pashkin}}, \bibinfo {author} {\bibfnamefont {Y.}~\bibnamefont {Nakamura}}, \
  and\ \bibinfo {author} {\bibfnamefont {J.~S.}\ \bibnamefont {Tsai}},\
  }\href@noop {} {\bibfield  {journal} {\bibinfo  {journal} {Phys Rev. Lett.}\
  }\textbf {\bibinfo {volume} {104}},\ \bibinfo {pages} {183603} (\bibinfo
  {year} {2010}{\natexlab{b}})}\BibitemShut {NoStop}%
\bibitem [{\citenamefont {Blencowe}(2005)}]{blencowe:2005}%
  \BibitemOpen
  \bibfield  {author} {\bibinfo {author} {\bibfnamefont {M.~P.}\ \bibnamefont
  {Blencowe}},\ }\href@noop {} {\bibfield  {journal} {\bibinfo  {journal}
  {Contemp. Phys.}\ }\textbf {\bibinfo {volume} {46}},\ \bibinfo {pages} {249}
  (\bibinfo {year} {2005})}\BibitemShut {NoStop}%
\bibitem [{\citenamefont {Schwab}\ and\ \citenamefont
  {Roukes}(2005)}]{schwab:2005}%
  \BibitemOpen
  \bibfield  {author} {\bibinfo {author} {\bibfnamefont {K.~C.}\ \bibnamefont
  {Schwab}}\ and\ \bibinfo {author} {\bibfnamefont {M.~L.}\ \bibnamefont
  {Roukes}},\ }\href@noop {} {\bibfield  {journal} {\bibinfo  {journal} {Phys.
  Today}\ }\textbf {\bibinfo {volume} {58}},\ \bibinfo {pages} {36} (\bibinfo
  {year} {2005})}\BibitemShut {NoStop}%
\bibitem [{\citenamefont {Aspelmeyer}\ and\ \citenamefont
  {Schwab}(2008)}]{aspelmeyer:2008}%
  \BibitemOpen
  \bibfield  {author} {\bibinfo {author} {\bibfnamefont {M.}~\bibnamefont
  {Aspelmeyer}}\ and\ \bibinfo {author} {\bibfnamefont {K.~C.}\ \bibnamefont
  {Schwab}},\ }\href@noop {} {\bibfield  {journal} {\bibinfo  {journal} {New J.
  Phys.}\ }\textbf {\bibinfo {volume} {20}},\ \bibinfo {pages} {095001}
  (\bibinfo {year} {2008})}\BibitemShut {NoStop}%
\bibitem [{\citenamefont {Xue}\ \emph {et~al.}(2007{\natexlab{a}})\citenamefont
  {Xue}, \citenamefont {Wang}, \citenamefont {xi~Liu},\ and\ \citenamefont
  {Nori}}]{Xue:2007p2711}%
  \BibitemOpen
  \bibfield  {author} {\bibinfo {author} {\bibfnamefont {F.}~\bibnamefont
  {Xue}}, \bibinfo {author} {\bibfnamefont {Y.~D.}\ \bibnamefont {Wang}},
  \bibinfo {author} {\bibfnamefont {Y.}~\bibnamefont {xi~Liu}}, \ and\ \bibinfo
  {author} {\bibfnamefont {F.}~\bibnamefont {Nori}},\ }\href {\doibase
  10.1103/PhysRevB.76.205302} {\bibfield  {journal} {\bibinfo  {journal} {Phys
  Rev B}\ }\textbf {\bibinfo {volume} {76}},\ \bibinfo {pages} {205302}
  (\bibinfo {year} {2007}{\natexlab{a}})}\BibitemShut {NoStop}%
\bibitem [{\citenamefont {Blencowe}\ and\ \citenamefont
  {Buks}(2007)}]{Blencowe:2007p285}%
  \BibitemOpen
  \bibfield  {author} {\bibinfo {author} {\bibfnamefont {M.~P.}\ \bibnamefont
  {Blencowe}}\ and\ \bibinfo {author} {\bibfnamefont {E.}~\bibnamefont
  {Buks}},\ }\href {\doibase 10.1103/PhysRevB.76.014511} {\bibfield  {journal}
  {\bibinfo  {journal} {Phys Rev B}\ }\textbf {\bibinfo {volume} {76}},\
  \bibinfo {pages} {014511} (\bibinfo {year} {2007})}\BibitemShut {NoStop}%
\bibitem [{\citenamefont {Buks}\ \emph {et~al.}(2007)\citenamefont {Buks},
  \citenamefont {Zaitsev}, \citenamefont {Segev}, \citenamefont {Abdo},\ and\
  \citenamefont {Blencowe}}]{Buks:2007p279}%
  \BibitemOpen
  \bibfield  {author} {\bibinfo {author} {\bibfnamefont {E.}~\bibnamefont
  {Buks}}, \bibinfo {author} {\bibfnamefont {S.}~\bibnamefont {Zaitsev}},
  \bibinfo {author} {\bibfnamefont {E.}~\bibnamefont {Segev}}, \bibinfo
  {author} {\bibfnamefont {B.}~\bibnamefont {Abdo}}, \ and\ \bibinfo {author}
  {\bibfnamefont {M.~P.}\ \bibnamefont {Blencowe}},\ }\href {\doibase
  10.1103/PhysRevE.76.026217} {\bibfield  {journal} {\bibinfo  {journal} {Phys
  Rev E}\ }\textbf {\bibinfo {volume} {76}},\ \bibinfo {pages} {026217}
  (\bibinfo {year} {2007})}\BibitemShut {NoStop}%
\bibitem [{\citenamefont {Regal}\ \emph {et~al.}(2008)\citenamefont {Regal},
  \citenamefont {Teufel},\ and\ \citenamefont {Lehnert}}]{Regal:2008p2560}%
  \BibitemOpen
  \bibfield  {author} {\bibinfo {author} {\bibfnamefont {C.~A.}\ \bibnamefont
  {Regal}}, \bibinfo {author} {\bibfnamefont {J.~D.}\ \bibnamefont {Teufel}}, \
  and\ \bibinfo {author} {\bibfnamefont {K.~W.}\ \bibnamefont {Lehnert}},\
  }\href@noop {} {\bibfield  {journal} {\bibinfo  {journal} {Nature Phys.}\
  }\textbf {\bibinfo {volume} {4}},\ \bibinfo {pages} {555} (\bibinfo {year}
  {2008})}\BibitemShut {NoStop}%
\bibitem [{\citenamefont {Teufel}\ \emph {et~al.}(2008)\citenamefont {Teufel},
  \citenamefont {Regal},\ and\ \citenamefont {Lehnert}}]{Teufel:2008p2398}%
  \BibitemOpen
  \bibfield  {author} {\bibinfo {author} {\bibfnamefont {J.~D.}\ \bibnamefont
  {Teufel}}, \bibinfo {author} {\bibfnamefont {C.~A.}\ \bibnamefont {Regal}}, \
  and\ \bibinfo {author} {\bibfnamefont {K.~W.}\ \bibnamefont {Lehnert}},\
  }\href@noop {} {\bibfield  {journal} {\bibinfo  {journal} {New J. Phys.}\
  }\textbf {\bibinfo {volume} {10}},\ \bibinfo {pages} {095002} (\bibinfo
  {year} {2008})}\BibitemShut {NoStop}%
\bibitem [{\citenamefont {Li}\ \emph {et~al.}(2008)\citenamefont {Li},
  \citenamefont {Wang}, \citenamefont {Xue},\ and\ \citenamefont
  {Bruder}}]{li:2008}%
  \BibitemOpen
  \bibfield  {author} {\bibinfo {author} {\bibfnamefont {Y.}~\bibnamefont
  {Li}}, \bibinfo {author} {\bibfnamefont {Y.~D.}\ \bibnamefont {Wang}},
  \bibinfo {author} {\bibfnamefont {F.}~\bibnamefont {Xue}}, \ and\ \bibinfo
  {author} {\bibfnamefont {C.}~\bibnamefont {Bruder}},\ }\href@noop {}
  {\bibfield  {journal} {\bibinfo  {journal} {Phys. Rev. B}\ }\textbf {\bibinfo
  {volume} {78}},\ \bibinfo {pages} {134301} (\bibinfo {year}
  {2008})}\BibitemShut {NoStop}%
\bibitem [{\citenamefont {Elste}\ \emph {et~al.}(2009)\citenamefont {Elste},
  \citenamefont {Girvin},\ and\ \citenamefont {Clerk}}]{elste:2009}%
  \BibitemOpen
  \bibfield  {author} {\bibinfo {author} {\bibfnamefont {F.}~\bibnamefont
  {Elste}}, \bibinfo {author} {\bibfnamefont {S.~M.}\ \bibnamefont {Girvin}}, \
  and\ \bibinfo {author} {\bibfnamefont {A.~A.}\ \bibnamefont {Clerk}},\
  }\href@noop {} {\bibfield  {journal} {\bibinfo  {journal} {Phys Rev. Lett.}\
  }\textbf {\bibinfo {volume} {102}},\ \bibinfo {pages} {207209} (\bibinfo
  {year} {2009})}\BibitemShut {NoStop}%
\bibitem [{\citenamefont {Hertzberg}\ \emph {et~al.}(2009)\citenamefont
  {Hertzberg}, \citenamefont {Rocheleau}, \citenamefont {Ndukum}, \citenamefont
  {Savva}, \citenamefont {Clerk},\ and\ \citenamefont
  {Schwab}}]{hertzberg:2009}%
  \BibitemOpen
  \bibfield  {author} {\bibinfo {author} {\bibfnamefont {J.~B.}\ \bibnamefont
  {Hertzberg}}, \bibinfo {author} {\bibfnamefont {T.}~\bibnamefont
  {Rocheleau}}, \bibinfo {author} {\bibfnamefont {T.}~\bibnamefont {Ndukum}},
  \bibinfo {author} {\bibfnamefont {M.}~\bibnamefont {Savva}}, \bibinfo
  {author} {\bibfnamefont {A.~A.}\ \bibnamefont {Clerk}}, \ and\ \bibinfo
  {author} {\bibfnamefont {K.~C.}\ \bibnamefont {Schwab}},\ }\href@noop {}
  {\bibfield  {journal} {\bibinfo  {journal} {Nature Phys.}\ }\textbf {\bibinfo
  {volume} {6}},\ \bibinfo {pages} {213} (\bibinfo {year} {2009})}\BibitemShut
  {NoStop}%
\bibitem [{\citenamefont {Kippenberg}\ and\ \citenamefont
  {Vahala}(2007{\natexlab{a}})}]{kippenberg:2007}%
  \BibitemOpen
  \bibfield  {author} {\bibinfo {author} {\bibfnamefont {T.~J.}\ \bibnamefont
  {Kippenberg}}\ and\ \bibinfo {author} {\bibfnamefont {K.~J.}\ \bibnamefont
  {Vahala}},\ }\href@noop {} {\bibfield  {journal} {\bibinfo  {journal} {Opt.
  Express}\ }\textbf {\bibinfo {volume} {15}},\ \bibinfo {pages} {17172}
  (\bibinfo {year} {2007}{\natexlab{a}})}\BibitemShut {NoStop}%
\bibitem [{\citenamefont {Marquardt}\ and\ \citenamefont
  {Girvin}(2009)}]{marquardt:2009}%
  \BibitemOpen
  \bibfield  {author} {\bibinfo {author} {\bibfnamefont {F.}~\bibnamefont
  {Marquardt}}\ and\ \bibinfo {author} {\bibfnamefont {S.~M.}\ \bibnamefont
  {Girvin}},\ }\href@noop {} {\bibfield  {journal} {\bibinfo  {journal}
  {Physics}\ }\textbf {\bibinfo {volume} {2}},\ \bibinfo {pages} {40} (\bibinfo
  {year} {2009})}\BibitemShut {NoStop}%
\bibitem [{\citenamefont {O'Connell}\ \emph {et~al.}(2010)\citenamefont
  {O'Connell}, \citenamefont {Hofheinz}, \citenamefont {Ansmann}, \citenamefont
  {Bialczak}, \citenamefont {Lenander}, \citenamefont {Lucero}, \citenamefont
  {Neeley}, \citenamefont {Sank}, \citenamefont {Wang}, \citenamefont {Weides},
  \citenamefont {Wenner}, \citenamefont {Martinis},\ and\ \citenamefont
  {Cleland}}]{oconnell:2010}%
  \BibitemOpen
  \bibfield  {author} {\bibinfo {author} {\bibfnamefont {A.~D.}\ \bibnamefont
  {O'Connell}}, \bibinfo {author} {\bibfnamefont {M.}~\bibnamefont {Hofheinz}},
  \bibinfo {author} {\bibfnamefont {M.}~\bibnamefont {Ansmann}}, \bibinfo
  {author} {\bibfnamefont {R.~C.}\ \bibnamefont {Bialczak}}, \bibinfo {author}
  {\bibfnamefont {M.}~\bibnamefont {Lenander}}, \bibinfo {author}
  {\bibfnamefont {E.}~\bibnamefont {Lucero}}, \bibinfo {author} {\bibfnamefont
  {M.}~\bibnamefont {Neeley}}, \bibinfo {author} {\bibfnamefont
  {D.}~\bibnamefont {Sank}}, \bibinfo {author} {\bibfnamefont {H.}~\bibnamefont
  {Wang}}, \bibinfo {author} {\bibfnamefont {M.}~\bibnamefont {Weides}},
  \bibinfo {author} {\bibfnamefont {J.}~\bibnamefont {Wenner}}, \bibinfo
  {author} {\bibfnamefont {J.~M.}\ \bibnamefont {Martinis}}, \ and\ \bibinfo
  {author} {\bibfnamefont {A.~N.}\ \bibnamefont {Cleland}},\ }\href@noop {}
  {\bibfield  {journal} {\bibinfo  {journal} {Nature}\ }\textbf {\bibinfo
  {volume} {464}},\ \bibinfo {pages} {697} (\bibinfo {year}
  {2010})}\BibitemShut {NoStop}%
\bibitem [{\citenamefont {Krommer}\ \emph {et~al.}(2000)\citenamefont
  {Krommer}, \citenamefont {Erbe}, \citenamefont {Tilke}, \citenamefont
  {Manus},\ and\ \citenamefont {Blick}}]{Krommer:2000p2946}%
  \BibitemOpen
  \bibfield  {author} {\bibinfo {author} {\bibfnamefont {H.}~\bibnamefont
  {Krommer}}, \bibinfo {author} {\bibfnamefont {A.}~\bibnamefont {Erbe}},
  \bibinfo {author} {\bibfnamefont {A.}~\bibnamefont {Tilke}}, \bibinfo
  {author} {\bibfnamefont {S.}~\bibnamefont {Manus}}, \ and\ \bibinfo {author}
  {\bibfnamefont {R.~H.}\ \bibnamefont {Blick}},\ }\href@noop {} {\bibfield
  {journal} {\bibinfo  {journal} {Europhys Lett}\ }\textbf {\bibinfo {volume}
  {50}},\ \bibinfo {pages} {101} (\bibinfo {year} {2000})}\BibitemShut
  {NoStop}%
\bibitem [{\citenamefont {Aldridge}\ and\ \citenamefont
  {Cleland}(2005)}]{Aldridge:2005p3614}%
  \BibitemOpen
  \bibfield  {author} {\bibinfo {author} {\bibfnamefont {J.~S.}\ \bibnamefont
  {Aldridge}}\ and\ \bibinfo {author} {\bibfnamefont {A.~N.}\ \bibnamefont
  {Cleland}},\ }\href@noop {} {\bibfield  {journal} {\bibinfo  {journal} {Phys
  Rev Lett}\ }\textbf {\bibinfo {volume} {94}},\ \bibinfo {pages} {156403}
  (\bibinfo {year} {2005})}\BibitemShut {NoStop}%
\bibitem [{\citenamefont {Almog}\ \emph {et~al.}(2007)\citenamefont {Almog},
  \citenamefont {Zaitsev}, \citenamefont {Shtempluck},\ and\ \citenamefont
  {Buks}}]{Almog:2007p2875}%
  \BibitemOpen
  \bibfield  {author} {\bibinfo {author} {\bibfnamefont {R.}~\bibnamefont
  {Almog}}, \bibinfo {author} {\bibfnamefont {S.}~\bibnamefont {Zaitsev}},
  \bibinfo {author} {\bibfnamefont {O.}~\bibnamefont {Shtempluck}}, \ and\
  \bibinfo {author} {\bibfnamefont {E.}~\bibnamefont {Buks}},\ }\href@noop {}
  {\bibfield  {journal} {\bibinfo  {journal} {Appl Phys Lett}\ }\textbf
  {\bibinfo {volume} {90}},\ \bibinfo {pages} {013508} (\bibinfo {year}
  {2007})}\BibitemShut {NoStop}%
\bibitem [{\citenamefont {Siddiqi}\ \emph {et~al.}(2004)\citenamefont
  {Siddiqi}, \citenamefont {Vijay}, \citenamefont {Pierre}, \citenamefont
  {Wilson}, \citenamefont {Metcalfe}, \citenamefont {Rigetti}, \citenamefont
  {Frunzio},\ and\ \citenamefont {Devoret}}]{Siddiqi:2004p1662}%
  \BibitemOpen
  \bibfield  {author} {\bibinfo {author} {\bibfnamefont {I.}~\bibnamefont
  {Siddiqi}}, \bibinfo {author} {\bibfnamefont {R.}~\bibnamefont {Vijay}},
  \bibinfo {author} {\bibfnamefont {F.}~\bibnamefont {Pierre}}, \bibinfo
  {author} {\bibfnamefont {C.}~\bibnamefont {Wilson}}, \bibinfo {author}
  {\bibfnamefont {M.}~\bibnamefont {Metcalfe}}, \bibinfo {author}
  {\bibfnamefont {C.}~\bibnamefont {Rigetti}}, \bibinfo {author} {\bibfnamefont
  {L.}~\bibnamefont {Frunzio}}, \ and\ \bibinfo {author} {\bibfnamefont
  {M.}~\bibnamefont {Devoret}},\ }\href@noop {} {\bibfield  {journal} {\bibinfo
   {journal} {Phys Rev Lett}\ }\textbf {\bibinfo {volume} {93}},\ \bibinfo
  {pages} {207002} (\bibinfo {year} {2004})}\BibitemShut {NoStop}%
\bibitem [{\citenamefont {Lupascu}\ \emph {et~al.}(2006)\citenamefont
  {Lupascu}, \citenamefont {Driessen}, \citenamefont {Roschier}, \citenamefont
  {Harmans},\ and\ \citenamefont {Mooij}}]{Lupascu:2006p3289}%
  \BibitemOpen
  \bibfield  {author} {\bibinfo {author} {\bibfnamefont {A.}~\bibnamefont
  {Lupascu}}, \bibinfo {author} {\bibfnamefont {E.~F.~C.}\ \bibnamefont
  {Driessen}}, \bibinfo {author} {\bibfnamefont {L.}~\bibnamefont {Roschier}},
  \bibinfo {author} {\bibfnamefont {C.}~\bibnamefont {Harmans}}, \ and\
  \bibinfo {author} {\bibfnamefont {J.}~\bibnamefont {Mooij}},\ }\href@noop {}
  {\bibfield  {journal} {\bibinfo  {journal} {Phys Rev Lett}\ }\textbf
  {\bibinfo {volume} {96}},\ \bibinfo {pages} {127003} (\bibinfo {year}
  {2006})}\BibitemShut {NoStop}%
\bibitem [{\citenamefont {Lee}\ \emph {et~al.}(2007)\citenamefont {Lee},
  \citenamefont {Oliver}, \citenamefont {Berggren},\ and\ \citenamefont
  {Orlando}}]{Lee:2007p3227}%
  \BibitemOpen
  \bibfield  {author} {\bibinfo {author} {\bibfnamefont {J.~C.}\ \bibnamefont
  {Lee}}, \bibinfo {author} {\bibfnamefont {W.~D.}\ \bibnamefont {Oliver}},
  \bibinfo {author} {\bibfnamefont {K.~K.}\ \bibnamefont {Berggren}}, \ and\
  \bibinfo {author} {\bibfnamefont {T.~P.}\ \bibnamefont {Orlando}},\
  }\href@noop {} {\bibfield  {journal} {\bibinfo  {journal} {Phys Rev B}\
  }\textbf {\bibinfo {volume} {75}},\ \bibinfo {pages} {144505} (\bibinfo
  {year} {2007})}\BibitemShut {NoStop}%
\bibitem [{\citenamefont {Metcalfe}\ \emph {et~al.}(2007)\citenamefont
  {Metcalfe}, \citenamefont {Boaknin}, \citenamefont {Manucharyan},
  \citenamefont {Vijay}, \citenamefont {Siddiqi}, \citenamefont {Rigetti},
  \citenamefont {Frunzio}, \citenamefont {Schoelkopf},\ and\ \citenamefont
  {Devoret}}]{Metcalfe:2007p1060}%
  \BibitemOpen
  \bibfield  {author} {\bibinfo {author} {\bibfnamefont {M.}~\bibnamefont
  {Metcalfe}}, \bibinfo {author} {\bibfnamefont {E.}~\bibnamefont {Boaknin}},
  \bibinfo {author} {\bibfnamefont {V.}~\bibnamefont {Manucharyan}}, \bibinfo
  {author} {\bibfnamefont {R.}~\bibnamefont {Vijay}}, \bibinfo {author}
  {\bibfnamefont {I.}~\bibnamefont {Siddiqi}}, \bibinfo {author} {\bibfnamefont
  {C.}~\bibnamefont {Rigetti}}, \bibinfo {author} {\bibfnamefont
  {L.}~\bibnamefont {Frunzio}}, \bibinfo {author} {\bibfnamefont {R.~J.}\
  \bibnamefont {Schoelkopf}}, \ and\ \bibinfo {author} {\bibfnamefont {M.~H.}\
  \bibnamefont {Devoret}},\ }\href@noop {} {\bibfield  {journal} {\bibinfo
  {journal} {Phys Rev B}\ }\textbf {\bibinfo {volume} {76}},\ \bibinfo {pages}
  {174516} (\bibinfo {year} {2007})}\BibitemShut {NoStop}%
\bibitem [{\citenamefont {Boaknin}\ \emph {et~al.}(2007)\citenamefont
  {Boaknin}, \citenamefont {Manucharyan}, \citenamefont {Fissette},
  \citenamefont {Metcalfe}, \citenamefont {Frunzio}, \citenamefont {Vijay},
  \citenamefont {Siddiqi}, \citenamefont {Wallraff}, \citenamefont
  {Schoelkopf},\ and\ \citenamefont {Devoret}}]{Boaknin:2007p3308}%
  \BibitemOpen
  \bibfield  {author} {\bibinfo {author} {\bibfnamefont {E.}~\bibnamefont
  {Boaknin}}, \bibinfo {author} {\bibfnamefont {V.~E.}\ \bibnamefont
  {Manucharyan}}, \bibinfo {author} {\bibfnamefont {S.}~\bibnamefont
  {Fissette}}, \bibinfo {author} {\bibfnamefont {M.}~\bibnamefont {Metcalfe}},
  \bibinfo {author} {\bibfnamefont {L.}~\bibnamefont {Frunzio}}, \bibinfo
  {author} {\bibfnamefont {R.}~\bibnamefont {Vijay}}, \bibinfo {author}
  {\bibfnamefont {I.}~\bibnamefont {Siddiqi}}, \bibinfo {author} {\bibfnamefont
  {A.}~\bibnamefont {Wallraff}}, \bibinfo {author} {\bibfnamefont {R.~J.}\
  \bibnamefont {Schoelkopf}}, \ and\ \bibinfo {author} {\bibfnamefont {M.~H.}\
  \bibnamefont {Devoret}},\ }\href@noop {} {\bibfield  {journal} {\bibinfo
  {journal} {arXiv:0702445v1}\ } (\bibinfo {year} {2007})}\BibitemShut
  {NoStop}%
\bibitem [{\citenamefont {Mancini}\ and\ \citenamefont
  {Tombesi}(1994)}]{Mancini:1994p2815}%
  \BibitemOpen
  \bibfield  {author} {\bibinfo {author} {\bibfnamefont {S.}~\bibnamefont
  {Mancini}}\ and\ \bibinfo {author} {\bibfnamefont {P.}~\bibnamefont
  {Tombesi}},\ }\href@noop {} {\bibfield  {journal} {\bibinfo  {journal} {Phys
  Rev A}\ }\textbf {\bibinfo {volume} {49}},\ \bibinfo {pages} {4055} (\bibinfo
  {year} {1994})}\BibitemShut {NoStop}%
\bibitem [{\citenamefont {Law}(1995)}]{LAW:1995p3039}%
  \BibitemOpen
  \bibfield  {author} {\bibinfo {author} {\bibfnamefont {C.~K.}\ \bibnamefont
  {Law}},\ }\href@noop {} {\bibfield  {journal} {\bibinfo  {journal} {Phys Rev
  A}\ }\textbf {\bibinfo {volume} {51}},\ \bibinfo {pages} {2537} (\bibinfo
  {year} {1995})}\BibitemShut {NoStop}%
\bibitem [{\citenamefont {Jacobs}\ \emph {et~al.}(1999)\citenamefont {Jacobs},
  \citenamefont {Tittonen}, \citenamefont {Wiseman},\ and\ \citenamefont
  {Schiller}}]{Jacobs:1999p2816}%
  \BibitemOpen
  \bibfield  {author} {\bibinfo {author} {\bibfnamefont {K.}~\bibnamefont
  {Jacobs}}, \bibinfo {author} {\bibfnamefont {I.}~\bibnamefont {Tittonen}},
  \bibinfo {author} {\bibfnamefont {H.}~\bibnamefont {Wiseman}}, \ and\
  \bibinfo {author} {\bibfnamefont {S.}~\bibnamefont {Schiller}},\ }\href@noop
  {} {\bibfield  {journal} {\bibinfo  {journal} {Phys Rev A}\ }\textbf
  {\bibinfo {volume} {60}},\ \bibinfo {pages} {538} (\bibinfo {year}
  {1999})}\BibitemShut {NoStop}%
\bibitem [{\citenamefont {Metzger}\ and\ \citenamefont
  {Karrai}(2004)}]{Metzger:2004p2871}%
  \BibitemOpen
  \bibfield  {author} {\bibinfo {author} {\bibfnamefont {C.}~\bibnamefont
  {Metzger}}\ and\ \bibinfo {author} {\bibfnamefont {K.}~\bibnamefont
  {Karrai}},\ }\href {\doibase 10.1038/nature03118} {\bibfield  {journal}
  {\bibinfo  {journal} {Nature}\ }\textbf {\bibinfo {volume} {432}},\ \bibinfo
  {pages} {1002} (\bibinfo {year} {2004})}\BibitemShut {NoStop}%
\bibitem [{\citenamefont {Gigan}\ \emph {et~al.}(2006)\citenamefont {Gigan},
  \citenamefont {Boehm}, \citenamefont {Paternostro}, \citenamefont {Blaser},
  \citenamefont {Langer}, \citenamefont {Hertzberg}, \citenamefont {Schwab},
  \citenamefont {Baeuerle}, \citenamefont {Aspelmeyer},\ and\ \citenamefont
  {Zeilinger}}]{Gigan:2006p1942}%
  \BibitemOpen
  \bibfield  {author} {\bibinfo {author} {\bibfnamefont {S.}~\bibnamefont
  {Gigan}}, \bibinfo {author} {\bibfnamefont {H.~R.}\ \bibnamefont {Boehm}},
  \bibinfo {author} {\bibfnamefont {M.}~\bibnamefont {Paternostro}}, \bibinfo
  {author} {\bibfnamefont {F.}~\bibnamefont {Blaser}}, \bibinfo {author}
  {\bibfnamefont {G.}~\bibnamefont {Langer}}, \bibinfo {author} {\bibfnamefont
  {J.~B.}\ \bibnamefont {Hertzberg}}, \bibinfo {author} {\bibfnamefont {K.~C.}\
  \bibnamefont {Schwab}}, \bibinfo {author} {\bibfnamefont {D.}~\bibnamefont
  {Baeuerle}}, \bibinfo {author} {\bibfnamefont {M.}~\bibnamefont
  {Aspelmeyer}}, \ and\ \bibinfo {author} {\bibfnamefont {A.}~\bibnamefont
  {Zeilinger}},\ }\href {\doibase 10.1038/nature05273} {\bibfield  {journal}
  {\bibinfo  {journal} {Nature}\ }\textbf {\bibinfo {volume} {444}},\ \bibinfo
  {pages} {67} (\bibinfo {year} {2006})}\BibitemShut {NoStop}%
\bibitem [{\citenamefont {Arcizet}\ \emph {et~al.}(2006)\citenamefont
  {Arcizet}, \citenamefont {Cohadon}, \citenamefont {Briant}, \citenamefont
  {Pinard},\ and\ \citenamefont {Heidmann}}]{Arcizet:2006p2873}%
  \BibitemOpen
  \bibfield  {author} {\bibinfo {author} {\bibfnamefont {O.}~\bibnamefont
  {Arcizet}}, \bibinfo {author} {\bibfnamefont {P.~F.}\ \bibnamefont
  {Cohadon}}, \bibinfo {author} {\bibfnamefont {T.}~\bibnamefont {Briant}},
  \bibinfo {author} {\bibfnamefont {M.}~\bibnamefont {Pinard}}, \ and\ \bibinfo
  {author} {\bibfnamefont {A.}~\bibnamefont {Heidmann}},\ }\href {\doibase
  10.1038/nature05244} {\bibfield  {journal} {\bibinfo  {journal} {Nature}\
  }\textbf {\bibinfo {volume} {444}},\ \bibinfo {pages} {71} (\bibinfo {year}
  {2006})}\BibitemShut {NoStop}%
\bibitem [{\citenamefont {Schliesser}\ \emph {et~al.}(2006)\citenamefont
  {Schliesser}, \citenamefont {Del'Haye}, \citenamefont {Nooshi},\ and\
  \citenamefont {Vahala}}]{Schliesser:2006p2558}%
  \BibitemOpen
  \bibfield  {author} {\bibinfo {author} {\bibfnamefont {A.}~\bibnamefont
  {Schliesser}}, \bibinfo {author} {\bibfnamefont {P.}~\bibnamefont
  {Del'Haye}}, \bibinfo {author} {\bibfnamefont {N.}~\bibnamefont {Nooshi}}, \
  and\ \bibinfo {author} {\bibfnamefont {K.}~\bibnamefont {Vahala}},\
  }\href@noop {} {\bibfield  {journal} {\bibinfo  {journal} {Phys Rev Lett}\
  }\textbf {\bibinfo {volume} {97}},\ \bibinfo {pages} {243905} (\bibinfo
  {year} {2006})}\BibitemShut {NoStop}%
\bibitem [{\citenamefont {Corbitt}\ \emph {et~al.}(2007)\citenamefont
  {Corbitt}, \citenamefont {Wipf}, \citenamefont {Bodiya}, \citenamefont
  {Ottaway}, \citenamefont {Sigg}, \citenamefont {Smith}, \citenamefont
  {Whitcomb},\ and\ \citenamefont {Mavalvala}}]{Mavalvala:2007p160801}%
  \BibitemOpen
  \bibfield  {author} {\bibinfo {author} {\bibfnamefont {T.}~\bibnamefont
  {Corbitt}}, \bibinfo {author} {\bibfnamefont {C.}~\bibnamefont {Wipf}},
  \bibinfo {author} {\bibfnamefont {T.}~\bibnamefont {Bodiya}}, \bibinfo
  {author} {\bibfnamefont {D.}~\bibnamefont {Ottaway}}, \bibinfo {author}
  {\bibfnamefont {D.}~\bibnamefont {Sigg}}, \bibinfo {author} {\bibfnamefont
  {N.}~\bibnamefont {Smith}}, \bibinfo {author} {\bibfnamefont
  {S.}~\bibnamefont {Whitcomb}}, \ and\ \bibinfo {author} {\bibfnamefont
  {N.}~\bibnamefont {Mavalvala}},\ }\href@noop {} {\bibfield  {journal}
  {\bibinfo  {journal} {Phys Rev Lett}\ }\textbf {\bibinfo {volume} {99}},\
  \bibinfo {pages} {160801} (\bibinfo {year} {2007})}\BibitemShut {NoStop}%
\bibitem [{\citenamefont {Thompson}\ \emph {et~al.}(2008)\citenamefont
  {Thompson}, \citenamefont {Zwickl}, \citenamefont {Jayich}, \citenamefont
  {Marquardt}, \citenamefont {Girvin},\ and\ \citenamefont
  {Harris}}]{Thompson:2008p72}%
  \BibitemOpen
  \bibfield  {author} {\bibinfo {author} {\bibfnamefont {J.~D.}\ \bibnamefont
  {Thompson}}, \bibinfo {author} {\bibfnamefont {B.~M.}\ \bibnamefont
  {Zwickl}}, \bibinfo {author} {\bibfnamefont {A.~M.}\ \bibnamefont {Jayich}},
  \bibinfo {author} {\bibfnamefont {F.}~\bibnamefont {Marquardt}}, \bibinfo
  {author} {\bibfnamefont {S.~M.}\ \bibnamefont {Girvin}}, \ and\ \bibinfo
  {author} {\bibfnamefont {J.~G.~E.}\ \bibnamefont {Harris}},\ }\href@noop {}
  {\bibfield  {journal} {\bibinfo  {journal} {Nature}\ }\textbf {\bibinfo
  {volume} {452}},\ \bibinfo {pages} {72} (\bibinfo {year} {2008})}\BibitemShut
  {NoStop}%
\bibitem [{\citenamefont {Tholen}\ \emph {et~al.}(2007)\citenamefont {Tholen},
  \citenamefont {Ergul}, \citenamefont {Doherty}, \citenamefont {Weber},
  \citenamefont {Gregis},\ and\ \citenamefont {Haviland}}]{Tholen:2007p1916}%
  \BibitemOpen
  \bibfield  {author} {\bibinfo {author} {\bibfnamefont {E.~A.}\ \bibnamefont
  {Tholen}}, \bibinfo {author} {\bibfnamefont {A.}~\bibnamefont {Ergul}},
  \bibinfo {author} {\bibfnamefont {E.~M.}\ \bibnamefont {Doherty}}, \bibinfo
  {author} {\bibfnamefont {F.~M.}\ \bibnamefont {Weber}}, \bibinfo {author}
  {\bibfnamefont {F.}~\bibnamefont {Gregis}}, \ and\ \bibinfo {author}
  {\bibfnamefont {D.~B.}\ \bibnamefont {Haviland}},\ }\href {\doibase
  10.1063/1.2750520} {\bibfield  {journal} {\bibinfo  {journal} {Appl Phys
  Lett}\ }\textbf {\bibinfo {volume} {90}},\ \bibinfo {pages} {253509}
  (\bibinfo {year} {2007})}\BibitemShut {NoStop}%
\bibitem [{\citenamefont {Dorsel}\ \emph {et~al.}(1983)\citenamefont {Dorsel},
  \citenamefont {McCullen}, \citenamefont {Meystre}, \citenamefont {Vignes},\
  and\ \citenamefont {Walther}}]{DORSEL:1983p2818}%
  \BibitemOpen
  \bibfield  {author} {\bibinfo {author} {\bibfnamefont {A.}~\bibnamefont
  {Dorsel}}, \bibinfo {author} {\bibfnamefont {J.~D.}\ \bibnamefont
  {McCullen}}, \bibinfo {author} {\bibfnamefont {P.}~\bibnamefont {Meystre}},
  \bibinfo {author} {\bibfnamefont {E.}~\bibnamefont {Vignes}}, \ and\ \bibinfo
  {author} {\bibfnamefont {H.}~\bibnamefont {Walther}},\ }\href@noop {}
  {\bibfield  {journal} {\bibinfo  {journal} {Phys Rev Lett}\ }\textbf
  {\bibinfo {volume} {51}},\ \bibinfo {pages} {1550} (\bibinfo {year}
  {1983})}\BibitemShut {NoStop}%
\bibitem [{\citenamefont {Marquardt}\ \emph {et~al.}(2006)\citenamefont
  {Marquardt}, \citenamefont {Harris},\ and\ \citenamefont
  {Girvin}}]{Marquardt:2006p103901}%
  \BibitemOpen
  \bibfield  {author} {\bibinfo {author} {\bibfnamefont {F.}~\bibnamefont
  {Marquardt}}, \bibinfo {author} {\bibfnamefont {J.~G.~E.}\ \bibnamefont
  {Harris}}, \ and\ \bibinfo {author} {\bibfnamefont {S.~M.}\ \bibnamefont
  {Girvin}},\ }\href@noop {} {\bibfield  {journal} {\bibinfo  {journal} {Phys
  Rev Lett}\ }\textbf {\bibinfo {volume} {96}},\ \bibinfo {pages} {103901}
  (\bibinfo {year} {2006})}\BibitemShut {NoStop}%
\bibitem [{\citenamefont {Yurke}\ \emph {et~al.}(1989)\citenamefont {Yurke},
  \citenamefont {Corruccini}, \citenamefont {Kaminsky}, \citenamefont {Rupp},
  \citenamefont {Smith}, \citenamefont {Silver}, \citenamefont {Simon},\ and\
  \citenamefont {Whittaker}}]{yurke:1989p2519}%
  \BibitemOpen
  \bibfield  {author} {\bibinfo {author} {\bibfnamefont {B.}~\bibnamefont
  {Yurke}}, \bibinfo {author} {\bibfnamefont {L.~R.}\ \bibnamefont
  {Corruccini}}, \bibinfo {author} {\bibfnamefont {P.~G.}\ \bibnamefont
  {Kaminsky}}, \bibinfo {author} {\bibfnamefont {L.~W.}\ \bibnamefont {Rupp}},
  \bibinfo {author} {\bibfnamefont {A.~D.}\ \bibnamefont {Smith}}, \bibinfo
  {author} {\bibfnamefont {A.~H.}\ \bibnamefont {Silver}}, \bibinfo {author}
  {\bibfnamefont {R.~W.}\ \bibnamefont {Simon}}, \ and\ \bibinfo {author}
  {\bibfnamefont {E.~A.}\ \bibnamefont {Whittaker}},\ }\href@noop {} {\bibfield
   {journal} {\bibinfo  {journal} {Phys Rev A}\ }\textbf {\bibinfo {volume}
  {39}},\ \bibinfo {pages} {2519} (\bibinfo {year} {1989})}\BibitemShut
  {NoStop}%
\bibitem [{\citenamefont {Yurke}\ and\ \citenamefont
  {Buks}(2006)}]{Yurke:2006p1911}%
  \BibitemOpen
  \bibfield  {author} {\bibinfo {author} {\bibfnamefont {B.}~\bibnamefont
  {Yurke}}\ and\ \bibinfo {author} {\bibfnamefont {E.}~\bibnamefont {Buks}},\
  }\href {\doibase 10.1109/JLT.2006.884490} {\bibfield  {journal} {\bibinfo
  {journal} {J Lightwave Technol}\ }\textbf {\bibinfo {volume} {24}},\ \bibinfo
  {pages} {5054} (\bibinfo {year} {2006})}\BibitemShut {NoStop}%
\bibitem [{\citenamefont {Bergeal}\ \emph {et~al.}(2008)\citenamefont
  {Bergeal}, \citenamefont {Vijay}, \citenamefont {Manucharyan}, \citenamefont
  {Siddiqi}, \citenamefont {Schoelkopf}, \citenamefont {Girvin},\ and\
  \citenamefont {Devoret}}]{bergeal:2008}%
  \BibitemOpen
  \bibfield  {author} {\bibinfo {author} {\bibfnamefont {N.}~\bibnamefont
  {Bergeal}}, \bibinfo {author} {\bibfnamefont {R.}~\bibnamefont {Vijay}},
  \bibinfo {author} {\bibfnamefont {V.~E.}\ \bibnamefont {Manucharyan}},
  \bibinfo {author} {\bibfnamefont {I.}~\bibnamefont {Siddiqi}}, \bibinfo
  {author} {\bibfnamefont {R.~J.}\ \bibnamefont {Schoelkopf}}, \bibinfo
  {author} {\bibfnamefont {S.~M.}\ \bibnamefont {Girvin}}, \ and\ \bibinfo
  {author} {\bibfnamefont {M.~H.}\ \bibnamefont {Devoret}},\ }\href@noop {}
  {\bibfield  {journal} {\bibinfo  {journal} {arXiv:0805.3452v1}\ } (\bibinfo
  {year} {2008})}\BibitemShut {NoStop}%
\bibitem [{\citenamefont {Castellanos-Beltran}\ \emph
  {et~al.}(2008)\citenamefont {Castellanos-Beltran}, \citenamefont {Irwin},
  \citenamefont {Hilton}, \citenamefont {Vale},\ and\ \citenamefont
  {Lehnert}}]{beltran:2008}%
  \BibitemOpen
  \bibfield  {author} {\bibinfo {author} {\bibfnamefont {M.~A.}\ \bibnamefont
  {Castellanos-Beltran}}, \bibinfo {author} {\bibfnamefont {K.~D.}\
  \bibnamefont {Irwin}}, \bibinfo {author} {\bibfnamefont {G.~C.}\ \bibnamefont
  {Hilton}}, \bibinfo {author} {\bibfnamefont {L.~R.}\ \bibnamefont {Vale}}, \
  and\ \bibinfo {author} {\bibfnamefont {K.~W.}\ \bibnamefont {Lehnert}},\
  }\href@noop {} {\bibfield  {journal} {\bibinfo  {journal} {Nature Phys.}\
  }\textbf {\bibinfo {volume} {4}},\ \bibinfo {pages} {928} (\bibinfo {year}
  {2008})}\BibitemShut {NoStop}%
\bibitem [{\citenamefont {Xue}\ \emph {et~al.}(2007{\natexlab{b}})\citenamefont
  {Xue}, \citenamefont {Wang}, \citenamefont {Sun}, \citenamefont {Okamoto},
  \citenamefont {Yamaguchi},\ and\ \citenamefont {Semba}}]{Xue:2007p2454}%
  \BibitemOpen
  \bibfield  {author} {\bibinfo {author} {\bibfnamefont {F.}~\bibnamefont
  {Xue}}, \bibinfo {author} {\bibfnamefont {Y.~D.}\ \bibnamefont {Wang}},
  \bibinfo {author} {\bibfnamefont {C.~P.}\ \bibnamefont {Sun}}, \bibinfo
  {author} {\bibfnamefont {H.}~\bibnamefont {Okamoto}}, \bibinfo {author}
  {\bibfnamefont {H.}~\bibnamefont {Yamaguchi}}, \ and\ \bibinfo {author}
  {\bibfnamefont {K.}~\bibnamefont {Semba}},\ }\href {\doibase
  10.1088/1367-2630/9/2/035} {\bibfield  {journal} {\bibinfo  {journal} {New J
  Phys}\ }\textbf {\bibinfo {volume} {9}},\ \bibinfo {pages} {35} (\bibinfo
  {year} {2007}{\natexlab{b}})}\BibitemShut {NoStop}%
\bibitem [{\citenamefont {Wang}\ \emph
  {et~al.}(2008{\natexlab{b}})\citenamefont {Wang}, \citenamefont {Semba},\
  and\ \citenamefont {Yamaguchi}}]{Wang:2007p2819}%
  \BibitemOpen
  \bibfield  {author} {\bibinfo {author} {\bibfnamefont {Y.~D.}\ \bibnamefont
  {Wang}}, \bibinfo {author} {\bibfnamefont {K.}~\bibnamefont {Semba}}, \ and\
  \bibinfo {author} {\bibfnamefont {H.}~\bibnamefont {Yamaguchi}},\ }\href@noop
  {} {\bibfield  {journal} {\bibinfo  {journal} {New J. Phys.}\ }\textbf
  {\bibinfo {volume} {10}},\ \bibinfo {pages} {043015} (\bibinfo {year}
  {2008}{\natexlab{b}})}\BibitemShut {NoStop}%
\bibitem [{\citenamefont {Drummond}\ and\ \citenamefont
  {Walls}(1980)}]{Drummond:1980p3191}%
  \BibitemOpen
  \bibfield  {author} {\bibinfo {author} {\bibfnamefont {P.~D.}\ \bibnamefont
  {Drummond}}\ and\ \bibinfo {author} {\bibfnamefont {D.~F.}\ \bibnamefont
  {Walls}},\ }\href@noop {} {\bibfield  {journal} {\bibinfo  {journal} {J.
  Phys. A: Math. Gen}\ }\textbf {\bibinfo {volume} {13}},\ \bibinfo {pages}
  {725} (\bibinfo {year} {1980})}\BibitemShut {NoStop}%
\bibitem [{\citenamefont {Babourina-Brooks}\ \emph {et~al.}(2008)\citenamefont
  {Babourina-Brooks}, \citenamefont {Doherty},\ and\ \citenamefont
  {Milburn}}]{BabourinaBrooks:2008p3350}%
  \BibitemOpen
  \bibfield  {author} {\bibinfo {author} {\bibfnamefont {E.}~\bibnamefont
  {Babourina-Brooks}}, \bibinfo {author} {\bibfnamefont {A.}~\bibnamefont
  {Doherty}}, \ and\ \bibinfo {author} {\bibfnamefont {G.~J.}\ \bibnamefont
  {Milburn}},\ }\href@noop {} {\bibfield  {journal} {\bibinfo  {journal} {New
  J. Phys.}\ }\textbf {\bibinfo {volume} {10}},\ \bibinfo {pages} {5020}
  (\bibinfo {year} {2008})}\BibitemShut {NoStop}%
\bibitem [{\citenamefont {Etaki}\ \emph {et~al.}(2008)\citenamefont {Etaki},
  \citenamefont {Poot}, \citenamefont {Mahboob}, \citenamefont {Onomitsu},
  \citenamefont {Yamaguchi},\ and\ \citenamefont {Zant}}]{etaki:2008}%
  \BibitemOpen
  \bibfield  {author} {\bibinfo {author} {\bibfnamefont {S.}~\bibnamefont
  {Etaki}}, \bibinfo {author} {\bibfnamefont {M.}~\bibnamefont {Poot}},
  \bibinfo {author} {\bibfnamefont {I.}~\bibnamefont {Mahboob}}, \bibinfo
  {author} {\bibfnamefont {K.}~\bibnamefont {Onomitsu}}, \bibinfo {author}
  {\bibfnamefont {H.}~\bibnamefont {Yamaguchi}}, \ and\ \bibinfo {author}
  {\bibfnamefont {H.~S. J. V.~D.}\ \bibnamefont {Zant}},\ }\href@noop {}
  {\bibfield  {journal} {\bibinfo  {journal} {Nature Phys.}\ }\textbf {\bibinfo
  {volume} {4}},\ \bibinfo {pages} {785} (\bibinfo {year} {2008})}\BibitemShut
  {NoStop}%
\bibitem [{\citenamefont {Orlando}\ and\ \citenamefont
  {Delin}(1991)}]{orlando}%
  \BibitemOpen
  \bibfield  {author} {\bibinfo {author} {\bibfnamefont {T.~P.}\ \bibnamefont
  {Orlando}}\ and\ \bibinfo {author} {\bibfnamefont {K.~A.}\ \bibnamefont
  {Delin}},\ }\href@noop {} {\emph {\bibinfo {title} {Foundations of Applied
  Superconductivity}}}\ (\bibinfo  {publisher} {Addison Wesley},\ \bibinfo
  {year} {1991})\BibitemShut {NoStop}%
\bibitem [{\citenamefont {Buks}\ and\ \citenamefont
  {Blencowe}(2006)}]{Buks:2006p346}%
  \BibitemOpen
  \bibfield  {author} {\bibinfo {author} {\bibfnamefont {E.}~\bibnamefont
  {Buks}}\ and\ \bibinfo {author} {\bibfnamefont {M.~P.}\ \bibnamefont
  {Blencowe}},\ }\href {\doibase 10.1103/PhysRevB.74.174504} {\bibfield
  {journal} {\bibinfo  {journal} {Phys Rev B}\ }\textbf {\bibinfo {volume}
  {74}},\ \bibinfo {pages} {174504} (\bibinfo {year} {2006})}\BibitemShut
  {NoStop}%
\bibitem [{\citenamefont {Gardiner}\ and\ \citenamefont
  {Collett}(1985)}]{Gardiner:1985p1483}%
  \BibitemOpen
  \bibfield  {author} {\bibinfo {author} {\bibfnamefont {C.~W.}\ \bibnamefont
  {Gardiner}}\ and\ \bibinfo {author} {\bibfnamefont {M.~J.}\ \bibnamefont
  {Collett}},\ }\href@noop {} {\bibfield  {journal} {\bibinfo  {journal} {Phys
  Rev A}\ }\textbf {\bibinfo {volume} {31}},\ \bibinfo {pages} {3761} (\bibinfo
  {year} {1985})}\BibitemShut {NoStop}%
\bibitem [{\citenamefont {Gardiner}\ and\ \citenamefont
  {Zoller}(2000)}]{gardiner}%
  \BibitemOpen
  \bibfield  {author} {\bibinfo {author} {\bibfnamefont {C.}~\bibnamefont
  {Gardiner}}\ and\ \bibinfo {author} {\bibfnamefont {P.}~\bibnamefont
  {Zoller}},\ }\href@noop {} {\emph {\bibinfo {title} {Quantum Noise, 2nd
  Ed.}}}\ (\bibinfo  {publisher} {Springer-Verlag, Berlin},\ \bibinfo {year}
  {2000})\BibitemShut {NoStop}%
\bibitem [{\citenamefont {Johansson}\ \emph {et~al.}(2006)\citenamefont
  {Johansson}, \citenamefont {Tornberg}, \citenamefont {Shumeiko},\ and\
  \citenamefont {Wendin}}]{Johansson:2006p2429}%
  \BibitemOpen
  \bibfield  {author} {\bibinfo {author} {\bibfnamefont {G.}~\bibnamefont
  {Johansson}}, \bibinfo {author} {\bibfnamefont {L.}~\bibnamefont {Tornberg}},
  \bibinfo {author} {\bibfnamefont {V.~S.}\ \bibnamefont {Shumeiko}}, \ and\
  \bibinfo {author} {\bibfnamefont {G.}~\bibnamefont {Wendin}},\ }\href
  {\doibase 10.1088/0953-8984/18/21/S14} {\bibfield  {journal} {\bibinfo
  {journal} {J Phys-Condens Mat}\ }\textbf {\bibinfo {volume} {18}},\ \bibinfo
  {pages} {S901} (\bibinfo {year} {2006})}\BibitemShut {NoStop}%
\bibitem [{\citenamefont {Dykman}\ and\ \citenamefont
  {Krivoglaz}(1980)}]{Dykman:1980p480}%
  \BibitemOpen
  \bibfield  {author} {\bibinfo {author} {\bibfnamefont {M.~I.}\ \bibnamefont
  {Dykman}}\ and\ \bibinfo {author} {\bibfnamefont {M.~A.}\ \bibnamefont
  {Krivoglaz}},\ }\href@noop {} {\bibfield  {journal} {\bibinfo  {journal}
  {Physica}\ }\textbf {\bibinfo {volume} {104A}},\ \bibinfo {pages} {480}
  (\bibinfo {year} {1980})}\BibitemShut {NoStop}%
\bibitem [{\citenamefont {Dykman}(2007)}]{Dykman:2007p1864}%
  \BibitemOpen
  \bibfield  {author} {\bibinfo {author} {\bibfnamefont {M.~I.}\ \bibnamefont
  {Dykman}},\ }\href@noop {} {\bibfield  {journal} {\bibinfo  {journal} {Phys
  Rev E}\ }\textbf {\bibinfo {volume} {75}},\ \bibinfo {pages} {011101}
  (\bibinfo {year} {2007})}\BibitemShut {NoStop}%
\bibitem [{\citenamefont {Caves}(1982)}]{Caves:1982p1311}%
  \BibitemOpen
  \bibfield  {author} {\bibinfo {author} {\bibfnamefont {C.~M.}\ \bibnamefont
  {Caves}},\ }\href@noop {} {\bibfield  {journal} {\bibinfo  {journal} {Phys
  Rev D}\ }\textbf {\bibinfo {volume} {26}},\ \bibinfo {pages} {1817} (\bibinfo
  {year} {1982})}\BibitemShut {NoStop}%
\bibitem [{\citenamefont {Clerk}(2004)}]{Clerk:2004p245306}%
  \BibitemOpen
  \bibfield  {author} {\bibinfo {author} {\bibfnamefont {A.~A.}\ \bibnamefont
  {Clerk}},\ }\href@noop {} {\bibfield  {journal} {\bibinfo  {journal} {Phys
  Rev B}\ }\textbf {\bibinfo {volume} {70}},\ \bibinfo {pages} {245306}
  (\bibinfo {year} {2004})}\BibitemShut {NoStop}%
\bibitem [{\citenamefont {Kippenberg}\ and\ \citenamefont
  {Vahala}(2007{\natexlab{b}})}]{Kippenberg:2007p1995}%
  \BibitemOpen
  \bibfield  {author} {\bibinfo {author} {\bibfnamefont {T.~J.}\ \bibnamefont
  {Kippenberg}}\ and\ \bibinfo {author} {\bibfnamefont {K.~J.}\ \bibnamefont
  {Vahala}},\ }\href@noop {} {\bibfield  {journal} {\bibinfo  {journal} {Opt
  Express}\ }\textbf {\bibinfo {volume} {15}},\ \bibinfo {pages} {17172}
  (\bibinfo {year} {2007}{\natexlab{b}})}\BibitemShut {NoStop}%
\bibitem [{\citenamefont {Marquardt}\ \emph {et~al.}(2008)\citenamefont
  {Marquardt}, \citenamefont {Clerk},\ and\ \citenamefont
  {Girvin}}]{Marquardt:2008p1309}%
  \BibitemOpen
  \bibfield  {author} {\bibinfo {author} {\bibfnamefont {F.}~\bibnamefont
  {Marquardt}}, \bibinfo {author} {\bibfnamefont {A.~A.}\ \bibnamefont
  {Clerk}}, \ and\ \bibinfo {author} {\bibfnamefont {S.~M.}\ \bibnamefont
  {Girvin}},\ }\href@noop {} {\bibfield  {journal} {\bibinfo  {journal} {J.
  Mod. Opt.}\ }\textbf {\bibinfo {volume} {55}},\ \bibinfo {pages} {3329}
  (\bibinfo {year} {2008})}\BibitemShut {NoStop}%
\bibitem [{\citenamefont {Marquardt}\ \emph {et~al.}(2007)\citenamefont
  {Marquardt}, \citenamefont {Chen}, \citenamefont {Clerk},\ and\ \citenamefont
  {Girvin}}]{Marquardt:2007p978}%
  \BibitemOpen
  \bibfield  {author} {\bibinfo {author} {\bibfnamefont {F.}~\bibnamefont
  {Marquardt}}, \bibinfo {author} {\bibfnamefont {J.~P.}\ \bibnamefont {Chen}},
  \bibinfo {author} {\bibfnamefont {A.~A.}\ \bibnamefont {Clerk}}, \ and\
  \bibinfo {author} {\bibfnamefont {S.~M.}\ \bibnamefont {Girvin}},\ }\href
  {\doibase 10.1103/PhysRevLett.99.093902} {\bibfield  {journal} {\bibinfo
  {journal} {Phys Rev Lett}\ }\textbf {\bibinfo {volume} {99}},\ \bibinfo
  {pages} {093902} (\bibinfo {year} {2007})}\BibitemShut {NoStop}%
\bibitem [{\citenamefont {Wilson-Rae}\ \emph {et~al.}(2007)\citenamefont
  {Wilson-Rae}, \citenamefont {Nooshi}, \citenamefont {Zwerger},\ and\
  \citenamefont {Kippenberg}}]{WilsonRae:2007p502}%
  \BibitemOpen
  \bibfield  {author} {\bibinfo {author} {\bibfnamefont {I.}~\bibnamefont
  {Wilson-Rae}}, \bibinfo {author} {\bibfnamefont {N.}~\bibnamefont {Nooshi}},
  \bibinfo {author} {\bibfnamefont {W.}~\bibnamefont {Zwerger}}, \ and\
  \bibinfo {author} {\bibfnamefont {T.~J.}\ \bibnamefont {Kippenberg}},\ }\href
  {\doibase 10.1103/PhysRevLett.99.093901} {\bibfield  {journal} {\bibinfo
  {journal} {Phys Rev Lett}\ }\textbf {\bibinfo {volume} {99}},\ \bibinfo
  {pages} {093901} (\bibinfo {year} {2007})}\BibitemShut {NoStop}%
\bibitem [{\citenamefont {Naaman}\ \emph {et~al.}(2008)\citenamefont {Naaman},
  \citenamefont {Aumentado}, \citenamefont {Friedland}, \citenamefont
  {Wurtele},\ and\ \citenamefont {Siddiqi}}]{Naaman:2008}%
  \BibitemOpen
  \bibfield  {author} {\bibinfo {author} {\bibfnamefont {O.}~\bibnamefont
  {Naaman}}, \bibinfo {author} {\bibfnamefont {J.}~\bibnamefont {Aumentado}},
  \bibinfo {author} {\bibfnamefont {L.}~\bibnamefont {Friedland}}, \bibinfo
  {author} {\bibfnamefont {J.~S.}\ \bibnamefont {Wurtele}}, \ and\ \bibinfo
  {author} {\bibfnamefont {I.}~\bibnamefont {Siddiqi}},\ }\href@noop {}
  {\bibfield  {journal} {\bibinfo  {journal} {Phys. Rev. Lett.}\ }\textbf
  {\bibinfo {volume} {101}},\ \bibinfo {pages} {117005} (\bibinfo {year}
  {2008})}\BibitemShut {NoStop}%
\bibitem [{\citenamefont {Ryvkine}\ \emph {et~al.}(2004)\citenamefont
  {Ryvkine}, \citenamefont {Dykman},\ and\ \citenamefont
  {Golding}}]{Dykman:2004p061102}%
  \BibitemOpen
  \bibfield  {author} {\bibinfo {author} {\bibfnamefont {D.}~\bibnamefont
  {Ryvkine}}, \bibinfo {author} {\bibfnamefont {M.~I.}\ \bibnamefont {Dykman}},
  \ and\ \bibinfo {author} {\bibfnamefont {B.}~\bibnamefont {Golding}},\
  }\href@noop {} {\bibfield  {journal} {\bibinfo  {journal} {Phys Rev E}\
  }\textbf {\bibinfo {volume} {69}},\ \bibinfo {pages} {061102} (\bibinfo
  {year} {2004})}\BibitemShut {NoStop}%
\bibitem [{\citenamefont {Dykman}\ \emph {et~al.}(2005)\citenamefont {Dykman},
  \citenamefont {Schwartz},\ and\ \citenamefont
  {Shapiro}}]{Dykman:2005p021102}%
  \BibitemOpen
  \bibfield  {author} {\bibinfo {author} {\bibfnamefont {M.~I.}\ \bibnamefont
  {Dykman}}, \bibinfo {author} {\bibfnamefont {I.~B.}\ \bibnamefont
  {Schwartz}}, \ and\ \bibinfo {author} {\bibfnamefont {M.}~\bibnamefont
  {Shapiro}},\ }\href@noop {} {\bibfield  {journal} {\bibinfo  {journal} {Phys
  Rev E}\ }\textbf {\bibinfo {volume} {72}},\ \bibinfo {pages} {021102}
  (\bibinfo {year} {2005})}\BibitemShut {NoStop}%
\bibitem [{\citenamefont {Serban}\ and\ \citenamefont
  {Wilhelm}(2007)}]{Serban:2007p3199}%
  \BibitemOpen
  \bibfield  {author} {\bibinfo {author} {\bibfnamefont {I.}~\bibnamefont
  {Serban}}\ and\ \bibinfo {author} {\bibfnamefont {F.~K.}\ \bibnamefont
  {Wilhelm}},\ }\href@noop {} {\bibfield  {journal} {\bibinfo  {journal} {Phys
  Rev Lett}\ }\textbf {\bibinfo {volume} {99}},\ \bibinfo {pages} {137001}
  (\bibinfo {year} {2007})}\BibitemShut {NoStop}%
\bibitem [{\citenamefont {Katz}\ \emph {et~al.}(2007)\citenamefont {Katz},
  \citenamefont {Retzker}, \citenamefont {Straub},\ and\ \citenamefont
  {Lifshitz}}]{Lifshitz:2007p040404}%
  \BibitemOpen
  \bibfield  {author} {\bibinfo {author} {\bibfnamefont {I.}~\bibnamefont
  {Katz}}, \bibinfo {author} {\bibfnamefont {A.}~\bibnamefont {Retzker}},
  \bibinfo {author} {\bibfnamefont {R.}~\bibnamefont {Straub}}, \ and\ \bibinfo
  {author} {\bibfnamefont {R.}~\bibnamefont {Lifshitz}},\ }\href@noop {}
  {\bibfield  {journal} {\bibinfo  {journal} {Phys Rev Lett}\ }\textbf
  {\bibinfo {volume} {99}},\ \bibinfo {pages} {040404} (\bibinfo {year}
  {2007})}\BibitemShut {NoStop}%
\bibitem [{\citenamefont {Kogan}(2008)}]{Kogan:2008p0972}%
  \BibitemOpen
  \bibfield  {author} {\bibinfo {author} {\bibfnamefont {O.}~\bibnamefont
  {Kogan}},\ }\href@noop {} {\bibfield  {journal} {\bibinfo  {journal}
  {arXiv:0805.0972v1}\ } (\bibinfo {year} {2008})}\BibitemShut {NoStop}%
\bibitem [{\citenamefont {Jacobson}(2003)}]{jacobson:2003}%
  \BibitemOpen
  \bibfield  {author} {\bibinfo {author} {\bibfnamefont {T.}~\bibnamefont
  {Jacobson}},\ }\href@noop {} {\bibfield  {journal} {\bibinfo  {journal}
  {arXiv:0308048 (unpublished)}\ } (\bibinfo {year} {2003})}\BibitemShut
  {NoStop}%
\bibitem [{\citenamefont {Mukhanov}\ and\ \citenamefont
  {Winitzki}(2007)}]{mukhanov:2007}%
  \BibitemOpen
  \bibfield  {author} {\bibinfo {author} {\bibfnamefont {V.}~\bibnamefont
  {Mukhanov}}\ and\ \bibinfo {author} {\bibfnamefont {S.}~\bibnamefont
  {Winitzki}},\ }\href@noop {} {\emph {\bibinfo {title} {Introduction to
  Quantum Effects in Gravity}}}\ (\bibinfo  {publisher} {Cambridge University
  Press},\ \bibinfo {year} {2007})\BibitemShut {NoStop}%
\bibitem [{\citenamefont {Hawking}(1974)}]{hawking:1974}%
  \BibitemOpen
  \bibfield  {author} {\bibinfo {author} {\bibfnamefont {S.~W.}\ \bibnamefont
  {Hawking}},\ }\href@noop {} {\bibfield  {journal} {\bibinfo  {journal}
  {Nature}\ }\textbf {\bibinfo {volume} {248}},\ \bibinfo {pages} {30}
  (\bibinfo {year} {1974})}\BibitemShut {NoStop}%
\bibitem [{\citenamefont {Unruh}(1981)}]{unruh:1981}%
  \BibitemOpen
  \bibfield  {author} {\bibinfo {author} {\bibfnamefont {W.~G.}\ \bibnamefont
  {Unruh}},\ }\href@noop {} {\bibfield  {journal} {\bibinfo  {journal} {Phys.
  Rev. Lett.}\ }\textbf {\bibinfo {volume} {46}},\ \bibinfo {pages} {1351}
  (\bibinfo {year} {1981})}\BibitemShut {NoStop}%
\bibitem [{\citenamefont {Painlev\'e}(1921)}]{painleve:1921}%
  \BibitemOpen
  \bibfield  {author} {\bibinfo {author} {\bibfnamefont {P.}~\bibnamefont
  {Painlev\'e}},\ }\href@noop {} {\bibfield  {journal} {\bibinfo  {journal} {C.
  R. Acad. Sci. (Paris)}\ }\textbf {\bibinfo {volume} {173}},\ \bibinfo {pages}
  {677} (\bibinfo {year} {1921})}\BibitemShut {NoStop}%
\bibitem [{\citenamefont {Garay}\ \emph {et~al.}(2000)\citenamefont {Garay},
  \citenamefont {Anglin}, \citenamefont {Cirac},\ and\ \citenamefont
  {Zoller}}]{garay:2000}%
  \BibitemOpen
  \bibfield  {author} {\bibinfo {author} {\bibfnamefont {L.~J.}\ \bibnamefont
  {Garay}}, \bibinfo {author} {\bibfnamefont {J.~R.}\ \bibnamefont {Anglin}},
  \bibinfo {author} {\bibfnamefont {J.~I.}\ \bibnamefont {Cirac}}, \ and\
  \bibinfo {author} {\bibfnamefont {P.}~\bibnamefont {Zoller}},\ }\href@noop {}
  {\bibfield  {journal} {\bibinfo  {journal} {Phys. Rev. Lett.}\ }\textbf
  {\bibinfo {volume} {85}},\ \bibinfo {pages} {4643} (\bibinfo {year}
  {2000})}\BibitemShut {NoStop}%
\bibitem [{\citenamefont {Volovik}(1999)}]{volovik:1999}%
  \BibitemOpen
  \bibfield  {author} {\bibinfo {author} {\bibfnamefont {G.~E.}\ \bibnamefont
  {Volovik}},\ }\href@noop {} {\bibfield  {journal} {\bibinfo  {journal} {JETP
  Lett.}\ }\textbf {\bibinfo {volume} {69}},\ \bibinfo {pages} {705} (\bibinfo
  {year} {1999})}\BibitemShut {NoStop}%
\bibitem [{\citenamefont {Schutzhold}\ and\ \citenamefont
  {Unruh}(2005)}]{schutzhold:2005}%
  \BibitemOpen
  \bibfield  {author} {\bibinfo {author} {\bibfnamefont {R.}~\bibnamefont
  {Schutzhold}}\ and\ \bibinfo {author} {\bibfnamefont {W.~G.}\ \bibnamefont
  {Unruh}},\ }\href@noop {} {\bibfield  {journal} {\bibinfo  {journal} {Phys.
  Rev. Lett.}\ }\textbf {\bibinfo {volume} {95}},\ \bibinfo {pages} {031301}
  (\bibinfo {year} {2005})}\BibitemShut {NoStop}%
\bibitem [{\citenamefont {Philbin}\ \emph {et~al.}(2008)\citenamefont
  {Philbin}, \citenamefont {Kuklewicz}, \citenamefont {Robertson},
  \citenamefont {Hill}, \citenamefont {K\"{o}nig},\ and\ \citenamefont
  {Leonhardt}}]{philbin:2008}%
  \BibitemOpen
  \bibfield  {author} {\bibinfo {author} {\bibfnamefont {T.~G.}\ \bibnamefont
  {Philbin}}, \bibinfo {author} {\bibfnamefont {C.}~\bibnamefont {Kuklewicz}},
  \bibinfo {author} {\bibfnamefont {S.}~\bibnamefont {Robertson}}, \bibinfo
  {author} {\bibfnamefont {S.}~\bibnamefont {Hill}}, \bibinfo {author}
  {\bibfnamefont {F.}~\bibnamefont {K\"{o}nig}}, \ and\ \bibinfo {author}
  {\bibfnamefont {U.}~\bibnamefont {Leonhardt}},\ }\href@noop {} {\bibfield
  {journal} {\bibinfo  {journal} {Science}\ }\textbf {\bibinfo {volume}
  {319}},\ \bibinfo {pages} {1367} (\bibinfo {year} {2008})}\BibitemShut
  {NoStop}%
\bibitem [{\citenamefont {Jacobson}(1991)}]{jacobson:1991}%
  \BibitemOpen
  \bibfield  {author} {\bibinfo {author} {\bibfnamefont {T.}~\bibnamefont
  {Jacobson}},\ }\href@noop {} {\bibfield  {journal} {\bibinfo  {journal}
  {Phys. Rev. D}\ }\textbf {\bibinfo {volume} {44}},\ \bibinfo {pages} {1731}
  (\bibinfo {year} {1991})}\BibitemShut {NoStop}%
\bibitem [{\citenamefont {Castellanos-Beltran}\ and\ \citenamefont
  {Lehnert}(2007)}]{beltran:2007}%
  \BibitemOpen
  \bibfield  {author} {\bibinfo {author} {\bibfnamefont {M.~A.}\ \bibnamefont
  {Castellanos-Beltran}}\ and\ \bibinfo {author} {\bibfnamefont {K.~W.}\
  \bibnamefont {Lehnert}},\ }\href@noop {} {\bibfield  {journal} {\bibinfo
  {journal} {Appl. Phys. Lett.}\ }\textbf {\bibinfo {volume} {91}},\ \bibinfo
  {pages} {083509} (\bibinfo {year} {2007})}\BibitemShut {NoStop}%
\bibitem [{\citenamefont {Haviland}\ and\ \citenamefont
  {Delsing}(1996)}]{haviland:1996}%
  \BibitemOpen
  \bibfield  {author} {\bibinfo {author} {\bibfnamefont {D.~B.}\ \bibnamefont
  {Haviland}}\ and\ \bibinfo {author} {\bibfnamefont {P.}~\bibnamefont
  {Delsing}},\ }\href@noop {} {\bibfield  {journal} {\bibinfo  {journal} {Phys.
  Rev. B}\ }\textbf {\bibinfo {volume} {54}},\ \bibinfo {pages} {6857}
  (\bibinfo {year} {1996})}\BibitemShut {NoStop}%
\bibitem [{\citenamefont {Unruh}\ and\ \citenamefont
  {Schutzhold}(2005)}]{unruh:2005}%
  \BibitemOpen
  \bibfield  {author} {\bibinfo {author} {\bibfnamefont {W.~G.}\ \bibnamefont
  {Unruh}}\ and\ \bibinfo {author} {\bibfnamefont {R.}~\bibnamefont
  {Schutzhold}},\ }\href@noop {} {\bibfield  {journal} {\bibinfo  {journal}
  {Phys. Rev. D}\ }\textbf {\bibinfo {volume} {71}},\ \bibinfo {pages} {024028}
  (\bibinfo {year} {2005})}\BibitemShut {NoStop}%
\bibitem [{\citenamefont {Hawking}(1976)}]{hawking:1976}%
  \BibitemOpen
  \bibfield  {author} {\bibinfo {author} {\bibfnamefont {S.~W.}\ \bibnamefont
  {Hawking}},\ }\href@noop {} {\bibfield  {journal} {\bibinfo  {journal} {Phys.
  Rev. D}\ }\textbf {\bibinfo {volume} {13}},\ \bibinfo {pages} {191} (\bibinfo
  {year} {1976})}\BibitemShut {NoStop}%
\bibitem [{\citenamefont {Unruh}\ and\ \citenamefont
  {Schutzhold}(2003)}]{unruh:2003}%
  \BibitemOpen
  \bibfield  {author} {\bibinfo {author} {\bibfnamefont {W.~G.}\ \bibnamefont
  {Unruh}}\ and\ \bibinfo {author} {\bibfnamefont {R.}~\bibnamefont
  {Schutzhold}},\ }\href@noop {} {\bibfield  {journal} {\bibinfo  {journal}
  {Phys. Rev. D}\ }\textbf {\bibinfo {volume} {68}},\ \bibinfo {pages} {024008}
  (\bibinfo {year} {2003})}\BibitemShut {NoStop}%
\bibitem [{\citenamefont {Blencowe}\ and\ \citenamefont
  {Vitelli}(2000)}]{blencowe:2000}%
  \BibitemOpen
  \bibfield  {author} {\bibinfo {author} {\bibfnamefont {M.~P.}\ \bibnamefont
  {Blencowe}}\ and\ \bibinfo {author} {\bibfnamefont {V.}~\bibnamefont
  {Vitelli}},\ }\href@noop {} {\bibfield  {journal} {\bibinfo  {journal} {Phys.
  Rev. A}\ }\textbf {\bibinfo {volume} {62}},\ \bibinfo {pages} {052104}
  (\bibinfo {year} {2000})}\BibitemShut {NoStop}%
\bibitem [{\citenamefont {Meschke}\ \emph {et~al.}(2006)\citenamefont
  {Meschke}, \citenamefont {Guichard},\ and\ \citenamefont
  {Pekola}}]{meschke:2006}%
  \BibitemOpen
  \bibfield  {author} {\bibinfo {author} {\bibfnamefont {M.}~\bibnamefont
  {Meschke}}, \bibinfo {author} {\bibfnamefont {W.}~\bibnamefont {Guichard}}, \
  and\ \bibinfo {author} {\bibfnamefont {J.~P.}\ \bibnamefont {Pekola}},\
  }\href@noop {} {\bibfield  {journal} {\bibinfo  {journal} {Nature}\ }\textbf
  {\bibinfo {volume} {444}},\ \bibinfo {pages} {187} (\bibinfo {year}
  {2006})}\BibitemShut {NoStop}%
\bibitem [{\citenamefont {Haviland}\ \emph {et~al.}(2000)\citenamefont
  {Haviland}, \citenamefont {Andersson},\ and\ \citenamefont
  {Agren}}]{haviland:2000}%
  \BibitemOpen
  \bibfield  {author} {\bibinfo {author} {\bibfnamefont {D.~B.}\ \bibnamefont
  {Haviland}}, \bibinfo {author} {\bibfnamefont {K.}~\bibnamefont {Andersson}},
  \ and\ \bibinfo {author} {\bibfnamefont {P.}~\bibnamefont {Agren}},\
  }\href@noop {} {\bibfield  {journal} {\bibinfo  {journal} {Journal Low Temp.
  Phys.}\ }\textbf {\bibinfo {volume} {118}},\ \bibinfo {pages} {773} (\bibinfo
  {year} {2000})}\BibitemShut {NoStop}%
\bibitem [{\citenamefont {Chow}\ \emph {et~al.}(1998)\citenamefont {Chow},
  \citenamefont {Delsing},\ and\ \citenamefont {Haviland}}]{chow:1998}%
  \BibitemOpen
  \bibfield  {author} {\bibinfo {author} {\bibfnamefont {E.}~\bibnamefont
  {Chow}}, \bibinfo {author} {\bibfnamefont {P.}~\bibnamefont {Delsing}}, \
  and\ \bibinfo {author} {\bibfnamefont {D.~B.}\ \bibnamefont {Haviland}},\
  }\href@noop {} {\bibfield  {journal} {\bibinfo  {journal} {Phys. Rev. Lett.}\
  }\textbf {\bibinfo {volume} {81}},\ \bibinfo {pages} {204} (\bibinfo {year}
  {1998})}\BibitemShut {NoStop}%
\bibitem [{\citenamefont {Thiemann}(2007)}]{thiemann:2007}%
  \BibitemOpen
  \bibfield  {author} {\bibinfo {author} {\bibfnamefont {T.}~\bibnamefont
  {Thiemann}},\ }\href@noop {} {\emph {\bibinfo {title} {Modern Canonical
  Quantum General Relativity}}}\ (\bibinfo  {publisher} {Cambridge University
  Press},\ \bibinfo {year} {2007})\BibitemShut {NoStop}%
\bibitem [{\citenamefont {Romero}\ \emph {et~al.}(2009)\citenamefont {Romero},
  \citenamefont {Garc\'{i}a-Ripoll},\ and\ \citenamefont
  {Solano}}]{romero:2009}%
  \BibitemOpen
  \bibfield  {author} {\bibinfo {author} {\bibfnamefont {G.}~\bibnamefont
  {Romero}}, \bibinfo {author} {\bibfnamefont {J.~J.}\ \bibnamefont
  {Garc\'{i}a-Ripoll}}, \ and\ \bibinfo {author} {\bibfnamefont
  {E.}~\bibnamefont {Solano}},\ }\href@noop {} {\bibfield  {journal} {\bibinfo
  {journal} {Phys. Rev. Lett.}\ }\textbf {\bibinfo {volume} {102}},\ \bibinfo
  {pages} {173602} (\bibinfo {year} {2009})}\BibitemShut {NoStop}%
\bibitem [{\citenamefont {Mathur}(2009)}]{mathur:2009}%
  \BibitemOpen
  \bibfield  {author} {\bibinfo {author} {\bibfnamefont {S.~D.}\ \bibnamefont
  {Mathur}},\ }\href@noop {} {\bibfield  {journal} {\bibinfo  {journal} {Class.
  Quant. Grav.}\ }\textbf {\bibinfo {volume} {26}},\ \bibinfo {pages} {224001}
  (\bibinfo {year} {2009})}\BibitemShut {NoStop}%
\bibitem [{\citenamefont {Hawking}(1975)}]{hawking:1975}%
  \BibitemOpen
  \bibfield  {author} {\bibinfo {author} {\bibfnamefont {S.~W.}\ \bibnamefont
  {Hawking}},\ }\href@noop {} {\bibfield  {journal} {\bibinfo  {journal}
  {Commun. Math. Phys.}\ }\textbf {\bibinfo {volume} {43}},\ \bibinfo {pages}
  {199} (\bibinfo {year} {1975})}\BibitemShut {NoStop}%
\bibitem [{\citenamefont {Hartle}\ and\ \citenamefont
  {Hawking}(1976)}]{hartle:1976}%
  \BibitemOpen
  \bibfield  {author} {\bibinfo {author} {\bibfnamefont {J.~B.}\ \bibnamefont
  {Hartle}}\ and\ \bibinfo {author} {\bibfnamefont {S.~W.}\ \bibnamefont
  {Hawking}},\ }\href@noop {} {\bibfield  {journal} {\bibinfo  {journal} {Phys.
  Rev. D}\ }\textbf {\bibinfo {volume} {13}},\ \bibinfo {pages} {2188}
  (\bibinfo {year} {1976})}\BibitemShut {NoStop}%
\bibitem [{\citenamefont {Boulware}(1976)}]{boulware:1976}%
  \BibitemOpen
  \bibfield  {author} {\bibinfo {author} {\bibfnamefont {D.~G.}\ \bibnamefont
  {Boulware}},\ }\href@noop {} {\bibfield  {journal} {\bibinfo  {journal}
  {Phys. Rev. D}\ }\textbf {\bibinfo {volume} {13}},\ \bibinfo {pages} {2169}
  (\bibinfo {year} {1976})}\BibitemShut {NoStop}%
\bibitem [{\citenamefont {Gibbons}\ and\ \citenamefont
  {Hawking}(1977)}]{gibbons:1977}%
  \BibitemOpen
  \bibfield  {author} {\bibinfo {author} {\bibfnamefont {G.~W.}\ \bibnamefont
  {Gibbons}}\ and\ \bibinfo {author} {\bibfnamefont {S.~W.}\ \bibnamefont
  {Hawking}},\ }\href@noop {} {\bibfield  {journal} {\bibinfo  {journal} {Phys.
  Rev. D}\ }\textbf {\bibinfo {volume} {15}},\ \bibinfo {pages} {2752}
  (\bibinfo {year} {1977})}\BibitemShut {NoStop}%
\bibitem [{\citenamefont {Parentani}(2000)}]{parentani:2000}%
  \BibitemOpen
  \bibfield  {author} {\bibinfo {author} {\bibfnamefont {R.}~\bibnamefont
  {Parentani}},\ }\href@noop {} {\bibfield  {journal} {\bibinfo  {journal}
  {Phys. Rev. D}\ }\textbf {\bibinfo {volume} {61}},\ \bibinfo {pages} {27501}
  (\bibinfo {year} {2000})}\BibitemShut {NoStop}%
\bibitem [{\citenamefont {Bombelli}\ \emph {et~al.}(1986)\citenamefont
  {Bombelli}, \citenamefont {Koul}, \citenamefont {Lee},\ and\ \citenamefont
  {Sorkin}}]{bombelli:1986}%
  \BibitemOpen
  \bibfield  {author} {\bibinfo {author} {\bibfnamefont {L.}~\bibnamefont
  {Bombelli}}, \bibinfo {author} {\bibfnamefont {R.~K.}\ \bibnamefont {Koul}},
  \bibinfo {author} {\bibfnamefont {J.}~\bibnamefont {Lee}}, \ and\ \bibinfo
  {author} {\bibfnamefont {R.~D.}\ \bibnamefont {Sorkin}},\ }\href@noop {}
  {\bibfield  {journal} {\bibinfo  {journal} {Phys. Rev. D}\ }\textbf {\bibinfo
  {volume} {34}},\ \bibinfo {pages} {373} (\bibinfo {year} {1986})}\BibitemShut
  {NoStop}%
\bibitem [{\citenamefont {Srednicki}(1993)}]{srednicki:1993}%
  \BibitemOpen
  \bibfield  {author} {\bibinfo {author} {\bibfnamefont {M.}~\bibnamefont
  {Srednicki}},\ }\href@noop {} {\bibfield  {journal} {\bibinfo  {journal}
  {Phys. Rev. Lett.}\ }\textbf {\bibinfo {volume} {71}} (\bibinfo {year}
  {1993})}\BibitemShut {NoStop}%
\bibitem [{\citenamefont {Eisert}\ \emph {et~al.}(2010)\citenamefont {Eisert},
  \citenamefont {Cramer},\ and\ \citenamefont {Plenio}}]{eisert:2010}%
  \BibitemOpen
  \bibfield  {author} {\bibinfo {author} {\bibfnamefont {J.}~\bibnamefont
  {Eisert}}, \bibinfo {author} {\bibfnamefont {M.}~\bibnamefont {Cramer}}, \
  and\ \bibinfo {author} {\bibfnamefont {M.~B.}\ \bibnamefont {Plenio}},\
  }\href@noop {} {\bibfield  {journal} {\bibinfo  {journal} {Rev. Mod. Phys.}\
  }\textbf {\bibinfo {volume} {82}},\ \bibinfo {pages} {277} (\bibinfo {year}
  {2010})}\BibitemShut {NoStop}%
\bibitem [{\citenamefont {Aharanov}\ \emph {et~al.}(1987)\citenamefont
  {Aharanov}, \citenamefont {Casher},\ and\ \citenamefont
  {Nussinov}}]{aharanov:1987}%
  \BibitemOpen
  \bibfield  {author} {\bibinfo {author} {\bibfnamefont {Y.}~\bibnamefont
  {Aharanov}}, \bibinfo {author} {\bibfnamefont {A.}~\bibnamefont {Casher}}, \
  and\ \bibinfo {author} {\bibfnamefont {S.}~\bibnamefont {Nussinov}},\
  }\href@noop {} {\bibfield  {journal} {\bibinfo  {journal} {Phys. Lett. B}\
  }\textbf {\bibinfo {volume} {191}},\ \bibinfo {pages} {51} (\bibinfo {year}
  {1987})}\BibitemShut {NoStop}%
\bibitem [{\citenamefont {Giddings}(1992)}]{giddings:1992}%
  \BibitemOpen
  \bibfield  {author} {\bibinfo {author} {\bibfnamefont {S.~B.}\ \bibnamefont
  {Giddings}},\ }\href@noop {} {\bibfield  {journal} {\bibinfo  {journal}
  {Phys. Rev. D}\ }\textbf {\bibinfo {volume} {46}},\ \bibinfo {pages} {1347}
  (\bibinfo {year} {1992})}\BibitemShut {NoStop}%
\bibitem [{\citenamefont {Hawking}(1988)}]{hawking:1988}%
  \BibitemOpen
  \bibfield  {author} {\bibinfo {author} {\bibfnamefont {S.~W.}\ \bibnamefont
  {Hawking}},\ }\href@noop {} {\bibfield  {journal} {\bibinfo  {journal} {Phys.
  Rev. D}\ }\textbf {\bibinfo {volume} {37}},\ \bibinfo {pages} {904} (\bibinfo
  {year} {1988})}\BibitemShut {NoStop}%
\bibitem [{\citenamefont {Frolov}\ \emph {et~al.}(1990)\citenamefont {Frolov},
  \citenamefont {Markov},\ and\ \citenamefont {Mukhanov}}]{frolov:1990}%
  \BibitemOpen
  \bibfield  {author} {\bibinfo {author} {\bibfnamefont {V.~P.}\ \bibnamefont
  {Frolov}}, \bibinfo {author} {\bibfnamefont {M.~A.}\ \bibnamefont {Markov}},
  \ and\ \bibinfo {author} {\bibfnamefont {V.~F.}\ \bibnamefont {Mukhanov}},\
  }\href@noop {} {\bibfield  {journal} {\bibinfo  {journal} {Phys. Rev. D}\
  }\textbf {\bibinfo {volume} {41}},\ \bibinfo {pages} {383} (\bibinfo {year}
  {1990})}\BibitemShut {NoStop}%
\bibitem [{\citenamefont {Page}(1993{\natexlab{a}})}]{page:1993}%
  \BibitemOpen
  \bibfield  {author} {\bibinfo {author} {\bibfnamefont {D.~N.}\ \bibnamefont
  {Page}},\ }\href@noop {} {\bibfield  {journal} {\bibinfo  {journal} {Phys.
  Rev. Lett.}\ }\textbf {\bibinfo {volume} {71}},\ \bibinfo {pages} {3743}
  (\bibinfo {year} {1993}{\natexlab{a}})}\BibitemShut {NoStop}%
\bibitem [{\citenamefont {Parikh}\ and\ \citenamefont
  {Wilczek}(2000)}]{parikh:2000}%
  \BibitemOpen
  \bibfield  {author} {\bibinfo {author} {\bibfnamefont {M.~K.}\ \bibnamefont
  {Parikh}}\ and\ \bibinfo {author} {\bibfnamefont {F.}~\bibnamefont
  {Wilczek}},\ }\href@noop {} {\bibfield  {journal} {\bibinfo  {journal} {Phys.
  Rev. Lett.}\ }\textbf {\bibinfo {volume} {85}},\ \bibinfo {pages} {5042}
  (\bibinfo {year} {2000})}\BibitemShut {NoStop}%
\bibitem [{\citenamefont {Hawking}(2005)}]{hawking:2005}%
  \BibitemOpen
  \bibfield  {author} {\bibinfo {author} {\bibfnamefont {S.~W.}\ \bibnamefont
  {Hawking}},\ }\href@noop {} {\bibfield  {journal} {\bibinfo  {journal} {Phys.
  Rev. D}\ }\textbf {\bibinfo {volume} {72}},\ \bibinfo {pages} {084013}
  (\bibinfo {year} {2005})}\BibitemShut {NoStop}%
\bibitem [{\citenamefont {Terno}(2005)}]{terno:2005}%
  \BibitemOpen
  \bibfield  {author} {\bibinfo {author} {\bibfnamefont {D.~R.}\ \bibnamefont
  {Terno}},\ }\href@noop {} {\bibfield  {journal} {\bibinfo  {journal} {Int. J.
  Mod. Phys. D}\ }\textbf {\bibinfo {volume} {14}},\ \bibinfo {pages} {2307}
  (\bibinfo {year} {2005})}\BibitemShut {NoStop}%
\bibitem [{\citenamefont {Jacobson}(1995)}]{jacobson:1995}%
  \BibitemOpen
  \bibfield  {author} {\bibinfo {author} {\bibfnamefont {T.~A.}\ \bibnamefont
  {Jacobson}},\ }\href@noop {} {\bibfield  {journal} {\bibinfo  {journal}
  {Phys. Rev. Lett.}\ }\textbf {\bibinfo {volume} {75}},\ \bibinfo {pages}
  {1260} (\bibinfo {year} {1995})}\BibitemShut {NoStop}%
\bibitem [{\citenamefont {Carlip}(2008)}]{carlip:2008}%
  \BibitemOpen
  \bibfield  {author} {\bibinfo {author} {\bibfnamefont {S.}~\bibnamefont
  {Carlip}},\ }\href@noop {} {\bibfield  {journal} {\bibinfo  {journal} {Class.
  Quant. Grav.}\ }\textbf {\bibinfo {volume} {25}},\ \bibinfo {pages} {154010}
  (\bibinfo {year} {2008})}\BibitemShut {NoStop}%
\bibitem [{\citenamefont {Dicke}(1953)}]{dicke:1953}%
  \BibitemOpen
  \bibfield  {author} {\bibinfo {author} {\bibfnamefont {R.~H.}\ \bibnamefont
  {Dicke}},\ }\href@noop {} {\bibfield  {journal} {\bibinfo  {journal} {Phys.
  Rev.}\ }\textbf {\bibinfo {volume} {93}},\ \bibinfo {pages} {99} (\bibinfo
  {year} {1953})}\BibitemShut {NoStop}%
\bibitem [{\citenamefont {Mollow}\ and\ \citenamefont
  {Glauber}(1967)}]{mollow:1967}%
  \BibitemOpen
  \bibfield  {author} {\bibinfo {author} {\bibfnamefont {B.~R.}\ \bibnamefont
  {Mollow}}\ and\ \bibinfo {author} {\bibfnamefont {R.~J.}\ \bibnamefont
  {Glauber}},\ }\href@noop {} {\bibfield  {journal} {\bibinfo  {journal} {Phys.
  Rev.}\ }\textbf {\bibinfo {volume} {160}},\ \bibinfo {pages} {1076} (\bibinfo
  {year} {1967})}\BibitemShut {NoStop}%
\bibitem [{\citenamefont {Travis}\ and\ \citenamefont
  {Cummings}(1968)}]{travis:1968}%
  \BibitemOpen
  \bibfield  {author} {\bibinfo {author} {\bibfnamefont {M.}~\bibnamefont
  {Travis}}\ and\ \bibinfo {author} {\bibfnamefont {F.~W.}\ \bibnamefont
  {Cummings}},\ }\href@noop {} {\bibfield  {journal} {\bibinfo  {journal}
  {Phys. Rev.}\ }\textbf {\bibinfo {volume} {170}},\ \bibinfo {pages} {379}
  (\bibinfo {year} {1968})}\BibitemShut {NoStop}%
\bibitem [{\citenamefont {Tucker}\ and\ \citenamefont
  {Walls}(1969)}]{tucker:1969}%
  \BibitemOpen
  \bibfield  {author} {\bibinfo {author} {\bibfnamefont {J.}~\bibnamefont
  {Tucker}}\ and\ \bibinfo {author} {\bibfnamefont {D.~F.}\ \bibnamefont
  {Walls}},\ }\href@noop {} {\bibfield  {journal} {\bibinfo  {journal} {Phys.
  Rev.}\ }\textbf {\bibinfo {volume} {178}},\ \bibinfo {pages} {2036} (\bibinfo
  {year} {1969})}\BibitemShut {NoStop}%
\bibitem [{\citenamefont {Walls}\ and\ \citenamefont
  {Barakat}(1970)}]{walls:1970}%
  \BibitemOpen
  \bibfield  {author} {\bibinfo {author} {\bibfnamefont {D.~F.}\ \bibnamefont
  {Walls}}\ and\ \bibinfo {author} {\bibfnamefont {R.}~\bibnamefont
  {Barakat}},\ }\href@noop {} {\bibfield  {journal} {\bibinfo  {journal} {Phys.
  Rev. A}\ }\textbf {\bibinfo {volume} {1}},\ \bibinfo {pages} {446} (\bibinfo
  {year} {1970})}\BibitemShut {NoStop}%
\bibitem [{\citenamefont {Lu}(1973)}]{lu:1973}%
  \BibitemOpen
  \bibfield  {author} {\bibinfo {author} {\bibfnamefont {E.~Y.~C.}\
  \bibnamefont {Lu}},\ }\href@noop {} {\bibfield  {journal} {\bibinfo
  {journal} {Phys. Rev. A}\ }\textbf {\bibinfo {volume} {8}},\ \bibinfo {pages}
  {1053} (\bibinfo {year} {1973})}\BibitemShut {NoStop}%
\bibitem [{\citenamefont {Agrawal}\ and\ \citenamefont
  {Mehta}(1974)}]{agrawal:1974}%
  \BibitemOpen
  \bibfield  {author} {\bibinfo {author} {\bibfnamefont {G.~P.}\ \bibnamefont
  {Agrawal}}\ and\ \bibinfo {author} {\bibfnamefont {C.~L.}\ \bibnamefont
  {Mehta}},\ }\href@noop {} {\bibfield  {journal} {\bibinfo  {journal} {J.
  Phys. A}\ }\textbf {\bibinfo {volume} {7}},\ \bibinfo {pages} {607} (\bibinfo
  {year} {1974})}\BibitemShut {NoStop}%
\bibitem [{\citenamefont {McNeil}\ and\ \citenamefont
  {Gardiner}(1983)}]{mcneil:1983}%
  \BibitemOpen
  \bibfield  {author} {\bibinfo {author} {\bibfnamefont {K.~J.}\ \bibnamefont
  {McNeil}}\ and\ \bibinfo {author} {\bibfnamefont {C.~W.}\ \bibnamefont
  {Gardiner}},\ }\href@noop {} {\bibfield  {journal} {\bibinfo  {journal}
  {Phys. Rev. A}\ }\textbf {\bibinfo {volume} {28}},\ \bibinfo {pages} {1560}
  (\bibinfo {year} {1983})}\BibitemShut {NoStop}%
\bibitem [{\citenamefont {Gerlach}(1976)}]{gerlach:1976}%
  \BibitemOpen
  \bibfield  {author} {\bibinfo {author} {\bibfnamefont {U.~H.}\ \bibnamefont
  {Gerlach}},\ }\href@noop {} {\bibfield  {journal} {\bibinfo  {journal} {Phys.
  Rev. D}\ }\textbf {\bibinfo {volume} {14}},\ \bibinfo {pages} {1479}
  (\bibinfo {year} {1976})}\BibitemShut {NoStop}%
\bibitem [{\citenamefont {Barnett}\ and\ \citenamefont
  {Knight}(1985)}]{barnett:1985}%
  \BibitemOpen
  \bibfield  {author} {\bibinfo {author} {\bibfnamefont {S.~M.}\ \bibnamefont
  {Barnett}}\ and\ \bibinfo {author} {\bibfnamefont {P.~L.}\ \bibnamefont
  {Knight}},\ }\href@noop {} {\bibfield  {journal} {\bibinfo  {journal} {J.
  Opt. Soc. Am. B}\ }\textbf {\bibinfo {volume} {2}},\ \bibinfo {pages} {467}
  (\bibinfo {year} {1985})}\BibitemShut {NoStop}%
\bibitem [{\citenamefont {Yurke}\ and\ \citenamefont
  {Potasek}(1987)}]{yurke:1987}%
  \BibitemOpen
  \bibfield  {author} {\bibinfo {author} {\bibfnamefont {B.}~\bibnamefont
  {Yurke}}\ and\ \bibinfo {author} {\bibfnamefont {M.}~\bibnamefont
  {Potasek}},\ }\href@noop {} {\bibfield  {journal} {\bibinfo  {journal} {Phys.
  Rev. A}\ }\textbf {\bibinfo {volume} {36}},\ \bibinfo {pages} {3464}
  (\bibinfo {year} {1987})}\BibitemShut {NoStop}%
\bibitem [{\citenamefont {Boyd}(2003)}]{boyd:2003}%
  \BibitemOpen
  \bibfield  {author} {\bibinfo {author} {\bibfnamefont {R.~W.}\ \bibnamefont
  {Boyd}},\ }\href@noop {} {\emph {\bibinfo {title} {Nonlinear Optics}}},\
  \bibinfo {edition} {2nd}\ ed.\ (\bibinfo  {publisher} {Academic Press},\
  \bibinfo {year} {2003})\BibitemShut {NoStop}%
\bibitem [{\citenamefont {Traux}(1985)}]{truax:1985}%
  \BibitemOpen
  \bibfield  {author} {\bibinfo {author} {\bibfnamefont {D.~R.}\ \bibnamefont
  {Traux}},\ }\href@noop {} {\bibfield  {journal} {\bibinfo  {journal} {Phys.
  Rev. D}\ }\textbf {\bibinfo {volume} {31}},\ \bibinfo {pages} {1988}
  (\bibinfo {year} {1985})}\BibitemShut {NoStop}%
\bibitem [{\citenamefont {Fuentes-Schuller}\ and\ \citenamefont
  {Mann}(2005)}]{fuentes:2005}%
  \BibitemOpen
  \bibfield  {author} {\bibinfo {author} {\bibfnamefont {I.}~\bibnamefont
  {Fuentes-Schuller}}\ and\ \bibinfo {author} {\bibfnamefont {R.~B.}\
  \bibnamefont {Mann}},\ }\href@noop {} {\bibfield  {journal} {\bibinfo
  {journal} {Phys. Rev. Lett.}\ }\textbf {\bibinfo {volume} {95}} (\bibinfo
  {year} {2005})}\BibitemShut {NoStop}%
\bibitem [{\citenamefont {Walls}\ and\ \citenamefont
  {Milburn}(2008)}]{walls:2008}%
  \BibitemOpen
  \bibfield  {author} {\bibinfo {author} {\bibfnamefont {D.~F.}\ \bibnamefont
  {Walls}}\ and\ \bibinfo {author} {\bibfnamefont {G.~J.}\ \bibnamefont
  {Milburn}},\ }\href@noop {} {\emph {\bibinfo {title} {Quantum Optics, 2nd
  Ed.}}}\ (\bibinfo  {publisher} {Springer},\ \bibinfo {year}
  {2008})\BibitemShut {NoStop}%
\bibitem [{\citenamefont {Kibble}\ and\ \citenamefont
  {Randjbar-Daemi}(1980)}]{kibble:1980}%
  \BibitemOpen
  \bibfield  {author} {\bibinfo {author} {\bibfnamefont {T.~W.~B.}\
  \bibnamefont {Kibble}}\ and\ \bibinfo {author} {\bibfnamefont
  {S.}~\bibnamefont {Randjbar-Daemi}},\ }\href@noop {} {\bibfield  {journal}
  {\bibinfo  {journal} {J. Phys. A}\ }\textbf {\bibinfo {volume} {13}},\
  \bibinfo {pages} {141} (\bibinfo {year} {1980})}\BibitemShut {NoStop}%
\bibitem [{\citenamefont {Horowitz}(1980)}]{horowitz:1980}%
  \BibitemOpen
  \bibfield  {author} {\bibinfo {author} {\bibfnamefont {G.~T.}\ \bibnamefont
  {Horowitz}},\ }\href@noop {} {\bibfield  {journal} {\bibinfo  {journal}
  {Phys. Rev. D}\ }\textbf {\bibinfo {volume} {21}},\ \bibinfo {pages} {1445}
  (\bibinfo {year} {1980})}\BibitemShut {NoStop}%
\bibitem [{\citenamefont {Anselmi}(2007)}]{anselmi:2007}%
  \BibitemOpen
  \bibfield  {author} {\bibinfo {author} {\bibfnamefont {D.}~\bibnamefont
  {Anselmi}},\ }\href@noop {} {\bibfield  {journal} {\bibinfo  {journal}
  {JHEP}\ }\textbf {\bibinfo {volume} {1}},\ \bibinfo {pages} {62} (\bibinfo
  {year} {2007})}\BibitemShut {NoStop}%
\bibitem [{\citenamefont {Unruh}(1984)}]{unruh:1984}%
  \BibitemOpen
  \bibfield  {author} {\bibinfo {author} {\bibfnamefont {W.~G.}\ \bibnamefont
  {Unruh}},\ }in\ \href@noop {} {\emph {\bibinfo {booktitle} {Quantum Theory of
  Gravity: Essays in Honor of the 60th Birthday if Bryce S. De Witt}}}\
  (\bibinfo  {publisher} {Bristol: Hilger},\ \bibinfo {year}
  {1984})\BibitemShut {NoStop}%
\bibitem [{\citenamefont {Manley}\ and\ \citenamefont
  {Rowe}(1956)}]{manley:1956}%
  \BibitemOpen
  \bibfield  {author} {\bibinfo {author} {\bibfnamefont {J.~M.}\ \bibnamefont
  {Manley}}\ and\ \bibinfo {author} {\bibfnamefont {H.~E.}\ \bibnamefont
  {Rowe}},\ }\href@noop {} {\bibfield  {journal} {\bibinfo  {journal} {Proc. of
  IRE}\ }\textbf {\bibinfo {volume} {44}},\ \bibinfo {pages} {904} (\bibinfo
  {year} {1956})}\BibitemShut {NoStop}%
\bibitem [{\citenamefont {Yurke}\ \emph {et~al.}(1986)\citenamefont {Yurke},
  \citenamefont {McCall},\ and\ \citenamefont {Klauder}}]{yurke:1986}%
  \BibitemOpen
  \bibfield  {author} {\bibinfo {author} {\bibfnamefont {B.}~\bibnamefont
  {Yurke}}, \bibinfo {author} {\bibfnamefont {S.~L.}\ \bibnamefont {McCall}}, \
  and\ \bibinfo {author} {\bibfnamefont {J.~R.}\ \bibnamefont {Klauder}},\
  }\href@noop {} {\bibfield  {journal} {\bibinfo  {journal} {Phys. Rev. A}\
  }\textbf {\bibinfo {volume} {33}},\ \bibinfo {pages} {4033} (\bibinfo {year}
  {1986})}\BibitemShut {NoStop}%
\bibitem [{\citenamefont {Brif}(1996)}]{brif:1996}%
  \BibitemOpen
  \bibfield  {author} {\bibinfo {author} {\bibfnamefont {C.}~\bibnamefont
  {Brif}},\ }\href@noop {} {\bibfield  {journal} {\bibinfo  {journal} {Phys.
  Rev. A}\ }\textbf {\bibinfo {volume} {54}},\ \bibinfo {pages} {5253}
  (\bibinfo {year} {1996})}\BibitemShut {NoStop}%
\bibitem [{\citenamefont {Nielson}\ and\ \citenamefont
  {Chuang}(2000)}]{nielson:2000}%
  \BibitemOpen
  \bibfield  {author} {\bibinfo {author} {\bibfnamefont {M.~A.}\ \bibnamefont
  {Nielson}}\ and\ \bibinfo {author} {\bibfnamefont {I.~L.}\ \bibnamefont
  {Chuang}},\ }\href@noop {} {\emph {\bibinfo {title} {Quantum Computation and
  Quantum Information}}}\ (\bibinfo  {publisher} {Cambridge University Press},\
  \bibinfo {year} {2000})\BibitemShut {NoStop}%
\bibitem [{\citenamefont {Page}(1993{\natexlab{b}})}]{page:1993-2}%
  \BibitemOpen
  \bibfield  {author} {\bibinfo {author} {\bibfnamefont {D.~N.}\ \bibnamefont
  {Page}},\ }\href@noop {} {\bibfield  {journal} {\bibinfo  {journal} {Phys.
  Rev. Lett.}\ }\textbf {\bibinfo {volume} {71}},\ \bibinfo {pages} {1291}
  (\bibinfo {year} {1993}{\natexlab{b}})}\BibitemShut {NoStop}%
\bibitem [{\citenamefont {Popescu}\ \emph {et~al.}(2006)\citenamefont
  {Popescu}, \citenamefont {Short},\ and\ \citenamefont
  {Winter}}]{popescu:2006}%
  \BibitemOpen
  \bibfield  {author} {\bibinfo {author} {\bibfnamefont {S.}~\bibnamefont
  {Popescu}}, \bibinfo {author} {\bibfnamefont {A.~J.}\ \bibnamefont {Short}},
  \ and\ \bibinfo {author} {\bibfnamefont {A.}~\bibnamefont {Winter}},\
  }\href@noop {} {\bibfield  {journal} {\bibinfo  {journal} {Nature Phys.}\
  }\textbf {\bibinfo {volume} {2}},\ \bibinfo {pages} {754} (\bibinfo {year}
  {2006})}\BibitemShut {NoStop}%
\bibitem [{\citenamefont {Drobn\'y}\ and\ \citenamefont
  {Bu\v{z}ek}(1994)}]{drobny:1994}%
  \BibitemOpen
  \bibfield  {author} {\bibinfo {author} {\bibfnamefont {G.}~\bibnamefont
  {Drobn\'y}}\ and\ \bibinfo {author} {\bibfnamefont {V.}~\bibnamefont
  {Bu\v{z}ek}},\ }\href@noop {} {\bibfield  {journal} {\bibinfo  {journal}
  {Phys. Rev. A}\ }\textbf {\bibinfo {volume} {50}},\ \bibinfo {pages} {3492}
  (\bibinfo {year} {1994})}\BibitemShut {NoStop}%
\bibitem [{\citenamefont {Groisman}\ \emph {et~al.}(2005)\citenamefont
  {Groisman}, \citenamefont {Popescu},\ and\ \citenamefont
  {Winter}}]{groisman:2005}%
  \BibitemOpen
  \bibfield  {author} {\bibinfo {author} {\bibfnamefont {B.}~\bibnamefont
  {Groisman}}, \bibinfo {author} {\bibfnamefont {S.}~\bibnamefont {Popescu}}, \
  and\ \bibinfo {author} {\bibfnamefont {A.}~\bibnamefont {Winter}},\
  }\href@noop {} {\bibfield  {journal} {\bibinfo  {journal} {Phys. Rev. A}\
  }\textbf {\bibinfo {volume} {72}},\ \bibinfo {pages} {032317} (\bibinfo
  {year} {2005})}\BibitemShut {NoStop}%
\bibitem [{\citenamefont {Chen}\ \emph {et~al.}(2007)\citenamefont {Chen},
  \citenamefont {Chen},\ and\ \citenamefont {Liang}}]{chen:2007}%
  \BibitemOpen
  \bibfield  {author} {\bibinfo {author} {\bibfnamefont {G.}~\bibnamefont
  {Chen}}, \bibinfo {author} {\bibfnamefont {Z.}~\bibnamefont {Chen}}, \ and\
  \bibinfo {author} {\bibfnamefont {J.}~\bibnamefont {Liang}},\ }\href@noop {}
  {\bibfield  {journal} {\bibinfo  {journal} {Phys Rev A}\ }\textbf {\bibinfo
  {volume} {76}},\ \bibinfo {pages} {055803} (\bibinfo {year}
  {2007})}\BibitemShut {NoStop}%
\bibitem [{\citenamefont {Lambert}\ \emph {et~al.}(2009)\citenamefont
  {Lambert}, \citenamefont {Chen}, \citenamefont {Johansson},\ and\
  \citenamefont {Nori}}]{lambert:2009}%
  \BibitemOpen
  \bibfield  {author} {\bibinfo {author} {\bibfnamefont {N.}~\bibnamefont
  {Lambert}}, \bibinfo {author} {\bibfnamefont {Y.}~\bibnamefont {Chen}},
  \bibinfo {author} {\bibfnamefont {R.}~\bibnamefont {Johansson}}, \ and\
  \bibinfo {author} {\bibfnamefont {F.}~\bibnamefont {Nori}},\ }\href@noop {}
  {\bibfield  {journal} {\bibinfo  {journal} {Phys Rev B}\ }\textbf {\bibinfo
  {volume} {80}} (\bibinfo {year} {2009})}\BibitemShut {NoStop}%
\bibitem [{\citenamefont {Fink}\ \emph {et~al.}(2009)\citenamefont {Fink},
  \citenamefont {Bianchetti}, \citenamefont {Baur}, \citenamefont {G\"{o}ppl},
  \citenamefont {Steffen}, \citenamefont {Filipp}, \citenamefont {Leek},
  \citenamefont {Blais},\ and\ \citenamefont {Wallraff}}]{fink:2009}%
  \BibitemOpen
  \bibfield  {author} {\bibinfo {author} {\bibfnamefont {J.~M.}\ \bibnamefont
  {Fink}}, \bibinfo {author} {\bibfnamefont {R.}~\bibnamefont {Bianchetti}},
  \bibinfo {author} {\bibfnamefont {M.}~\bibnamefont {Baur}}, \bibinfo {author}
  {\bibfnamefont {M.}~\bibnamefont {G\"{o}ppl}}, \bibinfo {author}
  {\bibfnamefont {L.}~\bibnamefont {Steffen}}, \bibinfo {author} {\bibfnamefont
  {S.}~\bibnamefont {Filipp}}, \bibinfo {author} {\bibfnamefont {P.~J.}\
  \bibnamefont {Leek}}, \bibinfo {author} {\bibfnamefont {A.}~\bibnamefont
  {Blais}}, \ and\ \bibinfo {author} {\bibfnamefont {A.}~\bibnamefont
  {Wallraff}},\ }\href@noop {} {\bibfield  {journal} {\bibinfo  {journal} {Phys
  Rev Lett}\ }\textbf {\bibinfo {volume} {103}},\ \bibinfo {pages} {083601}
  (\bibinfo {year} {2009})}\BibitemShut {NoStop}%
\bibitem [{\citenamefont {Schwinger}(1965)}]{schwinger:1965}%
  \BibitemOpen
  \bibfield  {author} {\bibinfo {author} {\bibfnamefont {J.}~\bibnamefont
  {Schwinger}},\ }\enquote {\bibinfo {title} {Quantum theory of angular
  momentum},}\ \ (\bibinfo  {publisher} {Academic, New York},\ \bibinfo {year}
  {1965})\BibitemShut {NoStop}%
\bibitem [{\citenamefont {Hepp}\ and\ \citenamefont {Lieb}(1973)}]{hepp:1973}%
  \BibitemOpen
  \bibfield  {author} {\bibinfo {author} {\bibfnamefont {K.}~\bibnamefont
  {Hepp}}\ and\ \bibinfo {author} {\bibfnamefont {E.}~\bibnamefont {Lieb}},\
  }\href@noop {} {\bibfield  {journal} {\bibinfo  {journal} {Ann. Phys.}\
  }\textbf {\bibinfo {volume} {76}},\ \bibinfo {pages} {360} (\bibinfo {year}
  {1973})}\BibitemShut {NoStop}%
\bibitem [{\citenamefont {Lambert}\ \emph {et~al.}(2004)\citenamefont
  {Lambert}, \citenamefont {Emary},\ and\ \citenamefont
  {Brandes}}]{lambert:2004}%
  \BibitemOpen
  \bibfield  {author} {\bibinfo {author} {\bibfnamefont {N.}~\bibnamefont
  {Lambert}}, \bibinfo {author} {\bibfnamefont {C.}~\bibnamefont {Emary}}, \
  and\ \bibinfo {author} {\bibfnamefont {T.}~\bibnamefont {Brandes}},\
  }\href@noop {} {\bibfield  {journal} {\bibinfo  {journal} {Phys Rev Lett}\
  }\textbf {\bibinfo {volume} {92}},\ \bibinfo {pages} {073602} (\bibinfo
  {year} {2004})}\BibitemShut {NoStop}%
\bibitem [{\citenamefont {Baumann}\ \emph {et~al.}(2010)\citenamefont
  {Baumann}, \citenamefont {Guerlin}, \citenamefont {Brennecke},\ and\
  \citenamefont {Esslinger}}]{baumann:2010}%
  \BibitemOpen
  \bibfield  {author} {\bibinfo {author} {\bibfnamefont {K.}~\bibnamefont
  {Baumann}}, \bibinfo {author} {\bibfnamefont {C.}~\bibnamefont {Guerlin}},
  \bibinfo {author} {\bibfnamefont {F.}~\bibnamefont {Brennecke}}, \ and\
  \bibinfo {author} {\bibfnamefont {T.}~\bibnamefont {Esslinger}},\ }\href@noop
  {} {\bibfield  {journal} {\bibinfo  {journal} {Nature}\ }\textbf {\bibinfo
  {volume} {464}},\ \bibinfo {pages} {1301} (\bibinfo {year}
  {2010})}\BibitemShut {NoStop}%
\bibitem [{\citenamefont {Tan}(2002)}]{tan:toolbox}%
  \BibitemOpen
  \bibfield  {author} {\bibinfo {author} {\bibfnamefont {S.~M.}\ \bibnamefont
  {Tan}},\ }\href@noop {} {\enquote {\bibinfo {title} {Quantum optics
  toolbox},}\ }\bibinfo {howpublished}
  {http://www.qo.phy.auckland.ac.nz/qotoolbox.html} (\bibinfo {year}
  {2002})\BibitemShut {NoStop}%
\end{thebibliography}%


